%% file: thesis.tex
\documentclass[10pt]{thesis_mw}

\usepackage{fancyhdr}
\usepackage[center,noinfo,dvips]{crop}
\usepackage[american,dutch]{babel}
\usepackage{parskip}
\usepackage{appendix}
\usepackage{comment}
\usepackage[usenames]{graphicx}
\usepackage{amsmath,amsfonts,amssymb,amsthm,dsfont,mathrsfs}
\usepackage{psfrag,xspace}
\usepackage{url}
\usepackage{wrapfig}
\usepackage{caption}
\usepackage[hidelinks,linktocpage]{hyperref}

\usepackage{subcaption}
\usepackage{tensor} 
\usepackage{array} 
\usepackage{mathtools,float}
\usepackage{mathrsfs}

\usepackage{authblk}
\usepackage{enumitem}  
  \usepackage{cite}

\newcommand{\ba}{\begin{eqnarray}}
\newcommand{\ea}{\end{eqnarray}}
\newcommand{\be}{\begin{equation}}
\newcommand{\ee}{\end{equation}}

\newcommand{\R}{{\mathbb R}}

\usepackage{chngcntr}
\counterwithout{footnote}{chapter}

\makeatletter
\newlength{\@testbar}%
\newlength{\@thusbar}%
\newif\ifthusstar
\newcommand*{\thus}{\@ifstar{\thusstartrue\@thus}{\@thus}}%
\newcommand*{\@thus}[1]{%
  \settowidth{\@testbar}{$\therefore$ #1}%
  \setlength{\@thusbar}{\minof{\@testbar}{\linewidth}}%
  \ifhmode\par\fi
  \vspace*{-1ex}\rule{\@thusbar}{0.3pt}\vspace*{-1ex}%
  \item\ifthusstar\else$\therefore$~\fi #1}%
\makeatother

\def\beq{\begin{equation}}
\def\eeq{\end{equation}}
\def\a{\alpha}
\def\d{\delta}
\def\e{\epsilon}
\def\k{\kappa}
\def\l{\lambda}
\def\r{\rho}
\def\z{\zeta}
\def\o{\omega}
\def\O{\Omega}
\def\L{\Lambda}
\def\S{\Sigma}
\def\R{{\cal R}}

\newtheorem*{hypothesis*}{Hypothesis}
\newcommand{\dd}{\mathrm{d}}
\newcommand{\lie}{\mathcal L}

\newcommand{\req}[1]{(\ref{#1})} 
\def \textAdS {{AdS}}

\def \textdS {{dS}}

\def \textMink {{Mink}}

\def \textBTZ {{BTZ}}

\def \textCFT {{CFT}}

\def\d{\partial}

\thesisfont{cmr} 
\thesissectionstyle{\rmfamily\bfseries}

\thesissectionsizes{\Large}{\large}{}

          \usepackage[explicit]{titlesec}
          \usepackage{lmodern}
          \newlength\chapnumb
          \newlength\chapnumbless
          \titleformat{\chapter}[block]
          {\normalfont\sffamily\scshape}{}{0pt} 
          {\parbox[b][1.5cm]{\chapnumb}{
             \fontsize{120}{110}\selectfont\thechapter}
            \parbox[b][1.5cm]{\dimexpr\textwidth-\chapnumb\relax}{
              \raggedleft
               {\hfill \linespread{0.8}\Huge#1\\}
              \rule{\dimexpr\textwidth-\chapnumb\relax}{0.4pt}} 
              }
         \titleformat{name=\chapter,numberless}[block]
          {\normalfont\sffamily\scshape}{}{0pt}
          {\parbox[b][0.5cm]{\chapnumbless}{
             \mbox{}}
            \parbox[b][0.5cm]{\dimexpr\textwidth-\chapnumbless\relax}{
              \raggedleft
              \hfill{\Huge#1}\\
              \rule{\dimexpr\textwidth-\chapnumbless\relax}{0.4pt}}}
\title{{  \sffamily Emergent Gravity \\  in a Holographic  Universe\\[5pt]} }
\thesiscolofon{\input{colofon}}

\author{Manus Renze Visser}
\placeofbirth{Edam-Volendam}

\theday{vrijdag}
\date{14 juni 2019}
\thetime{11.00}
\place{Aula der Universiteit}
 
\rectormagnificus{prof. dr. ir. K. I. J. Maex}
\promotor{\committeeentry{prof. dr. E. P. Verlinde}{Universiteit van Amsterdam}}

\copromotor{\committeeentry{dr. D. M. Hofman}{Universiteit van Amsterdam}}
\othercomitteemembers{
\committeeentry{prof. dr. J. de Boer}{Universiteit van Amsterdam},
\committeeentry{dr. A. Castro Anich}{Universiteit van Amsterdam},
\committeeentry{prof. dr. J. A. E. F. van Dongen}{Universiteit van Amsterdam},
\committeeentry{prof. dr. T. A. Jacobson}{University of Maryland},
\committeeentry{dr. J. P. van der Schaar}{Universiteit van Amsterdam},
}

\department{\small Faculteit der Natuurwetenschappen, Wiskunde en
Informatica}

\usepackage{pifont}
\usepackage[nottoc]{tocbibind}

\input{commands}

\newcommand{\committeeentry}[2]{\begin{tabular*}{0.78\textwidth}{p{0.42\textwidth}l}#1 & #2\end{tabular*}}

\begin{document}

\numberwithin{equation}{chapter}
    \selectlanguage{american}
    \frontmatter
        \maketitle
        \basedon{Publications}{\input{basedon}}
\setcounter{tocdepth}{2}
\tableofcontents
 

    \include{motto}

     \include{preface}

    \mainmatter

\include{chapter1}

       \include{chapter2}

\include{chapter3}

\include{chapter4}

       \summary{Summary and Outlook}{

\input{samenvatting}}
       \include{appendix}

\backmatter

	\bibliographystyle{utphys}
	\bibliography{diamonds,refs-enthigher,manus,publications,references,hints}
     \summary{Samenvatting}{\input{samenvattingNL}}
     \let\cleardoublepage\clearpage
     \acknowledgements{Dankwoord}{\input{dankwoord}}


\end{document}

%% file: colofon.tex
{    \small
\phantom{A} This research was   performed at the Institute for Theoretical Physics Amsterdam (ITFA) of the University of Amsterdam (UvA) and was   supported   by   the European Research Council (ERC) and  by the Spinoza grant, which is funded by the Netherlands Organisation for Scientific Research (NWO).
\vfill
\par\vskip 11cm

\noindent ISBN 978-94-6323-653-9 

\noindent\copyright{}  Manus Visser, 2019  

\noindent Cover design: Charlotte Kalshoven \\ 

   \begin{minipage}[b]{\textwidth}
All rights reserved. Without limiting the rights under copyright reserved above, no part of this book may be reproduced, stored in or introduced into a retrieval system, or transmitted, in any form or by any means (electronic, mechanical, photocopying, recording or otherwise) without the written permission of both the copyright owner and the author of the book.
      \end{minipage}    
\vfill
      }

%% file: commands.tex
\theoremstyle{plain}

\theoremstyle{definition}

\theoremstyle{remark}

\newcommand{\vev}[1]{\ensuremath{\langle #1 \rangle}\xspace}

\def\0{{\sst{(0)}}}
\def\1{{\sst{(1)}}}
\def\2{{\sst{(2)}}}
\def\3{{\sst{(3)}}}
\def\4{{\sst{(4)}}}
\def\5{{\sst{(5)}}}
\def\6{{\sst{(6)}}}
\def\7{{\sst{(7)}}}
\def\8{{\sst{(8)}}}

\let\a=\alpha   \let\d=\delta \let\e=\epsilon
\let\z=\zeta   \let\k=\kappa
\let\l=\lambda    \let\r=\rho

    \let\L=\Lambda

\def\sst#1{{\scriptscriptstyle #1}}

\def\<{ \langle }
\def\>{ \rangle }

\newcommand{\bea}{\begin{eqnarray}}
\newcommand{\eea}{\end{eqnarray}}
\newcommand{\ra}{\rightarrow}

%% file: basedon.tex
%

\setcounter{page}{5}

{\scshape This thesis is based on the following publications:}\\
(Note that the authors are sorted alphabetically in theoretical high-energy physics.) \\


\begin{itemize}
\item[\cite{Jacobson:2018ahi}]

\begin{minipage}[t]{\textwidth}
  Ted Jacobson and Manus Visser,
 
 ``{\it Gravitational Thermodynamics of Causal Diamonds in (A)dS,}''
 
 Submitted to SciPost Physics, \\
 { \ttfamily  arXiv:1812.01596 [hep-th]}. \\
  
  Presented in Chapter {\ref{ch2}}.  MV derived the Smarr formula and first law of causal diamonds in (A)dS space. Further,  he established the connection between entanglement equilibrium and stationarity of free energy, computed the York time for  diamonds in (A)dS and   worked out   various limiting cases of the first law. He wrote the majority of Sections \ref{sec:diamond}, \ref{sec:diamondsthermo}, \ref{negativetemp}, \ref{sec:freeenergy}, \ref{sec:cases} and Appendices \ref{sec:zerothlaw}--\ref{sec:conftrans}. 
  
  \end{minipage}
\end{itemize}
\bigskip

\begin{itemize}
\item[\cite{Bueno:2016gnv}]

\begin{minipage}[t]{\textwidth}
  Pablo Bueno, Vincent Min, Antony Speranza and Manus Visser,
  
  ``{\it Entanglement equilibrium for higher order gravity,}''
  
  Physical Review D   95 (2017) 4,  046003,\\
 {\ttfamily arXiv:1612.04374  [hep-th]}.
\\

Presented in Chapter {\ref{ch3}}. MV   extended the first law of causal diamonds to higher curvature gravity. He mainly contributed to 
Sections \ref{sec:firstlaw}, \ref{cons}  and Appendix \ref{app:W}.

\end{minipage} 
\end{itemize}
\bigskip

\begin{itemize}
\item[\cite{vanLeuven:2018pwv}]

\begin{minipage}[t]{\textwidth}
  Sam van Leuven, Erik Verlinde and Manus Visser,
 
 ``{\it Towards non-AdS Holography via the Long String Phenomenon,}''
 
 Journal of High Energy Physics 1806 (2018)  097, \\
 {\ttfamily arXiv:1801.02589 [hep-th]}.
  \\
  
  Presented in Chapter {\ref{ch4}}. MV     worked out the definitions of the three holographic quantities  and connected them to standard   quantities in 2d CFT. In particular, he wrote large parts of Sections \ref{sec:lessons},   \ref{sec:three-ex}, \ref{sec:vacenergy}  and  Appendix \ref{embedding}.

  \end{minipage}
\end{itemize}

 \bigskip
\par\vskip 1em
 
 \newpage 
 
{\scshape Other publications by the author:}
  \bigskip

\begin{itemize}
 \item[\cite{Bueno:2016ypa}]
 \begin{minipage}[t]{\textwidth}
   Pablo Bueno, Pablo Cano, Vincent Min and  Manus Visser, 
  
  {\it ``Aspects of general higher-order gravities,}"
   
   Physical Review D 95 (2017) no.~4, 044010, \\
   {\ttfamily arXiv:1612.03034 [hep-th]}.
    \end{minipage}
 \end{itemize}

  \bigskip

  \begin{itemize}
 \item[\cite{Brouwer:2016dvq}]
 \begin{minipage}[t]{\textwidth}
   Margot Brouwer, Manus Visser et al.,
  
  ``{\it First test of Verlinde's theory of Emergent Gravity using Weak Gravitational Lensing measurements,}''
  
   Monthly Notices of the Royal Astronomical Society 466 (2017) no.~3, \\
   {\ttfamily arXiv:1612.03034 [astro-ph.CO]}.
    \end{minipage}
 \end{itemize}

 \bigskip

  \begin{itemize}
 \item[\cite{Linnemann:2017hdo}]
 \begin{minipage}[t]{\textwidth}
   Niels Linnemann and Manus Visser, 
  
  ``{\it Hints towards the Emergent Nature of Gravity,}''
  
  Studies in History and Philosophy of Modern Physics B 64 (2018), 
   {\ttfamily arXiv:1711.10503 [physics.hist-ph]}.
    \end{minipage}
 \end{itemize}

  \bigskip

 \begin{itemize}
 \item[\cite{Pedraza:2018eey}]
 \begin{minipage}[t]{\textwidth}
   Juan Pedraza, Watse Sybesma and Manus Visser, 
  
  ``{\it Hyperscaling violating black holes with spherical and hyperbolic horizons,}''
  
  Classical and Quantum Gravity  36 (2019) no.~5, \\ 
   {\ttfamily arXiv:1807.09770 [hep-th]}.
    \end{minipage}
 \end{itemize}

 \bigskip

  \begin{itemize}
 \item[\cite{SVpapers1}]
 \begin{minipage}[t]{\textwidth}
   Sebastian de Haro, Jeroen van Dongen, Manus Visser and Jeremy Butterfield, 
  ``{\it Conceptual Analysis of Black  Hole  Entropy  in  String  Theory,}"   \\  
   Submitted to Studies in History and Philosophy of Modern Physics,  \\
     {\ttfamily arXiv:1904.03232 [physics.hist-ph]}. 

  \end{minipage}
  \end{itemize}
  
   \bigskip
   
  \begin{itemize}
 \item[\cite{SVpapers2}]
 \begin{minipage}[t]{\textwidth}
 Jeroen van Dongen,      Sebastian de Haro, Manus Visser and Jeremy Butterfield,
 ``{\it Emergence and Correspondence for String Theory Black Holes,}'' \\
   Submitted to Studies in History and Philosophy of Modern Physics, \\
   {\ttfamily arXiv:1904.03234 [physics.hist-ph]}.

    \end{minipage}
 \end{itemize}

\bigskip

  \begin{itemize}
 \item[\cite{Essay}]
 \begin{minipage}[t]{\textwidth}
 Ted Jacobson  and Manus Visser,\\
 ``{\it Spacetime Equilibrium at Negative Temperature and the Attraction of Gravity,}''\\ 
Honorable mention in  Gravity Research Foundation Essay Competition 2019, \\
{\ttfamily  arXiv:1904.04843 [gr-qc]}.

    \end{minipage}
 \end{itemize}

\bigskip




%% file: motto.tex

\thispagestyle{empty} 
\vspace*{1in}

\begin{minipage}{0.9\textwidth}
I have spent my life travelling across the universe, inside my mind. Through theoretical physics, I have sought to answer some of the great questions. At one point, I thought I would see the end of physics as we know it, but now I think the wonder of discovery will continue long after I am gone. We are close to some of these answers, but we are not there yet.\\

--- \textsc{Stephen Hawking}, \emph{Brief Answers to Big Questions}  (2018)
\end{minipage}

\newpage

\thispagestyle{empty} 
\vspace*{1in}

\begin{minipage}{0.9\textwidth}
\begin{flushright}
\emph{ \Large To Charlotte and Elias}\\
\bigskip
$\bigstar$ $\bigstar \qquad \qquad \qquad $\\
\bigskip
\emph{ \Large{The two stars 
in my life} }
\end{flushright}
\end{minipage}

%% file: preface.tex

\chapter{Preface and Outline}


\begin{quote}
Sir Donald Munger: ``Tell me, Commander, how far does your expertise extend into the field of diamonds?''\\
James Bond: ``Well, hardest substance found in nature, they cut glass, suggests marriage, I suppose it replaced the dog as the girl's best friend. That's about it.''\\
M: ``Refreshing to hear that there is one subject you're not an expert on!''
--- \emph{Diamonds are Forever}, film (1971)
\end{quote}
\begin{center}
  \vskip .2cm
 ***
  \vskip .2cm
\end{center}

This thesis is not about shiny   diamonds made out of carbon, but rather about the physics of \emph{causal diamonds}.  Causal diamonds appear in Einstein's theory of relativity: they are special regions that  extend both in space and time (which form one entity, called ``spacetime''). 
A causal diamond  is defined as the largest region of spacetime causally accessible to an observer with a finite lifetime \cite{Martinetti:2002sz}.  This means that it is the region with which an observer can exchange signals during his/her lifetime: he/she can receive  signals from \emph{and} send them to any other observer inside a causal diamond. A causal diamond is illustrated on the cover of this book and   in Figure~\ref{fig:zeno}, where time runs upwards and the spatial direction is horizontal (only one spatial direction is shown, the others are suppressed).   
This illustration   is a   symmetric version of a causal diamond: the observer is born  and dies at the same place,  respectively   the  past and future tip of the   diamond. In his/her lifetime the observer can move freely inside the diamond,   as long as the angle of his/her path  is more than  45 degrees  in the spacetime diagram (since otherwise he/she would move   faster than the speed of light).   A causal diamond is thus a spacetime representation of  everything a person can causally probe during his lifetime. The largest possible causal diamond  in spacetime is  that of  the observable  universe, with its   entire history and future included and the  Big Bang at its past tip. 
Before delving into the technical motivation for studying causal diamonds, we  will   first  tell a   story about Achilles and the tortoise moving inside a causal diamond,   to illustrate some interesting features of diamonds. 

\subsubsection*{Zeno's paradox   in relativity theory}

   \begin{figure}
	\centering
	\includegraphics
		[width=.42\textwidth]
		{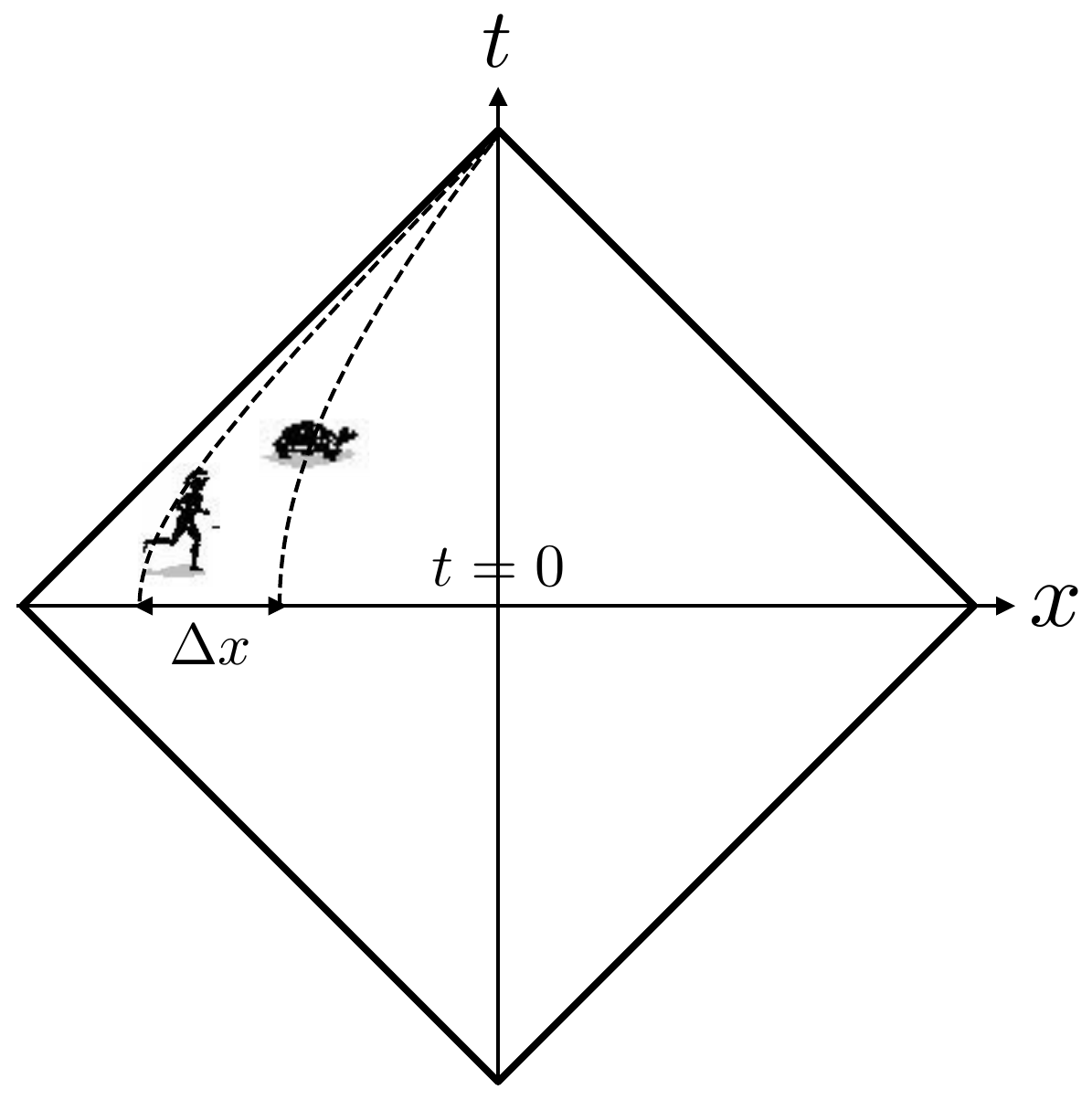}
\caption{\small   Achilles and the tortoise move with constant proper acceleration inside a causal diamond in flat spacetime (along the dashed trajectories). The tortoise gets a head start of   distance $\Delta x$ at $t=0$, but Achilles accelerates faster. 
Achilles  passes the tortoise at the future tip of the diamond, although this takes an infinite amount of  time in a particular coordinate system (the $(x,s)$-system in Figure \ref{fig:causaldiamondsandx} on p. \pageref{fig:causaldiamondsandx}). The proper time between the start and future tip    is finite for both trajectories, and   shorter for Achilles due to time dilation. See Appendix \ref{sec:accckv} for a derivation of the acceleration, velocity and proper time of these trajectories (which are orbits of a conformal Killing field).} 
	\label{fig:zeno}
\end{figure}

In a famous thought experiment   the Greek philosopher Zeno imagined a   running competition between the fleet-footed Achilles and the slow-moving tortoise, where the latter has been given a head start.\footnote{For a discussion of Zeno's paradoxes  we refer the reader to \cite{sep-paradox-zeno}, and references therein.}  
 It appears that Achilles can never overtake the tortoise, since after a certain time interval Achilles reaches the starting point of the tortoise, while in the mean time the tortoise has moved on. It   then takes Achilles some more time to run to the tortoise's next position, by which time the tortoise has crawled a bit further. Before Achilles has   arrived at this  new point, the tortoise has again advanced a tiny bit forward. Thus,  whenever  Achilles arrives at the tortoise's position, the tortoise has had enough time to move a   short distance ahead, and so Achilles seems to never catch up with the tortoise  (even though the distance between them becomes infinitesimally small).

 We will now formulate Zeno's paradox   in Einstein's theory of special relativity, assuming for the sake of argument that Achilles and the tortoise can move at a significant fraction of the speed of light (for example, because they  travel on    extremely fast spaceships). Suppose Achilles and the tortoise are uniformly accelerating   in flat spacetime (see Figure \ref{fig:zeno}). 
   They    start at rest inside the diamond at time $t=0$   and   distance $\Delta x$ from each other (in standard  flat coordinates). 
The tortoise has been given  a head start, just like in Zeno's standard   paradox. 
However, Achilles accelerates faster and   overtakes the tortoise at the future tip of the diamond. That is, they   meet at the future tip of the diamond, and when they continue outside the diamond,    Achilles   lies ahead of the tortoise. 

The paradox\footnote{A similar relativistic version of Zeno's thought experiment can be found in the book \emph{Expanding Universes} by Edwin Schr\"{o}dinger \cite{Schrodinger} (p. 20), but it came up independently in a discussion by the author and Erik Verlinde in 2014, during the preparation of a homework exercise on Rindler space for a special relativity course.}    arises due to the following fact: in a certain coordinate system it takes an \emph{infinite} amount of  time for both Achilles and the tortoise to reach the future tip (this time is called the ``conformal Killing time'', see Appendix~\ref{appyork}). In this coordinate system they never seem to arrive at their meeting point:   after a   finite coordinate time, Achilles has reached the tortoise's position, but the tortoise has moved ahead in the mean time. This happens   ad infinitum in   conformal Killing time, exactly as in Zeno's original paradox.  The paradox is resolved by pointing out that the coordinate time is not the physical time measured by   clocks which  follow   the same trajectories as Achilles and the tortoise. The   time measured by Achilles and the tortoise is   the proper time, which is finite along the trajectories   between the start and   future tip of the diamond. Thus, Achilles   passes the tortoise in a \emph{finite proper time}, despite the \emph{infinite coordinate time} it takes.\footnote{An interesting limiting case of this setup  is     \emph{Rindler space}, which is an infinite diamond in flat space (see Section \ref{largesmall:rindler}).  In   Rindler space     it takes an \emph{infinite proper time} for uniformly accelerating  observers  to meet, hence the tortoise can stay ahead of Achilles \emph{forever}!} 
  When they meet at the future tip  the tortoise has aged   more than Achilles, because his internal clock runs faster than Achilles'  clock (due to a relativistic effect called ``time dilation''), but that is the cost of engaging  in a relativistic    race with the fleet-footed Achilles.

%


This     thought experiment in relativity theory is relevant for the present thesis, since the trajectories of    Achilles and the tortoise coincide with the orbits   of the conformal Killing vector field that preserves   the diamond. This vector field plays an important role in Chapters \ref{ch2} and \ref{ch3} of the thesis. Whenever we write about the flow of the conformal Killing field, the reader can  imagine Achilles and the tortoise moving along that flow. We refer the reader to Appendix \ref{sec:accckv} for a computation of the acceleration, velocity and proper time of the orbits of the conformal Killing flow     in flat space.         Further, the thought experiment also illustrates the important notion of an \emph{event horizon}. The accelerating observers have an event horizon, which means that everything that happens outside the horizon cannot reach them (due to the finiteness of the speed of light). The future light cone of the diamond is called the ``future horizon'' (similarly for the ``past horizon''). \emph{All} timelike trajectories inside the diamond cannot receive messages from spacetime points outside the future horizon, nor can they send signals to points outside the past horizon.

\subsubsection*{The significance of causal diamonds}
 
In this thesis we are interested in causal diamonds\footnote{Causal diamonds also go under the less exciting names of ``double cones'' \cite{Hislop1982} and ``Alexandrov open sets'' \cite{Gibbons:2007nm,Penrose:1972ui}.} since they are well suited for studying universal aspects of spacetime and  gravity. Let us first mention that  causal diamonds  are the spacetime regions in which causal experiments can be performed.  A causal diamond can namely be defined as the intersection   of the causal future of a point $p$ and the causal past of point $q$: $C(p, q) = J^+ (p) \cap J^- (q)$.\footnote{Equivalently,   causal diamonds can be defined as the intersection of the future and past domain of dependence of a particular spatial region $\S$: $C (\S)= D^+ (\S) \cap D^- (\S)$.} An experiment that begins at  $p$ and ends at     $q$ on a timelike trajectory is in causal contact with every event in a   diamond.   A fundamental physical theory that describes local causal experiments thus necessarily should deal with  the physics inside a causal diamond, but arguably only one diamond at the time, since everything that happens outside   is operationally meaningless \cite{Bousso:2000nf}. 



 Second, causal diamonds have a null boundary and   hence define a covariant spacetime structure. The null boundary allows for a straightforward application of the covariant  entropy bound \cite{Bousso:1999xy}  (see Section \ref{intro:holographic}).  This bound implies that the entropy on the past and future null boundaries of the diamond is   bounded by the Bekenstein-Hawking entropy. Thus, causal diamonds have holographic properties, which deserve further study. 
 
 Third, apart from holographic properties event horizons typically obey thermodynamic laws. It is well known that    a temperature and an entropy  can be associated to black hole horizons, cosmological horizons and Rindler horizons. It is natural to ask whether other horizons also behave thermodynamically. The null boundary of a causal diamond is locally a Rindler horizon, hence  causal diamonds might inherit the  thermodynamic properties of Rindler horizons. Observers who accelerate uniformly  near the edge of the diamond   indeed measure an Unruh  temperature, because of the   boost Killing  symmetry near the edge. Causal diamonds are   an interesting case study for gravitational thermodynamics, since they can differ in size: Rindler space and the de Sitter static patch are   examples of large diamonds, but  perhaps small causal diamonds   also behave thermodynamically.
 
 Finally,   diamonds in maximally symmetric spaces are particularly appropriate     for studying gravitational thermodynamics, since they are conformally stationary.  
 The conformal Killing symmetry of maximally symmetric  diamonds implies that     the vacuum state restricted to these  diamonds is       thermal for conformal quantum field theories.
Just like black holes, maximally symmetric diamonds thus seem to behave as thermodynamic equilibrium states, at least for conformal fields.  In Chapters \ref{ch2} and \ref{ch3} we study the equilibrium properties of these  diamonds. 

\subsubsection*{Outline of this thesis}

In this thesis we   explore the thermodynamic, emergent and holographic aspects of gravity inside causal diamonds. These  three  topics roughly divide the thesis into three parts, in which we study the following   broad questions:

  \begin{enumerate}
 \item[(i)]  What are the gravitational thermodynamic  laws that govern causal diamonds in maximally symmetric spaces?    
 \\See Chapter  \ref{ch2}.
 \item[(iii)]What is the   universal microscopic principle from which (higher curvature) gravitational field equations can be derived?  
 \\See Chapter   \ref{ch3}.
 \item[(ii)] What   are the general features of   holographic microscopic theories for asymptotically non-Anti-de Sitter geometries?  
 \\See Chapter \ref{ch4}.
 \end{enumerate}
 
Chapter \ref{ch1} serves as an introduction to the rest of the thesis.
 Chapter \ref{ch2} is  devoted to the thermodynamics of causal diamonds in maximally symmetric spacetimes. We derive a Smarr relation and first law for causal diamonds  in (Anti-)de Sitter space, by employing Wald's Noether charge method and the conformal isometry of the diamonds.  In the semiclassical regime we deduce a universal principle of emergent  gravity from the first law, called ``entanglement equilibrium''  (see Section \ref{intro:EE5} for an introduction of this principle). We reformulate this principle in several ways, in particular as the stationarity of the diamond free energy in the vacuum.
 
In Chapter \ref{ch3} we generalize the first law of causal diamond to any higher curvature  theory of gravity. When quantum   matter fields are considered, we derive from  this first law an equilibrium condition on the entanglement entropy of   spherical regions in vacuum. For   small spherical regions we show that  the linearized higher derivative gravitational field equations are equivalent to this ``entanglement equilibrium'' condition. The existence of a  holographic duality is not   assumed,  nor necessary,   in this derivation. 

  
  Chapter \ref{ch4} examines how  holography  can be extended beyond asymptotically Anti-de Sitter (AdS) spacetimes.   We assume the existence of a  microscopic    theory on a closed codimension-two surface    that is holographically dual to Einstein gravity.    We probe the holographic dictionary   for  asymptotically non-AdS geometries,  such as   de Sitter   and Minkowski space and AdS space below its curvature radius, by conformally relating these geometries to $AdS_3 \times S^q$ spacetimes, and using  known results from AdS$_3$/CFT$_2$ holography. Concretely,  we derive three general  quantities for the microscopic theories of these spacetimes: the total number of degrees of freedom, the   typical excitation energy, and the number of excitations.   





\subsubsection{Assumptions, limitations and conventions}

\emph{Note on assumptions:} apart from Chapter \ref{ch1}   we {assume} in this thesis that gravity is emergent and holographic. In Section \ref{defEG} we explain what is meant by emergent gravity, and  {why} it is widely believed among physicists that gravity is emergent.  However, in the rest of the thesis we are interested in the ``how'' rather than the ``why"   of emergent gravity and holography. We  investigate the principles behind the emergence of gravity in Chapters~\ref{ch2}~and~\ref{ch3} and   analyze the dictionary of   holographic dualities in Chapter \ref{ch4}. These investigations  {do} provide further evidence or proof-of-concepts for  emergent gravity and holography in certain idealized setups, but are  {not}      theoretical  (let alone phenomelogical) proofs that gravity is emergent or holographic in the real world.\footnote{The title of the thesis is ``Emergent Gravity in \emph{a} Holographic Universe'' instead of   \emph{our} (holographic) universe. We believe, though, that the spacetimes we study (such as de Sitter space) are sufficiently close to our own universe, so that these investigations do teach us something about the real world.}


\emph{Note on spacetime solutions:} throughout the thesis we   only consider (variations of) spherically symmetric causal diamonds in static (i.e. non-rotating) spacetimes. In  Chapters~\ref{ch2}~and~\ref{ch3} we even restrict to maximally symmetric spacetimes, i.e. Minkowski   and (Anti)-de Sitter spacetime, whereas in Chapter \ref{ch4} we are also interested in  Schwarzschild black hole geometries.  Most equations in this thesis are valid in an arbitrary number of spacetime dimensions.
 
\emph{Note on theories of gravity:} Chapters   \ref{ch2} and \ref{ch4} are concerned with     general relativity   plus a cosmological constant, in some cases minimally coupled to matter fields.    Chapter~\ref{ch3} deals with  any  higher curvature theory of gravity minimally coupled to matter. Specifically,  the gravitational Lagrangian is an arbitrary, diffeomorphism-invariant function of the metric, Riemann tensor, and its covariant derivatives.

\emph{Note on conventions}: we set $k_{\rm B}=c=1$ (except in Section \ref{intro:emergentgrav}) and often also $\hbar =1$, but keep Newton's constant $G$ explicit throughout the thesis. We use metric signature $(-,+,+,\ldots)$, and denote the total number of spacetime dimensions with~$d$.   
We write the spacetime volume form as~$\epsilon$, and  occasionally we   denote it $\epsilon_a$ or $\epsilon_{ab}$, suppressing all but its first one or 
two abstract indices.

\emph{Note on language:} Throughout the thesis the pronoun ``we" is employed instead of ``I", since most work is based on collaborations. This is a common practice in   physics and in academia in general.

%

\begin{center}
 \vskip .5cm
 ***
 \vskip .5cm
\end{center}

%% file: chapter1.tex
\chapter{Introduction}
\label{ch1}

This chapter provides an introduction to the  concepts and topics  which are   relevant for this thesis.
In Section \ref{intro:emergentgrav} we start by defining emergent gravity and laying out the two main motivations behind the idea that gravity is emergent: black hole thermodynamics and the holographic principle.  Next, in Section \ref{intro:dSthermo} we discuss  in some detail the geometry and thermodynamics of the static patch of  de Sitter space. This serves as an illustration for a central  principle of quantum gravity   in this thesis due to \cite{Jacobson:2015hqa}: the generalized entropy of    causal diamonds is maximized in maximally symmetric  vacua. In  Section \ref{intro:EE5}  we show that this maximal  vacuum entropy principle is realized for first-order variations of the metric and quantum fields in   the de Sitter static patch  and  in small empty diamonds in Minkowski space. We also review Jacobson's derivation of the semiclassical Einstein equation   from this microscopic principle; in fact, he establishes an \emph{equivalence} between gravitational dynamics and this principle. Finally, in Section \ref{intro:AdS/CFT} we briefly describe some of the basic ingredients of the AdS$_3$/CFT$_2$ correspondence, in particular how the BTZ black hole entropy can be derived from the Cardy formula in 2d CFT. The first Section  \ref{intro:emergentgrav}  is written with a more philosophically-inclined readership in mind, whereas the latter three Sections \ref{intro:dSthermo}--\ref{intro:AdS/CFT}   are primarily for physicists and thus contain more technical details.\footnote{Section  \ref{intro:emergentgrav}  is largely based on the philosophy of physics paper  by Niels Linnemann and the author: ``Hints towards the Emergent Nature of Gravity'' \cite{Linnemann:2017hdo}.} 

\newpage
\section{Emergent gravity and holography} 
\label{intro:emergentgrav}

Gravity is notoriously harder to quantize than other interactions. It is well known that combining general relativity (GR) with quantum mechanics poses several difficulties, which are usually traced back to the perturbative non-renormalizability of general relativity. Proponents of quantum gravity approaches proper  --- such as loop quantum gravity, asymptotic safety and causal dynamical triangulation --- nevertheless expect that gravity is after all non-perturbatively renormalizable, and thereby possibly fundamental. In contrast, according to other approaches --- like string theory and emergent gravity accounts by Sakharov (1968) \cite{Sakharov1967}, Jacobson (1995, 2016) \cite{Jacobson:1995ab,Jacobson:2015hqa}, Hu (2009) \cite{Hu}, Padmanabhan (2010)  \cite{Padmanabhan:2009vy}, Verlinde (2011, 2017)  \cite{Verlinde:2010hp,Verlinde:2016toy}
---  the gravitational field cannot be fundamental but rather emerges from underlying novel microstructure. These approaches can be subsumed under the term ``emergent gravity'' (EG).

The idea that gravity and spacetime originate from some underlying microscopic reality, in which they do not exist, is a widely held belief in, for instance, the string theory community (cf. \cite{WittenReflections,Seiberg,Verlinde:2010hp}). In this section we   explain   \emph{why}  gravity is believed to be emergent (the arguments are often only implicitly contained in the   physics literature). First,   Section \ref{defEG}      describes  in more detail what is meant by the term ``emergent gravity''.  Second,  Sections \ref{intro:BHthermo}  and \ref{intro:holographic} contain  the    two main motivations behind  emergent gravity: black hole thermodynamics (or horizon thermodynamics in general) and the holographic principle. They are, of course, not fully independent motivations since the latter principle is itself motivated from black hole thermodynamics.



\subsection{Definition of emergent gravity}
\label{defEG}

Emergent gravity is roughly taken to be the view that gravity arises due to the ``collective action of the dynamics of more fundamental non-gravitational degrees of freedom'' \cite{HuggettWuethrich}. The aim of this subsection is to sharpen this rough characterization of emergent gravity   by providing answers to the following core questions: (A) ``emergence of what?'', (B) ``emergence from what?'' and (C) ``what kind of emergence?''. The first question asks what properties of gravitational physics are actually emergent. The second is concerned with the existence and nature of the underlying microstructure, and the third focuses on the emergence relation between the gravitational and non-gravitational degrees of freedom.

\subsubsection{(A) Emergent aspects of gravity}

\label{sec:definition}

Which aspects of gravity are emergent? Needless to say the answer to this question depends on what we take gravity to be. A gravitational theory has for instance been  defined by \cite{ThorneLee} as a spacetime theory that (approximately) reproduces Kepler's laws.\footnote{This definition seems to unfairly exclude  modifications of GR (such as higher curvature theories and bimetric theories, like TeVeS)  as gravitational theories.}  In any case, our current best theory of gravity is GR, which formally encompasses a differentiable  manifold with tensor fields on top. One of the tensor fields necessarily corresponds to a Lorentzian metric field $g_{ab}$.  The metric field satisfies the Einstein equation, $G_{ab} = \frac{8\pi G}{c^4} T_{ab}$, which relates the Einstein tensor $G_{ab}$ to the stress-energy tensor $T_{ab}$ obtained from all matter tensor fields. The matter fields themselves are subject to further (possibly coupled) dynamical equations.

One might think that ideally in an emergent model of GR all three aspects emerge together: the manifold, the metric field and its dynamics. The emergence of the manifold is, however, somewhat less important, since its physical significance is doubtful. Arguably, the (active) diffeomorphism invariance of fields in GR    renders the manifold as a mere mathematical scaffolding structure to define physical fields on (cf. the hole argument of \cite{NortonEarman}).  

In some approaches, such as the thermodynamic route to GR by \cite{Jacobson:1995ab}, the metric field is  given ab initio, and then its (Einstein) equation  of motion is derived. We think, however, that a genuine account of emergent gravity should  include the emergence of \emph{both} the metric field and its dynamics (rather than just the dynamics). This it not to say that thermodynamic accounts of gravity are not linked to emergent gravity. In fact,    we categorize it as an emergent gravity scenario, because the  thermodynamic interpretation of the dynamical  equations associated to the metric field $g_{ab}$  naturally leads to the emergent nature of $g_{ab}$ itself.

In the next part, we will further see that  not only the metric field and its dynamics are emergent  aspects of gravity, but    also the  graviton --- the spin-2 particle that mediates the gravitational force --- should count as ``gravitational'' (and  hence  as emergent). According to the spin-2 approach the metric field and its dynamics namely arise through the self-coupling of the graviton,  cf.  \cite{Deser:1969wk}. If we  viewed the spin-2 field as non-gravitational, then GR would emerge from quantized GR, which is however against the intuition behind emergent gravity.

\subsubsection{(B) Two types of microtheories}

\label{sec:microstructures}

In the EG literature,  gravity is unanimously taken to be emergent from (novel) microstructure underlying the metric degrees of freedom. For instance, Carlip \cite{Carlip} defines emergent gravity as: \begin{quote}
	``[...] the basic picture is that gravity, and perhaps space or spacetime themselves, are collective manifestations of very different underlying degrees of freedom.'' (p. 200) \end{quote} In a similar spirit, Sindoni \cite{Sindoni2012} writes: \begin{quote} ``As a provisional definition, we will intend as emergence of  a given theory  as a reorganization of the degrees of freedom of a certain underlying model in a way that leads to a regime in which the relevant degrees of freedom are qualitatively different from the microscopic ones.'' (p. 2)
\end{quote}  

\noindent Padmanabhan \cite{Padmanabhan:2009vy, Padmanabhan:2016bha}, Hu \cite{Hu}  and Verlinde \cite{Verlinde:2010hp} employ similar notions of EG. They tend to compare (or even equate) the idea of gravity having underlying microstructure (sometimes referred to as ``atoms of spacetime") to the sense in which thermodynamic or hydrodynamic systems have underlying microstructure. 

However, despite the apparent consensus in the literature, this definition of emergent gravity   is problematic. 
 In general, the  \emph{received definition of emergent gravity} in terms of \textit{very different} underlying degrees of freedom,  or \textit{novel} microstructure simply suffers from vagueness. It is vague both with respect to qualifiers like ``very" or ``novel", and the characterization of what there actually is  supposed to be over and above gravity (``different degrees of freedom" or ``microstructure"). In the following, we give three major instances of changes in degrees of freedom for which it is not a priori clear whether they should be counted as ``very different'' degrees of freedom or not. 
 
Firstly, the discretization of a theory of gravity changes its degrees of freedom in some sense. After all, the continuous version can only be reproduced upon coarse-graining again. But, according to the EG proponent, the degrees of freedom of a discretized theory  should   not be  ``very" or ``qualitatively" different from the degrees of freedom of the original theory. If that were the case, causal set theory would   count as an instance of emergent gravity, which would make the definition of EG too broad.

Secondly, quantization of a theory does to some extent change the theory's degrees of freedom. As quantization generically leads to discretization of classically continuous quantities --- such as energy, angular momentum, and possibly even spacetime structure ---  coarse-graining is typically part of restoring the classical limit (together with suppressing superposition effects). However, quantization does not suggest  a massive change of degrees of freedom, given that the quantum theory is constructed from the classical theory by translating classical  quantities into corresponding quantum ones, e.g. observables are promoted to operators. Presumably, the quantum degrees of freedom can therefore not be ``very'' different from the classical degrees of freedom.







	Thirdly,  in a quantum field theory (QFT) setting,   the renormalization group flow describes a change of theories under a coarse-graining operation. Coarse-graining here corresponds to integrating out high energy   modes.  If one starts  with a renormalizable theory which is well-behaved up to arbitrarily high energies, then coarse-graining will produce effective field theories which are only valid up to a certain level of energy. 
		Despite a change of its degrees of freedom  under renormalization (energy modes are integrated out), the resulting theory will not necessarily involve ``qualitatively" different degrees of freedom. Generally speaking, the   renormalization group flow only changes the coupling constants of a QFT.
		In the context of gravity, asymptotic safety claims that quantum general relativity (quantum GR\footnote{By ``quantum GR'' we denote any form of quantization approach turning GR into a quantum theory, including perturbative quantum gravity,  loop quantum gravity, causal dynamical triangulation, and more; but not, for instance, string theory and causal set theory.}) is subject to a renormalization group flow such that it  eventually runs into a UV fixed point in theory space. The only difference in the degrees of freedom of     low-energy and high-energy versions of quantum GR is then encoded in the UV cutoff of the corresponding theory.
 
The three examples above show that the received definition of EG is problematic, since it is unclear whether  discretization, quantization, and renormalization lead to ``qualitatively different'' degrees of freedom. This issue is important because we   want to distinguish EG from other quantum gravity accounts   that arise from discretizing, quantizing or renormalizing  GR. One could attempt to sharpen the characterization of the degrees of freedom underlying gravity by calling them  ``non-gravitational" (see also \cite{HuggettWuethrich}), but this remains unsatisfactory as long as one has no clear notion of  ``gravitational".  We already alluded in the previous part to the problem that we cannot straightforwardly give a definition of  ``gravitational'' in terms of GR alone, without running the risk of calling theories underlying GR non-gravitational, which are however not at all instances of emergent gravity. For example,   the spin-2 approach, loop quantum gravity, and causal set theory are possible quantum theories underlying GR that we would like to  exclude as cases of emergent gravity.  

We now propose to circumvent this issue by introducing a distinction between two types of microtheories, i.e. two types of theories underlying a  theory $T$:\footnote{We necessarily require microtheories to $T$ to reduce to $T$ in an appropriate limit, such as a classical or low-energy limit.}
  
\begin{quote}
We call a theory $M_1$ underlying a    theory $T$  a \emph{type I microtheory} to $T$ if and only if it is inspired 
 by $T$ (for instance through discretization, quantization or renormalization). An underlying theory $M_2$ to theory $T$ is called a \textit{type II microtheory} if and only if it  is not directly inspired or motivated\footnote{One might argue that  notions as ``inspired'' or ``motivated''  are   as vague as the notions ``qualitatively'' or ``very different'', criticized above. However,  despite their granted vagueness, the former notions allow for a more precise distinction between what is referred to as EG in the literature and what is not.} by  $T$. 
\end{quote}

\begin{quote}
The structure  linked to a microtheory is called \textit{microstructure}.
\end{quote}

\begin{quote}
A   theory is  called  \textit{emergent} if and only if there exists an underlying type II microtheory to it.
\end{quote}

\noindent 
 A type I microtheory to GR can now coherently be counted as gravitational, whereas a type II microtheory to GR is by definition non-gravitational since gravity emerges from it. Another advantage of this \textit{new definition of emergent gravity} in terms of type II microtheories is that the microstructure associated to these theories is automatically novel, because   it cannot be inspired by gravity. Examples of possible type I microtheories to GR are causal set theory, loop quantum gravity and other quantum gravity  approaches proper.   This is in line with the general idea of emergent gravity being something over and above quantum gravity proper (see \cite{Carlip, Padmanabhan:2009vy, Jacobson:1995ab}).  

 \subsubsection{(C) Different notions of emergence}

 With respect to the third question ---  ``what kind of emergence?'' ---  it is important to point out  that philosophers and physicists use the word ``emergent'' differently, and themselves not at all coherently either.\footnote{We would like to thank Karen Crowther  for sharing an unpublished note, in which she distinguished between the philosopher's, the physicist's and the philosopher of physics' usage of the word ``emergence''. This distinction goes back to \cite{Nickles}.}    \emph{Philosophers} in general tend to  define emergence as the failure of  reduction. A theory    is then called emergent from another  more fundamental  theory  if the former is not  derivable or explainable from the latter.

In contrast, \emph{philosophers of physics} mostly work with Butterfield's \cite{ButterfieldEmergence} notion of emergence in terms of supervenience on the one hand, and novelty and robustness on the other hand. 
 Supervenience of a theory  on another (standardly more fundamental) theory  means that there can be no change in the former theory  without a change in the latter theory. The problem with this notion of emergence, of course, lies in its deference of the difficulty in defining emergence to that of defining robustness and novelty (similar to how the philosopher's sense of emergence seems to ultimately depend on the definition of reducibility). Following \cite{ButterfieldEmergence}, we take ``novel'' properties to mean  features which are not contained in the microstructure, whereas ``robust" properties are properties that are not sensitive to (and somehow independent from) the specifics of the underlying structure.

Lastly, \emph{physicists} often use the word ``emergence'' in the sense of reducibility, that is derivability or explainability of a certain structure from another one. Take a string theorist claiming that he/she derived general relativity from perturbative string theory in a low-energy limit: he/she will probably call this an instance of emergent gravity. More concretely, Seiberg \cite{Seiberg}   takes space and time to be emergent if they ``will not be present in the fundamental formulation of the theory and will appear as approximate semiclassical notions  in the macroscopic world" (p. 163). Here, he identifies emergence of spacetime with its derivability (through semiclassical approximation) from a microscopic structure.

Which notion of emergence is applicable to emergent gravity? We should not forget that the notion of emergent gravity is  introduced by physicists --- not by philosophers. It is thus plainly wrong to presume that the notion of emergence in EG should be interpreted in  the philosopher's sense. Rather, it is   the physicist's notion which seems to apply here --- as physicists  expect GR to be reducible to more fundamental microscopic theories --- thereby excluding the philosopher's notion from the start.

Still, we do not take the physicist's reduction to exhaustively specify the relationship between the metric  and its    underlying microstructure. There is arguably an aspect of novelty and robustness in the emerging metric structure which need to be accounted for.\footnote{Some physicists might also   have this notion of emergence in mind. The previous distinction between the three senses of emergence should  be understood as a general labeling, not specific to the context of quantum gravity.}
 First, the non-gravitational degrees of freedom  (as defined in the previous section)  are qualitatively different  from (and hence \emph{novel} with respect to) the gravitational degrees of freedom, as the former are not inspired by the latter. Secondly, albeit more specifically, in thermodynamic approaches by Jacobson, Verlinde and Padmanabhan, gravity is viewed as \emph{robust}  with respect to the underlying microstructure, in the sense that it only depends on   coarse-grained notions such as entropy and temperature. Thus, the notion of emergence in emergent gravity  seems most suitably   characterized by the philosopher  of physics' sense. This also seems to capture best how advocates of EG models use the word ``emergence''.

In the following two subsections we follow a \emph{top-down strategy} for arguing that gravity is emergent.
The strategy is to look for analogies between   gravitational theories and  other theories said to have microstructure, such as thermodynamics or hydrodynamics.  A sufficiently strong analogy between, for example, gravity and thermodynamics in a relevant sense can be viewed as an important  hint that gravity is emergent.  
Certain features of gravity --- mainly its universality, perturbative non-renormalizability, its thermodynamic features and  its holographic nature --- allow for mounting such arguments from analogy.  Arguably, these gravitational features even remain ``mysteries'' in a fundamental take on the metric and appear to be in need of explanation,  which however  can be easily achieved in an emergent gravity framework. The two main motivations for emergent gravity --- black hole thermodynamics (Section \ref{intro:BHthermo}) and the holographic principle (Section \ref{intro:holographic}) --- are considered below, and  we refer the interested reader to  \cite{Linnemann:2017hdo} for a discussion of the universality    and   perturbative non-renormalizability of gravity as other hints for its  emergent nature. 

\subsection{Black hole thermodynamics}
\label{intro:BHthermo}

The connection between gravity and thermodynamics was originally discovered   in the context of black holes in the early 1970s.  In  \cite{Hawking:1971tu,Hawking:1971vc} Hawking showed that the area of the event horizon of a black hole can never decrease, assuming the null energy condition is satisfied (generalizing earlier work by Christodoulou \cite{Christodoulou:1970wf,Christodoulou:1972kt}).  This is reminiscent of the second law of thermodynamics which states that the entropy of a closed thermodynamic system can never decrease. Based on this similarity and on black hole  thought experiments,   Bekenstein \cite{Bekenstein:1972tm,Bekenstein:1973ur} proposed that black holes carry an entropy proportional to their  horizon area $A$:
\beq \label{Bekensteinentropy}
S_{\rm bh}=     k_{\rm B} \eta \frac{A}{\ell_{\rm P}^{d-2}}  \, ,
\eeq
where $\eta$ is a constant number of order unity, $k_{\rm B}$ is Boltzmann's constant,   and $\ell_{\rm P} := (\hbar G /c^3)^{\frac{1}{d-2}}$ is the Planck length in $d$ spacetime dimensions. 

In one of his thought experiments Bekenstein imagined dropping a     thermal object into a black hole, thereby decreasing the entropy outside   the black hole. For an outside observer there is no way to determine the interior entropy of the black hole, since a black hole in equilibrium is uniquely determined by three parameters: mass $M$,  angular momentum $J$ and charge  $Q$ (the so-called ``no-hair theorem''). In order to preserve the \emph{second law} of thermodynamics, he conjectured    that   the sum of the black hole entropy and the matter entropy outside of   the black hole --- called the ``generalized entropy'' ---  never decreases:
\beq \label{gensecondlaw4}
\Delta S_{\rm gen} := \Delta S_{\rm bh} + \Delta S_{\rm m} \ge 0 \, . 
\eeq
Bekenstein \cite{Bekenstein:1973ur} also derived   an identity for rotating charged black holes by varying their parameters: $\delta M = (\kappa/8\pi G) \delta A + \Omega_{\rm h} \delta J  + \Phi  \delta Q$. This variational relation is similar to the thermodynamic fundamental  identity: $dE = T dS + \mu_i d Q_i$, where $\mu_i$ is a chemical potential and $Q_i$ is the conjugate charge. The energy $E$ can be   identified with the mass $M$ of the black hole, and  if the two terms $\Omega_{\rm h} \d J  + \Phi \d Q$  are the analog of  the term $\mu_i dQ_i$, then
  $\kappa$ and $A$ are analogous to the temperature $T$ and the entropy~$S$, respectively.\footnote{Bekenstein obtained an expression for $\kappa$, but did not realize that this is an intensive variable (like the angular velocity $\Omega_{\rm h}$ of the horizon and the electrostatic potential~$\Phi$), nor that it had the interpretation of surface gravity. (We thank Ted Jacobson for this remark.)}  
 Bardeen, Carter and Hawking \cite{Bardeen:1973gs} extended the variational identity  to include matter terms: they derived a so-called \emph{first law} that relates       two nearby  stationary axisymmetric asymptotically flat solutions of the Einstein equation containing a black hole surrounded by matter.  In this thesis we restrict to    static    black holes, for which the first law  with an additional matter term is  given by
 \beq \label{firstlawBCH}
\delta M = \frac{\kappa}{8\pi G} \delta A + \int_\S \delta {T_a}^b \xi^a u_b dV  , 
\eeq
where $\kappa$ is the surface gravity of the black hole horizon; $\S$ is a spatial slice between the event horizon and asymptotic infinity, with $u^b$ the future pointing unit normal; and $\xi^a$ is the time translation Killing vector,  normalized as $\xi^a \xi_a  =-1$ near infinity. For the case in   which the matter outside the black hole is a perfect fluid,   the matter term in \eqref{firstlawBCH} can  be expressed in terms of   integrals over the change in particle number and entropy crossing a surface element, i.e.  $\int_\S \mu \delta dN + \int_\S T \delta dS$, where $\mu$ and $T$ are the redshifted chemical potential and temperature, respectively.\footnote{Bardeen, Carter and Hawking obtained the first law    by varying   a generalized Smarr formula \cite{Smarr:1972kt}, which for static black holes with surrounding stress-energy reads in general dimensions
\beq
 \frac{d-3}{d-2}M =  \frac{\kappa A }{8 \pi G}  +      \int_\S \left ( T_{ab} - \frac{1}{d-2}T g_{ab} \right) \xi^a u^b dV \,. 
\eeq
 The term between brackets is equal to  $R_{ab}/8\pi G$  due to the Einstein equation. This Smarr formula  for general relativity was derived in \cite{Bardeen:1973gs}   for $d=4$   from the Killing identity, but can  equivalently be obtained   from the Noether charge method  presented in Section \ref{sec:smarr}.} 
 
Another analogy between black holes and thermodynamics regards the \emph{zeroth law}. Bardeen, Carter and Hawking were able to prove, assuming the dominant energy condition, that $\kappa$ is constant on the event horizon of a stationary axisymmetric black hole, analogous to the temperature being constant in thermal equilibrium.  In total,  they formulated  four laws of black hole mechanics, which were strikingly similar to the standard four laws of thermodynamics. At the time they thought of this similarity   as a   mere formal analogy, since there was no evidence  that black holes emit thermal radiation.  This changed with the appearance of   \cite{Hawking:1974rv,Hawking:1974sw}, in which Hawking showed that   black holes have a temperature, equal to
\beq
T_{\rm H} = \frac{\hbar      }{ k_{\rm B} c}  \frac{\kappa}{2\pi}\, . 
\eeq
Hawking's discovery implied that  black holes are ``true'' thermodynamic systems. Further, together with the first law \eqref{firstlawBCH}, it fixed  the proportionality constant   between black hole entropy and the horizon area in~\eqref{Bekensteinentropy} to be $\eta=1/4$. Thus,
the formula for   black hole entropy (called the ``Bekenstein-Hawking formula'') is
\beq
S_{\rm BH} =  \frac{k_{\rm B}  c^3}{G \hbar}  \frac{A}{4}\, . 
\eeq
In this thesis we often set $k_{\rm B} = c  = \hbar=1$  (except in Chapter  \ref{ch2} where we keep $\hbar$ explicit). This beautiful formula combines   four  physical constants of nature, and   contains important information about the degrees of freedom of   quantum gravity. 

What does black hole thermodynamics imply about the   microstructure of the metric field and its dynamics? Thermodynamic phenomena are standardly seen to be underlied by non-trivial (type II) microstructure. For instance,   the   temperature of a gas is viewed as a measure of the averaged motion of the particles that make up the gas, and in a Boltzmannian view   entropy is a measure of the number of underlying microstates for a given macroscopic system. Since the laws of black hole mechanics are   thermodynamic laws, black hole metrics  as a consequence also seem to have underlying non-trivial  (type II) microstructure. This has for example   been verified by string theory calculations \cite{Strominger:1996sh} in which the Bekenstein-Hawking entropy was derived for five-dimensional extremal black holes by counting the   number of microscopic states of a D-brane configuration.   

Further, the laws of thermodynamics do not only apply to black hole horizons, but more generally to other types  of causal horizons, such as     cosmological horizons \cite{Gibbons:1976ue,Gibbons:1977mu} and acceleration horizons   \cite{Davies:1974th,Unruh:1976db,Jacobson:1995ab,Massar:1999wg,Jacobson:1999mi,Jacobson:2003wv}. It was even argued in \cite{Jacobson:1995ab} that they apply to local Rindler causal horizons   through each  point in any spacetime.  The connection between local causal horizons and thermodynamics  is a strong indication for the emergence of the metric and its dynamics for any spacetime.
 Jacobson  \cite{Jacobson:1995ab} namely  derived the Einstein field equation    from a Clausius relation   that governs all   local Rindler horizons in spacetime, and thereby he interpreted the Einstein equation as a thermodynamic equation of state. Thus, the gravitational field equations  of GR seem to be due to underlying non-trivial microstructure.

\subsection{The holographic principle}
\label{intro:holographic}

The holographic principle is generally viewed as an important guiding principle in the search for a theory of quantum gravity. It puts a precise limit on the number of fundamental degrees of freedom associated to spacetime regions, by stating that the maximal amount of information inside a given spacetime region cannot exceed the area of its boundary (measured in Planck units). 
In this subsection we are particularly interested in whether the holographic principle   entails anything about the nature of the putative microstructure of GR.  

\newpage

The \emph{holographic principle}, originally proposed by 't Hooft \cite{tHooft:1993dmi} and followed up by Susskind \cite{Susskind:1994vu}, is based on the following simple argument. Consider a  spacelike ball-shaped region  $\mathcal B$ with boundary $(d-2)$-surface $\mathcal S$  in a $d$-dimensional asymptotically flat spacetime. The  entropy and number of degrees of freedom inside $\mathcal S$ are maximized if the region contains a black hole whose horizon radius is equal to the radius of the ball. This is because, if the region $\mathcal B$ originally contained some matter, and  one tries to increase the entropy by putting in more and more matter, at some point it will collapse into a black hole. As we saw in the previous subsection,  thermodynamical arguments suggest that  black holes carry an entropy proportional to their horizon area. If one were to throw more matter into the black hole, then its size would grow until it does not fit    inside the surface $\mathcal S$ any more. Therefore, by  dropping the specific assumption of spherical symmetry and asymptotic structure, 't Hooft  concluded that the total number of microscopic degrees of freedom contained in any spatial region cannot exceed the area of the region's boundary (measured in Planck units). 

Later  Bousso \cite{Bousso:1999xy,Bousso:1999cb}  gave the holographic principle a covariant formulation. Firstly, the holographic principle   formulated above is violated for systems in which gravity is the dominant force, such as collapsing stars and large regions in cosmological spacetimes. In its most general form the holographic principle does not hold for spacelike regions, but   for  light sheets and their associated spacelike boundary. Secondly, a rendering of the holographic principle in terms of degrees of freedom already amounts to an interpretation of (black hole) entropy in a general-relativistic setting. Whether entropy in a gravitational context is due to a Boltzmannian-type counting procedure (as in string theory \cite{Strominger:1996sh}) or due to entanglement across the surface (see e.g. \cite{Sorkin:2014kta,Bombelli:1986rw,Srednicki:1993im,Frolov:1993ym,Solodukhin:2011gn,Jacobson:1994iw,Bianchi:2012br})  is however up to debate.
   The interpretation-neutral version of the covariant entropy bound now reads as follows \cite{Bousso:1999xy,Bousso:1999cb,Bousso:2002ju}
\begin{quote}
\emph{Covariant entropy bound:} The  entropy on any light sheet of a   spacelike ($d-2$)-surface $\mathcal S$ cannot exceed the Bekenstein-Hawking entropy  of $\mathcal S$ 
\end{quote}
\beq
S  \leq  \frac{k_{\rm B} c^3}{  G \hbar} \frac{A}{4} \, . 
\eeq
\noindent Here a light sheet of $\mathcal S$ is constructed by following light rays that emanate from $\mathcal S$, as long as they are not expanding. The covariant entropy bound can be shown to reduce to the original version of the holographic principle --- the so-called spacelike entropy bound ---    for spatial volumes and their associated boundary surfaces \cite{Bousso:2002ju}.

The holographic principle has led to important progress in quantum gravity research. Already 't Hooft put forward a much stronger statement  than the entropy bound,  namely that   all the gravitational degrees of freedom inside a region can be described by a quantum mechanical theory living on the region's boundary (which is now known as a \emph{holographic duality} or simply as  \emph{holography}). Such holographic duality relations have indeed been found, most notably the AdS/CFT correspondence \cite{Maldacena:1997re,Witten:1998qj,Aharony:1999ti}, and play a central role in quantum gravity research nowadays. We want to stress though that the holographic principle  (or, more precisely, the covariant entropy bound) should not be conflated with holographic duality relations. The latter is a conjecture about the equivalence of two, at first sight, unrelated theories, whereas the former is a bound on the gravitational entropy.  

Physicists often interpret the holographic relation as an emergent relation, in the sense that a theory of gravity in a given spacetime emerges from a quantum theory without gravity on the boundary. However, Dieks, Van Dongen and De Haro  \cite{Dieks}   among others worked out that emergence (also in our sense of underlying microstructure) and holographic duality are \textit{prima facie} independent notions. Only if the holographic duality is approximate at a certain level,   can it turn out that either the bulk or the boundary is more fundamental. In the latter case,  the bulk system can be seen to have underlying microstructure in terms of the (microscopic) boundary system. The jury is still out, though, on whether the AdS/CFT correspondence is an exact duality, because it is still only proven in a large $N$ limit (for semi-classical bulk gravity).

We take it, however, that the holographic principle \emph{itself} (and not just  the specific notion of holographic duality) already contains a hint for emergent gravity.  The holographic principle is not only concerned with the \emph{number} of fundamental degrees of freedom of quantum gravity, but it is also suggestive of the \emph{kind} of degrees of freedom that underly GR. As is well known, the holographic principle is in conflict with the conventional wisdom in statistical mechanics or quantum field theory that the number of degrees of freedom scales with the  volume size of a system. Conventional quantum field theories are interaction-local.\footnote{By interaction locality, we mean what is more usually dubbed `micro-causality', i.e. that spacelike separated operator field values   commute. For a scalar field $\phi$, this means that $[\phi(x), \phi(y)]=0$ provided that $x$ and $y$ are spacelike separated points.} Since degrees of freedom are linked to each spatial point, the information content of a spatial region grows with the volume. The holographic principle, on the other hand, implies that the entropy is bounded by the area of spacelike surfaces. Hence, the holographic principle implies that  the fundamental degrees of freedom of gravity are \emph{nonlocal}, and this  motivates regarding the underlying microstructure of GR as ``non-trivial'' (or of type II).  Thus, the holographic principle suggests that   spacetime is emergent from nonlocal underyling degrees of freedom.\footnote{Whether the nonlocality of the underlying microstructure is due to   quantum entanglement across the surface (as in induced gravity proposals \cite{Jacobson:1994iw}) or due to the holographic nature of the gravitational degrees of freedom (as in the  AdS/CFT holographic duality   \cite{Maldacena:1997re})  is still an open question. In the former case the area law for black hole entropy arises due to quantum correlations of UV degrees of freedom near the horizon, whereas in the latter case it is due to thermal correlations in a holographic boundary quantum field theory.}

\section{Thermodynamics of de Sitter space}
 \label{intro:dSthermo}
 
This thesis consists for a large part of a study of  the  thermodynamic  and holographic aspects of     de Sitter space. Cosmological observations suggest that both at very early and late   times our universe can be approximately described by de Sitter space. In the early stages our universe presumably went through an inflationary period, and can   be well approximated by the inflationary or flat patch of de Sitter space. Further, because of the observed accelerated expansion of the universe,   dark energy becomes dominant towards the future compared to other contributions to the   cosmological energy density    (i.e. matter and radiation). If dark energy is due to the cosmological constant, this means that the universe becomes closer and closer to de Sitter space at late times. In the far future the universe can hence be well described by the static patch of de Sitter space, where the horizon of our observable universe corresponds   to the cosmological horizon of a static observer in de Sitter space  (see Figure \ref{fig:penroseds} on p. \pageref{fig:penroseds}). Thus, a study of  de Sitter space is crucial for our understanding of the universe. In this thesis we will  be mainly concerned   with the static patch, and not  with  the inflationary patch  of de Sitter space. 
 
 \subsection{Classical geometry of de Sitter space}
 
 De Sitter  space (dS)\footnote{In Dutch last names which start with an article   or preposition, such as the names of Willem de Sitter and Johannes van der Waals, are written with a capital letter if the first name is omitted, like De Sitter and Van der Waals. However, in this thesis we deviate from this Dutch spelling rule and stick to the conventional abbreviations of the geometries   named after Willem de Sitter: dS and AdS (as opposed to the Dutch writing, DS and aDS). Another fun fact is that   ``sitter'' means  ``tailor'' or ``seamster'' in a certain old Dutch dialect.}  is     
  the       maximally symmetric  Lorentzian solution to 
 Einstein gravity plus a   positive cosmological constant $\L$, given by the Lagrangian
 \beq
L = \frac{1}{16 \pi G }   \left ( R - 2 \L  \right) \epsilon \, , 
 \eeq 
 where $\epsilon$ is the spacetime volume form.   Einstein's field equation takes the form
 \beq \label{fieldeqcosm}
 R_{ab} - \frac{1}{2} R g_{ab} + \L g_{ab} = 0 \, . 
 \eeq
Maximally symmetric spacetimes are locally characterized by the following condition for the Riemann tensor
\beq
R_{abcd} = \frac{R}{d (d-1)} \left ( g_{ac} g_{bd} - g_{ad} g_{bc} \right) , 
\eeq
with $d$ the number of spacetime dimensions and $R$ the Ricci scalar, which is constant throughout spacetime.
Inserting this   into the field equation  \eqref{fieldeqcosm} yields a   condition for the Ricci scalar:   $R =  \frac{2d}{d-2} \L$. For later convenience, we introduce the so-called de Sitter radius $L$, in terms of which the Ricci scalar and cosmological constant are given by
 \beq \label{dSradius}
 R  = \frac{d (d-1)}{L^2}   \qquad \text{and} \qquad \L =  \frac{(d-1)(d-2)}{2 L^2} \,.
  \eeq
  Note that de Sitter space has  constant positive curvature. 
    
  Alternatively, de Sitter space can be described as the maximally symmetric solution to Einstein gravity plus a perfect fluid. The Einstein equation is in this case
  \beq
  R_{ab} - \frac{1}{2} R g_{ab} = 8 \pi G T_{ab} \qquad \text{with} \qquad T_{ab} = (\rho + p) u_a u_b + p g_{ab} \, ,
  \eeq
  where $u^a$ is the velocity vector of the fluid, and $\rho$ and $p$ are the energy density and pressure of the perfect fluid, respectively. 
  Comparing with \eqref{fieldeqcosm} we see that the cosmological constant    in the field equation is equivalent to introducing a     stress-energy tensor   
  \beq \label{stresstensordS}
  T_{ab}^\L = -\frac{ \L}{8\pi G} g_{ab}  \qquad \text{and} \qquad \rho = - p =  \frac{ \L}{8\pi G}  \,. 
  \eeq
  Since the cosmological constant is positive for dS space, this maximally symmetric geometry can be viewed as a vacuum solution with   positive  constant   energy density and   negative pressure.
 We will use the two descriptions of the cosmological constant (a geometric term or a  perfect fluid term in the field equation), and hence of de Sitter space, interchangeably in this thesis. The perfect fluid description is particularly useful in finding the cosmological constant variation in the first law (Section \ref{sec:firstlawch2}). In Section \ref{sec:vacenergy} we give a microscopic interpretation of the vacuum energy of de Sitter space. 
  
  \begin{figure}
	\centering
	\includegraphics
		[width=.4\textwidth]
		{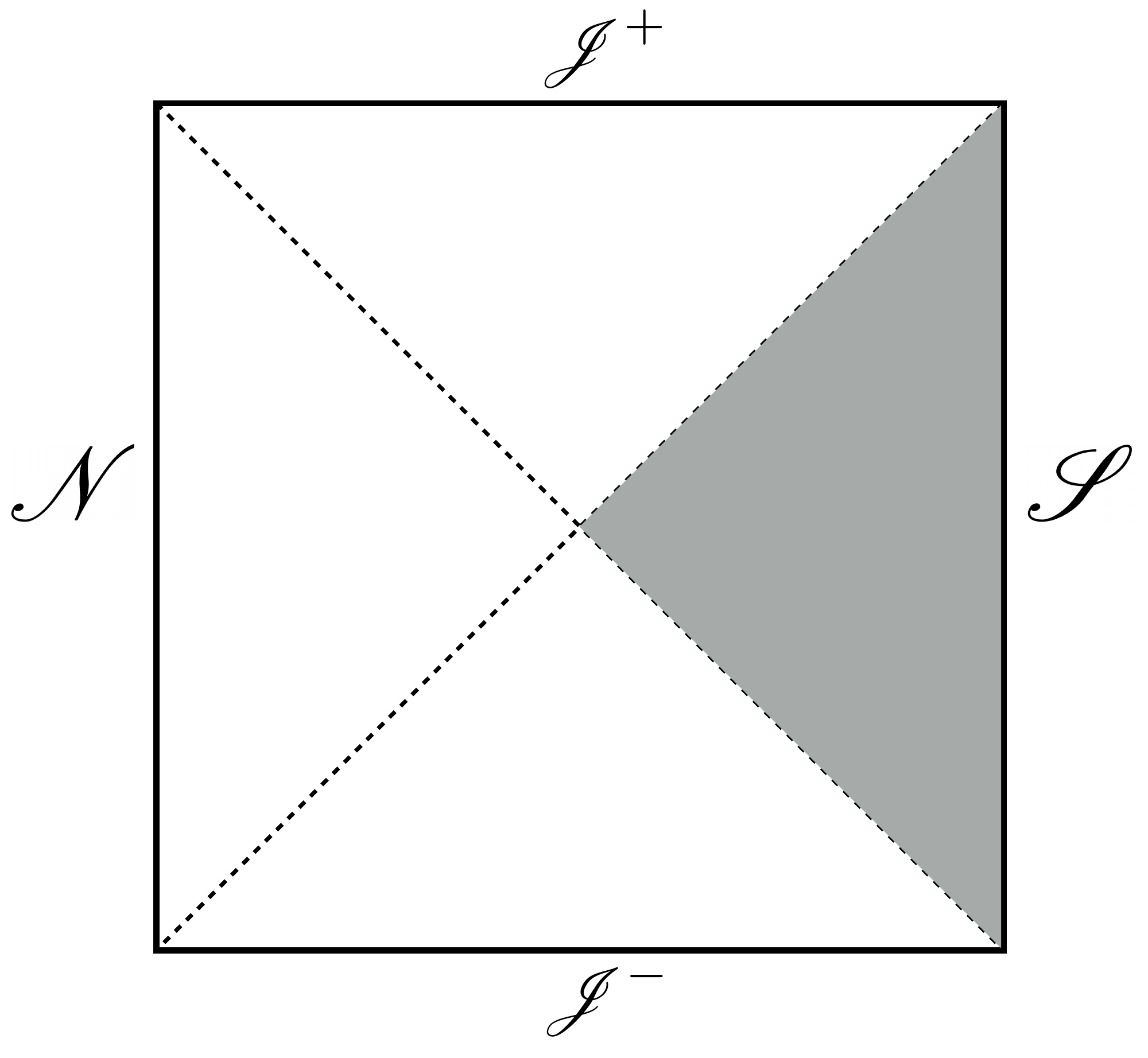}
\caption{\small   The Penrose diagram of $d$-dimensional de Sitter space for $d>2$. Spatial sections are $(d-1)$-spheres, so  points in the diagram represent   ($d-2$)-spheres, whose radius goes to zero at the northern and southern poles of the sections, denoted by $\mathscr{N}$ and $\mathscr{S}$ respectively. The causal diamond of a static observer sitting at one of the poles   is called the  \emph{static patch} (the southern   patch is shaded). The dotted lines    indicate the future and past horizons of   static observers at the poles.}
	\label{fig:penroseds}
\end{figure}

 In global coordinates the metric of dS space is
 \beq
 ds^2 = - dt^2 + L^2 \cosh^2 ({t/L}) d \Omega^{2}_{d-1} \, . 
 \eeq
 Note that equal time sections of this metric are $(d-1)$-spheres. The spatial sections of dS space have no asymptotic boundary, since they are compact.  The   topology of dS is   $\mathbb R \times S^{d-1}$, and its isometry group is $O(1,d)$.  The scale factor $L \cosh (t/L)$ is exponentially decreasing for $t <0$, reaches the minimum value $L$ at $t=0$, and is   exponentially increasing for $t>0$.   (See Figure  \ref{fig:penroseds} for the Penrose diagram of de Sitter space.) 
 
  \newpage
 In static   coordinates the metric   takes the spherically symmetric form
  \begin{equation}   \label{static1}
ds^2 = - [ 1 - (r/L)^2] dt^2 + [ 1 - (r/L)^2]^{-1}  dr^2 + r^2 d \Omega_{d-2}^2 \, ,
\end{equation}
where  $d \Omega_{d-2}^2$ is the   metric on a     unit ($d-2$)-sphere.  An observer sitting at $r=0$ is in causal contact with  phenomena in the region $r \le L$, and thus sees a cosmological horizon   at  $r=L$. In   global de Sitter space the horizon is located at the equator of the spacelike $(d-1)$-sphere.

 The observer-dependent horizon appears due to the rapid expansion of the de Sitter universe, so  that light emitted from behind the horizon can never reach the observer. The region $0 \le r \le L$     covers the    causal diamond of an observer 
whose worldline remains inside this region,
and is called the \emph{causal patch} or \emph{static patch} of dS space.  Notice that the metric \eqref{static1} has  a manifest timelike isometry with   Killing vector $\xi = \partial_t$, and the full symmetry group of the static patch is $\mathbb R \times  O(d-1)$, where $O(d-1)$ corresponds to the rotations that leave the horizon invariant. The rest of this thesis focuses on  the dS static patch --- and other maximally symmetric causal diamonds --- since it has thermodynamic properties, as we will see in the next subsection.\footnote{For   extensive reviews of de Sitter space, see \cite{Klemm:2004mb,Spradlin:2001pw,Anninos:2012qw,Balasubramanian:2001rb}.} 



 \subsection{De Sitter entropy and temperature}

 \subsubsection{Static patch of de Sitter space}
 Gibbons and Hawking   derived a  first law for the de Sitter static patch  \cite{Gibbons:1977mu} 
 \beq \label{firstlawdS1}
 \int_\S \delta {T_a}^{b}  \xi^a u_b dV= -\frac{ \kappa_{\rm c}}{8\pi G} \delta A_{\rm c} \, ,
 \eeq
 where 
 $\S$ is a spatial section   of the static patch, whose boundary is the bifurcation surface of the cosmological horizon, and    $u^b$ is its future pointing unit normal and $dV$ the proper volume element.   
Further,  $A_{\rm c}$ is the area of the bifurcation surface and $\kappa_{\rm c}$ is the surface gravity associated to the horizon, defined by the equation $\xi^a \nabla_a \xi^b = \kappa_{\rm c} \xi^b$,  evaluated on the  future cosmological horizon. Note there is no ADM mass term as in \eqref{firstlawBCH}, since spatial sections of dS space have no asymptotic boundary. 
   If the timelike Killing vector $\xi$ is normalized as $\xi^2 = -1$ at the origin of the static patch, i.e. $\xi = \partial_t$, then the surface gravity is given by $\kappa_{\rm c} = 1/L$.  In the case of a  point particle sitting at the origin of the static patch, the stress-energy tensor is $T_{ab}  = M u_a u_b \delta^{d-1} (r)$ and   the first law takes the form
 \beq \label{firstlawdS2}
 \delta A_{\rm c} = - 8 \pi G M L \, , 
 \eeq
 where we inserted $\kappa_{\rm c} =1/ L$. Thus, increasing the stress-energy inside the de Sitter static patch, decreases the cosmological horizon area.\footnote{We have kept  the cosmological constant     fixed in the first law \eqref{firstlawdS1}. A variation of $\L$ is included in the first law for the de Sitter static patch in Section \ref{sec:desitter}.}
 
 Gibbons and Hawking also computed the temperature as seen by an inertial observer moving along \emph{any} timelike geodesic in de Sitter space\footnote{For accelerating  observers that move on wordlines of  constant $r$,  along the flow of the timelike Killing vector $\xi$, the observed temperature is the Tolman temperature, i.e. the Gibbons-Hawking temperature    divided by the   norm of $\xi$. This can be   understood as  a combination of the Unruh effect (acceleration term) and    cosmological expansion effect (surface gravity term)  \cite{Narnhofer:1996zk,Deser:1998xb,Casadio:2010vq}
  $$ 
  T_{\rm{Tol}} = \frac{T_{\rm{GH}}}{\sqrt{- \xi \cdot \xi}} =  \frac{\hbar \sqrt{a^2 + \kappa_{\rm c}^2}}{2\pi}  \qquad \text{with}  \qquad a = \frac{   r/L^2}{\sqrt{1- r^2 /L^2}}  \, .
  $$
  Here $a = \sqrt{a^b a_b}$ is the proper acceleration of the observer. The acceleration 4-vector is defined as $a^b = u^c \nabla_c u^b$ and the velocity 4-vector with unit norm is $u^b = \xi^b / \sqrt{- \xi \cdot \xi}$. Note that $T=T_{\rm GH}$ for inertial observers at the poles, since $a=0$ at $r=0$.} 
 \beq \label{dStemp}
 T_{\rm{GH}} = \frac{\hbar  \kappa_{\rm c}}{2\pi} \, .
  \eeq
  By combining  this result with the first law \eqref{firstlawdS1}, they concluded that the entropy associated to the cosmological horizon is  proportional to the horizon area
  \beq \label{dSentropy}
  S_{\rm{GH}}  = \frac{A_{\rm c}}{4 G \hbar} \, .
  \eeq
This entropy can be interpreted as the ``lack of information of the observer about the regions which he cannot see'' \cite{Gibbons:1977mu}. The same   relations for the entropy and temperature appear in the context of black hole horizons, and therefore the close relationship between event horizons and thermodynamics seems to be universal. Unlike black hole horizons,  however, the cosmological event horizon is observer-dependent, and thus the particle production and horizon entropy of the static patch also depend on the observer. This makes the microscopic meaning of the cosmological horizon entropy  conceptually harder to grasp than that of black hole entropy.  
However, just like the entropy of specific black holes has been   computed in string theory \cite{Strominger:1996sh}, a sensible theory of quantum gravity should also give a microscopic derivation of the de Sitter entropy.

In our universe the current  horizon radius is related to the present Hubble constant, $L = c / H_0 \sim 10^{26}\,\text{m}$, and the   de Sitter  temperature and entropy are approximately equal to
\beq
T_{\rm GH} \sim 10^{-30} \, \text{K} \qquad \text{and} \qquad S_{\rm GH} \sim k_{\rm B} \cdot 10^{120} \, . 
\eeq
This temperature cannot be directly measured,  since it is much lower than the temperature of the cosmic microwave background (2.7 K). However, the existence of a temperature and entropy is  conceptually important because it shows that  the observable universe as a whole behaves thermodynamically.
  
 Another important comment concerns the minus sign  in the first law \eqref{firstlawdS1}. Due to this minus sign the entropy of the horizon decreases if the Killing energy increases. Hence the entropy of the de Sitter static patch  is   \emph{maximal}. In the next part, we will see that by putting a black hole inside de Sitter space one  still cannot increase the entropy of the static patch. 
 
 
The minus sign   raises the question how the first law for de Sitter spacetime can be interpreted as a ``first law'' of thermodynamics: $d E = T dS$.  Let us mention two interpretations of the minus sign, suggested in the literature. First, Spradlin, Strominger and Volovich \cite{Spradlin:2001pw} suggested that the minus sign results from varying with respect to the energy inside the horizon, rather than the energy outside the horizon, which is  the negative of the former: $d E = - \int_\S \delta {T_a}^{b} \xi^a d \S^b$. The Killing energy flux $T_{ab} \xi^a d\Sigma^b$ on the other side of the horizon, i.e. in the northern static patch, is negative since the Killing vector $\xi^a$  is past pointed outside the horizon. In fact, since global spatial section of dS are closed, the Killing energy outside   is exactly equal to the negative of the Killing energy inside the horizon. Therefore, according to Spradlin et al.  the first law should be defined with respect to the energy behind the horizon, just like the entropy is a measure for the number of degrees of freedom behind the horizon.

Second, Klemm and Vanzo \cite{Klemm:2004mb} argued   that the minus sign has to be included in the temperature rather than in the energy variation. Their proposal is that the de Sitter space static patch has a negative (absolute) temperature: $T = - T_{\rm GH}$. The motivation behind this proposal is that the entropy of the static patch decreases as the energy inside the same patch  increases, and hence the thermodynamic (or Gibbsian) temperature, defined by $1/T\equiv \partial S / \partial E$, is negative. Moreover, for thermodynamic systems with negative temperature there is an upper bound on the available energy per state and an upper bound on the entropy of the system. Both requirements are realized in de Sitter space, since the   dS 
entropy is finite and the   energy is bounded from above by the mass of the largest black hole that fits inside the cosmological horizon (called the Nariai black hole).   In this thesis we will argue that the negative temperature interpretation is   more sensible, especially when applied to causal diamonds of all sizes in maximally symmetric spaces.


  Another important feature  of de Sitter thermodynamics is that the specific heat 
   of the static patch is  {negative}, unlike   standard thermodynamic systems for which the specific heat is positive. The specific heat can be easily computed\footnote{Gibbons and Hawking wrote in their 1977 article on cosmological event horizons: ``Unlike the black-hole case, the surface gravity of the cosmological horizon decreases as the horizon shrinks" (p. 2751 \cite{Gibbons:1977mu}). This is wrong, since   $\kappa_{\rm c} =1/L$ increases as the de Sitter   radius $L$ decreases.}
\beq
  C = T_{\rm GH} \frac{dS_{\rm GH}}{dT_{\rm GH}} = -(d-2) \frac{A(L)}{4G \hbar}<0  \, ,
\eeq
  where we have inserted $S_{\rm GH}= A(L)/4G\hbar$ and $T_{\rm GH} = \hbar/2\pi L$. Note that the specific heat is independent of the sign of the temperature, so it is also negative if de Sitter space has   negative temperature. The horizon thermodynamics of the dS static patch is in that sense similar to that of an asymptotically flat black hole, since the latter also has negative specific heat. 
   In this thesis we will give an interpretation of this peculiar feature of horizon thermodynamics in terms of long strings. 
  
Thus, we have encountered   three important  research questions  which will be covered in this thesis:
\begin{enumerate} 
\item[(a)] What is the microscopic origin of   de Sitter entropy? \\See Section \ref{subsec:dS} for a long string interpretation.

\item[(b)] What is the physical interpretation of the minus sign in the first law for de Sitter spacetime (and, in fact, for any causal diamond in a maximally symmetric spacetime)? \\See  Section \ref{negativetemp} for the negative temperature interpretation.

\item[(c)] Why is the specific heat of the cosmological event horizon (and of the event horizon of a Schwarzschild black hole) negative?  \\See Section \ref{sec:bhentropy} for a long string interpretation.
\end{enumerate}

  
 \subsubsection{Schwarzschild-de Sitter space}
 
  The metric of a non-rotating black hole in asymptotically de Sitter space is  in static coordinates  \cite{Carter:1973,Gibbons:1977mu}\footnote{Schwarzschild-de Sitter space is a solution to \eqref{fieldeqcosm} with cosmological constant given in terms of $L$ by \eqref{dSradius}, but the space is not maximally symmetric.}
 \begin{align} \label{coordSdS}
& ds^2  = - f(r)dt^2 +   f(r)^{-1} dr^2 + r^2 d \Omega_{d-2}^2\, ,\nonumber \\
 &\text{with} \,\,\,  f(r) = 1 - \frac{r^2}{L^2}   - \frac{16 \pi G M }{(d-2) \Omega_{d-2} r^{d-3}} \, .
 \end{align}
 Here $M$ is the mass parameter and $\Omega_{d-2} = 2 \pi^{(d-1)/2} / \Gamma [(d-1)/2] $ is the volume of a unit $(d-2)$-sphere. For $d>3$ and $0 < M < M_{\rm{N}}$ the function  $f(r)$ 
  has
 two positive roots at $r= r_{\rm h}$ and $r=r_{\rm c}$. These values are the locations of the two event horizons for observers moving on world lines of constant $r$ between $r_{\rm h}$ and $r_{\rm c}$:   the black hole horizon at $r_{\rm h}$ and the cosmological horizon at $r_{\rm c}$  (with $r_{\rm c} \ge r_{\rm h}$).  
 The lower bound on the mass, $M=0$, corresponds  to the static patch of de Sitter space, for which $r_{\rm h}=0$ and $r_{\rm c} = L$. The upper bound  $M=M_{\rm{N}}$ corresponds to the case where the two horizons coincide, i.e. $r_{\rm h} = r_{\rm c} =r_{\rm{N}}$ (known as the \emph{Nariai solution} \cite{Nariai}). The horizon radius and mass of the Nariai solution can be computed by setting $f(r_{\rm{N}} )=0$ and $f'(r_{\rm{N}} )=0$:
 \beq \label{Nariai}
   r_{\rm{N}} = \sqrt{\frac{d-3}{d-1}} \,  L \qquad \text{and} \qquad M_{\rm{N}}  =\frac{d-2}{d-1} \frac{\Omega_{d-2}}{8 \pi G}  r_{\rm{N}}^{d-3}  \, . 
  \eeq
  For masses above the Nariai bound, $M > M_{\rm{N}}$, there is a  naked singularity.  Therefore, the largest physical black hole that fits inside the cosmological horizon is given by the Nariai black hole. 
  
  The Nariai solution can be obtained by defining the   coordinates $\tau = \epsilon \,  t$, $\rho = ( r- r_{\rm h})/\epsilon$ and taking the near-horizon limit $\epsilon \rightarrow 0, r_{\rm c} \rightarrow r_{\rm h}$, while keeping the ratio  $ \beta \equiv (r_{\rm c} - r_{\rm h})/\epsilon$ fixed
  \beq
  ds^2 = - \frac{\rho (\beta - \rho)}{\tilde L^2}  d \tau^2 + \frac{\tilde L^2 d \rho^2}{  \rho (\beta - \rho)} + (d-3) \tilde L^2 d \Omega_{d-2}^2 \, , 
  \eeq
 where $\tilde L = L / \sqrt{d-1}$ and $d>3$.    By changing coordinates $\tilde \tau =  \frac{\beta}{2 \tilde L} \tau, \tilde \rho =  \frac{ 2\tilde L }{\beta}   \rho- \tilde L$, the Nariai metric turns into that of $dS_2 \times S^{d-2}$:
  $$
  ds^2 =  - \left (1 -  \tilde \rho^2/\tilde L^2  \right) d \tilde \tau^2 + \left (1 - \tilde \rho^2 / \tilde L^2 \right)^{-1} \!\! d \tilde \rho^2 + (d-3) \tilde L^2 d \Omega_{d-2}^2 \, . 
  $$ 
 Note that the curvature radii of de Sitter and the sphere are not the same, for $d \neq 4$, and hence this is not  isomorphic to $d$-dimensional de Sitter space. Although the original coordinate distance  $r$ between the black hole horizon and the cosmological horizon is zero for the Nariai solution, the geodesic distance between the two horizons is finite and equal to $\pi \tilde L$. 
 
\begin{figure}
	\centering
	\includegraphics
		[width=.9\textwidth]
		{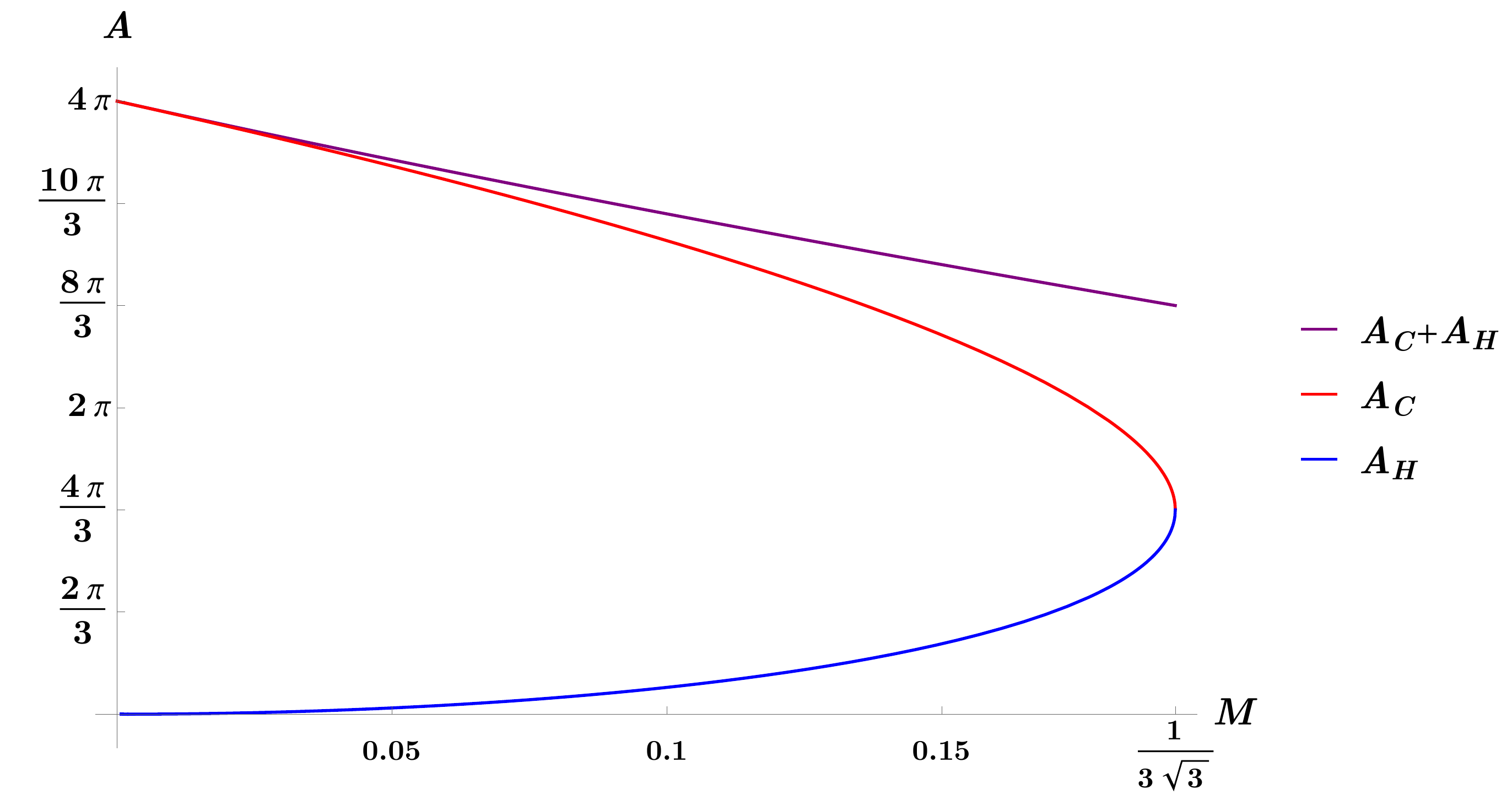}
\caption{\small   The cosmological and black hole horizon area, $A_{\rm c}$ and $A_{\rm h}$, as a function of the mass $M$ in Schwarzschild-de Sitter space. The plot is for     $d=4$ and we have set $L = G =1.$ The sum of the horizon areas, $A_{\rm c} + A_{\rm h}$, decreases monotonically with increasing $M$, and is hence always less than the cosmological horizon area of pure dS space ($A_{\rm c}=4\pi, M=0$). The  cosmological and black hole horizon  coincide for the Nariai solution ($M = 1/3\sqrt{3}$) and their area is in that case given by  $A_{\rm c} = A_{\rm h}= 4\pi/3.$}
	\label{fig:areasds}
\end{figure}

Let us return to the generic Schwarzschild-de Sitter case. As  $M$   increases from zero to $M_{\rm N}$, the black hole horizon increases monotonically and the cosmological horizon decreases monotonically. This can be seen by expressing the de Sitter  radius   and  the mass parameter    in terms of the 
 horizon radii (which satisfy $f(r_{\rm c}) = f(r_{\rm h}) =0$)
  \beq 
  L^2 = \frac{ r_{\rm h}^d r_{\rm c} - r_{\rm h} r_{\rm c}^d}{r_{\rm h}^{d-2} r_{\rm c} -r_{\rm h}  r_{\rm c}^{d-2}} \qquad \text{and} \qquad \frac{16 \pi G M}{(d-2) \Omega_{d-2}} = \frac{ r_{\rm h}^2 - r_{\rm c}^2 }{r_{\rm h}^{2} r_{\rm c}^{3-d} - r_{\rm h}^{3-d} r_{\rm c}^2} \, . 
  \eeq
 Expanding these expressions around $r_{\rm h} =0$,  we find to leading order away from pure de Sitter space
    \beq \label{seriesLandM}
L^2 = r_{\rm c}^2 \left ( 1+ \frac{r_{\rm h}^{d-3}}{r_{\rm c}^{d-3}}+ \dots \right) \qquad \text{and} \qquad      \frac{16 \pi G M}{(d-2) \Omega_{d-2}} =  r_{\rm h}^{d-3} + \dots \, . 
  \eeq
  The black hole horizon radius thus increases with $M$ (at least perturbatively around pure dS), and the cosmological horizon radius of  Schwarzschild-de Sitter spacetime is \emph{smaller} than the   horizon radius of the dS static patch, by an amount 
   \beq
 \delta r_{\rm c} =  r_{\rm c} - L = - \frac{8 \pi G M}{(d-2) \Omega_{d-2} L^{d-2}} + \dots .
 \eeq
 Amusingly, the infinitesimal horizon change can   be expressed in terms of Newton's potential: $\delta r_{\rm c} = \Phi (L) L$ \cite{Verlinde:2016toy}. 
From the first equation in \eqref{seriesLandM} we can  now  compute the cosmological horizon area perturbatively around the de Sitter static patch 
    \beq \label{cosmoareaSdS}
 A(r_{\rm c}) = A(L) -  A(r_{\rm h}) \frac{d-2}{2}  \frac{r_{\rm c}}{r_{\rm h}} + \dots\, . 
  \eeq
  By also using the second equation in \eqref{seriesLandM} we recover  the first law  \eqref{firstlawdS2} of   de Sitter space, $\delta A_{\rm c} =  - 8 \pi G M L$, where $M$ is the mass of the black hole.
Since $r_{\rm c} \ge r_{\rm h}$  equation \eqref{cosmoareaSdS} implies that the sum of the black hole and cosmological horizon areas is less than the area of the pure dS cosmological horizon  \cite{Bousso:2000nf,Bousso:2000md}\footnote{In \cite{Maeda:1997fh} an even stronger  inequality was   proven for any asymptotically de Sitter spacetime  containing a black hole: $A_{\rm h} + A_{\rm c} + \sqrt{A_{\rm h}   A_{\rm c}} \le A(L)$. }
  \beq \label{dSareabound}
A_{\rm h} + A_{\rm c}  \le A(L) \, . 
  \eeq
\noindent This holds for any physical value of $M$, and the inequality is saturated for $r_{\rm h}=0$ (see Figure \ref{fig:areasds}).      By the Bekenstein-Hawking area-entropy relation, we see that the total entropy of Schwarzschild-de Sitter space is hence  less than that of pure de Sitter. In other words, one cannot increase the entropy of the dS static patch by putting a black hole inside, since the sum of the horizon areas $A_{\rm h} + A_{\rm c}$ decreases. It is also impossible to increase the de Sitter entropy by putting matter inside, since matter always has less entropy than a black hole for the same mass. 
Thus, the entropy of the de Sitter vacuum is  \emph{maximal}, as we   anticipated from the first law for the static patch.\footnote{The minimum amount of entropy in a de Sitter universe with a cosmological horizon is given by the entropy of the Nariai solution:
$
S_{\rm N} =2 \frac{A_{\rm N}}{4 G \hbar} = 2 \left ( \frac{d-3}{d-1} \right)^{(d-2)/2} \!\! S_{\rm dS}.
$
This is still a significant fraction of the   pure de Sitter entropy $S_{\rm dS}= \frac{A(L)}{4 G \hbar}$. The fraction is 2/3 for $d=4$ and the fraction attains its maximum for an infinite number of   dimensions, i.e.   $2/e \approx 0.736$, so it is bounded.
}

\subsubsection{Entropy bounds for cosmological spacetimes}
There is a more general argument why the total entropy in a de Sitter universe cannot exceed the de Sitter entropy \cite{Banks:2000fe,Bousso:2000nf}.
This argument holds for   spacetimes that approach de Sitter space in the asymptotic future, and follows from applying the generalized second law  to cosmological horizons. It is based on the following thermodynamic process.
 Consider a matter system   in a 
  universe that is asymptotically de Sitter in the future.  The matter system is surrounded by the cosmological horizon of an observer, who moves relative to the matter system into the asymptotic de Sitter region.  For the observer, the matter   falls across the cosmological horizon, and hence the matter entropy $S_{\rm m}$ is lost for the observer. However, when  matter crosses the horizon, the area $A_{\rm c}$ of the cosmological horizon increases. According to the generalized second law \cite{Bekenstein:1972tm,Bekenstein:1973ur,Bekenstein:1974ax}, the Gibbons-Hawking entropy $A_{\rm c}/4 G \hbar$ associated to   the cosmological horizon grows at least enough to compensate for the lost matter entropy.
 To be specific, 
the entropy of the initial state is the sum of the entropy of the cosmological horizon  and the matter entropy (which may include black hole entropy) 
  \beq \label{gendSinitial}
S_{\rm gen} = S_{\rm GH}  + S_{\rm m}  \, . 
\eeq
  The entropy of the final state of the observable universe is that of empty de Sitter space
  \beq \label{dSfinal}
  S_{\rm dS} = \frac{A_{\rm dS}}{4 G \hbar } = \frac{\Omega_{d-2}}{4 G \hbar} \left ( \frac{(d-1)(d-2)}{2 \L}\right)^{\frac{d-2}{2}}  \!\! . 
  \eeq
 The    generalized second law of thermodynamics states that the initial (generalized)  entropy \eqref{gendSinitial} cannot exceed the entropy of the final state \eqref{dSfinal}
\beq
S_{\rm gen} \le S_{\rm dS} \, . 
 \eeq
Using   \eqref{dSentropy} and \eqref{dSfinal}, this puts a non-trivial bound on the  matter entropy in terms of  the change of the horizon area 
 \beq
S_{\rm m} \le \frac{1}{4 G \hbar} \left ( A_{\rm{dS}} - A_{\rm c} \right).
\eeq
 Bousso called this the \emph{D-bound}, where `D' stands for the difference between the two horizon areas \cite{Bousso:2000nf,Bousso:2000md}.  Since $S_{\rm m} \ge 0$, the D-bound implies in particular that $A_{\rm{dS}} \ge A_{\rm c}$, i.e. the horizon   of empty de Sitter space must have a  larger area than a cosmological horizon that encloses matter (which is indeed the case for Schwarzschild-de Sitter space, see eq. \eqref{cosmoareaSdS}).\footnote{For Schwarzschild-de Sitter space the D-bound evaluates to $S_{\rm m} \le \frac{d-2}{8 G \hbar} A(r_{\rm h})  \frac{r_{\rm c}}{r_{\rm h}}$ in the regime $r_{\rm h} \ll r_{\rm c}$. This bound is   satisfied if we take the matter entropy to be equal to the Bekenstein-Hawking entropy associated to the black hole, i.e. $S_{\rm m} = A(r_{\rm h})/4 G\hbar.$}

In an even more general setting, for any universe with a positive cosmological constant $\L$ --- including those that do not evolve to de Sitter space in the future --- Bousso conjectured that the observable entropy is bounded by the de Sitter entropy 
$
S_{\rm gen} \le S_{\rm dS} 
$
\cite{Bousso:2000nf}.
Here, $S_{\rm gen}$ denotes the   matter and horizon entropy associated to a causal diamond.  This is called the \emph{$N$-bound}, because   de Sitter entropy can be interpreted as the total number of degrees of freedom in the observable universe, also known as the   $\L-N$ correspondence: 
$
N = S_{\rm dS} = 3 \pi / \L G \hbar \, 
$
for $d=4$  \cite{Banks:2000fe}. The $N$-bound was proven by Bousso \cite{Bousso:2000nf} by applying the covariant entropy bound and the D-bound, under the assumption of spherical symmetry. 

To conclude, entropy bounds suggest that the entropy of a causal diamond in a cosmological spacetime cannot be greater than the entropy of   de Sitter space.

 \section{Einstein equation from   entanglement equilibrium}
\label{intro:EE5}

The fact that the de Sitter entropy   is the maximal entropy implies that to first order its variation vanishes in the vacuum.  We will show more explicitly in Section \ref{sec:statgenent} that Bekenstein's \emph{generalized entropy} \cite{Bekenstein:1973ur}, defined as the sum of the Bekenstein-Hawking entropy and the matter entropy,   is stationary in the de Sitter vacuum:
\beq \label{statgenprinciple}
\delta S_{\rm{gen}} =0 \,   \qquad \text{with} \qquad  S_{\rm{gen}} :=  S_{\rm BH}   +  S_{\rm   m} \,.
\eeq
This turns out to be a   general principle of quantum gravity.   Note that the static patch of de Sitter space is the causal diamond for any observer whose worldline starts and ends at the southern pole at $\mathscr{J}^-$ and $\mathscr{J}^+$, respectively (see Figure \ref{fig:penroseds} on p. \pageref{fig:penroseds}). Hence one might wonder whether this principle also applies to other causal diamonds in   de Sitter space, or to causal diamonds in general. In Chapter~\ref{ch2}  we   show that the  generalized entropy   is stationary for  causal diamonds of \emph{any size} in a maximally symmetric spacetime (i.e.   in (Anti-)de Sitter space and Minkowski space).\footnote{For  finite causal diamonds  in de Sitter space  the generalized entropy is only stationary when the volume of the maximal slice of the diamond is kept fixed.
} Below, in Section \ref{sec:areavolume} and Section \ref{EEthesis}, we derive  equation \eqref{statgenprinciple}, as a warmup exercise,   for small diamonds in Minkowski spacetime from the semiclassical Einstein equation. 

Moreover, it was proposed by Jacobson \cite{Jacobson:2015hqa} that  the semiclassical Einstein equation can be derived from the principle \eqref{statgenprinciple} (thus establishing an \emph{equivalence} between the entropy   principle and the semiclassical Einstein equation). 
He identified the generalized entropy with   entanglement entropy --- see Section \ref{EEthesis} for the motivation behind this identification --- and conjectured that  the total entanglement entropy  in small balls is \emph{maximized} at fixed volume in the vacuum state, for a simultaneous variation of the geometry and quantum fields away from the vacuum. 
This conjecture is called the ``maximal vacuum entanglement hypothesis'', or   \emph{entanglement equilibrium} hypothesis   for short, since the vacuum is an equilibrium state if its entropy is maximized \cite{Jacobson:2015hqa}. It implies that  the total entanglement entropy vanishes for   first-order variations, and   is negative for finite variations of the geometry and the quantum fields (in the entire dissertation we   focus   on first-order variations, but see \cite{Jacobson:2017hks} for an analysis of second-order variations). 
 Jacobson   showed that  the semiclassical Einstein equation is equivalent to the stationarity of  the total entanglement entropy  at fixed volume in small local diamonds everywhere in spacetime.   We start this section by deriving \eqref{statgenprinciple} for de Sitter space from the quantum corrected first law of the cosmological horizon, and then  review the link    between the semiclassical Einstein equation  and    entanglement equilibrium.



\subsection{Stationarity of generalized entropy}
\label{sec:statgenent}

Gibbons and Hawking  \cite{Gibbons:1977mu} 
 showed that the de Sitter vacuum is a thermal state with respect to   the Hamiltonian that generates  time translations in the static patch. The reduced density matrix of the vacuum restricted to a   slice $\S$ of the static patch is 
\beq
\rho_{\rm vac} =\frac{1}{Z} e^{- K / T_{\rm GH}} ,
\eeq
where $K$ an operator called the ``modular Hamiltonian'', and  the Gibbons-Hawking temperature is given by \eqref{dStemp}:  $ T_{\rm GH} = \hbar \kappa_{\rm c} / 2\pi$.
This is the de Sitter analog of the Davies-Unruh effect in the Rindler wedge of Minkowski spacetime \cite{Davies:1974th,Unruh:1976db}.
For the static patch, the modular Hamiltonian $K$ is equal to the Hamiltonian $H^{\rm m}_\xi$ generating time translations on the static patch  
\beq
 K =  H^{\rm m}_\xi := \int_\S    {T_{a}}^b    \xi^a u_b dV  \, . 
\eeq
For infinitesimal variations of the density matrix, the variation of the vacuum expectation value of the Hamiltonian is related  to the variation of the matter entropy $S_{\rm m}$ by
\beq \label{firstlawdSmatter}
\delta   \langle K    \rangle = T_{\rm GH} \delta S_{\rm   m} \,, 
\eeq
where $\delta  \langle K    \rangle = \text{Tr}\, (K \delta \rho_{\rm vac})$, and the matter entropy is defined as the  von Neumann entropy associated to the thermal density matrix: $S_{\rm   m}  = - \text{Tr}\, ( \rho_{\rm vac} \log \rho_{\rm vac}).$ This is   the usual Clausius relation for a  thermal  state.  If the variation is to a global pure state, the matter entropy variation is purely entanglement entropy, which is why this is sometimes known as the ``first law of entanglement''.

We will now derive the stationarity of generalized entropy \eqref{statgenprinciple} from the ``quantum corrected'' first law of de Sitter space.  For quantum matter fields on a classical background spacetime  the stress-energy tensor in the first law \eqref{firstlawdS1}   is replaced by its expectation value 
 \beq
 \int_\S \delta \langle {T_{a}}^b \rangle \xi^a u_b dV = - \frac{\hbar \kappa_{\rm c}}{2\pi  }  \frac{\delta A_{\rm c}}{4 G \hbar} \, .
 \eeq
 The left-hand side of this equation can be expressed in terms of the matter entropy variation through the Clausius relation \eqref{firstlawdSmatter} for the quantum matter. The right-hand side is equal to (minus) the Gibbons-Hawking temperature  \eqref{dStemp} times the variation of the Gibbons-Hawking entropy \eqref{dSentropy}, so that the first law becomes
\beq
T_{\rm GH} \delta S_{\rm   m} = - T_{\rm GH}  \delta S_{\rm   GH} \, .
\eeq
Recall that $S_{\rm   GH} = A_{\rm c} / 4 G \hbar$ is the usual entropy-area relation for the cosmological horizon, originally proposed by Bekenstein and Hawking for black hole horizons. By putting both entropy variations on one side of the equation, we find that   the matter entropy adds to the de Sitter horizon entropy, forming the statement that Bekenstein's generalized entropy is stationary
\beq
 T_{\rm GH}  \delta S_{\rm{gen}}  =  T_{\rm GH}  \delta (S_{\rm{GH}} +  S_{\rm   m} ) =0 \,  .
\eeq
 The generalized entropy was   introduced by Bekenstein \cite{Bekenstein:1972tm,Bekenstein:1973ur,Bekenstein:1974ax}, in the context of black holes in order to restore the second law of thermodynamics. When throwing some thermal matter into a black hole, the matter entropy is lost, but the horizon area of the black hole increases, in such a way that the generalized second law is satisfied, i.e. $ S_{\rm BH}^{\rm old} + S_{\rm m}  \le S^{\rm new}_{\rm BH}$, where $S_{\rm BH}$ is the Bekenstein-Hawking entropy. The generalized second law thus states that the  generalized entropy defined in \eqref{statgenprinciple}  never decreases: $\Delta S_{\rm gen} \ge 0$, see also equation \eqref{gensecondlaw4}.   
 


\subsection{Area and volume variations in Einstein gravity}
\label{sec:areavolume}

The Einstein equation, $G_{ab} = 8\pi G T_{ab}$, relates the curvature of spacetime to the stress-energy of matter fields. An intuitive way of describing this is by comparing the     area and volume of spatial ball-shaped regions in curved spacetime (i.e. in the presence of   stress-energy) with those in flat spacetime. The curvature causes an \emph{excess} in the proper volume of a ball, when the area is kept fixed, whereas the surface area of a ball has a \emph{deficit} in curved spacetime relative to flat spacetime (both when the volume and the geodesic length are kept fixed).   We will verify these statements --- which date back to work by Pauli, Riemann and Feynman \cite{Pauli,Feynman:1996kb,Feynman:1963uxa} --- in this subsection for small geodesic balls. We will also show that the Einstein equation is equivalent to the statement that the surface area deficit of small balls  is proportional to the energy density at the center of the ball.  In this subsection we  will closely follow  Section  II and Appendix A of   \cite{Jacobson:2015hqa}. 

The basic idea is to use a Riemann normal coordinate expansion of the metric around any point $O$ in spacetime. A geodesic ball-shaped region $\S$ centered at $O$  is constructed by generating a congruence of spacelike geodesics through $O$, which are orthogonal to a   timelike unit vector $u^\mu$.  The size of the ball is set by the geodesic radius $\ell$, which is taken to be much smaller than the local curvature radius, i.e. $\ell \ll L_{\text{curvature}}$. The Riemann normal coordinates (RNC)  with origin $O$ are defined by: $x^\mu = r n^\mu$, where $r$ is the geodesic radius and $n^\mu$ is a unit tangent vector to a geodesic starting at $O$. The ball $\S$ lies by definition within the $x^0=0$ surface, so that $n^\mu = n^i \delta^\mu_i$ where the index $i$ runs only over the spatial coordinates. 
In RNC coordinates the metric at $O$ is the $d$-dimensional Minkowski metric. In a small neighborhood of $O$ the metric can be expanded      in powers of the curvature (and its derivatives) at $O$. 
In particular, the induced metric on $\S$ in RNC coordinates  is  
\beq
h_{ij} = \delta_{ij} - \frac{1}{3} r^2 R_{ikjl} n^k n^l + \mathcal{O}(r^3),
\eeq
where $R_{ikjl}$ are the spatial components of the spacetime Riemann tensor evaluated at $r=0$. When the  spatial  metric is expressed as a sum of the flat metric and a perturbation, $h_{ij} = \delta_{ij} + \gamma_{ij}$, then   to first order in the perturbation the volume density is
\beq
\sqrt{h} = \sqrt{\delta} \left ( 1 + \frac{1}{2} \delta^{ij} \gamma_{ij} \right) = \left ( 1- \frac{1}{6} r^2 {{R_{ik}}^i}_l n^k n^l + \mathcal{O}(r^3)\right) r^{d-2} \, . 
\eeq
The proper volume of the ball is the integral of the volume density over the geodesic radius and spherical coordinates, which to lowest nontrivial order in $\ell/L_{\text{curvature}}$ is given by:
\beq
V = \int  d\Omega_{d-2} \int_0^\ell \sqrt{h} dr = \left ( 1 - \frac{\ell^2 }{6 (d+1)} \mathcal{R}\right) \frac{\Omega_{d-2} \ell^{d-2}}{d-1} \,.
\eeq
Here  $\mathcal R = {R_{ij}}^{ij}$  is   evaluated at $O$ and, by the Gauss-Codacci equation,  it  is equal to the induced Ricci scalar on $\S$, since the extrinsic curvature of $\S$ at $O$ vanishes. Further, the integral of the area element on the unit sphere was evaluated using  $\int d \Omega_{d-2} n^k n^l = \Omega_{d-2} \delta^{kl} / (d-1)$. The volume variation at fixed geodesic length is thus
\beq  \label{VatfixedlRiemann1}
\delta V \big|_\ell = - \frac{\Omega_{d-2} \ell^{d+1}}{6 (d^2-1)} \mathcal R \, . 
\eeq
To connect  with the Einstein equation, Jacobson noted that the spatial Ricci scalar at $O$ is related to the $00$-component of the Einstein curvature tensor at $O$, as follows:
\beq \label{curvatureRiemann}
\mathcal R = {R_{ij}}^{ij} = R - 2 {R_0}^0 = 2 \left (R_{00} - \frac{1}{2} R g_{00} \right) = 2 G_{00} \ . 
\eeq
Only the  third equality makes use of RNC, since it follows from  $g_{00} (O) =-1$. Inserting  the curvature relation \eqref{curvatureRiemann} into  the volume variation \eqref{VatfixedlRiemann1} gives
\beq \label{VatfixedlRiemann2}
\delta V |_\ell = - \frac{\Omega_{d-2} \ell^{d+1}}{3 (d^2-1)}   G_{00} \, . 
\eeq
The area variation at fixed geodesic radius can be derived from this equation by taking the derivative with respect to the geodesic length:
\beq \label{AatfixedlRiemann}
\delta A \big|_\ell = \frac{d}{d\ell} \delta V \big |_\ell= - \frac{\Omega_{d-2} \ell^d}{3 (d-1)} G_{00} \, . 
\eeq
In order to derive the stationarity of generalized entropy \eqref{statgenprinciple}, we need to compute  the area variation at fixed volume, which is given by the following combination of variations (see Section \ref{sec:constrained} for an explanation)
\beq  \label{areavariationatfixedvolume10}
\delta A \big|_V = \delta A  \big|_\ell  - k\delta V \big|_\ell   \, ,
\eeq
where $k$ is the trace of the extrinsic curvature of the boundary  $\partial \S$ as embedded in $\S$. Since the background geometry is flat space, the extrinsic trace is equal to $k = (d-2)/\ell$. 
Inserting \eqref{VatfixedlRiemann2} and \eqref{AatfixedlRiemann} into   \eqref{areavariationatfixedvolume10} yields 
\beq  \label{AatfixedVRiemann}
\delta A \big|_V 
= - \frac{\Omega_{d-2} \ell^d}{  d^2 -1} G_{00} \, .
\eeq
 By inserting the $00$-component of the Einstein equation, evaluated at the center $O$ of the ball, we find that the area variation is proportional to the energy density  at $O$ (just like all   other variations in this subsection)
\beq  \label{AatfixedVRiemann2}
\delta A \big |_V = - \frac{8 \pi G  \Omega_{d-2} \ell^d }{d^2 -1} T_{00} \, .
\eeq
 This is a consequence, and nice manifestion, of the Einstein equation. If this identity holds everywhere in  spacetime   and for all timelike unit vectors, then equations \eqref{AatfixedVRiemann} and \eqref{AatfixedVRiemann2} together imply the full tensorial Einstein equation (i.e. all its components). In contrast to   Wheeler's loose (but succinct) summary of general relativity --- \emph{matter tells spacetime how to curve} \cite{Wheeler:1998vs}\footnote{We left out the first part of Wheeler's summary --- \emph{spacetime tells matter how to move} --- since we are only concerned here with the backreaction of matter fields on the metric, i.e. the Einstein equation, not  with the action of the metric field on matter, i.e. the geodesic equation.} --- this is a    precise  (but long-winded) formulation of the Einstein equation:\footnote{For a different formulation    in terms of the motion of freely falling test particles, see \cite{Baez:2001qy}.}

\begin{quote}
\emph{The Einstein equation is equivalent to the fact that the   area deficit of a small geodesic ball, compared to flat space  at fixed volume,  is proportional to the energy density at the center of the ball, if this holds at any spacetime point  and for all boosted frames.}
\end{quote}

The statement holds similarly for other variations in this setting, such as the volume excess at fixed area, or for the volume deficit at fixed geodesic length, etc.   
For future reference we also give the volume variation at fixed area, and  the variation of the geodesic length at fixed area and volume 
\begin{align}
\delta V \big |_A 
&= - \frac{\ell}{d-2} \delta A \big |_V  \label{volumeatfixedA}
= + \frac{\Omega_{d-2} \ell^{d+1}}{(d-2)(d^2-1)} G_{00}\, ,  \\
\delta \ell \big |_A &= -  \frac{\ell}{d-2} \frac{\delta A \big |_\ell}{A(\ell)} = + \frac{\ell^3}{3 (d-1)(d-2)} G_{00} \,,  \label{radiusatfixedA}\\
\delta \ell \big |_V &= - \frac{1}{A(\ell)} \delta V \big |_\ell \,\,\,=   + \frac{\ell^3}{3 (d-1)(d+1)} G_{00} \, .   \label{radiusatfixedV}
\end{align}
Note that all  these variations have an excess,  whereas the  previous variations  \eqref{VatfixedlRiemann2}, \eqref{AatfixedlRiemann} and \eqref{AatfixedVRiemann} have a deficit. 
In particular, the variation of the surface area at fixed proper volume has a deficit, while, conversely, the variation of the proper volume at fixed area has an excess.

\subsubsection{Example: constant energy distribution}

To gain more intuition for the geometric variations, we compute them explicitly  for an energy distribution with constant energy density. This can be achieved by taking the ball small enough so that the energy density inside it is effectively constant. The energy inside the  ball is   $M c^2  = T_{00} V(\ell)$, where we take the volume $V(\ell) = \Omega_{d-2} \ell^{d-1} / (d-1)$ to be the volume of a ball in flat space. The energy density $\rho = T_{00}$ is related to the $00$-component of the Einstein tensor through the Einstein equation: $G_{00} = \frac{8 \pi G}{c^4} T_{00} $, with the factors of the speed of light restored. In terms of the mass $M$ the  
area and volume variations at fixed geodesic radius are
\beq \label{areaandvolumeformass}
\delta A |_\ell = -\frac{8 \pi G M \ell}{3c^2} \qquad \text{and} \qquad \delta V |_\ell = - \frac{8\pi G M \ell^2}{3 (d+1)c^2}\, . 
\eeq
From these expressions we can compute all the other geometric variations above. In particular,  the area variation at fixed volume is
\beq
 \delta A \big |_V =  - \frac{ 8 \pi G M \ell }{(d+1)c^2}   \,, 
\eeq
and
the  variation of the geodesic radius   at fixed area is 
\beq
\delta \ell \big |_A =  \frac{8\pi G M }{3(d-2) \Omega_{d-2} \ell^{d-4} c^2} \overset{(d=4)}{=} \frac{G M }{3 c^2}  \, .
\eeq
This expression for the radius excess in $d=4$ was already known to  Feynman (see Section 11.2 of \cite{Feynman:1996kb} and Section 42-3 of \cite{Feynman:1963uxa}; and    Appendix A of  \cite{Jacobson:2015hqa}). He pointed out that the Einstein equation can be interpreted as the fact that, for all timelike directions, the radius excess of a small ball is proportional to the  mass inside it. 

Feynman   calculated that the radius excess for the Earth is approximately equal to $\delta \ell |_A = 1.5 \, \text{mm}$, and for the Sun $\delta \ell |_A =  0.5 \, \text{km}, $ assuming that the energy density is independent of the radius. Note that this radius excess is equal to the   actual radius of the ball in curved space minus the   radius computed from the actual surface area through the standard Euclidean formula, i.e. $\delta \ell |_A = \ell - \sqrt{A/4\pi}$. Similarly,   the   area deficit at fixed geodesic radius   for the Earth is roughly  $\delta A |_\ell = -237 \, \text{m}^2$, and for the Sun  $\delta A |_\ell = - 8602 \, \text{km}^2$, where we have inserted into the first equation in \eqref{areaandvolumeformass} the equatorial radius for the Earth and the radius to the photosphere for the Sun. The area deficit is equal to the difference between the actual area in curved space and the area in flat space computed from the actual  geodesic radius, i.e. $\delta A |_\ell = A - 4\pi \ell^2.$



\subsubsection{Example: area and volume in de Sitter space and flat space}
Here we check the equations \eqref{VatfixedlRiemann2}, \eqref{AatfixedlRiemann} and \eqref{AatfixedVRiemann}   for the   example where the perturbed spacetime is the de Sitter static patch.\footnote{We thank Maulik Parikh for suggesting this example.}
 Recall the metric for the dS static patch  
\beq
ds^2 = - [1- \tilde r^2/L^2] dt^2 + [1- \tilde r^2/L^2]^{-1}d\tilde r^2 + \tilde  r^2 d \Omega_{d-2}^2 \, .
\eeq
For $L \to \infty$ the metric reduces to that of Minkowski space. We will compare the area and volume of a ball of coordinate radius $R$ in flat space with the area and volume of a ball of coordinate radius $\tilde R$ in the dS static patch. In order to match with the Riemann normal coordinate results above we evaluate the area and volume differences for small balls in de Sitter space, i.e. in the limit $\tilde R \ll L$. When comparing the two geometries  we   keep the geodesic radius of the ball fixed, which in de Sitter space  is given by $\tilde \ell = L \arcsin (\tilde R/L)$. Setting this equal to the geodesic radius in flat space, $\ell = R$, yields a relation between the two coordinate radii
\beq
\tilde \ell = \ell \quad \Rightarrow \quad \tilde R = L \sin (R/L) = R - \frac{R^3}{6 L^2} + ...
\eeq
 To first order in $R/L$, the difference between the area  of the ball in dS space and in flat space, at fixed geodesic radius, is equal to
\beq \label{areadSandflat}
\Delta A \big |_\ell = A (\tilde R) - A(R)=  - \frac{(d-2) \Omega_{d-2} R^d}{6 L^2}  =  - \frac{ \Omega_{d-2} R^d}{3 (d-1)  }    \L \,.
\eeq
This is consistent with the  Riemann normal coordinate expression \eqref{AatfixedlRiemann}, since the $00$-component of the Einstein tensor evaluated at the center of the ball is: $G_{00} (\tilde r=0) = - \L g_{00} (\tilde r=0)=\L$.
Further, the proper volume of a ball in the dS static patch is
\beq
V (\tilde R) = \int_{0}^{\tilde R} \frac{\tilde r^{d-2} d\tilde r d \Omega_{d-2}}{\sqrt{1- \tilde r^2 /L^2}} = \frac{\Omega_{d-2}\tilde R^{d-1}}{d-1} \left ( 1 + \frac{(d-1)}{2 (d+1)} \frac{\tilde R^2}{L^2}  + ... \right) \,. 
\eeq
To first order in $R/L$, the difference between the volume in dS and flat space at fixed geodesic radius is
\beq  \label{volumedSandflat}
\Delta V \big |_\ell = V(\tilde R) - V(R)=-  \frac{(d-2)\Omega_{d-2} R^{d+1}}{6 (d+1) L^2} = -  \frac{\Omega_{d-2} R^{d+1}}{3 (d^2-1) }   \L \,.
\eeq
Hence,  by combining   \eqref{areadSandflat} and \eqref{volumedSandflat},  we find the area variation at fixed volume  
\beq
\Delta A \big |_V = \Delta A \big |_\ell - \frac{d-2}{R} \Delta V \big |_\ell = - \frac{\Omega_{d-2} R^{d}}{d^2-1} \L \,.
\eeq 
The latter expression can be written in terms of the stress-energy tensor of de Sitter space evaluated at the center: $T_{00} (\tilde r=0)= \L / 8 \pi G$ (see eq. \eqref{stresstensordS}).  Thus, we recover \eqref{AatfixedVRiemann2} for this specific example. 


\subsection{Entanglement entropy of a causal diamond}
\label{EEthesis}

In the previous subsection we have seen that the Einstein equation is equivalent to the statement \eqref{AatfixedVRiemann2} that the area deficit of small balls is proportional to the energy density. In the present subsection, we will reinterpret this statement in the \emph{semiclassical} regime, i.e. by considering quantum matter fields on a   classical background, as the stationarity of generalized entropy  \eqref{statgenprinciple} for first order variations of the vacuum state of a conformal field theory (CFT) inside the ball. Thus, it is shown that  the    stationarity of generalized entropy for a CFT vacuum in all small balls   at every point in spacetime  is \emph{equivalent} to  the Einstein equation  at the center of the balls. 


In the semiclassical Einstein equation, i.e. $G_{ab} = 8\pi G   \langle T_{ab} \rangle$, the curvature tensor is kept classical   but the stress-energy tensor is replaced by its quantum expectation value.   
Inserting this semiclassical gravitational equation into the area variation at fixed volume \eqref{AatfixedVRiemann} yields a ``quantum corrected'' variational identity 
\beq \label{qcidentity}
\frac{1}{4 G \hbar} \delta A \big |_V  +  \frac{2\pi}{\hbar} \frac{ \Omega_{d-2} \ell^d }{d^2 -1} \delta \langle T_{00 } \rangle = 0 \, .
\eeq
Note that we have divided the identity by the reduced Planck constant $\hbar$. Moreover, we have rearranged the terms  such that the identity becomes equivalent to the statement that generalized entropy is stationary in the vacuum state of a CFT. 
This is because the first term in \eqref{qcidentity} is the variation of the Bekenstein-Hawking  entropy $S_{\rm BH}$ and, as we will show below, the second term  is the variation of the matter entanglement entropy $S_{\rm m}$ of a CFT vacuum restricted to a small causal diamond in flat space. Together, the   Bekenstein-Hawking  entropy and the matter entanglement entropy form the generalized entropy, see \eqref{statgenprinciple}. 

We will now introduce and evaluate the   entanglement entropy variation $\delta S_{\rm m}$. The reduced density matrix of the vacuum state of  any quantum field theory, restricted to  a causal diamond in flat space, can be formally expressed as
\beq \label{vacuumdensitymatrixdiamond}
\rho_{\rm vac} = \frac{1}{Z} e^{-K  / T } \,, 
\eeq
where $K$ is the modular Hamiltonian and $T $ is  a number that is factored out for convenience and which has the interpretation of a temperature in a number of   cases.  For example, for the infinite diamond that coincides with Rindler space, $K$ is the Lorentz boost Hamiltonian and $T$ is the Unruh temperature \cite{Unruh:1976db}. The ``temperature'' of the diamond 
can   be chosen to be   
\beq \label{chosenT}
T = \frac{\hbar  }{2 \pi} \, . 
\eeq
The matter  entanglement entropy of the diamond is defined as the von Neumann entropy associated to the reduced density matrix \eqref{vacuumdensitymatrixdiamond}
\beq
S_{\rm m} = - \text{Tr} \,  \rho_{\rm vac} \log \rho_{\rm vac} \, . 
\eeq
For first order varations of the density matrix, $\rho_{\rm vac} \to \rho_{\rm vac} + \delta \rho_{\rm vac}$,  the variation of the entanglement entropy is related to the  variation of the expectation value of the modular Hamiltonian \cite{Blanco:2013joa, Bhattacharya2012}
\beq \label{firstlawofentanglement3}
\delta S_{\rm m} = \frac{2\pi}{\hbar} \delta \langle K \rangle \, . 
\eeq
 This is the same as equation \eqref{firstlawdSmatter}. Thus, in order to evaluate $\delta S_{\rm m}$  we need to know the quantum expectation value of $K$ in the perturbed state.

The modular Hamiltonian is  a non-local operator for general quantum field theories in the causal diamond. Only for conformal field theories   is $K$ a simple integral of a local operator (the stress-energy tensor). This can be derived from two facts: (1) a causal diamond in flat space is Weyl equivalent to    the Rindler wedge in flat space (see Appendix \ref{sec:conftrans}); (2)      the reduced density matrix of the vacuum  state of any Poincar\'{e} invariant QFT  restricted to the Rindler wedge is thermal, i.e. it takes the form \eqref{vacuumdensitymatrixdiamond} where  $K$ is the Hamiltonian generating Lorentz boosts \cite{Bisognano:1976za}. For   field theories that are conformally invariant   it follows from these two facts that the reduced density matrix of the vacuum in a diamond in flat space is also thermal \cite{Hislop1982,Haag:1992hx,Martinetti:2002sz,Casini:2008cr,Casini:2011kv}. For the CFT vacuum   in a diamond the temperature is again given by \eqref{chosenT} and the modular Hamiltonian   $K$ is equal to the Hamiltonian $H^{\rm m}_\zeta$ generating evolution along the   conformal Killing flow that preserves the diamond, i.e. 
\beq  \label{modularHam5}
K = H^{\rm m}_\zeta := \int_\Sigma {T_{a}}^b \zeta^a    u_b dV \, ,
\eeq
with $\zeta$   the conformal Killing vector\footnote{A conformal Killing vector is a vector field that satisfies $\mathcal L_\zeta g_{ab}= 2 \alpha g_{ab}$, where $\mathcal L_\zeta$ is the Lie derivative along $\z$ and $\alpha= \nabla \cdot \z /d$. For the conformal Killing vector \eqref{ckv6} and $g_{ab} = \eta_{ab}$, one can readily check that $\alpha =- 2t/\ell$. Thus, at the maximal slice $t=0$ the   conformal Killing vector is an ``instantaneous'' Killing vector.} whose flow remains   inside the diamond    (see Figure \ref{fig:causaldiamond} on p. \pageref{fig:causaldiamond} for a depiction of the flow) \cite{Faulkner:2013ica,Jacobson:2015hqa} 
\beq \label{ckv6}
\zeta = \frac{1}{2 \ell} \left [ (\ell^2 - r^2 -t^2)\partial_t - 2 rt \partial_r \right] \, . 
\eeq
Here, $\ell$ is the  radius of the ball and $(t,r)$ are the standard coordinates in Minkowski spacetime.
The normalization of $\zeta$ is chosen such that the associated surface gravity, defined through $\nabla_a \zeta^2 = - 2 \kappa \zeta_a$, is equal to unity: $\kappa =1.$ This normalization agrees with the chosen temperature \eqref{chosenT}.\footnote{There is no preferred normalization for the conformal Killing vector $\zeta$ of a diamond, like there is for the timelike Killing vector of Schwarzschild spacetime, i.e. $\xi = \partial_t$, whose norm is set equal to one   at asymptotic spatial infinity. For a different normalization of $\zeta$ the temperature would be given by   Hawking's formula: $T = \hbar \kappa /2\pi$, where $\kappa$ is the surface gravity of   $\zeta$.} For a more detailed description of the conformal Killing flow (in particular in (Anti-)de Sitter space), see Section \ref{sec:diamond} and Appendices \ref{sec:ckv} and \ref{appkill}.

Next, for perturbation of the vacuum state  the typical   length scale of an excitation is assumed to be much larger than the radius of the ball, i.e. $L_{\rm excitation} \gg \ell$. Under this assumption the expectation value of the stress-energy tensor in the perturbed state can be treated as a constant inside the ball, and the variation  of the expectation value of the   Hamiltonian \eqref{modularHam5} is hence given by
\beq \label{modHamflatspace3}
\delta \langle H^{\rm m}_\z \rangle = \delta \langle T_{00} \rangle \int d\Omega_{d-2} \int_0^\ell dr \,  r^{d-2}  \frac{\ell^2 - r^2}{2\ell}=  \frac{\Omega_{d-2} \ell^d}{d^2-1} \delta \langle T_{00} \rangle \,. 
\eeq
Recall that the ball $\S$  lies in the $t=0$ surface in flat space.
The right-hand side of this equation matches precisely with the  second term in  \eqref{qcidentity}, up to a factor $2\pi /\hbar$. By combining this equation  \eqref{modHamflatspace3} with the ``first law of entanglement'' \eqref{firstlawofentanglement3}, we find that the second term in the quantum corrected variational identity \eqref{qcidentity} is equal to the variation of the matter entanglement entropy $\delta S_{\rm m}$.
Thus, the variational identity  can be interpreted as the statement that  the generalized entropy variation at fixed volume is zero to first order for small causal diamonds in Minkowski space  \cite{Jacobson:2015hqa}
 \beq \label{genentropynew}
\delta S_{\rm gen} \big |_V =  \delta S_{\rm BH} \big |_V+ \delta S_{\rm m} =  0    \, . 
\eeq
Recall that the variational identity \eqref{qcidentity}   is equivalent to the  semiclassical Einstein equation, if the former holds for any spacetime point  and for all timelike unit vectors at the center of the ball. The equivalence between this variational identity and the entropy principle \eqref{genentropynew} thus implies that:

\begin{quote}
\emph{For  conformal matter the semiclassical Einstein equation is equivalent to 
 the stationarity of the vacuum generalized entropy  at fixed volume in all small     diamonds centered at any spacetime point.}
\end{quote}

A number of remarks are in order regarding this  statement:

 First, in  \cite{Jacobson:2015hqa} Jacobson identified the generalized entropy with   entanglement entropy, and reformulated \eqref{genentropynew} as the statement that the total entanglement entropy is in equilibrium in the vacuum. This identification is based on the idea that both the Bekenstein-Hawking entropy and the matter entropy are entanglement entropies of the microscopic degrees of freedom, which are entangled across the boundary of the ball. We   discuss both entropy terms separately below, but  their sum --- the generalized entropy --- is presumably more physical   since it appears to be  UV cutoff independent.

 Regarding the matter entropy, $S_{\rm m}$ is indeed the entanglement entropy of a global pure   state, i.e. the vacuum state of the CFT.\footnote{The CFT vacuum state is a thermofield double state, $| \text{TFD} \rangle \!=\! \frac{1}{\sqrt{Z}} \sum_n e^{- K / 2T} | n \rangle_L \otimes | n \rangle_R$. The TFD state has the special property that the entanglement entropy of the reduced density matrix associated to system $L$ (or $R$) is equal to   thermodynamic entropy in the canonical ensemble (if $K$ is a local Hamiltonian that generates a geometric flow, like \eqref{modularHam5}).} If the variation of the quantum fields is to a global pure state, then $\delta S_{\rm m}$ is  also  purely entanglement entropy. However,  if the   perturbed global state is   mixed, then the von Neumann entropy of the perturbed reduced density matrix     captures the entropy of classical correlations (in addition  perhaps  to the entropy of quantum correlations, depending on whether the mixed state is entangled). Thus, the variation $\delta S_{\rm m}$ might not be pure entanglement entropy.  
 
 Regarding the gravitational entropy, there is numerous literature  \cite{Sorkin:2014kta,Bombelli:1986rw,Srednicki:1993im,Frolov:1993ym,Solodukhin:2011gn,Jacobson:1994iw,Jacobson:2012yt,Jacobson:2012ek,Bianchi:2012br,Bianchi:2012ev} on the identification  of Bekenstein-Hawking entropy $S_{\rm BH}$ with  the vacuum entanglement entropy of   quantum gravitational  degrees of freedom.  This is based on three assumptions: (1) the density of states is dominated by short-distance  degrees of freedom near the boundary, in which case the entanglement entropy   scales  to leading order with the boundary area \cite{Sorkin:2014kta,Bombelli:1986rw,Srednicki:1993im}
 \beq \label{eqn:SEE5}
S_\text{ent} =   c_0 \frac{A}{\epsilon^{d-2}} +\{\text{subleading divergences}\} + S_\text{finite} \, ,
\eeq
where  $c_0$ is a dimensionless constant and   $\epsilon$ is a short-distance cutoff (an artificial UV regulator at this stage); (2) the entanglement entropy is rendered finite by a \emph{physical}  short-distance cutoff on the quantum gravity degrees of freedom; (3) the short-distance cutoff is proportional to the Planck length,  and yields a universal result in the sense that   to leading order   the entanglement entropy is equal to the Bekenstein-Hawking formula   by   identifying $c_0/\epsilon^{d-2} $ with $1/ 4G$ \cite{Frolov:1993ym,Jacobson:1994iw,Jacobson:2012yt}.  This identification  is of course sensitive to the details of   UV  physics,    in particular the constant $c_0$ depends on the UV regulator. There is various evidence \cite{Callan:1994py,Susskind:1994sm,Cooperman:2013iqr} though that the   \emph{generalized entropy} is 
independent of the UV regulator. This is because the  UV divergences in the area term in the matter  entanglement entropy could be absorbed in the renormalization of the gravitational constant $G$,\footnote{The area divergences in the entanglement entropy   only match the matter loop divergences that renormalize $G$ if the \emph{same} short-distance cutoff $\epsilon$   is used to regulate the matter loop divergences  and to distinguish the area   term    from  the  finite term in \eqref{eqn:SEE5}.} making the generalized entropy
 \beq \label{EEisgen}
 S_{\rm gen} =  \frac{A}{4 G^{(\epsilon)}} + S_{\rm m}^{(\epsilon)}  
 \eeq
independent of  the short-distance cutoff $\epsilon.$    The equality between the    entanglement entropy \eqref{eqn:SEE5}  (neglecting the subleading divergences)  and the generalized entropy \eqref{EEisgen} only holds in the ``induced gravity'' scenario where the bare gravitational coupling constant vanishes \cite{Jacobson:1994iw}. 
 Conclusive evidence for $S_{\rm ent} =  S_{\rm gen}$, however, would require a better understanding of  quantum gravity effects at the Planck scale.\footnote{To avoid the suggestive link with entanglement, the quantum gravity principle \eqref{genentropynew} could also be referred to as ``stationarity of generalized entropy" or ``entropy equilibrium''. It namely  remains to be seen what generalized entropy precisely counts or represents in a complete  theory of quantum gravity. Having said that, in this dissertation we   submit    to the interpretation of  \eqref{genentropynew}  in terms of entanglement.}

Another remark concerns the fact that the second term in the variational identity \eqref{qcidentity} is proportional to the  Hamiltonian variation $\delta \langle H^{\rm m}_\zeta \rangle $. This  proportionality is not a coincidence. Actually, there exists a  variational identity  for causal diamonds of \emph{any size} in flat space --- the so-called ``first law of causal diamonds'' --- which relates the area and volume variation of the maximal slice to the matter Hamiltonian variation. The identity \eqref{qcidentity} is just a special  case of this first law, in     that it treats the stress-energy tensor variation  as a constant on the ball (which follows from the assumption   $\ell \ll L_{\rm excitation}$). The first law of causal diamonds in flat space was derived  in    Appendix D of   \cite{Jacobson:2015hqa}.   One of the main results of Chapter~\ref{ch2} is to extend this first law to any maximally symmetric spacetime (see Section \ref{sec:firstlawch2}).  Further, we show that the first law in the  semiclassical regime is   equivalent to the stationarity of generalized entropy  at the maximally symmetric vacuum at fixed volume (see Section \ref{sec:EE}).

Next,  for non-conformal matter the semiclassical Einstein equation  can also be derived from the same principle \eqref{genentropynew}. However, the derivation depends on a nontrivial conjecture about the    modular Hamiltonian variation $\delta \langle K \rangle$ for the vacuum state of non-conformal field theories (which has been checked in \cite{Casini:2016rwj,Speranza:2016jwt}).  One immediate consequence of this extension to non-conformal matter is that the cosmological constant also appears in the derived Einstein equation.  In Section \ref{sec:EE}  we review this derivation  in light of the new results in Chapter \ref{ch2}.

Further, the fixed volume requirement is an input of the derivation of the Einstein equation without a clear physical motivation. What does it mean microscopically to keep the volume of the maximal slice fixed? And why is the volume fixed and not, for example, the geodesic length? Note that    the generalized entropy variation would not have been obtained had we used the area variation at fixed geodesic radius rather than at fixed volume; more precisely, the variational identity   of the form $\delta A |_\ell \sim - \delta \langle T_{00} \rangle$ is equivalent to   $\delta S_{\rm BH} |_\ell + ((d+1)/3 )\delta S_{\rm m}=0$, and not to the stationary of generalized entropy \eqref{statgenprinciple}. 
In Chapter \ref{ch2} we explain that the volume variation in  the first law of causal diamonds appears due to the variation of the gravitational Hamiltonian, thus keeping the volume fixed means keeping this gravitational energy fixed. Further, we show that the stationarity of generalized entropy at fixed volume can be reformulated as the extremization of a free conformal energy with the volume not fixed. The latter   principle can thus feature as a different input assumption in the derivation of the Einstein equation, with the advantage that the volume is not required to be fixed. 


Finally, why can the full non-linear Einstein equation  be derived from a variational identity that   holds only for first-order variations? The reason is that in Riemann normal coordinates the non-linear and linearized Einstein equation, evaluated at the center of the ball, are \emph{equal} to leading order in $\ell / L_{\rm curvature}$. That is, the RNC expansion of the linearized Einstein tensor around flat space is: $\delta G_{00} = G_{00} (O) + \mathcal O(r)$.  Since we  only have been dealing with the linearized constraint, one could question
whether it gives a good approximation to the non-linear field equations at all points within 
the small ball.  This
requires estimating the size of the nonlinear corrections to this field equation.   When 
integrated over the ball, the corrections to the curvature in RNC 
are of order $\ell^2/L^2_{\rm curvature}$.  Since we took the ball
size to be much smaller than the radius of curvature, these terms are already suppressed
relative to the linear order terms in the field equation.  For Einstein gravity the \emph{full non-linear} semiclassical field equations can thus be obtained from the stationarity of generalized entropy. In Chapter \ref{ch3} we show that this not hold for higher derivative theories of gravity: only the \emph{linearized} semiclassical gravitational field equations are equivalent to the entanglement equilibrium principle (see especially Section \ref{sec:equations}).

To summarize,  the following questions will be addressed in the rest of this thesis:

\begin{itemize}

\item[(d)] Is the generalized entropy stationary for  all maximally symmetric   diamonds?\\ See Section \ref{sec:thermo}, and in particular Section \ref{sec:EE}, for a discussion.

\item[(e)] Can the Einstein equation  be derived  for non-conformal matter   and with cosmological constant?\\This has been addressed by Jacobson in Section V of \cite{Jacobson:2015hqa}, but see   Sections \ref{sec:qc}--\ref{sec:EE}    for a revised answer.

\item[(f)] Can the Einstein equation  be derived from a different quantum gravity principle in which  the volume is not   fixed? \\See Section \ref{sec:freeenergy} for a new principle: the stationarity of  the free conformal energy of a causal diamond.  
 
\item[(g)] Can the gravitational fields equations with arbitrary higher curvature corrections be derived  from an entanglement equilibrium condition?\\This is the main focus of Chapter \ref{ch3}.


\end{itemize}




\section{The AdS$_3$/CFT$_2$ correspondence}
\label{intro:AdS/CFT}

The canonical     example of holography is the AdS/CFT correspondence \cite{Maldacena:1997re,Witten:1998qj,Aharony:1999ti}, which is a conjectured duality between quantum gravity in $d$-dimensional asymptotically  Anti-de Sitter (AdS) space  and a conformal field theory (CFT) in $(d-1)$-dimensional flat space. A particularly well-understood   case of the AdS/CFT duality is for $d=3$, i.e. a three-dimensional bulk and a two-dimensional boundary theory.  Einstein gravity in three dimensions is rather simple since it has  no propagating degrees of freedom, but it is still complex and interesting enough to study its holographic properties, since it allows   for     nontrivial black hole solutions with AdS asymptotics. In this section we   briefly review some basic ingredients of the 3d/2d holographic dictionary, which serves as background knowledge for Chapter~\ref{ch4} (for an extensive review of the AdS$_3$/CFT$_2$ correspondence, see \cite{Kraus:2006wn}). 

\subsection{The 3d/2d holographic dictionary}


A crucial ingredient of the 3d/2d holographic dictionary is the relationship  between the central charge and   scaling dimension  in     CFT$_2$, on the one hand, and the curvature radius and   energy of an asymptotically    AdS$_3$ spacetime, on the other hand. The central charge $c$ of a   CFT is defined in terms of the normalization of the two-point function of the stress tensor, i.e. $\langle T(z) T(w) \rangle = \frac{c/2}{(z-w)^4}$, and  it measures the number of field theoretic degrees of freedom of    the CFT. In order to avoid    gravitational anomalies, we     assume that the left- and right-moving central charges are equal: $c = \bar c.$  The scaling dimension is defined as the sum of the left- and right-moving conformal weights of a primary operator, i.e. $\Delta = h + \bar h$, and it determines   the energy of a state through the operator-state correspondence.  For a CFT that lives on a cylinder $\mathbb R \times S^1$, with $L$   the radius of the circle, the total energy $E$ of a primary state is given by: $E L = \Delta - c/12.$
The holographic dictionary states that the central charge is related to the curvature radius $L$ (also called AdS radius), and the scaling dimension is related to the ADM energy $E$ of the dual spacetime  (and to the AdS radius  $L$)
 \begin{align}
&\text{central charge:} \quad  \quad \,\,   \frac{c}{12} = \frac{2\pi L}{16 \pi G_{3}}  \,,\label{BrownHen5}\\
&\text{scaling dimension:}   \quad       \Delta -   \frac{c}{12 } =  E L        \,.
\label{excnumber5}
 \end{align}
Equation  (\ref{BrownHen5}) is just a rewriting of the  Brown-Henneaux formula for the central charge, $c = 3 L / 2 G_{3}$, where $G_3$ is the three-dimensional Newton's constant \cite{Brown:1986nw}. 
Equation  (\ref{excnumber5}) identifies    the energy of a   CFT state on the circle, whose radius is equal to the AdS radius, with the ADM energy of the asymptotically AdS spacetime.\footnote{This identification  between the energies and the radii  follows from the following set of rules. The   CFT which is dual to gravity in asymptotically global AdS spacetime lives on a cylinder. The metric on the cylinder could be any Weyl equivalent metric of    asymptotically AdS spacetime, i.e. $\lim_{r \to \infty} g_{\rm AdS} =\Omega^2 g_{\rm CFT}$. Within this class of Weyl equivalent metrics, a particular  representative is singled out by the restriction that the CFT time is the same as global time in asymptotically AdS. 
This assumption is realized if the Weyl factor is $\Omega=r/L$, and hence the CFT metric is: $d  s^2  = - dt^2 +   L^2 d \phi^2$. Note that for this choice of the CFT metric the radius of the circle is equal to the AdS radius~$L$. Further,   the 
CFT Hamiltonian and the (asymptotic) ADM Hamiltonian are both  generators of time translations along $t$   and   can hence be identified.}
 The vacuum state  satisfies $\Delta = 0$ and has a negative energy   $- c/12 L$, which is the Casimir energy of a 2d~CFT on a circle of size $L$. 
Since  the   CFT lives on a circle, it has an energy gap between the ground state and the first excited state, which is equal to $1/L$.  The energy gap puts an IR cutoff   on the large-distance degrees of freedom, since the    length scale of an excitation cannot exceed the system's size. Hence, the scaling dimension \eqref{excnumber5} counts the energy of the state in terms of the IR cutoff scale.


Next, we describe the holographic dictionary between particular   states in the CFT and  three-dimensional solutions to Einstein gravity with a negative cosmological constant. We   only consider   static (i.e. non-rotating) asymptotically AdS$_3$ spacetimes. In the CFT this amounts to  setting  $  h = \bar h$, since  the   angular momentum   is given by   the difference between the right- and left-moving conformal weights:  $J=   h - \bar h$. The holographic dictionary between CFT states and gravitational solutions is encoded in the following 
   metric:   
    \begin{equation}
   ds^2  = - \left ( \frac{r^2}{L^2} - \frac{\Delta - c/12}{c/12} \right ) dt^2 + \left (\frac{r^2}{L^2} - \frac{\Delta - c/12}{c/12}  \right )^{-1} dr^2 + r^2 d\phi^2 \, . 
 \end{equation}
The central charge $c$  is fixed by the Brown-Henneaux formula (\ref{BrownHen5}). For different choices of the scaling dimension $\Delta$, the metric describes different asymptotically AdS$_3$ spacetimes (which are   \emph{locally} equivalent, since the latter two spacetimes are quotients of pure AdS$_3$ \cite{Banados:1992gq}):  
    \begin{align} \label{threecases}
&\text{Global empty AdS$_3$:}  \qquad \qquad   \,   \Delta = 0 \, ,  \nonumber\\
&\text{AdS$_3$ with a conical defect:} \quad  \frac{ \Delta - c/12}{c/12} =  - \frac{1}{N^2}      \, , \\
&\text{BTZ black hole:}  \qquad \qquad \quad \,\, \,\,\,  \frac{ \Delta - c/12}{c/12} =  \frac{r_{\rm h}^2}{L^2}    \, .  \nonumber
 \end{align}
By comparing equations  (\ref{excnumber5})  and (\ref{threecases}) the  ADM  energy $E$ of these three   spacetimes can be determined.  The first case, $\Delta =0$,  corresponds to the vacuum state in the CFT, which is dual to   empty AdS$_3$.   In the second case, $\Delta \le c/12$, the scaling dimension is associated to a $\mathbb Z_N$ twist operator in the CFT with integer value $N$, which corresponds to a conical defect  in AdS$_3$ with deficit angle $2 \pi (1-1/N)$. The last case, $\Delta \ge c/12$, describes a thermal state in the CFT which is dual to   a non-rotating BTZ black hole in AdS, whose horizon radius is denoted by $r_{\rm h}$ \cite{Banados:1992wn}. 
  Note that there is a mass gap between   global empty AdS and the horizonless black hole  solution $r_{\rm h}=0$, since the latter solution is identical to   a conical defect in AdS with $N = \infty$ (or to Poincar\'{e} AdS, but \emph{not} to global AdS). 

The special case $r_{\rm h} = L$ (or $\Delta = c/6$) corresponds to the critical point of a first order phase transition between thermal AdS (i.e.  AdS spacetime with thermal radiation) and the BTZ black hole: the Hawking-Page phase transition \cite{Hawking:1982dh}. The temperature of a black hole at this critical point is $T_{\rm HP} = 1/(2\pi L)$.\footnote{For higher-dimensional Schwarzschild-AdS black holes the horizon radius at the critical point of the Hawking-Page phase transition remains $r_{\rm h}=L$, and the Hawking-Page temperature is $T_{\rm HP} = (d-2)/(2\pi L)$, where $d$ is the number of bulk spacetime dimensions.} Above this temperature the black hole dominates the canonical ensemble, i.e. its free energy is lower than that of thermal AdS, whereas below the Hawking-Page temperature thermal AdS  is thermodynamically preferred and the free energy of the black hole is larger than that of thermal AdS. This phase transition thus makes a natural distinction between large and small BTZ black holes.\footnote{This is different from the standard distinction between large and small Schwarzschild-AdS black holes, based on the sign of the specific heat $C  = \partial  M / \partial T$. The specific heat of Schwarzschild-AdS   is negative for small black holes, with horizon radius $r_{\rm h} < r_0$, and positive for large black holes, with $r_{\rm h} > r_0$, where   $r_0 = L \sqrt{(d-3)/(d-1)}$. At $r_{\rm h} = r_0$ the specific heat diverges and the temperature is $T_0 = \sqrt{(d-3)(d-1)} / (2\pi L)$ (note that $ T_0 < T_{\rm HP}$). Below this  temperature black hole solutions do not exist. For BTZ black holes   the specific heat is  always positive, since $r_0=0$, hence the usual distinction between small and large black holes does not apply in $d=3$.}   Further, in the CFT the Hawking-Page phase transition can be interpreted as the   transition between a confining ($T < T_{\rm HP}$) and deconfining phase ($T> T_{\rm HP}$) \cite{Witten:1998zw}. 

\subsection{Black hole entropy from the Cardy formula}

In unitary, modular invariant 2d CFTs with a gapped spectrum the thermodynamic properties of a given macrostate, such as entropy and temperature,  can be derived from the central charge   and the scaling dimension. In particular,  the thermodynamic entropy   is given by the Cardy formula  (assuming $c = \bar c$ and $h = \bar h$) \cite{Cardy:1986ie}
 \begin{equation}
 \label{cardy}
 S_{\rm Cardy} = 4 \pi \sqrt{\frac{c}{12} \left(\Delta - \frac{c}{12} \right)} \, .
 \end{equation}
Cardy's original derivation  of this formula is only valid in the  regime  $\Delta \gg c/6$, with $c$ a fixed but not necessarily large number.      It was recently shown in  \cite{Hartman:2014oaa} that the Cardy formula holds in the entire regime $\Delta \geq c/6$, for  2d CFTs with a large central charge $c\gg1$ and a sparse light spectrum.     Both derivations are   based on the modular invariance of the partition function on the torus. 
The Cardy formula matches the Bekenstein-Hawking entropy of a BTZ black hole through the holographic dictionary (\ref{BrownHen5}) and (\ref{excnumber5}) (or equivalently \eqref{threecases}) \cite{Strominger:1997eq}
\begin{equation}
S_{\rm BH}= \frac{2\pi r_{\rm h}}{4 G_3}\, .
\end{equation}
The matching works just as well for rotating BTZ black holes. Note that the derivation only holds for $\Delta \ge c/6$, hence $r_{\rm h} \ge L$, since the Cardy formula does not apply for lower scaling dimensions. Hence, the entropy of small black holes   cannot be derived in the standard way using the   Cardy formula.     

The temperature of a CFT   macrostate follows from the Cardy formula using the   usual thermodynamic definition
\beq
T_{\rm CFT} = \left ( \frac{\partial S_{\rm Cardy}}{\partial E} \right)^{-1} =  \frac{1}{2 \pi L} \sqrt{\frac{\Delta - c/12}{c/12}} \, . 
\eeq
The length scale   $L$ ensures that the temperature has the correct dimensionality. Using the dictionary \eqref{BrownHen5} and \eqref{threecases} we recover   the Hawking temperature of   non-rotating BTZ black holes
\beq
T_{\rm H} =  \frac{r_{\rm h}}{2\pi L^2} \, .
\eeq
 To conclude, the   entropy and temperature  are   derived quantities in 2d CFT; they  are uniquely determined  by   two   numbers $c$ and $\Delta$ (and in general  also by $\bar c$ and $\bar h$).  We emphasize that these derivations of black hole entropy and Hawking temperature do not rely on all of the microscopic details of the CFT. Only two general   quantities  associated to the theory and the state --- the number of field degrees of freedom and the energy measured in the IR cutoff --- are needed to reproduce the horizon entropy and   temperature.
 
  A natural question is whether these quantities can be generalized to the microscopic theories on other holographic surfaces than those close to the AdS boundary. In Chapter \ref{ch4} we introduce three    microscopic holographic quantities on a    spherical $(d-2)$-surface, that are generalizations of the central charge, scaling dimension and energy gap in the CFT. This is similar in spirit to holographic renormalization \cite{Henningson:1998gx, Akhmedov:1998vf,Alvarez:1998wr, Balasubramanian:1999jd, Skenderis:1999mm,deBoer:1999tgo} or the   $T\bar T$ deformation of a 2d CFT \cite{Smirnov:2016lqw,McGough:2016lol}. However, we use a different approach to derive these three microscopic quantities, namely a Weyl equivalence between certain    non-AdS geometries and $AdS_3\times S^q$ spacetimes. We derive these   holographic quantities for sub-AdS regions,  de Sitter space  and Minkowski space in general dimensions, and connect them to properties of symmetric product   2d CFTs dual to the   AdS$_3$ spacetimes that appear  after  a  Weyl transformation.  

%% file: chapter2.tex
 \chapter{Gravitational Thermodynamics of Causal Diamonds}
\label{ch2}


\section{Introduction}
\label{sec:introdiamond}

Horizon thermodynamics was first discovered in the context of black holes \cite{Bekenstein:1973ur,Bardeen:1973gs,Hawking:1974sw}, but the principles are far more universal than that. The case of cosmological horizons was quickly understood \cite{Gibbons:1976ue,Gibbons:1977mu} and, rather less quickly insofar as entropy is concerned, 
that of acceleration horizons as well \cite{Davies:1974th,Unruh:1976db,Jacobson:1995ab,Massar:1999wg,Jacobson:1999mi,Jacobson:2003wv}. 
Most recently, in the setting of AdS/CFT duality, the gravitational thermodynamics of ``entanglement wedges" has been discovered \cite{Casini:2011kv,Blanco:2013joa}. 
An entanglement wedge is the domain of dependence of a partial Cauchy surface of the bulk, whose boundary consists of a subregion ${\cal R}$ of a conformal boundary Cauchy slice together with the minimal area bulk surface that meets the conformal boundary at $\partial {\cal R}$. The area $A$ of the minimal surface corresponds to the CFT entanglement entropy of the subregion ${\cal R}$ to leading order in Newton's constant, 
via the Ryu-Takayanagi formula, which is nothing but the 
Bekenstein-Hawking entropy $A/4\hbar G$ \cite{Ryu:2006bv,Ryu:2006ef}. In particular, a link has been established between fundamental properties of CFT entanglement entropy and the bulk Einstein equation, drawing a connection between 
the behavior of quantum information and gravitational dynamics in this holographic setting \cite{Lashkari:2013koa,Faulkner:2013ica,Swingle:2014uza,Faulkner:2017tkh}.

In the examples just described, the thermodynamic system extends to a boundary of spacetime. 
Quasilocal relations analogous to the laws of
thermodynamics have been found for various sorts of `apparent' horizons, but of course these find application only
when such horizons are present \cite{Ashtekar:2004cn}. If the lesson from all we have learned is that 
something fundamentally  statistical
underlies gravitational dynamics, then
that something should be at play {\it everywhere} in spacetime. This viewpoint 
was the motivation for introducing the notion of ``local causal horizon," with which it was possible to derive the Einstein equation from the Clausius relation applied to the area-entropy changes of all such horizons in spacetime \cite{Jacobson:1995ab}.  
Essential to that argument was the fact that the near vicinity of any spacetime point looks like part of a slightly deformed Minkowski spacetime, so that one can identify an approximate local boost Killing field with respect to which the relevant notion of energy can be defined. While that derivation was reasonably 
plausible on physical grounds, the ``system" under consideration was not sharply defined. It was 
taken to be a very small, near horizon subsystem, but no precise definition was given, and possible
effects of entanglement with neighboring regions were not addressed. 

To localize and isolate it,  the thermodynamic system was defined in \cite{Jacobson:2015hqa} as the spacetime inside 
a causal diamond, i.e.\ the domain of dependence of a spatial ball.\footnote{The thermal behavior of conformal quantum fields -- without gravity -- inside a causal diamond in flat space  was studied earlier, for example in Refs.~\cite{Hislop1982,Haag:1992hx,Martinetti:2002sz,Casini:2008cr,Casini:2011kv}.} 
For such a system, rather than describing a time dependent physical process, one can just consider variations,
comparing the equilibrium state to nearby states.
A link was established between the semiclassical Einstein equation, and the stationarity of the 
total entanglement entropy of the diamond with respect to variations of the geometry and quantum fields
away from the vacuum state at fixed volume, taking the diamond much smaller than the local curvature length scale
of the background spacetime. 

A {\it classical} ingredient of this argument is a ``first law of causal diamonds,"
which relates variations, away from  
flat spacetime, of the area, volume, and matter energy-momentum tensor inside. 
This relation is similar to the first law of black hole mechanics, and was derived in an Appendix of \cite{Jacobson:2015hqa} using the same methods as employed by Wald for black holes \cite{Wald:1993nt}. 
A diamond in Minkowski spacetime plays the 
role of equilibrium state in this first law.  
Unlike a black hole, however, a causal diamond is only {\it conformally stationary}.
That is, it does not admit a timelike Killing vector,  but it admits a timelike conformal Killing vector,
for which the null boundary is a conformal Killing horizon \cite{Dyer:1979,Dyer:2004}, with a well-defined surface gravity.\footnote{It was shown   in \cite{Jacobson:1993pf} that the surface gravity $\kappa$ of a conformal Killing horizon has a conformal invariant definition (see  Appendix \ref{sec:zerothlaw} for an explanation, and for a proof of the zeroth law). This fact can be viewed as a hint that conformal Killing horizons have thermodynamic properties, since $\kappa$ is identical to the surface gravity of a conformally related true Killing horizon (for which $\hbar \kappa/2\pi$ is the Hawking temperature).}
Remarkably, this is sufficient to derive the first law, but an additional contribution arises,
namely, the gravitational Hamiltonian, which is proportional to the spatial volume of the ball.

In this chapter we generalize the first law of causal diamonds   to       (Anti-)de Sitter spacetime -- i.e. it applies to any maximally symmetric space -- and we include variations of the cosmological constant and matter stress-energy, both using
a fluid description as done for black holes by Iyer \cite{Iyer:1996ky}.  
Treated this way, the matter can contribute to a first order variation
 away from a maximally symmetric spacetime.\footnote{Matter 
described by a typical field theory action 
could not contribute, since the matter fields would vanish in the maximally
 symmetric background spacetime, and the action is at least quadratic in the fields.}
 Motivation for varying the cosmological constant is discussed in 
 Section~\ref{sec:varyingLambda}.  The first law we obtain for Einstein gravity minimally coupled to fluid matter takes the form
 \beq\label{1}
 \d H_\z =  -\frac{\k}{8\pi G}\,\d A \, ,
 \eeq
 where $\z$ is the conformal Killing vector of the unperturbed, maximally symmetric diamond (computed explicitly in Appendix \ref{sec:ckv}),
 $H_\z$ is the total Hamiltonian 
 for gravity and matter along the flow of $\z$,  and 
 $\k$ is the surface gravity (defined as positive) with respect to $\z$. 
Much as for black holes, if we multiply and divide by Planck's constant $\hbar$, 
 the quantity on the right-hand side of \eqref{1} takes the form $-T_{\rm H}\d S_{\rm BH}$, where $T_\text{H}=\hbar\k/2\pi$ is the Hawking temperature and $S_{\rm BH}=A/4\hbar G$ is the Bekenstein-Hawking entropy. Unlike for black holes, however, there is a {\it minus sign} in front, indicating that the temperature is {\it negative}.
This minus sign is  familiar from the limiting case in which the diamond consists of the entire static patch of de Sitter (dS) spacetime, bounded by the de Sitter horizon \cite{Gibbons:1977mu}. It was   argued in the dS case that the temperature of the static patch is therefore negative \cite{Klemm:2004mb}, and further arguments in favor of this interpretation were given recently in \cite{Cvetic:2018dqf}. 

The left-hand side of \eqref{1} consists of minimally coupled matter, cosmological constant, and gravitational terms, i.e. 
$\d H_\zeta = \d H_\z^{\text{\~{m}}} + \d H_\z^\L + \d H_\z^\text{g}$. 
The contribution of the fluid matter with arbitrary equation of state is 
\beq  \label{matter1}
\d H^\text{\~m}_\zeta =\int_\S \d   (  {T_a}^{b}  )^\text{\~m} \z^a u_b\, dV  \, ,
\eeq
where $(  {T_a}^{b}  )^\text{\~m}= g^{bc} T_{ac}^\text{\~m}$ is the Hilbert stress-energy tensor with one index raised by the inverse metric
and $\Sigma$ is the maximal slice of the unperturbed diamond, with future pointing unit normal vector $u^a$
 and proper volume element $dV$. 
The cosmological constant can be thought of as a perfect fluid with stress-energy tensor  $T^\L_{ab}=-\L/(8\pi G)g_{ab}$.
Since it is maximally symmetric, it 
 may be  nonzero in the background. Its Hamiltonian variation 
is given by
\beq \label{firstdefoftheta}
\d H^{\L}_\z=\frac{V_\zeta}{8\pi G}\,\d\L \, , \qquad \text{with} \qquad V_\zeta:= \int_\S |\zeta| dV  ,
\eeq
where $|\zeta| :=\sqrt{-\z\cdot\z} $ is the norm of the conformal Killing vector. 
The quantity $V_\zeta$ is called the ``thermodynamic volume" \cite{Kastor:2009wy,Dolan:2010ha,Cvetic:2010jb}, and is well known in the context of the first law for black holes, when extended to include variations of $\L$. Finally,  the gravitational term takes the form
\beq\label{dHg}
\d H_\z^\text{g} = -\frac{\k k}{8\pi G} \d V  , 
\eeq
where $k$ is the trace of the (outward) extrinsic curvature of $\partial\S$ as embedded in $\S$, and $V$ is the proper volume of  the ball-shaped spacelike region $\S$. This term arises because $\z$ fails to be a true Killing vector. In the special case of the static patch of dS space,
$\z$ is a true Killing vector, and indeed \eqref{dHg} vanishes, since we have $k=0$.
That this gravitational Hamiltonian variation is proportional to the maximal volume variation suggests 
a connection with the {\it York time Hamiltonian} \cite{York:1972sj}, which  generates evolution   along a constant mean curvature foliation and   is proportional to the  proper  volume of  slices of this foliation (see  Section~\ref{sec:Yorktime}). 
 In fact, we show in Appendix \ref{appyork} that slices of 
the constant mean curvature foliation of a maximally symmetric diamond coincide with slices of  constant conformal Killing time.
 
 Moreover, we extend the first law of causal diamonds to the semiclassical regime, i.e. by considering quantum matter fields on a fixed classical background. For  any type of quantum matter in a small diamond we show that  the semiclassical first law  can be written in terms of Bekenstein's generalized entropy\footnote{For conformal matter this expression for the first law actually holds  for any sized diamond, whereas for  non-conformal matter the derivation  depends on an assumption \eqref{conja} about the    modular Hamiltonian variation for small diamonds  which was conjectured in   \cite{Jacobson:2015hqa} and tested in \cite{Casini:2016rwj,Speranza:2016jwt}.}
\beq
-\frac{\kappa k}{8\pi G} \delta V + \frac{V_\zeta}{8\pi  G} \delta \L = T \delta S_{\rm{gen}} \, ,
\eeq 
where the temperature is minus the Hawking temperature, i.e. $T= - \hbar \kappa/2\pi$, and the generalized entropy is defined as the sum of the Bekenstein-Hawking entropy and the matter entanglement entropy, i.e. $S_{\rm{gen}} := S_{\rm{BH}} + S_{\rm \tilde{m}}$.
At fixed volume $V$ and cosmological constant  $\L$ this implies that the generalized entropy is stationary in a  maximally symmetric vacuum. This coincides with the   \emph{entanglement equilibrium} condition of  \cite{Jacobson:2015hqa}, which was shown in that paper to be equivalent to the semiclassical Einstein equation. We also argue in Section \ref{sec:freeenergy} that the entanglement equilibrium condition is equivalent to the stationarity of a free energy   at fixed cosmological constant, but \emph{without} fixing the volume. We thus find that the semiclassical Einstein equation is also equivalent to the   free energy extremization.


This  chapter is organized as follows.  In Section \ref{sec:diamond} we describe our setup of       causal diamonds in  maximally symmetric spaces in more detail. The Smarr formula and  first law for causal diamonds are derived in Section   \ref{sec:diamondsthermo} using Wald's Noether charge formalism. In Section \ref{sec:thermo} we give  a thermodynamic interpretation to the first law, and we derive the entanglement equilibrium condition from the semiclassical first law. In Section  \ref{sec:remarks} we comment on various aspects of the first law, and in Section \ref{sec:cases} we describe   a number of limiting cases of the first law for maximally symmetric causal diamonds: de Sitter static patch, flat space,  Rindler space, AdS-Rindler space and the Wheeler-DeWitt patch of AdS.
We end with a summary of results    in Section \ref{sec:discussion}.  
The Appendices \ref{sec:accckv}--\ref{sec:conftrans} are devoted to  establishing several properties of   conformal Killing fields in (A)dS    and in flat space, and  of bifurcate conformal Killing horizons in general.


\section{Maximally symmetric causal diamonds}
\label{sec:diamond}

In this section we discuss causal diamonds and their conformal Killing vectors in maximally symmetric spaces. 
It suffices to write the equations for the case of positive curvature, i.e.\ de Sitter space, since the 
negative curvature (Anti-de Sitter) and flat cases can be obtained from these by sending the curvature length scale $L$ to $iL$, or to infinity, respectively. The line element for a static patch of de Sitter space in $d$ spacetime dimensions is 
\begin{equation} \label{staticpatch}
ds^2 = - [ 1 - (r/L)^2] dt^2 + [ 1 - (r/L)^2]^{-1}  dr^2 + r^2 d \Omega_{d-2}^2 \, ,
\end{equation}
where 
we use units with $c=1$.
In terms of retarded and advanced time coordinates, 
 \begin{equation}  \label{uandv}
 u = t - r_* \,  \qquad \text{and} \qquad v = t + r_* \, ,
 \end{equation}
with $r_*$ the ``tortoise coordinate"  defined by 
\begin{equation}
dr_* = \frac{dr}{1-(r/L)^2} \,,  \qquad r = L\tanh(r_*/L) \,,
\end{equation}
the line element \eqref{staticpatch} takes the form
\beq
 \begin{aligned}\label{linehyperbolic}
 ds^2 &= -[ 1 - (r/L)^2] \,du dv + r^2 d \Omega_{d-2}^2   \\
 &= \text{sech}^2(r_*/L) \left [ - du dv + L^2\sinh^2(r_*/L) \, d  \Omega_{d-2}^2 \right] . 
 \end{aligned}
 \eeq
Note that $r_*=r+O(r^3)$, so in particular $r=0=r_*$ at the origin.
For dS the cosmological horizon, $r=L$, corresponds to $r_* = \infty$. For AdS,
we have $r=L \tan (r_*/L)$, so $r=\infty$ corresponds to $r_* = L \pi  / 2$. 
In the flat space limit, $L \to \infty$, the tortoise and radial coordinates coincide.

\begin{figure}
	\centering
	\includegraphics
		[width=.35\textwidth]
		{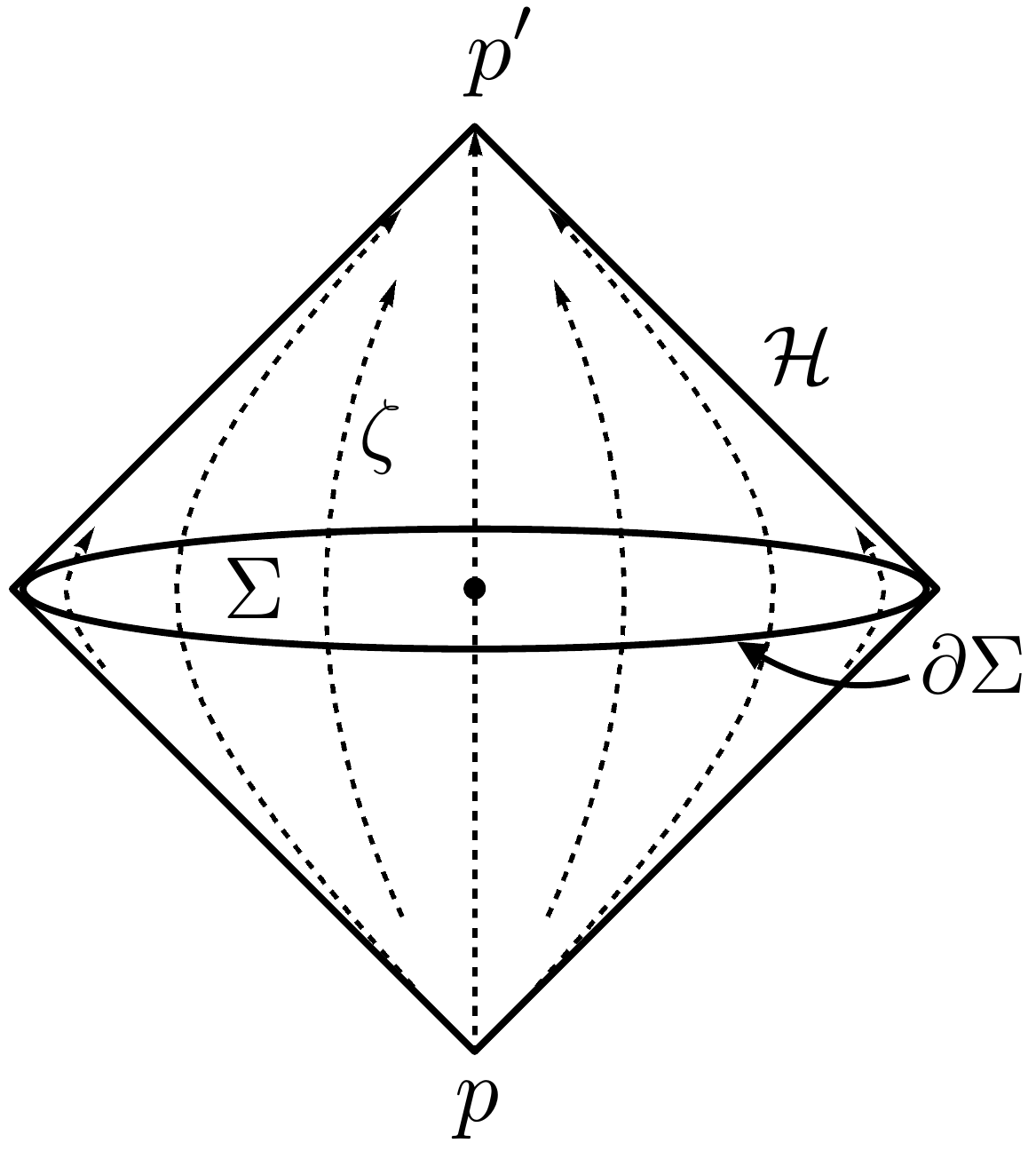}
\caption{\small   A   causal diamond in a maximally symmetric spacetime for a  ball-shaped spacelike region $\Sigma$. The past and future vertices of the diamond are denoted by $p$ and $p'$, respectively, and $\mathcal H$ is the null boundary. The dashed arrows are the flow lines of  the conformal Killing vector   $\zeta$, whose flow sends the boundary of the diamond into itself, and vanishes at $\partial \Sigma$, $p$ and $p'$.}
	\label{fig:causaldiamond}
\end{figure}

A spherical causal diamond in
a maximally symmetric space 
can be defined as the domain of dependence of a spherical spacelike region $\S$  with vanishing extrinsic curvature. Equivalently, it can be described as the intersection of the future of some point $p$,
with the past of another point $p'$ (see Figure \ref{fig:causaldiamond}). 
All such diamonds are equivalent, once the geodesic proper time 
between the {\it vertices} $p$ to $p'$ has been fixed. 
The intersection
of the future light cone of $p$ and the past light cone of $p'$ is called the {\it edge} of the diamond.
The edge is the boundary $\partial\Sigma$ of a $(d-1)$-dimensional ball-shaped region $\Sigma$. 
The symmetries of such diamonds are rotations about the $pp'$ line, reflection across $\Sigma$, and 
a conformal isometry to be discussed shortly. 

Fixing a causal diamond, 
we place the origin of the above coordinate system at the center, and choose the $t$ coordinate so that    
$\Sigma$ lies in the $t=0$ surface. The geodesic joining the vertices is then the line $r=0$,
and the diamond is the intersection of the two regions $u>- R_*$ and $v<R_*$, 
for some $R_*$.  The vertices are located at the points
$p= \{ u =v= - R_*\}$ and $p'=\{u=v=R_*\}$, and the edge of the diamond is the $(d-2)$-sphere
$ \partial\Sigma= \{  v = - u =R_* \}$,  
with coordinate radius  $r_* = R_*$ and area radius   $r = R = L\tanh (R_*/L)$.

 The unique conformal isometry that preserves the causal diamond 
 is generated by the conformal Killing vector   
 \begin{equation}  \label{ckv1}
   \zeta
   = \frac{L}{\sinh (R_*/L)} \Big[\big(\cosh (R_*/L) - \cosh (u/L) \big)\partial_u 
   + \big(\cosh (R_*/L)- \cosh (v/L) \big) \partial_v \Big]      \, .
   \end{equation}
A derivation of this fact is given in Appendix \ref{sec:ckv}. 
The flow generated by $\zeta$ 
sends the boundary of the diamond into itself, and leaves fixed the vertices and the edge.
In the interior $\zeta$ is the sum of two future null vectors, so it is 
timelike and future directed. It is the null  tangent to the past and future null boundaries,
$\mathcal H = \{ v = R_*,   u = -R_*\}$,  so those boundaries are conformal Killing horizons.
The intersection of these null boundaries, i.e. the edge $\partial \Sigma$,   is  therefore referred to as the {\it bifurcation surface} of $\mathcal H$.\footnote{The  conformal Killing vector also acts   outside the   diamond, and remains null on the  continuation of the null boundaries of the diamond. In total there are four conformal Killing horizons $\{u=\pm R_*, v=\pm R_* \}$, which divide the maximally symmetric spacetime up into six regions (see \cite{DeLorenzo:2017tgx} for an analysis of the flow of the conformal Killing field in Minkowski spacetime).}

In terms of the $t$ and $r$ coordinates introduced above, the conformal Killing vector  reads
  \begin{equation}  \label{ckv2}
 \zeta  \!=\! \frac{L^2}{R} \! \!\left[    \Big ( 1 \!-\!  \frac{\sqrt{1- (R/L)^2}}{\sqrt{1- (r/L)^2}}   \cosh(t/L)  \Big)    \partial_t -   \frac{r}{L} \sqrt{\left ( 1 \!-\! (R/L)^2 \right)  \left (1 \!-\! (r/L)^2\right) }   \sinh(t/L)  \partial_r  \!     \right].
 \end{equation}
Note that if the boundary of the diamond  coincides with the cosmological horizon, i.e. if $R=L$, then $\zeta= L\partial_t$.
That is, the conformal Killing symmetry becomes the usual time translation of the entire static patch of dS, which is  a causal diamond with   infinite  time duration  but finite spatial width. 
In Appendix \ref{sec:embedding} we use the  two-time embedding formalism of dS and AdS  to derive an expression for $\zeta$  
in terms of the generators of the conformal group.\footnote{Note that maximally symmetric spaces are conformally flat, so they admit the $O(2,d)$   group of conformal isometries.} 
We show that $\zeta$ can be written as a linear combination of a time translation of the surrounding static patch and a    conformal transformation (which is a special conformal transformation in the case of flat space) -- see equation (\ref{ckvdsemb}).

The surface gravity $\kappa$ of any conformal Killing vector with a bifurcation surface ${\cal B}$ can be defined 
exactly as for a true Killing vector: If we contract the conformal Killing equation, $\nabla_{(a}\zeta_{b)}=\alpha g_{ab}$
with $m^a m^b$, for $m^a$ any tangent vector to ${\cal B}$, the left-hand side vanishes, since $\zeta^a = 0$ on ${\cal B}$.  Thus we learn that 
(as for the particular example of $\zeta^a$ under study) $\alpha=0$ on ${\cal B}$. It follows that  
$\partial_a\zeta^b$ is a generator
of Lorentz transformations in the two-dimensional normal plane at each point of ${\cal B}$. Like for true Killing vectors, 
we may therefore define the surface gravity $\kappa$ at $\cal{B}$ (or rather its absolute value)  by $2\kappa^2 = (\partial_a\zeta^b)(\partial_b\zeta^a)$.
Other definitions for $\kappa$, which are equivalent for a true Killing vector, are not equivalent for a conformal
Killing vector. The definition that is invariant under conformal rescalings of the metric \cite{Jacobson:1993pf}, 
$\nabla_a \zeta^2 = - 2 \kappa \zeta_a$, is also constant along the generators
of the conformal Killing horizon, and coincides with the definition just given at ${\cal B}$. We establish these 
general properties in Appendix \ref{sec:zerothlaw}, along with the zeroth law, i.e.\ the fact that $\kappa$ is constant on  $\cal{H}$. The normalization of $\zeta^a$ in \eqref{ckv1} or \eqref{ckv2} has been chosen so that $\kappa=1$ at the future horizon ($v = R_*$) and $\kappa = - 1$ at the past horizon ($u= - R_*$). In the rest of the thesis we take $\kappa$ to be positive and keep it explicit,  to indicate where it appears if a different normalization is chosen.\footnote{If the conformal Killing vector were normalized such that $\zeta^2 =-1$ at $t=r=0$, then the surface gravity would be $\kappa =(R/L^2)   (1 - \sqrt{1-(R/L)^2}  )^{-1}$. This is the choice usually made in the case of the static patch of de Sitter space, where $R=L$ and $\kappa_{\text{dS}} =1/L $. In the flat space limit ($L\to \infty$) this surface gravity becomes $\kappa_{\rm{flat}} = 2/R$, and for an infinite diamond in AdS ($R/L \to \infty$, the ``Wheeler-DeWitt patch'') it reduces to $\kappa_{\rm{WdW}}= 1/L,$ where $L$ is the AdS radius.}

 The Lie derivative of the metric with respect to a conformal Killing vector in general has the form
 \begin{equation}\label{lie}
 \mathcal L_\zeta g_{ab} = \nabla_a\zeta_b+\nabla_b\zeta_a = 2 \alpha g_{ab} \, ,   \qquad \alpha =(\nabla_c\zeta^c)/d \, . 
  \end{equation}
In addition, $\zeta$ has some special properties that will be important for us.
Under the reflection symmetry of the diamond, $\zeta \rightarrow -\zeta$, so $\alpha\rightarrow-\alpha$.
It follows that $\alpha$ vanishes on $\Sigma$, so $\zeta$ 
acts ``instantaneously" as a true Killing vector on $\Sigma$:
\begin{equation} \label{liezero}
\mathcal L_\zeta g_{ab}  \big|_\Sigma =  0 \, .
\end{equation}
It also follows that $\nabla_a\alpha$ is normal to $\Sigma$, so we have
\beq\label{dLg}
\nabla_c \left (\mathcal L_\zeta g_{ab}  \right)  \big|_\Sigma =  - 2 \dot\alpha\, u_c g_{ab}\, ,\qquad \dot\alpha=u^c\nabla_c \alpha \, , 
 \eeq
 where  $u^c$ is the  future pointing unit normal to $\Sigma$, given by $u^c \partial_c \!=\!  \left (  1 \!-\!(r/L)^2  \right)^{-1/2} \! \partial_t $. 
 Computation of $\alpha$ using the expression  \eqref{ckv2} easily yields
 \begin{align}
 \alpha &= -\frac{L}{R} \sqrt{ \left ( 1-(R/L)^2 \right)  \left ( 1-(r/L)^2 \right)} \sinh ( t/L) \,  ,  \label{alpha} \\
  \dot\alpha\big |_{t=0}&= -\frac{1}{R} \sqrt{1-(R/L)^2}  = \frac{-1}{L \sinh(R_*/L)}\, . \label{alphadot}
 \end{align}
Notice that $\dot\alpha$ is {\it constant} on $\Sigma$. This property
will be crucial for the existence of a geometric form of the first law of causal diamonds.
Further, since 
\beq\label{k}
k:=\frac{d-2}{R} \sqrt{1 - (R/L)^2}
\eeq
is the trace of the extrinsic curvature of the surface $\partial \Sigma$ as embedded in $\Sigma$,
we may write 
\begin{equation}   \label{dotalpha}
 \dot\alpha \big |_\Sigma= -\frac{\kappa k}{d-2} \, ,
 \end{equation}
allowing for a normalization of $\zeta$ with surface gravity $\kappa$ rather than unity. 

While we established  \eqref{dotalpha} by explicit computation, 
it can also be derived by examining derivatives along the
horizon at $\partial\Sigma$, and using the properties that (i) the trace of the extrinsic curvature of 
$\Sigma$ vanishes, and (ii) $\dot\alpha$ is constant on $\Sigma$. 
However, we have not found an underlying geometric reason for the constancy of
$\dot\alpha$ on $\Sigma$. It probably requires maximal symmetry of the spacetime, 
since we checked that $\dot\alpha$ is not constant for the case of a causal diamond in time cross hyperbolic space, $\mathbb R \times \mathbb H^{d-1}$, which also admits a diamond preserving conformal Killing vector.
 In Appendix \ref{appyork} the constancy of $\dot\alpha$ on $\S$ 
is established in a different way.

 \section{Mechanics of causal diamonds in (A)dS}
 \label{sec:diamondsthermo}
 
  In this section we first derive a Smarr formula for causal diamonds in (A)dS, by equating the Noether charge to the integral of the Noether current.  This method of obtaining the Smarr formula is quite general, and it illustrates the origin of the ``thermodynamic volume" term. As an aside, we also show how a finite Smarr formula for AdS black holes can be obtained by subtracting the (divergent) empty AdS Smarr formula from the black hole one.
We then move on to our main objective, which is to  derive the first law of  causal diamonds  in (A)dS. 
First we employ the usual dimensional scaling argument to deduce from the Smarr formula 
a first law for variations between maximally symmetric diamonds. 
Next we employ the Noether current method, used by Wald for the case of black holes\cite{Wald:1993nt,Iyer:1994ys},
as was done for variations of Minkowski space diamonds in Appendix D of \cite{Jacobson:2015hqa}.
Here  we  extend that derivation to  (Anti-)de Sitter space, and include fluid matter (allowing in particular for a variable cosmological constant) as in \cite{Iyer:1996ky}, obtaining a first law that applies to arbitrary variations to nearby solutions.

\subsection{Smarr formula for causal diamonds}
\label{sec:smarr}

We start with deriving a Smarr-like formula for causal diamonds in (A)dS space. We will obtain this relation using a slightly unusual but very general method: equating  the Noether current associated with diffeomorphism symmetry to the exterior derivative of the Noether charge and integrating over the ball $\Sigma$.
Liberati and Pacilio \cite{Liberati:2015xcp} used the same Noether method to derive a Smarr formula for Lovelock black holes. Throughout this section we will employ Wald's Noether charge formalism \cite{Wald:1993nt}.
See e.g.  \cite{Iyer:1994ys,Wald:1999wa} for further details about this formalism. 

To every   Lagrangian $d$-form depending on the dynamical fields $\phi$ there is an associated {\it symplectic potential} $(d-1)$-form $\theta$, defined through
\beq \label{dL}
\delta L = E \, \delta \phi + d \theta (\phi, \delta \phi) \, , 
\eeq
where $E$ is the equation of motion $d$-form, and  tensor indices are suppressed. 
For  a  variation     $\delta \phi =  \mathcal L_\chi \phi$ induced by the flow of a vector field $\chi$, 
there is an associated {\it Noether current} $(d-1)$-form, 
\beq\label{jchi}
j_\chi := \theta (\phi, \mathcal L_\chi \phi) - \chi \cdot L \, .
\eeq 
When  $L$ is diffeomorphism covariant, the variation $\delta_\chi L$ 
produced by $\delta \phi =  \mathcal L_\chi \phi$ is equal to $\mathcal L_\chi L $,
which implies that the Noether current is closed for all $\chi$
when the equation of motion $E=0$ holds, and so is an exact form,
\beq\label{j=dQ}
j_\chi = d Q_\chi.  
\eeq
The $(d-2)$-form $Q_\chi$ is constructed from the dynamical fields together with $\chi$ and its
first derivative, and is called the {\it Noether charge} form.

The Smarr formula comes from the integral version of the identity \eqref{j=dQ},
\beq\label{integral}
\oint_{\partial \R} Q_\chi = \int_\R j_\chi,
\eeq
where $\R$ is a $(d-1)$-dimensional submanifold with boundary $\partial \R$.
For a black hole with bifurcate Killing horizon, $\chi$ can be taken as the horizon generating 
Killing vector, and $\R$ can be taken as a hypersurface 
extending from the bifurcation surface to spatial infinity. 
Since $\mathcal L_\chi \phi=0$  when $\chi$ is a Killing symmetry of all the dynamical fields, the first term of \eqref{jchi}   vanishes.
 For vacuum Einstein gravity, without a cosmological constant, 
 the second term  of \eqref{jchi}
vanishes on shell, so \eqref{integral} reduces to the statement that the Noether charge 
of the horizon is equal to that of the sphere at spatial infinity, both orientations being taken outward (toward larger radius).
This yields the Smarr formula \cite{Smarr:1972kt},
\beq \label{smarr1}
\frac{d-3}{d-2} M -\Omega_{\mathcal H} J=\frac{\kappa}{8\pi G} A,
\eeq
where $M$ is the mass, $J$ is the angular momentum, and $\Omega_{\mathcal H}$ is the 
angular velocity of the horizon, and $A$ is the horizon area.  

For a maximally symmetric causal diamond in Einstein gravity, with or without a cosmological constant, 
we can instead choose the region $\R$ to be the ball $\Sigma$,
and choose the vector field $\chi$ to be the conformal Killing vector $\zeta$ of the 
diamond. The left-hand side of \eqref{integral} is then just a single integral, 
which is the same as the horizon integral 
in the black hole case (having flipped the orientation of the latter),\footnote{\label{sign}The orientation is chosen to be outward, toward  larger radius, according to Stokes' theorem. 
The minus sign arises because the   conformal Killing vector is future  pointing on the inside  and past pointing on the outside of the diamond, which is opposite to the   black hole case. 
}
\beq \label{noetherarea}
\oint_{\partial\S}Q_\z = -\frac{\k}{8\pi G}\, A. 
\eeq
On the other hand,  the contribution from the integral of the Noether current 
on the right-hand side of \eqref{integral} no longer vanishes: since $\z$ is not a Killing vector,
the symplectic potential term in the Noether current \eqref{jchi} is nonzero and, if the cosmological constant is
nonvanishing, the Lagrangian no longer vanishes on shell so the second term in the Noether current is also nonzero.
To evaluate the contribution from the first term in the Noether current, we note that the symplectic potential for Einstein gravity is given by \cite{Burnett:1990,Iyer:1994ys}
\beq
\theta(g,\d g) = \frac{1}{16\pi G} \epsilon_a (g^{ab}g^{ce}-g^{ae}g^{bc})\nabla_e\d g_{bc} \, ,
\eeq
where $\epsilon_{a}  = \epsilon_{a a_2 \cdots a_{d}}$ is the volume form with the first index displayed and the remaining $(d-1)$ indices   suppressed. 
Setting $\d g_{bc}={\cal L}_\z g_{bc}=2\a g_{bc}$, evaluating on $\S$, 
and using \eqref{dLg}, we obtain
\beq
\theta(g,{\cal L}_\z g)|_\S = \frac{(d-1)\dot\alpha}{8\pi G}\,u\cdot\e,
\eeq
and together with \eqref{dotalpha} this yields
\beq \label{thetaterm}
\int_\S\theta(g,{\cal L}_\z g) = -\frac{d-1}{d-2}\frac{\k k}{8\pi G}\,V,
\eeq
where $V=\int_\S u\cdot\e$ is the proper volume of the ball. To evaluate the contribution from the 
second term in the Noether current, we note that the off-shell Lagrangian is 
\beq
L = \frac{R-2\L}{16\pi G}\,\e \, .
\eeq
On shell we have $R-2\L =4\L/(d-2)$, so the on-shell Lagrangian is 
\beq \label{onshellL}
L^{\rm on-shell} = \frac{\L}{(d-2)4\pi G}\,\e,
\eeq
and the second term in the integral of the Noether current is thus 
\beq \label{zLterm}
-\int_\S \z\cdot L =  -  \frac{\L}{(d-2)4\pi G}\,V_\zeta,
\eeq
where
\beq \label{Theta1}
V_\zeta:=    \int_\S \z\cdot \e  \, . 
\eeq
Since $\z$ is orthogonal to $\S$, $V_\zeta$ is just the proper volume of $\S$ weighted locally 
by the norm of the conformal Killing vector, given in \eqref{firstdefoftheta}. 
For the case of a black hole in asymptotically Anti-de Sitter spacetime,
a quantity close to $V_\zeta$ was first identified in \cite{Kastor:2009wy} 
as the variable thermodynamically conjugate to $\L$. (See subsection \ref{AdSBHs} for a discussion of 
that case.) 
For a true Killing vector it is commonly called the {\it thermodynamic volume} \cite{Dolan:2010ha,Cvetic:2010jb}, and we will use that term 
also
in the conformal Killing case. 
 For the  conformal Killing vector (\ref{ckv2}) in dS space $V_\zeta$ is easily found to be
given by
\begin{equation} \label{thetads1}
V_\zeta  
=  \frac{\kappa  L^2 }{R} \left ( V_R^{\text{flat}}  -   \sqrt{1- (R / L)^2 }   \,\, V_R \right)   \, .
\end{equation}
where $V_R^{\text{flat}}=R^{d-1}\Omega_{d-2}/(d-1)$ is the volume of a sphere of radius $R$ in Euclidean space and 
$V_R$ is the proper volume of a ball of  radius $R$ in dS space. 
It follows from the definition \eqref{Theta1} that $V_\zeta$ is positive in both dS and AdS space, although that is not
obvious from the expression in \eqref{thetads1}.  In the flat space limit $R/L \to 0$ it becomes 
$V_\zeta^{\text{flat}}= \kappa R V_R^{\text{flat}}/(d+1)$.

 Combining the two terms (\ref{thetaterm}) and (\ref{zLterm}), the integral of the Noether current is thus
given by 
\beq
\int_\S j_\z =  -\frac{{(d-1)}{\k k}\,V  +  2\L V_\zeta}{(d-2)8 \pi G} \, ,
\eeq
 so \eqref{integral} yields the Smarr formula,
\beq  \label{smarr}
(d-2){\kappa} A= {{(d-1)}{\k k}\,V +  2{V_\zeta} \L } \, . 
\eeq
 At the cosmological horizon of de Sitter space the extrinsic curvature trace $k$ vanishes, hence the Smarr formula reduces to a   relation between the horizon area and the cosmological constant.  In flat space the cosmological constant is zero, so that the formula turns into the trivial connection between the area and the volume. For generic sizes of the causal diamond,  and for a positive cosmological constant, equation (\ref{smarr}) can be checked explicitly by using the formulas (\ref{k}) and (\ref{thetads1}) for $k$ and $V_\zeta$, respectively, and the expression for the cosmological constant of dS space: $\Lambda = +(d-1)(d-2)/ 2L^2$. 
 
\subsubsection{Smarr formula for AdS black holes}
\label{AdSBHs}
 
 As an aside from the main topic of this chapter, in this subsection we discuss briefly how the Smarr
formula for asymptotically AdS black holes \cite{Kastor:2009wy} can be derived using \eqref{integral}. 
In that setting $\chi$ is replaced by the horizon generating  Killing field $\xi$, and the domain of integration $\S$ is from the black hole horizon to infinity. This does not yet yield a meaningful Smarr formula, since 
both $\oint_\infty Q_\xi$ and $\int_\S j_\xi$ diverge.
However, these divergences are the same as those that arise for pure AdS, so by subtracting the 
pure AdS Smarr formula from the AdS black hole Smarr formula, one obtains a finite relation: 
\beq\label{AdSSmarr}
\oint_{\infty} (Q_\xi - Q_\xi^{\text{AdS}} ) - \oint_{\mathcal H} Q_\xi = \int_{\Sigma} j_\xi - \int_{\Sigma'} j_\xi^{\text{AdS}} \, ,
\eeq
where the Noether charge integrals are both outward oriented. The domain of integration $\S$ in the black hole integral on the right extends from the horizon to infinity, while in the pure AdS integral the domain $\S'$ extends across the entire spacetime.

The first integral on the left-hand side of \eqref{AdSSmarr}
 is proportional to 
the (AdS background subtracted) Komar mass and angular momentum, and for Einstein gravity the horizon   integral is proportional to the surface gravity times the horizon area. Using (\ref{onshellL}) and $\theta (g, \mathcal L_\xi g) =0$ for Killing vectors, we find that the  Noether current   is   $ - \L \xi \cdot \epsilon / ((d-2) 4 \pi G)$. Thus the Smarr formula for AdS black holes is \cite{Kastor:2009wy}
\beq
\frac{d-3}{d-2} M - \Omega_{\mathcal H} J= \frac{\kappa A }{8\pi G}  - \frac{ 2  \bar V_\xi \Lambda }{(d-2) 8 \pi G} \, ,
\eeq
where   
\beq
\bar V_\xi := \int_\S \xi \cdot \epsilon - \int_{\S'} \xi^{\text{AdS}}\cdot \epsilon^{\text{AdS}}
\eeq 
is the background subtracted thermodynamic volume.\footnote{ In the literature $\bar V_\xi$ is usually denoted by $\Theta$. Moreover, the  background subtracted thermodynamic volume is   expressed in  \cite{Kastor:2009wy} in terms of  surface integrals of  the Killing potential $(d-2)$-form $\o_\xi$, defined through $\xi \cdot \epsilon = d \o_\xi$, which can be solved at least locally for $\o_\xi$  
because $\xi \cdot \epsilon$ is closed for Killing vectors. Thus,   $\bar V_\xi = \oint_{\infty} (\o_\xi - \o_\xi^{\text{AdS}}) - \oint_\mathcal{H} \o_\xi$, where the orientation is outward (toward larger radius) at both $\infty$ and ${\cal H}$. This agrees with the expression  (22) in \cite{Kastor:2009wy}, up to a minus sign in the definition of $\omega_\xi$. \label{notetheta}}
The relative sign between the area term and cosmological constant term is the same as in the Smarr formula for causal diamonds \eqref{smarr}. For AdS-Schwarzschild, however, the quantity $\bar V_\xi$ is  negative (it is minus the `flat' volume excluded by the black hole, i.e. $\bar V_\xi = - V^{\text{flat}}_{r_{\mathcal H}}$), whereas for causal diamonds  $V_\zeta :=  \int_{\Sigma} |\zeta| dV$  is positive.

\subsection{First law of causal diamonds}
\label{sec:firstlawch2}

 From the Smarr formula one can derive a variational identity analogous to the first law for black holes  using a simple scaling argument (see e.g. \cite{Kastor:2009wy}). In fact, Smarr \cite{Smarr:1972kt} originally derived the relation (\ref{smarr1}) for stationary black holes from the first law of black hole mechanics by using Euler's theorem for homogeneous functions, applied to the black hole mass considered as a function of the horizon area,  angular momentum, and charge.   In the context of causal diamonds, the  area is a function  of the volume  and the cosmological constant   alone, $A(V,\L)$, since $V$ and $\L$ determine a unique diamond up to isometries. It follows from dimensional analysis that  $\lambda^{d-2} A(V,\L) = A (\lambda^{d-1} V, \lambda^{-2} \L)$, where $\lambda$ is a dimensionless scaling parameter. For a   function with this property 
 Euler's theorem     implies
 \beq
 (d-2) A = (d-1)   \left (   \frac{\partial A}{ \partial V}  \right)_{\!\!\L}  \! V - 2 \left (  \frac{\partial A}{\partial \L}  \right)_{\!\!V}  \! \L \, . 
 \eeq 
 Comparing this with the Smarr formula (\ref{smarr}) we find that 
 \beq \label{firstlaw1a}
 \left ( \frac{\partial A}{ \partial V} \right)_{\!\!\L} = k \, , \qquad \left (  \frac{\partial A}{\partial \L}  \right)_{\!\!V}= - \frac{V_\zeta}{\kappa} \, ,
 \eeq
 which yields the   first law for causal diamonds in (A)dS,
 \beq\label{1st1stlaw}
\kappa\, \delta A =  \kappa k\, \delta V - V_\zeta \,\delta \Lambda \, . 
 \eeq
Notice that an increase of the cosmological constant at fixed volume leads to a decrease of the area. This is because  the spatial curvature is increased inside the ball.  
 The fact that the coefficient of $\delta V$ is the extrinsic curvature $k$ can be understood 
by considering a variation of the radius of a ball 
in a fixed, maximally symmetric space. If the proper radius increase is $\d\ell$, the volume increase is $\d V=A\d\ell$, while the area increase is $\d A =  kA\d\ell$, hence $\delta A =k \delta V$.

The first law (\ref{firstlaw1a})   involves only variations of the parameters that characterize  the maximally symmetric causal diamond, and matter fields are not included in this approach because there are no maximally symmetric solutions with  matter (except the cosmological constant). 
The first law can be extended to allow for variations away from maximal symmetry,  
thereby   
 permitting
variations of  
 the matter stress tensor,  
as has been done both  for black holes \cite{Bardeen:1973gs} and 
for de Sitter space \cite{Gibbons:1977mu}. 
We next derive
 such an extended first law by varying the identity \eqref{integral}, 
as was done for vacuum black holes in \cite{Wald:1993nt},   but including matter stress-tensor
variations   as in  Refs. \cite{Iyer:1996ky,Gao:2001ut}. The  variations we consider are  arbitrary variations of the dynamical fields $\phi$ to nearby solutions, while keeping   the manifold,   the  vector field $\zeta^a$ and the surface   $\S$  of the unperturbed  diamond fixed.

The variation of the Noether current \eqref{jchi} is given on shell 
by 
\beq \label{varnoethcurrent}
\delta j_\chi = \omega (\phi, \delta \phi, \mathcal L_\chi \phi) + d (\chi \cdot \theta (\phi, \delta \phi)),
\eeq
where $\omega (\phi, \delta_1 \phi, \delta_2 \phi) := \delta_1 \theta (\phi, \delta_2 \phi) - \delta_2 \theta (\phi, \delta_1 \phi)$ is the \emph{symplectic current} $(d-1)$-form. 
The variation of the integral identity \eqref{integral} thus yields
\beq\label{omega-id}
\int_\Sigma \omega (\phi, \delta \phi, \mathcal L_\chi \phi)   =   \oint_{\partial \Sigma}  \left [ \delta Q_\chi - \chi \cdot \theta (\phi, \delta \phi) \right ] . 
\eeq
This relation holds provided the background equations for all the fields and the linearized constraint equations associated with the diffeomorphism generated by $\chi$ are satisfied on the hypersurface $\Sigma$.\footnote{The fact that \eqref{omega-id} 
invokes only the (linearized) initial value constraint equations (as opposed to    linearized dynamical field equations), is explained in the Appendix of \cite{Iyer:1995kg} and 
Appendix B of \cite{Faulkner:2013ica}.} The left-hand side of \eqref{omega-id} is the symplectic form on the (covariant) phase space of solutions
which, by Hamilton's equations,\footnote{The variation of a Hamiltonian $H$ for a general dynamical system is related 
to the symplectic form $\omega$ on phase space and the flow vector field 
$X_H$ of the background solution via Hamilton's equations, 
$dH(v) = \omega(v,X_H)$, where $v$ is any tangent vector on phase space. 
In the present case $v$ corresponds to   $\delta \phi$, $X_H$ to $\mathcal L_\chi \phi$, and  $dH(v)$ is written as $\delta H$ \cite{Lee:1990nz}. }
is equal to the variation of the Hamiltonian, 
\beq\label{dH}
\delta H_\chi =   \int_\Sigma \omega (\phi, \delta \phi, \mathcal L_\chi \phi) \,. 
\eeq
Equation \eqref{omega-id} thus yields the on-shell identity relating the Hamiltonian variation 
to the Noether charge variation and the symplectic potential, 
\begin{equation} \label{varid}
\delta H_\chi = \oint_{\partial \Sigma} \left [ \delta Q_\chi - \chi \cdot \theta (\phi, \delta \phi) \right]\, .
\end{equation}
If $\chi$ is a true Killing vector of the background metric and matter fields, 
then \eqref{dH} implies $\delta H_\chi=0$, so the variational identity 
reduces to a relation between the boundary integrals.
This is how the first law of black hole mechanics arises \cite{Wald:1993nt,Iyer:1994ys}:
taking $\Sigma$ to be a hypersurface bounded by 
the black hole horizon and spatial infinity,  the identity relates the variation
of the horizon Noether charge to the variations of total energy and angular momentum. 

A special case for the first law of causal diamonds 
is the first law of a static patch of dS space \cite{Gibbons:1977mu},
which in vacuum is just the statement that the variation of the area of the de Sitter horizon vanishes. 
The first law derived by Gibbons and Hawking allowed for variations in the Killing energy of matter,
but matter contributions do not appear in $\delta H_\chi$ if the matter is described by fields 
 that appear quadratically 
in the Lagrangian 
 and vanish in the de Sitter background.
 However, for matter described by a diffeomorphism invariant fluid theory
first order variations of the matter stress tensor can arise, 
and because the fields are potentials which do not share the background 
Killing symmetry enjoyed by the stress tensor, 
a volume contribution containing the matter Killing energy 
appears in the variational relation \cite{Bardeen:1973gs, Iyer:1996ky}. 
In the  derivation of the first law for causal diamonds below we also allow for variations of fluid matter fields.

We consider the case where   
the  gravitational theory is general relativity, the matter sector consists of  minimally coupled fluids with arbitrary equation of state, the background metric
is pure dS, and the vector field $\chi$ is the conformal Killing vector $\zeta$ of a causal diamond.\footnote{Many steps in the derivation below remain valid for other conformally flat solutions. First of all, causal diamonds in conformally flat spacetimes still allow for a unique conformal Killing field whose flow preserves the diamond. Moreover, the diamonds  still have a reflection symmetry around $\Sigma$, so that $\mathcal L_\zeta g_{ab} = 0 $ on~$\Sigma$. However,   $\dot{\alpha}$ might not be constant in other spacetimes, so all the equations up to (\ref{symplecticonSigma}) hold, but not \eqref{metrichamfinal}, since  the  relation \eqref{dotalpha} for $\dot \alpha$ might be specific to maximally symmetric spacetimes.}
One of the fluids describes the cosmological constant, with equation of state $p = - \rho = - \Lambda/(8\pi G)$.
Since $\zeta$ is zero at the edge $\partial \Sigma$ of the diamond, the second term on the right-hand side in \eqref{varid} vanishes.
The surface integral of $\delta Q_\zeta$ in this case  is 
%
\begin{equation} \label{noether}
 \oint_{\partial \Sigma} \delta Q_\zeta = - \frac{\kappa }{8\pi G} \delta A \, ,
\end{equation}
where $\kappa$ is the surface gravity and $A$ is the area of the bifurcation surface $\partial \Sigma$.\footnote{The minus sign appears for the same reason as in (\ref{noetherarea}),  which is explained in footnote \ref{sign}.}
With this result, the identity \eqref{varid} takes the form 
\begin{equation} \label{dA=-dH}
\delta H_\zeta = -\frac{\kappa }{8\pi G} \delta A \, . 
\end{equation}
 For the present field content  the variation of the total Hamiltonian splits into a (nonvanishing) term associated to the background metric and one associated to the matter fields 
\begin{equation} \label{separation}
\delta H_\zeta = \delta H_\zeta^{\text{g}} + \delta H_\zeta^{\text{m}} \, .
\end{equation}
In the following we will first evaluate the gravitational term and then the matter term. The result will be that the gravitational term is proportional to minus the
variation of the volume of $\Sigma$,   the matter term contains a term proportional to the thermodynamic volume times the variation of the cosmological constant as well as 
 the variation of the canonical Killing energy for the other fluids.

We evaluate $\delta H^\text{g}$ through its relation to the symplectic form (\ref{dH}). For general relativity the symplectic current  takes the form  \cite{Crnkovic:1986ex, Burnett:1990}  
\begin{equation} \label{symp}
\omega  (g, \delta_1 g, \delta_2 g)= \frac{1}{16\pi G} \epsilon_a P^{abcdef} \left ( \delta_2 g_{bc} \nabla_d \delta_1 g_{ef}   -  \delta_1 g_{bc} \nabla_d \delta_2  g_{ef}    \right),   
\end{equation}
with
\begin{equation}
P^{abcdef}= g^{ae} g^{bf} g^{cd} -  \frac{1}{2} g^{ad} g^{be} g^{cf} - \frac{1}{2} g^{ab} g^{cd} g^{ef}  - \frac{1}{2} g^{bc} g^{ae} g^{fd}  + \frac{1}{2} g^{bc} g^{ad} g^{ef}  .
\end{equation}
Note that in (\ref{dH}) the symplectic current is evaluated on the Lie derivative of the fields  along $\zeta$. If $\zeta$ were  a Killing vector, the metric contribution $\delta H_\zeta^\text{g}$  would  hence  vanish, as it does when deriving the first law of black hole mechanics \cite{Wald:1993nt}. 
However, since  for a diamond $\zeta$ is only a {\it conformal} Killing vector,   
$\delta H_\zeta^{\text{g}}$ makes a nonzero contribution to the first law.
When 
$\delta_1g=\delta g$ and $\delta_2 g=\mathcal L_\zeta g$, 
the first term in  (\ref{symp})  is zero at $\Sigma$, 
 since  $\mathcal L_\zeta g|_\S=0$ \eqref{liezero}.
Using \eqref{dLg} the second term yields 
\begin{equation}
\omega (g, \delta g, \mathcal L_\zeta g)  \big |_\Sigma = - \frac{(d-2)\dot\alpha}{ 16\pi G}  \epsilon_a  \! \left (h^{ab} u^c - h^{bc} u^a  \right) \delta g_{bc} \,, 
\end{equation}
where $h_{ab}:= g_{ab} + u_a u_b$ is the induced metric on $\Sigma$ and $u_a$ is the unit normal to $\S$. 
Only the pullback of $\omega$ to $\S$ is relevant in the integral in \eqref{dH}.
 Using
\beq\label{epullback}
\epsilon_a\big |_\Sigma= - u_a (u \cdot \epsilon),
\eeq
  this pullback can be simplified as  
\begin{equation}  \label{symplecticonSigma}
\omega (g, \delta g, \mathcal L_\zeta g)  \big |_\Sigma =\frac{ (d-2)\dot\alpha}{16\pi G} (u\cdot \epsilon) \, h^{bc} \delta h_{bc} = 
\frac{ (d-2)\dot\alpha}{8\pi G} \delta (u\cdot \epsilon) \, ,
\end{equation}
where pullback of all forms to $\Sigma$ is implicit.
%
The metric contribution to $\delta H_\zeta$ is therefore equal to
\begin{equation} \label{metrichamfinal}
\delta H_\zeta^{\text{g}} =  \frac{d-2}{8\pi G} \int_\Sigma  \dot{\alpha}\,\delta (u\cdot\epsilon) 
=  - \frac{\kappa k}{8 \pi G} \delta V \, , 
\end{equation}
where  $V= \int_\Sigma   u \cdot \epsilon $ is the proper volume of $\S$, and in the last equality we used (\ref{dotalpha})
and  the fact that 
$\dot\alpha$   is constant over $\Sigma$. 
The constancy of $\dot\alpha$ is hence crucial for arriving at 
an intrinsic geometric quantity, the variation of the proper volume.

Combining \eqref{dA=-dH}, \eqref{separation} and \eqref{metrichamfinal},  we find the extended first law for causal diamonds, which includes a variation of the matter Hamiltonian,
\beq   \label{finalfirstlaw'}
  \delta H^{\text{m}}_\zeta  = \frac{\kappa}{8\pi G} \left ( -\delta A + k \delta V \right) \,.
\eeq
Next, we compute the matter Hamiltonian variation  explicitly  through its relation with the symplectic form. 

The precise on-shell relation between the symplectic current $\omega^\text{m}$ and the   Noether current $j^\text{m}_\chi$ for matter fields is \cite{Iyer:1996ky}
\begin{equation} \label{onshellcurrents}
\omega^{\text{m}} (\phi, \delta \phi, \mathcal L_\chi \phi) = \delta j^\text{m}_\chi + \frac{1}{2}  \chi \cdot \epsilon  \,  T^{ab} \delta g_{ab} - d (\chi \cdot \theta^\text{m} (\phi, \delta \phi)) \, .
\end{equation}
Here, $T^{ab}$ is the  Hilbert  stress-energy tensor   defined through the matter Lagrangian.\footnote{In particular, the variation of the matter Lagrangian $d$-form with respect to the matter fields $\psi $ and the metric $g_{ab}$ is given by: $\delta L^\text{m}  = E^\text{m} \delta \psi + \epsilon \frac{1}{2} T^{ab} \delta g_{ab} + d \theta^\text{m} (\phi, \delta \phi)$, where $E^\text{m}=0$ are the matter equations of motion, $T^{ab}$ is the stress-energy tensor, and $\theta^\text{m}$ is the symplectic potential associated to $L^\text{m}$ \cite{Iyer:1996ky}.}
Compared to the equivalent identity \eqref{varnoethcurrent} for all the dynamical fields, we see that in the identity above for the matter sector the term involving the stress-energy tensor is new. This term arises from the metric variation of the matter Lagrangian. A similar identity exists for the pure metric sector, with the extra term being $-(1/16\pi G) \chi \cdot \epsilon \, G^{ab}\delta g_{ab}$, so that  (\ref{varnoethcurrent})   holds when the pure metric and matter sectors are combined and the metric equation of motion is imposed.

Further, the Noether current for matter fields is on shell given by  \cite{Iyer:1996ky}
\beq \label{matterNoethercurrent}
j_\chi^\text{m} = dQ_\chi^\text{m}  - T_a{}^b  \chi^a \epsilon_b \, .
\eeq
The stress-energy term appears on the right-hand side because only the full Noether current $j_\chi = j^\text{g}_\chi + j^\text{m}_\chi$ is an exact form on shell.
Inserting (\ref{matterNoethercurrent}) into the variational identity (\ref{onshellcurrents}) and using Hamilton's equations (\ref{dH}), we find that the   matter Hamiltonian variation is
\beq \label{matterHamvariation}
\delta H^\text{m}_\chi = \oint_{\partial \S} \left [ \delta Q_\chi^\text{m} - \chi \cdot \theta^\text{m} (\phi, \delta \phi )\right]   + \int_\Sigma \left[ - \delta 
(T_a{}^b  \chi^a \epsilon_b)  + \frac{1}{2}     \chi \cdot \epsilon \, T^{ab} \delta g_{ab}  \right]  .
\eeq
This equality is true for an arbitrary smooth vector field $\chi$ on spacetime, and holds provided the field equations   and the linearized equations of motion are satisfied for the matter fields, i.e. $E^\text{m} = \delta E^\text{m} = 0$. 

We now specialize to the conformal Killing vector  $\zeta$ that preserves  a   diamond in (A)dS (the analysis below is actually valid in any conformally flat spacetime). Since $\zeta=0$  at   $\partial \Sigma$, the second term in the boundary integral in (\ref{matterHamvariation}) vanishes.  In addition, the Noether charge variation also does not contribute at the bifurcation surface $\partial \S$. This is because for a generic diffeomorphism invariant Lagrangian the Noether charge ($d-2$)-form can be expressed as $Q_\zeta= W_c (\phi) \zeta^c + X^{cd} (\phi)\nabla_{[c} \zeta_{d]}$ \cite{Iyer:1994ys}.\footnote{Here we have fixed the ambiguity in the definition of the Noether charge, coming from the freedom to shift the symplectic potential $\theta$  by an exact form $d Y(\phi, \delta \phi)$, such that $Y(\phi, \mathcal L_\zeta \phi)=0$. If one were to allow for a nonzero $Y$ form, then the first law would not  be modified, since the Noether charge variation (together with symplectic potential) associated to matter fields in (\ref{matterHamvariation}) cancels anyway in the variational identity (\ref{varid}) against an identical term on the  right-hand side. The cancellation  of $\oint_{\partial \S}[\delta Q^\text{m}_\chi - \chi \cdot \theta^\text{m} (\phi, \delta \phi)]$ in the first law  was pointed out by Iyer in \cite{Iyer:1996ky}.} The first term vanishes at $\partial \S$, and the second term involves a form $X^{cd}$, which is purely constructed from derivatives of the Lagrangian with respect to the Riemann tensor (and its covariant derivatives). For minimally coupled matter fields, this form does not receive contributions from the matter sector, so $Q^\text{m}_\zeta=0$ at $\partial \Sigma$ for  the present field content (and also  $\delta Q^\text{m}_\zeta$ vanishes at $\partial \Sigma$).

 The matter Hamiltonian variation on the maximal slice $\Sigma$ in a diamond  is therefore given by
\begin{equation} 
\delta H^\text{m}_\zeta =   \int_\Sigma\left [  -  
\delta   ( T_a{}^b \zeta^a \epsilon_b  ) + \frac{1}{2}   \zeta \cdot \epsilon \, T^{ab} \delta g_{ab} \right] \,  ,
\end{equation}
 which can be rewritten, using $\delta \epsilon_b = \frac{1}{2} \epsilon_b   g^{cd} \delta g_{cd}$,  as
a sum of stress tensor and metric variation terms,
\beq \label{newhamvar}
\delta H^\text{m}_\zeta = \int_\S \left [   - \delta {T_a}^b  + \frac{1}{2} ( {\delta_a}^b T^{cd} - {T_a}^b g^{cd}) \delta g_{cd}  \right]  \! \zeta^a \epsilon_b\, .
\eeq
Notice that the trace part drops out of the second term.\footnote{We should have been able to anticipate this feature, but have not yet found a way to do so.}
 In a maximally symmetric background  the tracefree part of the stress tensor must vanish, hence the  
matter Hamiltonian variation takes the form
\beq \label{newhamvar2}
\delta H^\text{m}_\zeta = -\int_\S \delta {T_a}^b  \zeta^a \epsilon_b\, ,
\eeq
which receives contributions from all types of matter.\footnote{Using \eqref{epullback}, the integrand becomes  
$\delta {T_a}^b  \zeta^a u_b\,u\cdot\e$, and the unfamiliar minus sign in \eqref{newhamvar2} disappears.} 
%

 Since the cosmological constant can be obtained from a field or fields covariantly coupled to the metric, its
variation falls within the class of ``matter" variations to which the first law \eqref{finalfirstlaw'} applies, and so 
may be included in \eqref{newhamvar2}.
To separate out this contribution, we split the matter stress tensor as 
\beq
T_{ab}  =T_{ab}^\text{\~m}  + T_{ab}^\L \, , 
\eeq
 where $T_{ab}^\text{\~m}  $ is the stress-energy tensor of  
 matter other than the cosmological constant,
 and $T^\L_{ab}=-(\L/8\pi G) g_{ab}$  is the ``vacuum" energy-momentum tensor corresponding to the 
 cosmological constant.
The contribution of the $\Lambda$ term to the variation of the Hamiltonian is thus  
\beq  \label{Theta}
\delta H^\Lambda_\zeta = \frac{V_\zeta \,  \delta \Lambda}{8 \pi G}   \,   ,
\eeq
where $V_\zeta$ is the thermodynamic volume  defined in \eqref{Theta1}.
 Note that   $V_\zeta$ is not varied, 
since the metric variation was already separated out
 in \eqref{newhamvar}.

In conclusion, by inserting   $\delta H^\text{m}_\zeta  = \delta  H^\text{\~m}_\zeta + \delta H^\L_\zeta$ 
and (\ref{Theta}) into   \eqref{finalfirstlaw'}, we arrive at the final form of the  first law of causal diamonds, 
%
%
%
 \begin{equation}\label{firstlawcc1}
  \delta   H^\text{\~m}_\zeta 
   = \frac{1 }{8 \pi G } \left ( - \kappa \, \delta A +   \kappa k   \,  \delta V  - V_\zeta  \,   \delta \Lambda  \right)   .
   \end{equation} 
We remind the reader of what all these symbols represent:  
$    H^\text{\~m}_\zeta $ is the conformal Killing energy of 
matter 
other than the cosmological constant $\Lambda$, $\kappa$ is the surface gravity, $A$ is the area
of the edge $\partial\S$, $k$ is the trace of the (outward) extrinsic curvature of $\partial \S$ as embedded in the maximal slice $\S$, $V$ is
the proper volume of the maximal slice, and $V_\zeta $ is the proper volume weighted locally by the norm of the conformal Killing vector $\zeta$.

The derivation above also goes through for causal diamonds in AdS.
Note that the form of the first law is the same for (A)dS as for Minkowski space, except that all the quantities should now be evaluated in (A)dS.  Hence, we  have established a    variational identity  in general relativity which holds for spherical regions of any size in maximally symmetric spacetimes. 



 \newpage

\section{Thermodynamics of causal diamonds in (A)dS}
\label{sec:thermo}

 As is well known   black holes admit a true thermodynamic interpretation.
In this section we will     explore to what extent the same is true for causal  diamonds in (A)dS. 
We will also relate the first law of causal diamonds to the entanglement equilibrium proposal in \cite{Jacobson:2015hqa}.

\subsection{Negative temperature}
\label{negativetemp}

Like the first law of black hole mechanics, and its generalizations mentioned in the introduction, 
the first law of causal diamonds \eqref{dA=-dH}, 
\beq \label{newfirstlaw}
 \delta H_\zeta = - \frac{\kappa}{8\pi G} \delta A,
\eeq
admits a thermodynamic interpretation. 
What is unusual, however, is the minus sign on the right-hand side. 
The $\kappa  \, \delta A$ term   is  usually  identified with a $T_\text{H} \delta S_{\text{BH}} $ term, where $S_{\text{BH}} = A/4\hbar G$ is   the Bekenstein-Hawking entropy and $T_\text{H} = \kappa \hbar / 2\pi$ is the  Hawking  temperature.
However, in the present context this identification calls for a negative temperature\footnote{The conceptual possibility of negative absolute temperature was  discussed for the first time by Afanassjewa  in 1925 \cite{Ehrenfest-Afanassjewa1925}. In 1951  Purcell and Pound  \cite{PurcellPound} prepared and measured a nuclear spin system  at negative temperature in an external magnetic field.  Subsequently, the thermodynamic and statistical mechanical implications of negative temperature were studied in detail by Ramsey   \cite{Ramsey1956}. We thank Jos Uffink for  bringing the work of Afanassjewa to our attention \cite{Uffink}. 
See \cite{NegativeTemp} for a recent review of negative temperature.}
\beq \label{eqnegativetemp}
T = - T_{\rm H},
\eeq
%
because an increase of conformal Killing energy in the diamond is associated with a {\it decrease} of horizon entropy.\footnote{It was suggested in Ref.~\cite{Spradlin:2001pw} that in global de Sitter space 
 this minus sign (which was there attached to the entropy rather than to the temperature) 
results from imposing the first law with the energy inside the horizon, rather than the energy outside the horizon
which is the negative of the former. The sensibility of this proposal is debatable, since the opposite sign
of the energy results from the fact that the Killing vector is past-oriented outside the horizon. 
Moreover, in flat spacetime or AdS the energy outside the horizon is not the negative of that inside, and yet we still encounter the same minus sign.}
This negative temperature interpretation has previously been suggested by Klemm and Vanzo \cite{Klemm:2004mb} in the special case of the static patch of de Sitter space, and was recently advocated  in the context of multiple Killing horizons in \cite{Cvetic:2018dqf}, where the cosmological event horizon was also assigned a negative (Gibbsian) temperature.\footnote{The arguments in \cite{Cvetic:2018dqf} appealed to the fact that the surface gravity of the cosmological horizon is negative. Although not stated in \cite{Cvetic:2018dqf} (nor elsewhere in the literature that we are aware of), this holds only on the {\it past} cosmological horizon (if we take the Killing vector to be future pointing on the past and future horizons).
The surface gravity of the {\it future} cosmological horizon is {\it positive}. 
(The surface gravity is positive (negative) if the Killing vector  
is stretched (shrunk) with respect to affine parameter along the Killing flow on the Killing horizon.)
A varied diamond can be viewed as the result of a physical process in which a perturbation has passed   through the past horizon, and  entered what would otherwise have been a maximally symmetric diamond. The surface gravity   
of the {\it past} horizon should thus be expected to play the role of temperature in the first law.} 
 Negative temperature typically requires of a system that  i) its energy spectrum is bounded above and  ii) the Hilbert space is finite-dimensional. Klemm and Vanzo have argued that these requirements are indeed satisfied for the de Sitter space static patch.\footnote{The proposal that  the  Hilbert space of an observer's patch in asymptotic de Sitter space  is finite-dimensional  is due to Banks and Fischler \cite{Banks:2000fe,Fischler:2000}.} 
   Their arguments can actually be applied to all causal diamonds: i) the mass inside is bounded above by the mass of the largest black hole that fits inside 
 a diamond with a given boundary area
   and ii) the entropy associated to the horizon is finite due to the holographic principle or covariant entropy bound   \cite{Bousso:1999xy,Bousso:2002ju}.  
 It therefore 
   seems feasible that causal diamonds  have a negative temperature in quantum gravity.

Using    the   negative temperature \eqref{eqnegativetemp},   the first law \eqref{newfirstlaw} can   be written as  
 \beq\label{dH=TdS}
 \delta  H_\zeta  = T \delta S_{\text{BH}},
 \eeq 
%
which is a standard thermodynamic relation between energy, temperature and entropy. 
 As a special case, we find that the static patch of de Sitter space  has a negative temperature (see Section \ref{sec:desitter}). This is in apparent conflict with the positive Gibbons-Hawking temperature for dS space, computed using quantum field theory on a fixed background \cite{Gibbons:1977mu}.  In Section \ref{sec:qc} below we 
shall
propose a resolution to this apparent conflict involving the quantum corrections to the first law, but first we want to further discuss the   thermodynamic interpretation of the leading order classical quantities in this relation.





Instead of writing the  cosmological constant term   \eqref{Theta}
 in $\d H_\z$ as an energy variation, we can also take it to the right-hand side of the first law  and write it as  the thermodynamic volume times the   variation of the pressure, i.e. $V_\zeta \, \delta p$. This is because 
the cosmological constant can be interpreted both as an energy density,  
$\rho =  \Lambda/ 8\pi G$,  and as a pressure $p = - \Lambda / 8 \pi G$.
In this way 
\eqref{dH=TdS} can be expressed as  
 \beq \label{firstlawpressure}
 \delta  H^{\rm g +  \rm  \tilde m}_\zeta  = T \delta S_{\text{BH}} + V_\zeta \,  \delta p\,,
 \eeq
 where $\rm g$ labels the gravitational contribution  \eqref{dHg} to the Hamiltonian variation, and $\rm \tilde m$ labels the matter contribution \eqref{matter1} other than the cosmological constant. 
This form of the first law suggests that $H^{\rm g +  \rm  \tilde m}_\zeta  $ 
is an enthalpy, rather than 
an energy, just like the ADM mass for black holes \cite{Kastor:2009wy}. The matter Hamiltonian vanishes in the background, so the enthalpy of causal diamonds in Minkowski  and (A)dS space is $H_\zeta^{\rm g}$, which is defined above only through its variation. We leave it to future work to evaluate $H_\zeta^{\rm g}$ itself.


Through a Legendre transformation, $U = H^{\rm g +  \rm  \tilde m}_\zeta  - p \, V_\zeta$, the first law can be written in the standard form
\beq
\delta U = T \delta S_{\rm BH} - p \,  \delta V_\zeta \, , 
\eeq
where $U$ 
 plays the role of
the internal energy associated to causal diamonds. 
Using the equation of state $p=-\rho$, the contribution of the $\L$ term 
to $U$ may be expressed as $\rho \, V_\zeta$, which is the redshifted vacuum energy associated to the cosmological constant.
(See Section \ref{sec:desitter} for a similar discussion for the special case of the de Sitter static patch.)



\subsection{Quantum corrections}

\label{sec:qc}



 The first law can be extended into the semiclassical regime by considering quantum matter fields 
 (instead of classical fields) on a classical background spacetime.
The ``quantum corrected'' first law of causal diamonds reads
 \beq \label{quantumfirstlaw1}
 \delta \langle H^{\rm \tilde m}_\zeta \rangle  + \delta H_\zeta^{\rm g + \Lambda}= T \delta S_{\rm BH} \, ,
 \eeq
 which can be derived (along the lines of Section \ref{sec:firstlawch2}) from the semiclassical Einstein equation, where the stress-energy tensor is replaced by its quantum expectation value but the metric is kept classical. Our aim in this section is to show how this first law can be written in terms of the variation of Bekenstein's generalized entropy \cite{Bekenstein:1973ur} --- defined in \eqref{genent} --- which at the same time explains why the negative temperature $T$ is  consistent with the positive Gibbons-Hawking temperature $T_{\rm H}$.
 We will first restrict the discussion to conformal matter and then generalize it to any quantum matter.

\subsubsection{Conformal matter}

In the particular case of the   vacuum state of a conformal quantum field theory,   the   matter Hamiltonian $H^{\rm \tilde m}_\zeta$ associated to a spherical region $\Sigma$ in flat space or (A)dS is equal to the so-called modular Hamiltonian $K$,
i.e.\
\beq \label{modHam1}
H^{\rm \tilde m}_\zeta : = \int_\Sigma ({T_a}^b)^{\rm \tilde m} \zeta^a u_b dV  =   K \, , 
\eeq
where  $K$ 
is  the operator implicitly defined   via 
the reduced density matrix of the vacuum restricted 
to the region $\Sigma$, $\rho_{\rm vac} = e^{-K/T_H}/Z$.\footnote{It is well known that the reduced density matrix of the vacuum  in   Rindler space is thermal with respect to the Lorentz boost Hamiltonian \cite{Haag:1992hx}. The thermal behavior of conformal quantum fields in a global vacuum state inside   maximally symmetric causal diamonds  can be derived from the Weyl equivalence between these diamonds and Rindler space. In Appendix \ref{appyork}
we show explicitly that all maximally symmetric diamonds are Weyl equivalent to conformal Killing time cross hyperbolic space, $\mathbb R \times \mathbb{H}^{d-1}$, and therefore to each other. It is also known that a diamond in flat space is Weyl equivalent to 
Rindler space, see Appendix~\ref{sec:conftrans}.
 Therefore, the reduced density matrix of the conformal vacuum on a diamond in (A)dS and flat space is thermal. For the special case of diamonds in flat space this was discussed before in \cite{Hislop1982,Haag:1992hx,Martinetti:2002sz,Casini:2008cr,Casini:2011kv}.}
For infinitesimal variations of the reduced density matrix, the variation of the expectation value of the modular Hamiltonian 
is equal 
to the variation of 
the matter entropy $S_{\rm \tilde m}$,
by
%
\beq\label{dent}
\delta \langle K \rangle=T_{\rm H}\delta S_{\rm \tilde m},
\eeq
with the {\it positive} sign for the temperature $T_{\rm H}$.  
 If the variation is to a global pure state $\delta S_{\rm \tilde m}$ is purely entanglement entropy,
which is why this is known as the ``first law of entanglement''.

Initially it appears that this opposite sign for the temperature indicates an inconsistency:  if $\delta \langle H^{\rm \tilde m}_\zeta \rangle$ were added to the rest of the conformal energy variation $\delta H_\zeta$, and at the same time $\delta S_{\rm \tilde m}$ were added to $\delta S_{\rm BH}$ in the classical first law \eqref{dH=TdS}, then --- because of the mismatch in the signs of the temperatures --- the  first law would   no longer be valid for any temperature. However, in gravitational thermodynamics, it would  be incorrect to add both the energy term and the entropy term. The derivation of the {\it gravitational} first law follows from diffeomorphism invariance and the gravitational field equation, and the matter entropy does not enter in the derivation \cite{Wald:1993nt,Iyer:1996ky}. On the other hand, this first law must be consistent with the {\it thermodynamic} first law, so it must also be possible to take into account the matter entropy. 

The way this works is perhaps easiest to understand in the setting of an asymptotically flat  black hole \cite{Bardeen:1973gs}. The gravitational first law indicates that a matter energy variation increases the variation of the total ADM mass $M$, for a fixed value of the horizon area. On the other hand, the energy variation of a thermal fluid can be expressed in terms of its {\it entropy} and {\it particle number} variations. If that is done, the fluid registers {\it explicitly} in the first law only via those variations, yet it also registers {\it implicitly} in the $\delta M$ term, to which it contributes via the constraints. When the first law is expressed in this way, the total entropy variation appears explicitly, and is related to the total energy variation. When expressed instead using the energy of the fluid rather than its entropy, the first law is ``purely gravitational," and in particular refers only to the horizon area contribution to the entropy. It is quite peculiar to gravitational thermodynamics \cite{Martinez:1996vy} that the first law has simultaneously these two different meanings, one gravitational and one thermodynamical. And, for the same reason, the fluid contribution  does not enter explicitly {\it both} as an energy contribution and as an entropy contribution, unlike in ordinary thermodynamics.


It thus seems that the correct procedure is to add {\it only} the matter energy variation   
$\delta \langle H^{\rm \tilde m}_\zeta \rangle$
to the classical first law,  as we anticipated in \eqref{quantumfirstlaw1}, where  the classical matter Hamilonian variation $\delta H^{\rm \tilde m}$ is   \emph{replaced} by its expectation value. 
If desired, one can use \eqref{modHam1} and \eqref{dent} to express $\delta \langle H^{\rm \tilde m}_\zeta \rangle$  in terms of the 
 matter
entropy variation. 
  Thanks to the opposite sign of the temperature in \eqref{dent}, this can
then be combined with $\d S_{\rm BH}$ in \eqref{quantumfirstlaw1} 
to form
the variation of the {\it generalized entropy},
%
\beq \label{genent}
S_{\rm gen} := S_{\rm BH} + S_{\rm \tilde m} \, ,
\eeq
%
in terms of which the quantum corrected first law  of causal diamonds becomes
\beq \label{quantumfirstlaw2}
\d H^{\rm g+\Lambda}_\z  = T \d S_{\rm gen} \,. 
\eeq
In fact, this appears to be a more satisfactory formulation of the first law because, 
unlike the matter entropy and Bekenstein-Hawking entropy separately,
the  generalized entropy is plausibly invariant under a change of the 
UV cutoff  
(see the appendix of \cite{Bousso:2015mna} for a discussion of this idea).

In this way, we see that the opposite sign of the temperature in \eqref{dH=TdS} and \eqref{dent} is precisely what is needed in order for the Bekenstein-Hawking entropy and matter entropy to combine as $S_{\rm gen}$. At least for conformal fields, this resolves the apparent conflict alluded to before between the negative temperature in the first law for causal diamonds and the positive temperature in the first law of entanglement.  The negative and positive temperature  seem compatible with each other, since   the first law of entanglement \eqref{dent} and the quantum corrected first law of causal diamonds \eqref{quantumfirstlaw2} are valid simultaneously.


\subsubsection{Non-conformal matter}
\label{ncm}

We now show how the generalized entropy variation can   be obtained in the first law for generic quantum fields. For non-conformal fields the matter Hamiltonian is not equal to the modular Hamiltonian, and hence the    term $   \delta \langle H_\zeta^{\rm \tilde  m} \rangle$ cannot be directly related to the entanglement entropy variation. 
  However, in \cite{Jacobson:2015hqa} it was postulated that,  
  for causal diamonds that are small compared to the local curvature scale, the length scale of the quantum state, and any  length scale in the quantum field theory 
  defined by a relevant deformation of a conformal field theory,
  this term is in fact  related to the variation of the expectation value of the modular Hamiltonian 
 via
\beq \label{conja}
 \delta \langle H_\zeta^{\rm \tilde  m} \rangle  = \delta \langle K \rangle  -  V_\zeta \,  \delta X \, .  
\eeq
Here  $X$ is a spacetime scalar that can depend on the size of the diamond but is  invariant under Lorentz boosts that leave the center of the diamond fixed.  The thermodynamic volume $V_\zeta$ has been factored out for later convenience.\footnote{For small diamonds the conformal Killing energy variation    can be approximated by $\delta \langle H^{\rm \tilde m}_\zeta \rangle = V_\zeta \delta \langle T_{00}^{\rm \tilde m} \rangle$ and the thermodynamic volume is to first order given by $V_\zeta = \kappa  \, \Omega_{d-2} R^d /(d^2 -1)$ (see   Section \ref{item:small}). Inserting these approximations into \eqref{conja} yields the actual conjecture   (21) in \cite{Jacobson:2015hqa}.} The modular Hamiltonian $K$ is here defined for the vacuum   of a quantum field theory restricted to a ball-shaped region, and the variation denotes a perturbation of   the vacuum state.
The  assumption \eqref{conja} was checked in \cite{Casini:2016rwj,Speranza:2016jwt}, and it was found in particular that   $\delta X$ may depend on the radius $R$  of the ball, and can dominate at small $R$ (depending on the conformal dimension of the operator that deforms the CFT).  

With the postulate \eqref{conja} and  the first law of entanglement \eqref{dent}, the quantum first law \eqref{quantumfirstlaw1} can be written in terms of the generalized entropy variation as
\beq
\d H^{\rm g+\lambda}_\z - V_\zeta \,  \d X 
= T \d S_{\rm gen} \, . 
\eeq
Here, we  have   denoted the local cosmological constant in a small maximally symmetric diamond by $\lambda$, in order to  distinguish it from the total cosmological constant variation to be introduced below.
The variations $\delta H^{\rm g}$ and $\delta H^{\l}$ are  explicitly given by \eqref{dHg} and \eqref{firstdefoftheta}, respectively, so we can further rewrite this first law as 
\beq \label{genentlambda}
- \frac{\kappa k}{8\pi G} \d V + \frac{V_\zeta}{8\pi G} (\d\lambda - 8\pi G \d X ) = T \d S_{\rm gen} \, . 
\eeq
 Now, since the  $\delta X$ contribution from non-conformal matter 
 appears together with the local cosmological constant variation
 $\d \l$ in this way, we may combine them into one
 net variation,
%
\beq \label{effectivecc}
 \delta  \Lambda :=  \delta \lambda  - 8 \pi G \delta X ,
\eeq
%
in terms of which \eqref{genentlambda} is expressed as
\beq \label{genentlambda2}
- \frac{\kappa k}{8\pi G} \d V +\frac{V_\zeta}{8\pi G}    \d     \Lambda  = T \d S_{\rm gen} \, . 
\eeq
%
%
The first law 
 including non-conformal quantum fields
 is thus also expressed in terms of the generalized entropy variation,
and the  temperature in this first law is still negative. 
We conclude that the assignment of a  negative temperature to the diamond remains consistent when  extended to the semiclassical realm. 
 

\subsection{Entanglement equilibrium}
\label{sec:EE}

 If the proper volume $V$ and   cosmological constant   $  \L$ are held fixed in \eqref{genentlambda2}, 
 then the generalized entropy   is stationary in a maximally symmetric vacuum,
\beq \label{ee}
\delta S_{\rm{gen}} \big |_{V,   \L} = \delta S_{\rm BH} \big |_{V,  \L} +  \delta S_{\rm \tilde m}= 0 \, ,
\eeq
 There is no need to fix   $V$ and $\L$  in the matter entropy variation, 
because there is no first order metric variation of the matter entropy, since the zeroth order matter state is the vacuum.
 The condition $\d\L=0$ means that $\d\l$ is chosen to cancel the 
change in the effective local cosmological constant,  $-8\pi G\d X$. 

 In \cite{Jacobson:2015hqa}, it was shown, assuming the conjecture \eqref{conja}, that the semiclassical Einstein equation holds if and only if the generalized entropy is stationary at fixed volume in small local diamonds everywhere in spacetime.\footnote{In that context, the area term in the generalized entropy was assumed to have the form $\eta A$, and the gravitational constant $G$ was found to be given by $1/4\hbar\eta$.} The validity of the latter property was called the ``maximal vacuum entanglement hypothesis", but we  shall refer to it as the \emph{entanglement equilibrium} hypothesis. Here we have deduced the entanglement equilibrium statement \eqref{ee} from the conjecture \eqref{conja} together with the quantum corrected first law of causal diamonds \eqref{quantumfirstlaw1}, which itself was derived from the semiclassical gravitational equations of motion. In this sense  the semiclassical Einstein equation is equivalent to the quantum corrected first law for small, and therefore maximally symmetric diamonds. However, the variations in the entanglement equilibrium setting \cite{Jacobson:2015hqa} and in this chapter are viewed somewhat differently, so the
precise relation between the two results is not immediately clear. 
We shall now explain how they may be brought into alignment. 

In the setting of \cite{Jacobson:2015hqa}, an arbitrary spacetime and 
matter state, $(g, |\psi\ra)$, are considered, in every small causal diamond, 
as a variation of a maximally symmetric spacetime (MSS) and vacuum 
$(g_\lambda,|0\ra)$, with an initially arbitrary $\lambda$. The idea is that 
any spacetime is locally close to an ``equilibrium" state, and that all maximally symmetric states qualify as equilibria. The generalized entropy $S_{\rm{gen}}$ in the diamond is then compared to that of the diamond with the same volume in the MSS, the difference being 
\beq\label{dSEE}
\delta S_{\rm{gen}} |_{V,\lambda} = S_{\rm{gen}}|_{V(g,|\psi\ra)} - S_{\rm{gen}}|_{V(g_\lambda,|0\ra)}.
\eeq
The notation suggests ``fixed $\lambda$'', but at this stage
$\lambda$ is just an arbitrary background value for the comparison. 
The entanglement equilibrium assumption amounts to the postulate 
that there exists {\it some} $\lambda$, in each diamond, 
for which the stationary condition $\delta S_{\rm gen} |_{V,\lambda}=0$ holds. 
When applied to all diamonds this condition, together with energy-momentum conservation and the Bianchi identity, implies that 
$\lambda$ for each diamond is determined,
up to one overall spacetime constant $\Lambda$,
by the $\delta X$ of the state,  and it 
implies that the Einstein equation holds for that $\Lambda$.

To bring this more in line with the variational
relations of the present chapter, instead of setting the
difference \eqref{dSEE} to zero, we may first reckon the 
diamond entropy of $(g, |\psi\ra)$ relative to that of a diamond in flat spacetime.
In the notation of \cite{Jacobson:2015hqa} this yields
\begin{equation}\label{dSflat}
\delta S_{\rm{gen}} |_V = \eta \delta A |_V + \frac{2\pi }{\hbar  } \delta \langle K \rangle = \frac{\Omega_{d-2} R^d}{d^2-1} \left [ - \eta G_{00} + \frac{2\pi}{\hbar} (\delta \langle T_{00} \rangle + \delta X) \right].
\end{equation}
Now, rather than postulating that this variation vanishes, we postulate that it
is the same as would be obtained by varying from the Minkowski vacuum to 
a MSS vacuum with {\it some} cosmological constant $\l$ \eqref{1st1stlaw}, 
\begin{equation}\label{dSMSS}
\delta_{\rm MSS} S_{\rm{gen}}|_V = \eta\, \delta_{\rm MSS} A |_V 
= -\frac{\Omega_{d-2} R^d}{d^2-1}\eta\, \l.
\end{equation}
The equality of \eqref{dSflat} and \eqref{dSMSS} 
implies the relation
\beq
G_{00} +  \lambda\, g_{00} = \frac{2\pi}{ \hbar \eta} (\delta \langle T_{00} \rangle - \delta X\, g_{00}),
\eeq
(since $g_{00} =-1$ in Riemann normal coordinates at the center of the diamond)
and the validity of this relation for all small diamonds in spacetime implies the
tensor equation
\beq
G_{ab} +\L\, g_{ab} = \frac{2\pi}{\hbar \eta} \delta \langle T_{ab} \rangle,
\eeq
where\footnote{Eq.~\eqref{EEL}
agrees with the result in Ref.~\cite{Jacobson:2015hqa} 
after correcting the sign error there of the $\delta X$ term in (25), 
which appears due to an error in  equation (24).}    
\beq\label{EEL}
\Lambda : =  \lambda + \frac{2\pi}{\hbar \eta} \delta X. 
\eeq
With the identification $\eta=1/4\hbar G$, the relation
\eqref{EEL} matches that found in \eqref{effectivecc},
when it is recognized that $\l$ and $\L$ here refer to a
maximally symmetric {\it comparison} spacetime, whereas 
in \eqref{effectivecc} the background spacetime is implicit 
and $\d\l$ and $\d\L$ are part of the one overall variation that is made.
Adding $\lambda$ to the comparison spacetime yields the same change of entropy as including a variation $\delta\lambda = -\lambda$ in the variation being considered, and similarly for $\Lambda$,
so that the appropriate identification is 
$\l=-\d\l$ and $\L=-\d\L$. 
This establishes how the first law derived in this chapter 
from the Einstein equation is related to the entanglement equilibrium 
postulate used in \cite{Jacobson:2015hqa} to derive the 
Einstein equation.

\subsection{Free conformal energy}
\label{sec:freeenergy}



In this section we address two questions concerning the entanglement equilibrium proposal. First, an essential ingredient in the derivation of the Einstein equation in \cite{Jacobson:2015hqa} 
was  the fixed volume requirement. It 
 is desirable
to understand this requirement better, and to see whether the Einstein equation can somehow be derived without that condition. Second, in standard thermodynamics the stationarity of entropy at fixed energy 
follows from the stationarity of free energy at fixed temperature. Hence, 
it is natural to ask
whether we can identify a thermodynamic potential (e.g. free energy) in our setting 
whose extremization corresponds to an
equilibrium condition. Moreover, a characterization of the equilibrium state in terms of free energy extremization, instead of entropy extremization, has the advantage that the fixed volume constraint is relaxed.
We address 
these questions below in reverse order.

Let us start with the free energy for classical matter configurations. The classical first law \eqref{dH=TdS} implies that the  \emph{free conformal energy}\footnote{The term ``conformal'' here refers to the fact that the Hamiltonian $H_\zeta$ generates evolution along the conformal Killing vector $\zeta$,  and not to a conformal symmetry of the matter fields (as for CFTs). } 
\beq \label{freenenergy}
F = H_\zeta - T S_{\rm BH}
\eeq
is stationary at fixed temperature, i.e. $\delta F = - S_{\rm BH} \delta T= 0$.\footnote{We remind the reader that the temperature of a vacuum causal diamond is always $T=-\hbar/2\pi$.}
This means that   causal diamonds in flat space and (A)dS are   equilibrium states. Whether the free energy is minimized or maximized   does not follow from the first law, but can only be determined from the second order variation of the free energy.\footnote{We expect that maximally symmetric causal diamonds have \emph{maximum} free energy. This is because local thermodynamic stability requires that entropy be maximized at fixed energy, which for systems with negative temperature (such as our diamonds) implies that free energy is maximized  (see e.g. \cite{Wisniak}). We thank Batoul Banihashemi for pointing this out.}
We leave the study of second order variations   for future work (see also \cite{Jacobson:2017hks}). 

In the semiclassical regime,   the quantum corrected free conformal energy can    be defined by replacing the matter Hamiltonian by its quantum expectation value
 \beq \label{quanfreeenergy1}
 F_{\rm quan} =  \langle H_\zeta^{\rm \tilde m} \rangle + H_\zeta^{\rm g + \L}  - T S_{\rm BH} \, .
 \eeq
  The extremization of the quantum free energy  at fixed temperature in the vacuum, i.e. $\delta F_{\rm quan}  = 0$, follows from the quantum  first law of causal diamonds  \eqref{quantumfirstlaw1}. 
This free energy extremization can be shown to be equivalent to the entanglement equilibrium hypothesis, i.e.   the stationarity of generalized entropy  in the vacuum at fixed $V$ and $\L$. For small diamonds and for non-conformal matter this follows from combining the relations \eqref{firstdefoftheta},  \eqref{dHg}, \eqref{dent}, \eqref{conja} with the quantum free energy  extremization:
 \beq \label{statfreeenergy1}
 \delta F_{\rm quan}  \big |_{V, \L} = 0 \qquad \Longleftrightarrow \qquad \delta S_{\rm gen} \big |_{V, \L} = 0 \, .
 \eeq
Note that keeping the   volume  and  (effective) cosmological constant  fixed is the same as keeping the conformal energy fixed (assuming, for non-conformal mattter, that  $\Lambda$ defined in \eqref{effectivecc}   can be treated as the energy density of some perfect fluid). Therefore, the entanglement equilibrium proposal is just the   statement that generalized entropy is extremal at fixed conformal energy (as it should be for an equilibrium state).

In standard thermodynamical equilibrium, not only is entropy extremal at fixed energy, but also energy is extremal at fixed entropy. Let us now see to what extent the latter is true for causal diamonds.
For conformal matter, and for any sized diamond, if the generalized entropy is kept fixed and the conformal energy is allowed to vary, then the Hamiltonian associated to $\rm{g}$ and $\L$ is extremal in the vacuum 
    \beq \label{statfreeenergy2}
 \delta F_{\rm quan} \big |_{S_{\rm{gen}}} = 0 \qquad \Longleftrightarrow \qquad \delta    H_\zeta^{\rm g + \L}  \big |_{S_{\rm{gen}}}= 0 \, . 
 \eeq
 For non-conformal matter, we must restrict to small diamonds if we want to hold the generalized entropy fixed. There is a potential obstruction to this, however, since it is not clear whether there exists a Hamiltonian such that 
 $
\delta H^{    \L}_\zeta =  (V_\zeta /8\pi G) \,  \d     \L ,
$
where $\L$ is the effective cosmological constant \eqref{effectivecc} in a small diamond.\footnote{It is not clear to us whether this Hamiltonian exists, since $    \L$ includes both a   local cosmological constant and a piece  from   the non-conformal   matter. The latter contribution    seems to spoil the derivation of   equation \eqref{Theta}.} 
But if instead  we keep the Bekenstein-Hawking entropy fixed, rather than the full generalized entropy, then we need not  restrict to small diamonds and the \emph{total} Hamiltonian is extremal in the vacuum, i.e.   
  \beq \label{statfreeenergy3}
 \delta F_{\rm quan} \big |_{S_{\rm{BH}}} = 0 \qquad \Longleftrightarrow \qquad \delta \big (\langle H_\zeta^{\rm \tilde m} \rangle +   H_\zeta^{\rm g + \L} \big) \big |_{S_{\rm{BH}}}= 0 \, .  
 \eeq
This  equilibrium condition states that the total conformal Killing energy is extremal in the vacuum if the dimension of the Hilbert space is fixed (if we interpret the Bekenstein-Hawking entropy as the logarithm of that dimension).
It would be interesting to explore this energy relation 
further. 
  
Next, we return to the role of the fixed volume constraint in  \eqref{statfreeenergy1}.  
We note the remarkable fact that  any variation of the free energy at fixed 
 $\L$ is equal to the variation of the free energy at fixed $V$ and $\L$ composed with some variation induced by a diffeomorphism that changes the volume. This is because such a diffeo-induced variation leaves  the free energy  unchanged, as explained in Section  \ref{sec:constrained}. To be explicit, for small diamonds the former variation can be expressed as
 \beq  \label{statfreeenergy4}
 \frac{1}{T_H} \delta F_{\rm quan} \big |_\L= \frac{1}{4G\hbar}  ( \delta A - k \, \delta V ) \big |_\Lambda +    \delta S_{\rm \tilde m}    \, ,
 \eeq
 whereas
 the latter variation is given by
 \beq  \label{statfreeenergy5}
 \frac{1}{T_H} \delta F_{\rm quan} \big |_{V,\L}= \frac{1}{4G\hbar}    \delta A   \big |_{V,\Lambda} +    \delta S_{\rm \tilde m}    \, ,
\eeq
 where the volume variation comes from the variation of $H^{\rm{g}}_\zeta$ in \eqref{quanfreeenergy1}.
As shown in Section  \ref{sec:constrained},  the two free energy variations are \emph{equivalent}, i.e. 
\begin{equation}
\delta F_{\rm quan} \big |_{\Lambda} \sim \delta F_{\rm quan} \big |_{V,\Lambda} \, ,
\end{equation}
where the equivalence is  modulo diffeo-induced variations (which do not affect the free energy).
This implies that
  the stationarity of the free energy at fixed $\L$ is equivalent  to the stationarity of the free energy at fixed $V$ and $\L$,  and hence --- using the equivalence \eqref{statfreeenergy1} ---  equivalent to the entanglement equilibrium postulate. 
  Therefore, the derivation of the Einstein equation in \cite{Jacobson:2015hqa} can be rephrased  without the requirement to  fix the volume. One can thus  take as the input assumption   that the free energy is stationary at fixed $\L$, rather than that the generalized entropy is stationary
  at fixed $V$ and  $\L$ (see our Gravity Research Foundation Esssay \cite{Essay}).

\section{Further remarks on the first law}
\label{sec:remarks}

In this section we collect several comments on aspects of the first law of causal diamonds.

\subsection{Role of maximal volume}

 When we evaluated the variation of the gravitational Hamiltonian $\delta H_\zeta^\text{g}$ in 
 \eqref{metrichamfinal}, we chose to carry out the integral over the maximal slice of the unperturbed diamond. Because the symplectic current is conserved, the value would have been the same had we chosen any other
 slice bounded by $\partial\Sigma$, although it would not have been given in the same way by the volume variation. The slice $\Sigma$ therefore has a somewhat preferred status. Furthermore, although we described this as the variation of the volume of ``the slice that was the maximal slice in the unperturbed diamond," we could just as well describe it as the variation of the volume of the maximal slice itself. This is because the volume change due to the variation of the location of the maximal slice itself vanishes, precisely because that slice is maximal to begin with. This is satisfying, since it allows the first law to be stated in a manifestly ``gauge-invariant" fashion --- i.e.\ independently of how the spacetime interior
  of the varied diamond is identified with that of the original diamond ---   and the second and higher order variations of the maximal slice are unambiguously defined.

\subsection{Fixed volume and fixed area variations}
\label{sec:constrained}

For variations that fix the volume, 
\eqref{finalfirstlaw'} becomes a relation between the area variation at fixed volume and the 
variation of the matter Hamiltonian,
\begin{equation} \label{vararea}
\frac{\kappa}{8\pi G} \delta A \big |_V= - \delta H_\zeta^\text{m}.
 \end{equation}
%
That is, the presence of positive conformal Killing energy 
matter produces an area deficit at fixed volume.
Similarly, 
for variations in which 
the area  is fixed we obtain the relation
 \begin{equation}  \label{varvolume}
 \frac{\kappa k }{8\pi G} \delta V \big |_A= \delta H_\zeta^\text{m}. 
 \end{equation}
Hence, the presence of positive conformal Killing energy 
matter produces a volume excess at fixed area.

Importantly, the first law \eqref{vararea} and free energy variation \eqref{statfreeenergy4} at fixed volume are \emph{equivalent} to  those \eqref{finalfirstlaw'} and \eqref{statfreeenergy5} without holding the volume fixed. 
This is because  
(i) it is always possible to compose any variation with a variation induced by a diffeomorphism, such that the volume $V$ is unchanged under the complete variation;  and (ii) for \emph{all} diffeo-induced variations,
both sides of the first law equation 
vanish, and, similarly, the free energy variation  vanishes. That the combination $\delta A - k\delta V$ vanishes in such a 
variation
is shown below;\footnote{The combination of variations $\delta A - k\delta V$  can thus be interpreted as the part of the area change that would remain if one were to compose  a generic variation with a diffeo-variation  restoring the volume  to its original value.} and since  matter can be present only after the field variation away from maximal symmetry,
  a diffeo-induced field variation affects  $\delta H_\zeta^\text{m}$ and $\delta S_{\rm \tilde m}$
only at the next variational order. Thus, we are free to add to the first law  \eqref{finalfirstlaw'}  a diffeo-induced variation  that restores the volume $V$ to its original value, so that the first law takes the form \eqref{vararea}.
 Similarly, one can also freely add a diffeo-induced variation that restores the area $A$ to its original value, such that  to the first law becomes \eqref{varvolume}.

 To establish vanishing of $\delta A - k\delta V$ under diffeo-variations at first order,  
we first note that the effect of the diffeomorphism on the fields with a fixed ball is the same as the effect   of the opposite diffeomorphism on the ball with fixed fields. Now consider the latter setup: a diffeo-flow of the ball with fixed fields.  The flow carries the points on the original ball surface to a new ball surface. Let $n$ be a field of unit vectors normal to the original ball surface as well as to all the layers of the onion between there and the final ball. Then using Stokes' theorem, the difference in areas of the initial and final surface is $\int  \!d( n \cdot \varepsilon)$, where $\varepsilon$ is the spatial volume form and the integral is over the region between the ball surfaces. Since $d(n \cdot \varepsilon) = (D \cdot n) \varepsilon$ --- where $D$ is the covariant derivative compatible with the induced metric on $\S$ ---  the area change is $\int (D  \cdot n) \varepsilon $ over the region between the surfaces. For a small change of the ball $D \cdot n$ is approximately equal to its original value, the trace of the extrinsic curvature $k$ of the original ball surface, so in that limit the area change $\delta A$ becomes equal to $k\int \! \varepsilon = k \, \d V$, hence $\delta A - k\delta V$ vanishes. 
\label{notediff}

\subsection{Varying the cosmological constant}
\label{sec:varyingLambda}

In a thermodynamic interpretation, causal diamonds in maximally symmetric spacetimes are all ``equilibrium states" from which variations can be made. The diamonds differ only in size, and in the cosmological constant of the background. 
It is natural to allow also $\Lambda$ to vary, since it is evidently an equilibrium state variable, and there are circumstances under which it might vary.  For instance, there may be mechanisms by which it can decay. Also, in the context of the AdS/CFT correspondence, the negative cosmological constant is controlled by the number of stacked D-branes, which could in principle change \cite{Kastor:2009wy}. Another reason to consider variable $\Lambda$ arises in formulating the principle of vacuum entanglement  equilbrium for non-conformal matter fields, see Section \ref{ncm} and Ref.~\cite{Jacobson:2015hqa}. Consistency with the Bianchi identity made it necessary to allow for an initially undetermined local cosmological constant in small causal diamonds, which ended up being related to the part of the entanglement entropy variation not captured by the energy-momentum tensor. It is thus of interest to include variations of $\Lambda$ in the first law. Ref.~\cite{Kubiznak:2016qmn} provides an extensive review of black hole thermodynamics extended to include variable $\L$, a.k.a. ``black hole chemistry".  


There are many ways to accommodate a cosmological constant variation in the first law. 
In the literature this has been done for the first law for black holes and for holographic entanglement entropy by employing various methods, see e.g. \cite{Kastor:2009wy,Urano:2009xn,Kastor:2014dra,Caceres:2016xjz,Couch:2016exn,Kubiznak:2016qmn}. In Section \ref{sec:firstlawch2} we treated the cosmological constant as a perfect fluid, and made use of Iyer's 
 generalized derivation 
of the first law to allow for matter fields which are non-stationary yet have a stationary stress-energy tensor \cite{Iyer:1996ky}.   In this approach the cosmological constant term in the first law comes from the variation of the stress-energy tensor of the fluid. Yet another way of introducing a cosmological constant is    to promote it to a dynamical scalar field, and to add it to the Lagrangian together with a $(d-1)$-form field $B$ as: $ \L(dB - \epsilon)  / (8\pi G)$ \cite{Henneaux:1989zc}. The $B$ field equation implies that $\L$ is constant, while the $\L$ field equation implies $dB=\epsilon$.  The addition to the symplectic potential due to this Lagrangian is $\theta (\phi, \delta \phi) =  \Lambda \delta B / (8\pi G)$, where $\phi = (\Lambda, B)$.  Moreover, the additional term in the symplectic current is given on shell by $\omega(\phi,  \delta \phi, \mathcal L_\zeta \phi) = [  \delta \Lambda \,  \zeta  \cdot \epsilon + d (\delta \Lambda  \,  \zeta \cdot B) ] /(8\pi G)$. When integrated over $\Sigma$ this gives precisely $V_\zeta \delta \Lambda /(8\pi G)$, as in \eqref{Theta}, since $\zeta$ vanishes at the edge  $\partial \Sigma$.


\newpage

\subsection{Gravitational field Hamiltonian and York time}
\label{sec:Yorktime}

\begin{figure}
	\centering
	\includegraphics
		[width=.46\textwidth]
		{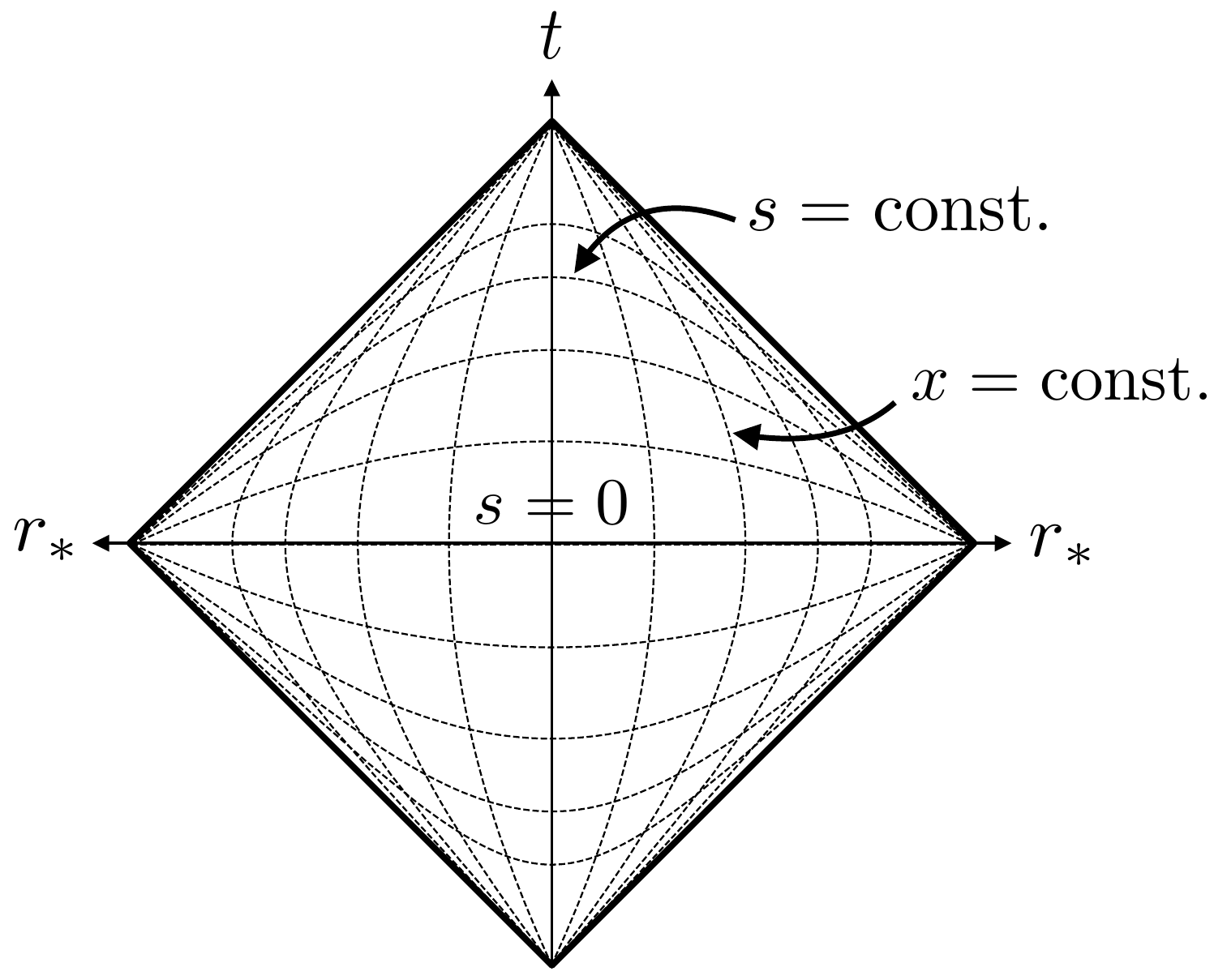}
\caption{\small   The $(x,s)$ coordinate chart of a maximally symmetric causal diamond. The coordinate $s  \in (-\infty, \infty) $ is the conformal Killing time,  defined as the function that vanishes on the maximal slice $\S$ and satisfies
$\zeta \cdot ds =1$. The coordinate $x  \in [0, \infty)$ is spherically symmetric and satisfies $\zeta\cdot dx = 0$ and $|dx|=|ds|$.
Constant $s$ and $x$ lines are plotted at equal coordinate intervals of $0.5$. 
See Appendix \ref{appyork} for a demonstration that $ds$ and $dx$ are everywhere orthogonal, and for the  
line element in these coordinates.}
	\label{fig:causaldiamondsandx}
\end{figure}

 We have seen that the gravitational contribution to the variation of the Hamiltonian 
$H_\zeta$ generating evolution along the conformal Killing flow of the background maximally symmetric diamond 
is proportional to the volume variation.
This ``volume as Hamiltonian'' is reminiscent of a ``York time" Hamiltonian for general relativity \cite{York:1972sj}, 
which generates evolution along  a foliation by spacelike hypersurfaces with 
constant mean curvature $K$ (i.e.\ along a ``CMC" foliation), using $K$ as the time parameter,
 and with an arbitrary shift vector field. (Mean curvature can be defined as $K:=\nabla_a u^a$, where $u^a$ is the future pointing unit normal to a spacelike hypersurface.) Such a Hamiltonian 
is proportional to the spatial volume of the CMC slices. 

The similarity is not accidental. 
It arises from the fact that (i) the conformal Killing vector $\zeta^a$ is orthogonal to $\S$, which is a CMC
surface, and (ii) $\z^a\nabla_a K$ is constant on $\S$. Actually, these two properties hold on {\it all} leaves of the CMC foliation:
 as shown in Appendix \ref{appyork}, surfaces of constant conformal Killing parameter $s$ --- defined by $\zeta^a \nabla_a s =1$ with the initial condition $s=0$ on $\Sigma$  --- 
coincide with surfaces of constant $K$  everywhere in the diamond,
and $\z^a$ is everywhere orthogonal to these surfaces (see Figure  \ref{fig:causaldiamondsandx} for an illustration).
More specifically, $K$ and $s$ are related  by 
\beq   \label{Kands2}
K 
= (d-1) \dot \alpha |_{s=0} \sinh s \,, 
\eeq
where $\alpha = \nabla_a \zeta^a /d$ and $\dot \alpha = u^a \nabla_a \alpha$.  
In particular, $K$ vanishes at the extremal surface $s=0$, and its first derivative with respect to $s$ at $s=0$ is given by
\beq  \label{Kands1}
\frac{dK}{ds} \Big |_{s=0}  =  (d-1)\dot\alpha|_{s=0} = -\frac{d-1}{d-2}\kappa k \, ,
\eeq
where  \eqref{dotalpha} is used in the last equality.\footnote{Note that $K$ decreases as $s$ increases, and is hence negative to the future of the slice $\Sigma$ (and positive to the past of $\Sigma$).} 
Equation  \eqref{Kands1} establishes that York time  and conformal Killing time are proportional, to first order about the maximal slice, for a maximally symmetric diamond. This indicates, as we will now argue, that the variation   $\delta H_\zeta^{\rm g}$  agrees, up to the constant  \eqref{Kands1}, with the York time Hamiltonian variation $\delta H_{\rm Y}$.

In the context of the first law,  the perturbed spacetime is not the maximally symmetric one. When the metric is varied, the definition of York time varies, so the surface on which we should be computing the volume varies, as does the rate of time flow. 
Nevertheless, since the $t=0$ surface has vanishing $K$, the volume variation induced by varying the surface vanishes. Also, the field variation is already first order, so the change of flow rate of $K$ makes a higher order contribution to $\delta H_{\rm Y}$. It follows that
\beq
\delta H_\z^{\rm g} = \frac{dK}{ds} \Big |_{s=0} \delta H_{\rm Y} \, ,  \qquad \text{with} \qquad  \delta H_{\rm Y} = \frac{d-2}{d-1} \frac{\delta V}{8\pi G} \, .
\eeq
Therefore, the gravitational Hamiltonian variation \eqref{metrichamfinal} is  indeed
equal to a constant times the York time Hamiltonian variation. The York time Hamiltonian variation would be  equal to the negative of the proper volume variation, i.e. $\delta H_{\rm Y} = - \delta V$, if one used the  time  variable 
$
 t_{\rm{Y}}
= - \frac{d-2}{d-1} \frac{ K}{8 \pi G}  
$
instead of $K$.
 This is precisely the time variable that York originally introduced for general relativity  in $d=4$  \cite{York:1972sj}.  
 In the literature the minus sign in the time variable is often omitted, in which case the Hamiltonian is equal to   the volume (see e.g. \cite{FischerMoncrief}).

   \vspace*{.5cm}
 
\section{Limiting cases: small and large     diamonds}
\label{sec:cases}

In this section we comment on various    limiting cases of the first law of causal diamond mechanics (\ref{firstlawcc1}). Since the law applies to arbitrary sized diamonds in (A)dS, it has a wide domain of applicability.  Here we apply it to the static patch of de Sitter spacetime, small diamonds in any maximally symmetric spacetime, flat and AdS Rindler spacetimes,  and to the so-called ``Wheeler-DeWitt patch" in AdS,
tying these limiting cases together into one framework.


 \subsection{De Sitter static patch}
\label{sec:desitter}

\begin{figure}
	\centering
	\includegraphics
		[width=.35\textwidth]
		{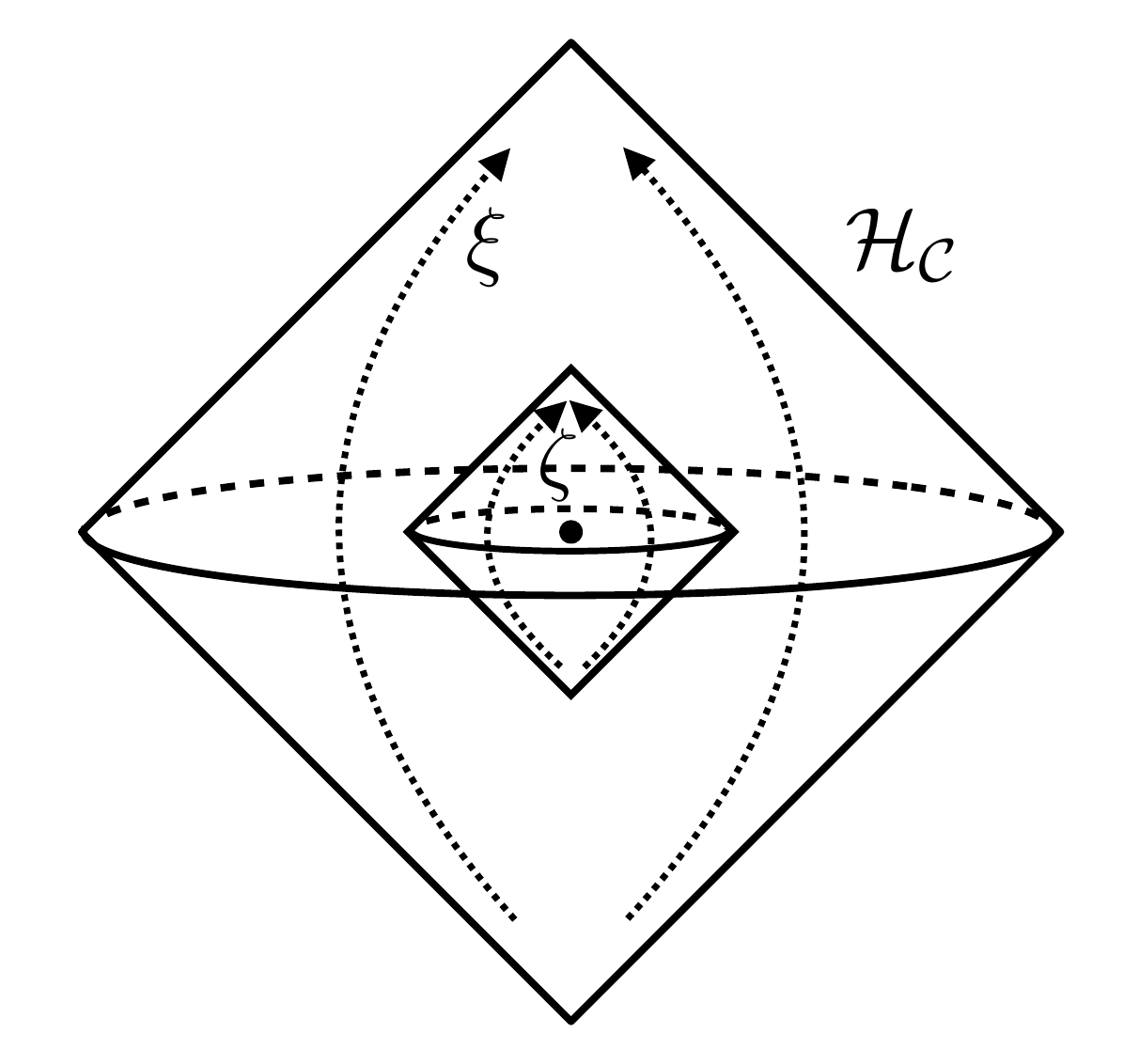}
\caption{\small   A   causal diamond   in the  de Sitter space static patch. The conformal Killing vector~$\zeta$ turns  into the timelike   Killing vector $\xi$ if the boundary of the diamond coincides with the cosmological horizon $\mathcal H_C$.}		
	\label{fig:dScausaldiamond}
\end{figure}

If the boundary of a causal diamond in dS space coincides with the cosmological horizon, i.e. if $R=L$, then the conformal Killing vector \eqref{ckv2} becomes the time translation Killing vector of the static patch,\footnote{In this section we use the letter $\xi$ for Killing vectors and retain $\z$ for conformal Killing vectors.} 
 \begin{equation}  \label{ckv2dS}
\xi^{\text{dS}}= L \partial_t,
 \end{equation}
normalized so that the surface gravity is unity  (see Figure \ref{fig:dScausaldiamond}).
Since this is a true Killing vector, the variation of the gravitational part of the Hamiltonian (\ref{dH}) vanishes.
This is consistent with \eqref{metrichamfinal}, because the Sitter horizon has extremal area on the $(d-1)$-sphere, so  $k=0$. The first law \eqref{firstlawcc1} thus reduces to 
\beq\label{firstlawdSstatic}
  \delta H_\xi^\text{\~m} =- \frac{1}{8 \pi G} \left ( \kappa  \, \delta A +  V_\xi \,  \delta \Lambda \right) \, . 
\eeq
The thermodynamic volume \eqref{thetads1} in this case reduces to 
$V_\xi^{\text{dS-static-patch}}=\kappa LV^{\text{flat}}_L$.
The relation \eqref{firstlawdSstatic}  with $\d\L=0$ was established long ago by 
Gibbons and Hawking  \cite{Gibbons:1977mu}, and 
was  generalized to include a variation of the cosmological constant in \cite{Sekiwa:2006qj,Urano:2009xn}.

 If one assigns a negative temperature $T= - \kappa \hbar /2\pi$ and pressure $p = - \Lambda/ 8\pi G$ to the dS static patch, the first law   can be turned into a proper thermodynamic relation\footnote{For $\delta   H_\xi^\text{\~m} =0$, one could  assign a positive temperature   to the dS static patch, i.e. $T=T_{\rm H}>0$, but then the pressure should also be positive according to the first law, i.e. $p = \L/8\pi G>0$, in contradiction to the usual sign of the pressure associated to the cosmological constant in dS space.} 
\beq \label{dSthermo1}
\delta   H_\xi^\text{\~m}   = T  \delta S_{\text{BH}} + V_\xi \, \delta p \, . 
\eeq
Since this thermodynamic relation is of the form $d H = T dS + V^{\text{th}} dp$, where $H$ is the enthalpy of the system  and $V^{\text{th}}$ is the thermodynamic volume,      $H_\xi^\text{\~m}$ coincides with the enthalpy instead of 
 simply the internal energy.   The matter Hamiltonian vanishes in the background, hence we observe that the enthalpy of  the static patch of dS space is zero. 
 
 Through a Legendre transformation, $U = H - pV^{\text{th}}$,   the first law can be rewritten in the standard form $dU = T dS - p\,dV^{\text{th}}$. But what is the internal energy $U$ of de Sitter space? 
 The common lore,   cf. e.g.\ \cite{Gibbons:1977mu},  is  that the energy of   dS space is zero, because it has no asymptotic infinity.  However,  we find that its (vacuum) internal energy  is nonzero and given by $U_{\text{vac}} = \rho \, V_\xi$,  the redshifted vacuum energy, where we used the equation of state $p = - \rho$ and the fact that $H=0$ for dS. 
The first law can thus be expressed as 
\beq
\delta U= T \delta S_{\text{BH}} - p \, \delta V_\xi \, , 
\eeq
where $U = U_{\text{vac}} + H_\xi^\text{\~m}$ is the internal energy.
Finally, we note that the Gibbs free energy of   dS  space is $G = H - TS = -TS$, and the Helmholtz free energy is $F = U_{\text{vac}} - TS$.  As usual, the former is extremized in a fixed $(T,p)$ ensemble, whereas the latter  is extremized in a fixed $(T, V_\xi)$ ensemble.  The free energy computed in \cite{Gibbons:1977mu} from the on-shell Euclidean action agrees with the Gibbs free energy, and not the Helmholtz free energy, 
because the Euclidean action was extremized there at fixed period (i.e. fixed temperature) 
and fixed cosmological constant.\footnote{If one takes the timelike Killing vector  to be $\xi^{\text{dS}}= \partial_t$, so that   $\xi^2 = -1$ at the center of the diamond,  then  temperature and  pressure are not independent in the dS static patch. That is because the surface gravity is set by the dS radius in this case, i.e. $\kappa_{\text{dS}}=1/L$, and the pressure is determined by the cosmological constant  $\L = (d-1)(d-1)/2L^2$. Hence,  by fixing the temperature one also fixes the pressure, and vice versa, 
 when using this normalization of the Killing field to define the temperature.}

\subsection{Small diamonds and Minkowski space}
\label{item:small}
 
\begin{figure*}[t!]
    \centering
    \begin{subfigure}[t]{0.38\textwidth}
        \centering
        \includegraphics[width=\textwidth]
		{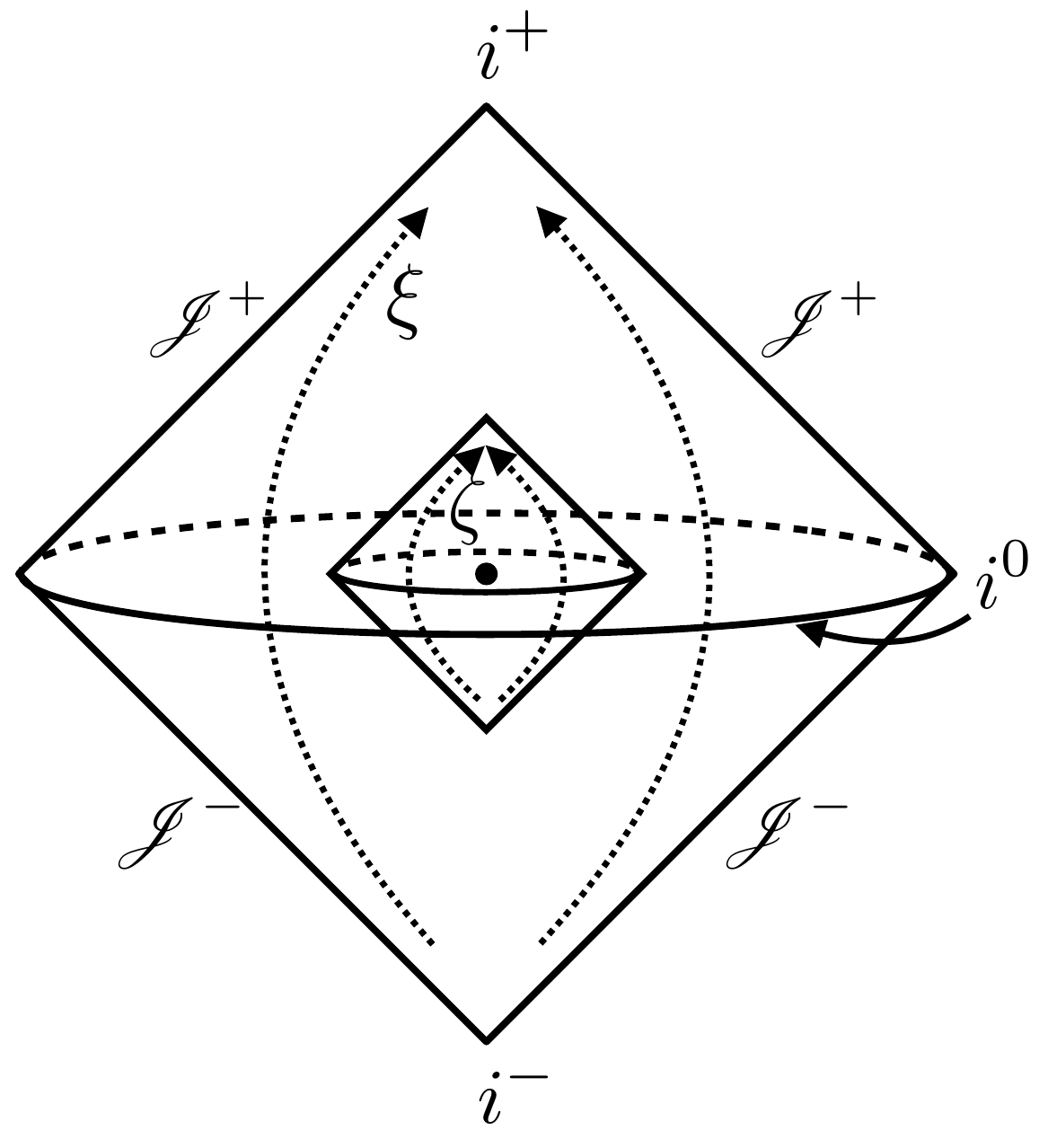}
        \caption{ }
        \label{fig:mink1}
    \end{subfigure}%
    ~~~~~~~~~~~~~~~
    \begin{subfigure}[t]{0.38\textwidth}
        \centering
        \includegraphics[width=\textwidth]
		{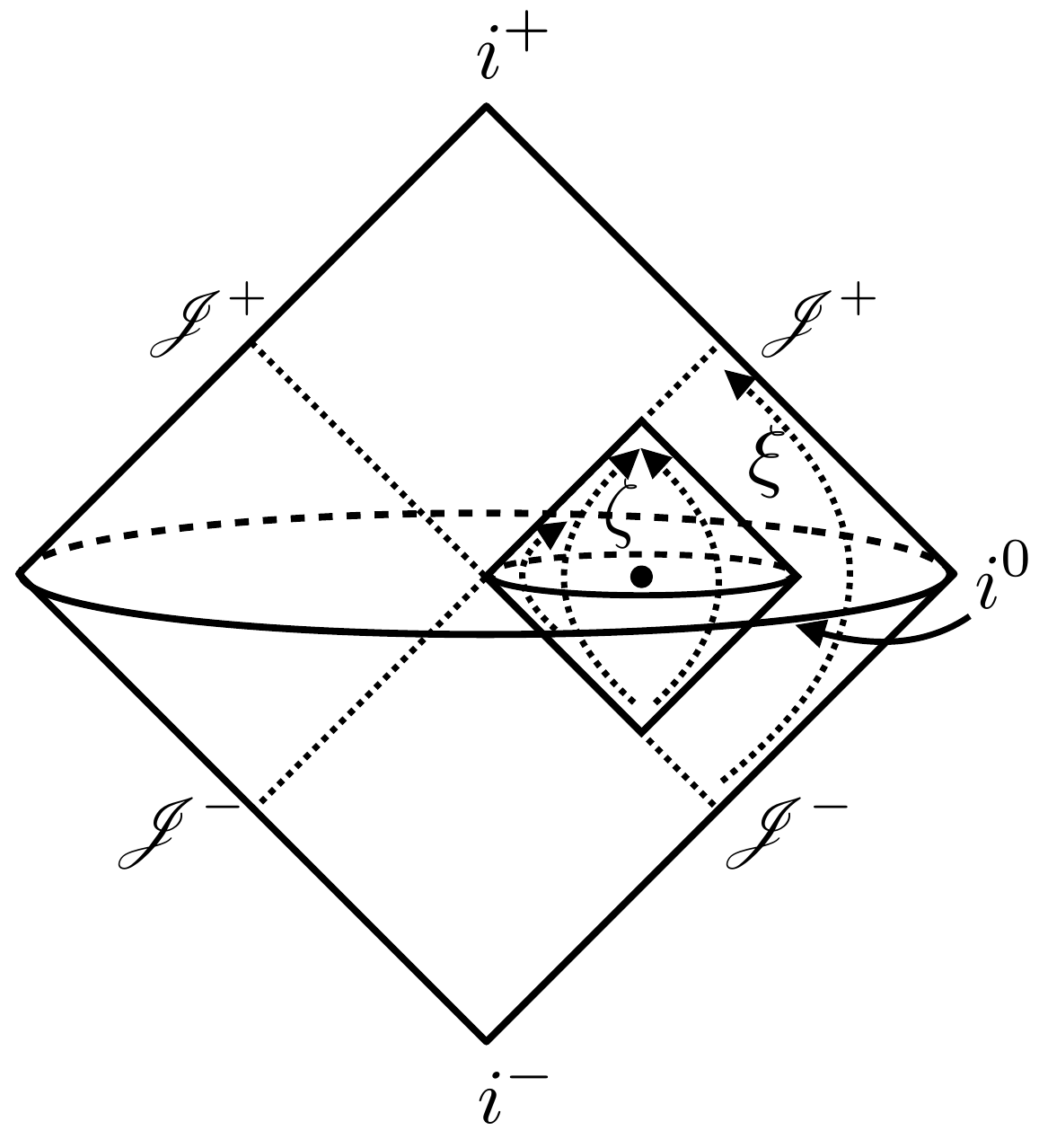}
        \caption{ }
        \label{fig:mink2}
    \end{subfigure}
    \caption{ \small {(a)  A causal diamond associated to a ball  at the center of Minkowski space. If one the normalizes   the conformal Killing vector that preserves the diamond such that $\zeta^2 = -1$ at the center of the  ball, then it becomes identical to the timelike Killing vector $\xi=\partial_t$ in the infinite-size limit. (b)  A causal diamond whose edge touches the bifurcation surface of two Rindler horizons in Minkowski space. In the infinite-size limit the diamond coincides with the right Rindler wedge, and the conformal Killing vector $\zeta$ becomes the boost Killing vector $\xi$.  }}
\end{figure*}

In the small radius limit $R \ll L$
the mean curvature \eqref{k} and
the thermodynamic volume \eqref{thetads1} are 
given up to second order in $R/L$ by
\begin{align}
   k&=\frac{d-2}{R}\left(1 - \frac12\frac{R^2}{L^2} + \dots \right)\label{smallk} \, , \\
 V_\zeta   &=   \frac{ \kappa \, \Omega_{d-2} R^{d}}{d^2-1} \left ( 1 +\frac{d}{d+3}  \frac{R^2}{L^2}  + \dots \right)  \, .  \label{smallTheta} 
 \end{align}
To first order in $R/L$, the first law (\ref{firstlawcc1}) thus reduces to the one that would be
found in flat spacetime, 
\begin{equation}\label{flatlaw}
 \delta H^\text{\~m}_\zeta =  -\frac{\kappa}{8 \pi G} \left [     \delta A - \frac{d-2}{R}    \,  \delta V^\text{flat} + \frac{\Omega_{d-2} R^{d}}{d^2-1} \delta \Lambda \right] \, .
\end{equation}
This identity, without the cosmological constant term, is the one derived in \cite{Jacobson:2015hqa}, both by Riemann normal coordinate expansion, and by varying the Noether current for the conformal Killing vector that preserves a causal diamond in flat space. That conformal Killing vector can be recovered from (\ref{ckv2}) in the limit  $L \rightarrow \infty$, and is given by 
\begin{equation} \label{ckvflat1}
\zeta^{\text{flat}} = \frac{1}{2R} \left [  \left (  R^2 -  t^2 - r^2 \right) \partial_t - 2 t r\partial_r  \right] \,.
\end{equation}
(See Figure   \ref{fig:mink1} for a conformal diagram of a diamond in Minkowski space.) As a side remark, if the variation  
 $\delta T_{ab}^{\text{\~m}}u^a u^b$ is constant  on $\Sigma$,\footnote{Note that since $\S$ has vanishing extrinsic curvature, 
 $m^c\nabla_c u^a=0$ for any vector $m^c$ tangent to $\S$. Thus constancy on $\S$ of the scalar
 $\delta T_{ab}^{\text{\~m}}u^a u^b$ is equivalent to the condition 
 $(m^c\nabla_c \delta T_{ab}^{\text{\~m}})u^au^b=0$.} 
  then the  conformal Killing energy becomes proportional to the thermodynamic volume
\begin{equation} \label{modHamsmall}
\delta  H_\zeta^\text{\~m} =\delta  T_{ab}^{\text{\~m}}  u^a u^b\, V_\zeta   \, , 
 \end{equation}
where $\delta T_{ab}^{\text{\~m}}  u^a u^b$ is evaluated at the center of the ball.

\subsection{Rindler space}
 \label{largesmall:rindler}
 
From the first law (\ref{flatlaw}) in flat space  one can   derive a first law for Rindler space in the infinite-diamond limit. This is because a Rindler wedge  is an infinite diamond in flat space, as can be seen from the Penrose diagram   in Figure \ref{fig:mink2}.  More precisely, the right  Rindler wedge can   be obtained by inflating   the causal diamond whose   edge touches the origin of flat space and whose center is located at $\{ t=0, x^1 = R, x^2=0, \dots, x^{d-1} = 0 \}$. In the infinite $R$ limit, 
the  null boundaries of the diamond  coincide with 
the Rindler horizon.\footnote{Instead of increasing the size of the diamond to infinity, one could also directly relate the  diamond nestled in the corner of the right Rindler wedge  to the entire wedge itself  through   a conformal map. See Appendix \ref{sec:conftrans} for further details.}
 Moreover, 
 by replacing
 the coordinate $x^1$ by   $x^1 - R$ in $\zeta^{\text{flat}}$ and 
 then taking the limit $R\to \infty$,
  the conformal Killing vector (\ref{ckvflat1}) becomes the boost Killing vector of Rindler space,
\begin{equation}
\xi^{\text{Rindler}} =   x^1 \partial_t + t \partial_{  x^1} \, . 
\end{equation}
Although  the first law was originally derived for causal diamonds which are centered at the origin of flat space, it actually applies to any causal diamond associated to a spherical region whose center is located at  $\{ t=t_0, x^i=x^i_0 \}$. Hence, in particular it applies to the causal diamond described above which is centered at $x^1 = R$, and also to the infinite-size version of this diamond. Thus, by taking the limit $R\rightarrow \infty$ of (\ref{flatlaw}), we obtain the first law for a Rindler horizon 
 (setting $\delta \L=0$)
\begin{equation} \label{rindlerfirstlaw1}
 \delta H^{\rm \tilde{m}}_\xi =  -\frac{\kappa}{8 \pi G}     \delta A  \, .
\end{equation}
This is the \emph{equilibrium variation version} of the first law  for a Rindler horizon  in flat space. It compares the areas of the bifurcation surface in two infinitesimally nearby configurations, the  Rindler horizon and the   perturbed horizon  due to the presence of matter.  Notice that, as defined by this limiting process, there is no outer boundary term in the first law for the Rindler wedge.

In the literature, on the other hand,   a  \emph{physical process version} of the first law  for Rindler space  was stated for the first time in \cite{Jacobson:1999mi}, and subsequently proven using the Einstein equation in \cite{Jacobson:2003wv, Amsel:2007mh,Bianchi:2013rya}.
This   version  of the first law describes how the horizon area changes if one  throws an infinitesimal amount of Killing energy through the     horizon. Mathematically, the physical process version  states that
\begin{equation}  \label{rindlerfirstlaw2}
\delta E_\xi = \frac{\kappa}{8\pi G} \delta A  \, ,
\end{equation}
where $\delta A$ is the horizon area change and $\delta E_\xi = \int_{\mathcal H} \delta T_{ab} \xi^a d \Sigma^b$ is the  flux of Killing energy across the horizon.  
The positive sign in the process version (\ref{rindlerfirstlaw2}) arises because of two  implicit boundary conditions: the  generators of the horizon are parallel at future null infinity (due to the teleological nature of the horizon)  and the area  is fixed at the bifurcation surface.  When matter with   positive Killing energy crosses the future horizon, the generators of the horizon converge because of   the attractive nature of gravity. In order to satisfy both boundary conditions, the generators must be initially expanding and hence the area of the horizon  cross-section \emph{increases} towards the future.\footnote{Ref. \cite{Bianchi:2013rya} 
also interpreted this in terms of an equilibrium variation version of the first law,
which refers to  the variation of the asymptotic horizon area rather than the variation of the bifurcation surface as in our version.}
 
Although the equilibrium   version \eqref{rindlerfirstlaw1} and     process version \eqref{rindlerfirstlaw2} of the first law are conceptually distinct, they are,  of course, consistent with each other.  One can  
obtain a different 
sign in the physical process version   by choosing a different boundary condition. 
If one  were to fix  the asymptotic area, instead of the initial edge area, then since the horizon area 
would still \emph{increase} towards the future, if positive Killing energy crosses the horizon, the  area of the bifurcation surface must be \emph{smaller} in the perturbed configuration when compared to the unperturbed configuration (i.e. the original Rindler space).
  This is consistent with  the equilibrium version, where    the area of the bifurcation surface is allowed to vary, and decreases when matter is added to the right Rindler wedge. Thus, the sign difference between (\ref{rindlerfirstlaw1}) and (\ref{rindlerfirstlaw2}) arises due to  different boundary conditions.\footnote{The `negative sign' also appears in the process version of the past horizon (if the area of the bifurcation surface is allowed to vary). Suppose one would throw some positive Killing energy through the past horizon 
into
the right Rindler wedge. If the 
 area of the horizon at  past null infinity is fixed to agree with 
that of the original Rindler wedge, 
then the area of the bifurcation surface must \emph{decrease}, since the flux of Killing energy converges the horizon generators.}

\begin{figure*}[t!]
    \centering
    \begin{subfigure}[t]{0.38\textwidth}
        \centering
        \includegraphics[width=\textwidth]
		{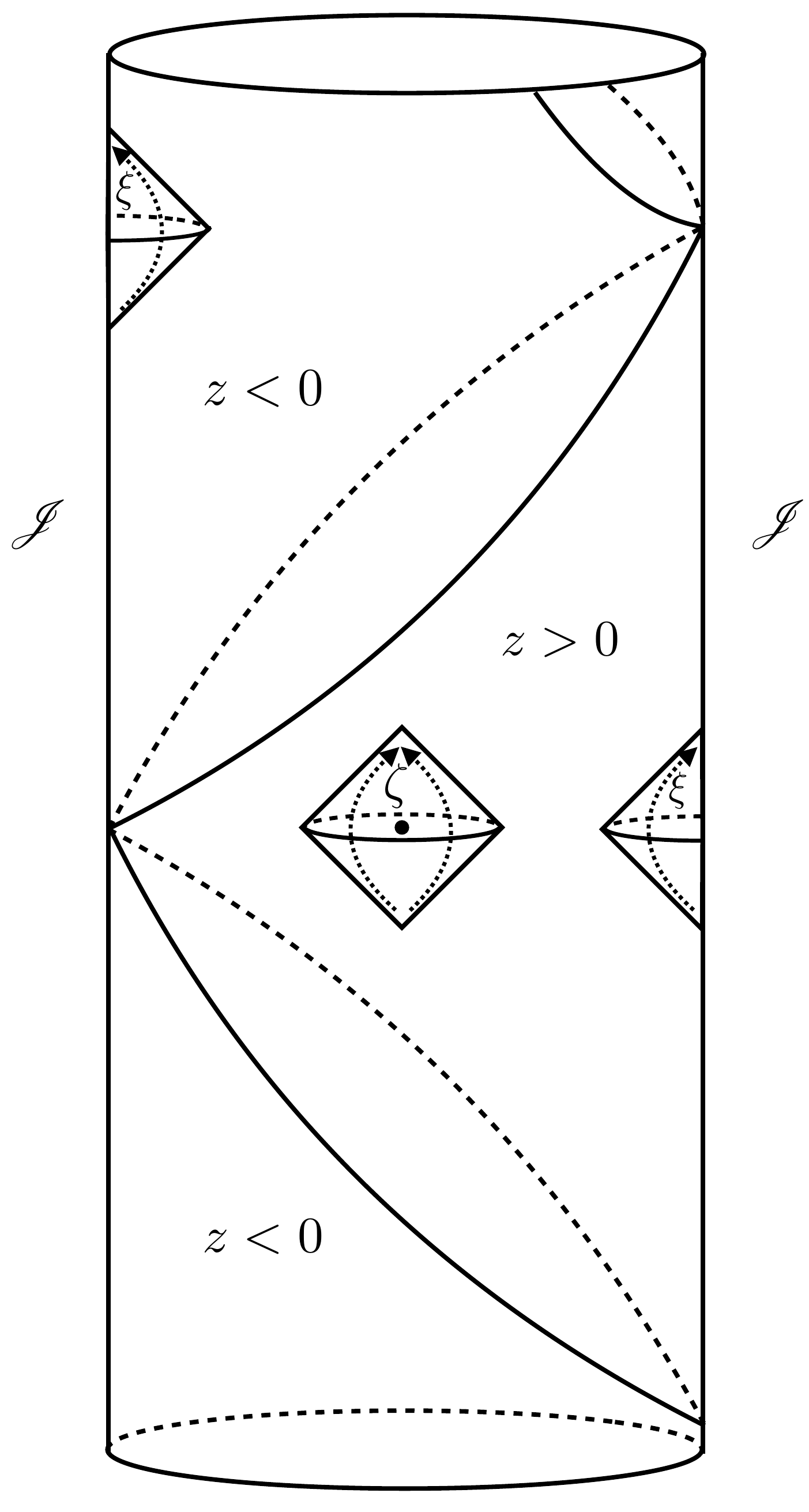}
        \caption{ }
        \label{fig:adsrindler}
    \end{subfigure}%
    ~~~~~~~~~~~~~~~~~~
    \begin{subfigure}[t]{0.38\textwidth}
        \centering
        \includegraphics[width=\textwidth]
		{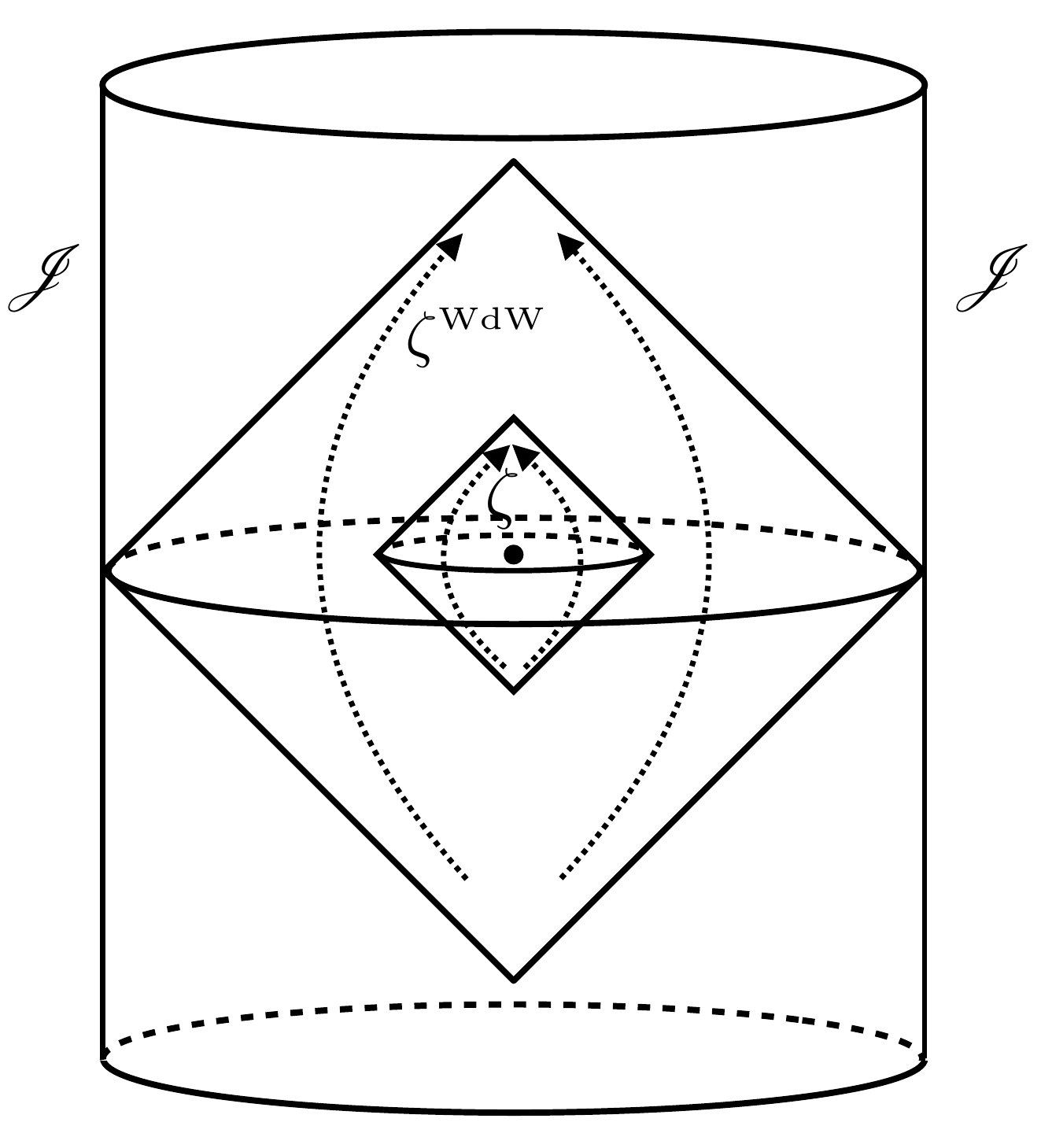}
        \caption{ }
        \label{fig:wdw}
    \end{subfigure}
    \caption{ \small { 
 (a) A causal diamond centered at the origin of AdS space can be mapped to two equally large AdS-Rindler wedges in two different Poincar\'{e} patches. The situation is  depicted for a coordinate radius $R< L /\sqrt{3}$ of the centered ball, since for larger values $R>L/\sqrt{3}$ the diamond overlaps with the right Rindler wedge. (b) A small causal diamond and infinite size diamond in AdS space. The infinite size diamond is called the Wheeler-DeWitt patch of global AdS, and is also preserved by a conformal Killing vector.}}
\end{figure*}

\subsection{AdS-Rindler space}

In empty AdS space there also exists a Rindler wedge, which  admits a boost Killing vector. Unlike in Minkowski  space, accelerating observers in AdS start and end on the boundary at a finite global Killing time. 
In the Penrose diagram the AdS-Rindler wedge therefore has the shape of a half diamond rather than a full diamond (see Figure \ref{fig:adsrindler}). The vertices of the diamond are located at the AdS boundary, and are separated by an infinite proper time but a finite global  Killing  time. Due to the presence of the conformal boundary the first law of causal diamonds (\ref{firstlawcc1}) is not valid for AdS-Rindler space. In addition to the area variation one should  take   the boundary term $\int_\infty [\delta Q_\xi - \xi \cdot \theta (g, \delta g)]$ at spatial infinity into account    in the variational identity \eqref{varid}. This boundary term has been shown in \cite{Papadimitriou:2005ii,Hollands:2005wt} to be equal to the gravitational energy variation $ \delta E_{\zeta}^\text{g}  = \int_\infty \delta T_{ab}^{\rm{g}} \zeta^a_{\text{flat}} d \Sigma^b$, where $ T_{ab}^{\rm{g}}$ is the holographic stress-energy tensor, and $\zeta^{\text{flat}}$ is the boundary conformal Killing vector  given by (\ref{ckvflat1}). The first law for AdS-Rindler space can subsequently be derived along the lines of Section \ref{sec:firstlawch2} using Wald's variational identity, but now applied  to the boost Killing vector.

The variational identity was   established in \cite{Blanco:2013joa} for general relativity, and later generalized in \cite{Faulkner:2013ica} to an arbitrary higher-derivative theory of gravity. The bulk modular energy term $\delta H_\xi^{\tilde{\rm{m}}}$ was included   in the first law in \cite{Swingle:2014uza,Jafferis:2015del}, and it  has been extended to allow for variations of the cosmological constant in \cite{Kastor:2014dra,Caceres:2016xjz}. 
For general relativity the full first law reads 
\beq
 \delta \bar E_{\zeta}^{\text{g}}= \frac{1}{8\pi G} \left ( \kappa \, \delta A  + \bar V_\xi \,  \delta \L \right  )+ \delta H_\xi^{  \tilde{\rm{m} }}  \, .
\eeq
 The proper volume term is   absent  since   the gravitational Hamiltonian variation $\delta H^{\rm{g}}_\xi$ vanishes for a true 
  Killing vector.  In the energy variation $ \delta \bar E_{\zeta}^{\text{g}} : = \int_\infty [\delta Q_\xi  - \delta_\L Q^{\text{AdS}}_\xi - \xi \cdot \theta (g, \delta g)] $ we have subtracted the boundary term in the AdS background due to a variation of $\L$, such that it stays finite when the cosmological constant is   varied.\footnote{We note that $  \theta (g, \delta_\L g) = 0$  since the symplectic potential for general relativity is linear in $\nabla_c \delta g_{ab}$, and we have $\delta_\L g_{ab} = - (\delta  \L  /\L) g_{ab}$ (see also   Appendix C of \cite{Caceres:2016xjz}).} From the RHS   of the equation we have also   subtracted the outer boundary integral of the Noether charge variation $\delta_\L Q_\xi^{\text{AdS}}$,  which for general relativity is given by
  \beq
  \delta_\L Q_\xi^{\text{AdS}}  = - \frac{d-2}{2 \L}  Q_\xi^{\text{AdS}}  \delta \L
  = \frac{ \delta \L }{8\pi G}\,  \omega_\xi^{\text{AdS}} \, ,
  \eeq
  where $\omega_\xi^{\text{AdS}}$ is the Killing potential form in AdS.
This first equality is due to the scaling $Q_\xi^{\text{AdS}} \sim L^{d-2}$,  and the second equality follows from the definition of the Noether charge  $j_\xi = d Q_\xi$ and the Killing potential form $\xi \cdot \epsilon = d \o_\xi$, together with  equations \eqref{jchi} and  \eqref{onshellL}. This amounts to replacing the thermodynamic volume by the background subtracted thermodynamic volume:
   $\bar V_\xi = \int_\infty (\omega_\xi - \omega_\xi^{\text{AdS}}) - \int_{\mathcal H} \omega_\xi$ (similar to the expression in  footnote~\ref{notetheta}).  Since $\omega_\xi = \omega_\xi^{\text{AdS}}$ for the AdS-Rindler wedge, the boundary term at infinity cancels and only the horizon surface integral remains.
 More explicitly, in  \cite{Kastor:2014dra} the background subtracted thermodynamic volume was  found to be equal to: $\bar V_\xi 
  = -  \kappa \,  A \, L^2 / (d-1)$, 
  where $A$ is the area of the bifurcation surface of the AdS-Rindler horizon.

Although our first law of causal diamonds does not directly apply  to AdS-Rindler space, there exists a map from the causal diamond centered at the origin of empty AdS to the AdS-Rindler wedge. This map consists of shifting the  diamond   towards the boundary in the  $z$ direction, where $z$ is the radial Poincar\'{e} coordinate.
 At the level of the (conformal) isometries, the conformal Killing vector which preserves the diamond transforms   under this shift  into the boost Killing vector of AdS-Rindler space.
In   Appendix~\ref{sec:embedding} we 
 find
the conformal Killing vector that preserves a diamond in AdS explicitly in Poincar\'{e} coordinates (see equation (\ref{ckvpoin}))
 \begin{equation}   \label{adsckv1}
\zeta^{\text{AdS}} = \frac{1}{2R} \left [  \sqrt{L^2 + R^2 }  (2 z \partial_t + 2 t \partial_z ) -   (L^2 + t^2 + \vec x^2 + z^2)  \partial_t  -  2 t x^i \partial_i  -  2 t z \partial_z   \right]  .
\end{equation}
Now, by shifting the radial coordinate as $z \rightarrow z + \sqrt{L^2 + R^2}$, the conformal Killing vector above turns into the boost Killing vector of an AdS-Rindler wedge \cite{Faulkner:2013ica}
\begin{equation} \label{boostads}
\xi^{\text{AdS-Rindler}}    = \frac{1}{2R}\left [  \left ( R^2 - t^2 - \vec x^2 - z^2 \right)  \partial_t - 2 t x^i \partial_i - 2 t z \partial_z  \right]  . 
\end{equation}
At the boundary ($z=0$) this reduces to the conformal Killing vector $\zeta^{\text{flat}}$ in  (\ref{ckvflat1}).
The  boost Killing vector   becomes null on the Killing horizon $\mathcal H = \left \{ z^2 = (R \pm t)^2 - \vec x^2 \right \}$, and   vanishes at the vertices $  \left \{  t= \pm R, \, z=x^i=0   \right \}$ and  hemisphere $\mathcal B = \left \{  t= 0, \, z^2 +\vec x^2 =    R^2    \right \} $. Notice, though, that there are two solutions to these quadratic equations, one for which $z$ takes positive values and the other for which $z$ takes negative values. Thus,   one part of the causal diamond is mapped to the AdS-Rindler wedge in the $z>0$ Poincar\'{e} patch, and the other part is shifted to the equally large Rindler wedge   in the $z<0$ Poincar\'{e} patch (see Figure \ref{fig:adsrindler}).

\subsection{Wheeler-DeWitt patch in  AdS}
 
 The first law applies also  
 to   causal diamonds  whose spatial slice is an entire slice of AdS. 
 The spacetime region covered by such an infinite diamond is   commonly known as the ``Wheeler-DeWitt patch'' of AdS (see Figure \ref{fig:wdw}). In this  limit, i.e. $R/L \rightarrow \infty$,  the   vector field $\zeta^{\text{AdS}}$ (\ref{adsckv1})     takes the following simple form in Poincar\'{e} coordinates
 \beq \label{wdwckv}
 \zeta^{\text{WdW}} = z \partial_t  + t \partial_z \, . 
 \eeq
This limiting value  also follows directly from expression \eqref{ckvpoin} for $\zeta^{\text{AdS}}$ in terms of the generators of the conformal group.  
By taking the limit $R/L \to \infty$ of \eqref{firstlawcc1}, we obtain the first law  for the Wheeler-DeWitt patch 
\begin{equation} \label{firstlawwdw}
  \delta H^\text{\~m}_{\zeta} = -  \frac{\kappa}{8\pi G} \left [  \delta A  - \frac{d-2}{L} \delta V  \right] .
\end{equation}
In this limit the proper volume  $V  \approx A\, L / (d-2)$ of the $t=0$ timeslice in   AdS   is divergent, but its variation can be finite. 
Here we  have  restricted to fixed cosmological constant, i.e.  $\delta \L = 0$. 
 Note that the proper volume variation is still present  in the first law for the Wheeler-DeWitt patch, because the conformal Killing vector \eqref{wdwckv} is not a true Killing vector.

\section{Discussion}
\label{sec:discussion}

In this chapter we explored aspects of the gravitational thermodynamics of causal diamonds
in maximally symmetric spacetimes and their first order variations. Our starting point was the 
notion that the maximally symmetric diamonds behave as thermodynamic equilibrium states.
This is initially motivated by the examples offered by the static patch of de Sitter spacetime and the Rindler wedge of Minkowski spacetime, which are special cases of causal diamonds 
admitting a true Killing field.
Other maximally symmetric causal diamonds, in particular finite ones, admit only a conformal Killing vector. Since neither general relativity nor ordinary matter are conformally invariant, it is not 
at all clear from the outset that the presence of a conformal Killing symmetry should be adequate to
support the interpretation of a physical equilibrium state. However, it seems in all respects to be sufficient.
This can be traced   to   three important facts: (i) the conformal Killing vector is an ``instantaneous" Killing vector at the maximal volume slice, which (ii) behaves at the edge as a boost-like Killing vector, with a well-defined surface gravity; and (iii) in a maximally symmetric diamond  the vacuum of a conformal matter field restricted to the diamond is thermal with respect to the Hamiltonian that generates the conformal Killing flow. 

We first established a classical Smarr formula and   a ``first law" variational identity for causal diamonds
in maximally symmetric spacetime, i.e.\ in either Minkowski, de Sitter, or Anti-de Sitter spacetime.
Since we include a cosmological constant and variations thereof, the ``thermodynamic volume"~\cite{Kastor:2009wy,Dolan:2010ha,Cvetic:2010jb}  plays a role, generalized here to the case of a conformal Killing vector. The name is appropriate for this quantity since it is thermodynamically conjugate to the cosmological constant which is a type of pressure. It 
is defined, given a (conformal) Killing vector $\z$,
by $V_\z=\int_\S \z\cdot\e$, which might well be called the ``redshifted volume". For finite causal diamonds
it appears as such, while for infinite asymptotically Anti-de Sitter diamonds or black hole spacetimes it diverges.
We reviewed how finite relations have nevertheless been obtained 
by subtracting the same divergence from both sides of an equation. 

We then analyzed the thermodynamic interpretation of the first law,
finding that the gravitational temperature of a diamond is 
{\it minus} the Hawking temperature associated with the horizon of the conformal Killing vector.
The idea that a negative temperature should be assigned to the static patch of de Sitter  spacetime has been floated before \cite{Klemm:2004mb}, but not followed up. 
We found the that the notion appears sound, and indeed is required by the 
thermodynamics of causal diamonds in general. 
The consistency with the {\it positive} Gibbons-Hawking temperature 
hinges on the fact that, in gravitational thermodynamics, 
the matter contribution to the first law enters {\it either} as an energy contribution,  
{\it or} as an entropy contribution, unlike in ordinary thermodynamics. 
In the latter form,
the matter entropy combines with the Bekenstein-Hawking area entropy to comprise the 
generalized entropy. We showed that it works this way also for quantum corrections.
In establishing this for nonconformal matter we invoked a conjecture, 
previously postulated in \cite{Jacobson:2015hqa} and checked in \cite{Casini:2016rwj,Speranza:2016jwt},
to relate the conformal boost energy to the entanglement entropy. 

We next reformulated the entanglement equilibrium proposal of \cite{Jacobson:2015hqa}, 
replacing the role of generalized entropy maximization by 
a free energy extremization, and showed that in this way, 
the need to fix the volume is no longer present.
We  find  that extremization of the free conformal energy $H_\zeta - T S_{\text{BH}}$ 
at fixed cosmological constant is    
equivalent to the stationarity of the generalized entropy at fixed cosmological constant and fixed volume.
Therefore, the Einstein equation is implied by stationarity of the free energy at fixed cosmological constant, 
without fixing the volume. 

In the final section we considered limiting cases of causal diamonds, such as small diamonds, 
the de Sitter static patch, Rindler space, and AdS-Rindler space, and showed how our general result
for the first law appears and reduces to known results in these different settings. In particular, we gained some interesting perspective on the first law in Rindler space, viewed as the infinite size limit of a causal diamond.  

In one of the appendices we established a link between conformal Killing time and York time in a maximally
symmetric causal diamond. The latter time is defined by the mean curvature on a foliation by constant mean curvature slices. We found that these slices coincide with the foliation orthogonal to the conformal Killing vector, and that the conformal Killing time parameter also labels these surfaces.



%% file: chapter3.tex
\chapter{Higher Curvature Gravity from Entanglement Equilibrium} 
\label{ch3}


\section{Introduction} \label{sec:introhigher}

Black hole entropy 
remains one of the best windows into the nature 
of quantum gravity available to dwellers of the infrared. 
Bekenstein's original motivation for introducing it
was to avoid gross violations of the 
second law of thermodynamics by sending matter
into the 
black hole, decreasing the entropy of the exterior   
\cite{Bekenstein:1972tm, Bekenstein:1973ur}.  
The subsequent discovery by Hawking that black holes radiate thermally at 
a temperature $T_{\rm H}=\kappa/2\pi$, with $\kappa$ the surface gravity,
fixed the value of the entropy in terms of the area to be
$
S_\text{BH} = A/4G,
$
and suggested  a deep connection to  quantum properties
of gravity \cite{Hawking:1974rv}.

The appearance of area in $S_\text{BH}$ is somewhat mysterious from
a classical perspective; however, an intriguing explanation emerges by considering 
the entanglement entropy of quantum fields outside the horizon \cite{Sorkin:2014kta,
Bombelli:1986rw, Srednicki:1993im, Frolov:1993ym}.  Entanglement entropy is UV divergent,
and upon regulation it takes the form
\beq \label{eqn:SEE}
S_{\rm ent} = c_0 \frac{A}{\epsilon^{d-2}} +\{\text{subleading divergences}\} + S_\text{finite} \, ,
\eeq
with $\epsilon$ a regulator and $c_0$ a constant. 
Identifying the coefficient $c_0/\epsilon^{d-2}$ with $1/4G$ would allow $S_\text{BH}$
 to be attributed to the leading divergence in the entanglement entropy. 
The subleading divergences 
could similarly be associated with higher curvature gravitational couplings,  
 which change the expression for the black hole entropy 
to  the Wald entropy \cite{Wald:1993nt}.

To motivate these identifications, one must assume that the quantum gravity theory
is UV finite (as occurs in string theory), yielding a finite 
entanglement entropy,  cut off 
near the Planck length, $\epsilon \sim \ell_{\rm P}$.  
Implementing this cutoff would seem to depend on a detailed knowledge  
of the UV theory, 
inaccessible from the vantage of low energy effective field theory.  Interestingly, this issue can be resolved
 within the effective theory by the renormalization of  the gravitational
couplings by matter loop divergences.  
There is mounting evidence that 
these precisely match the entanglement entropy divergences, making 
the {\it generalized entropy}
\beq \label{eqn:Sgen}
S_\text{gen} = S_\text{Wald}^{(\epsilon)}+ S^{(\epsilon)}_{\rm m}
\eeq  
independent of $\epsilon$ \cite{Susskind:1994sm, 
Cooperman:2013iqr, Solodukhin:2011gn, Bousso:2015mna}. Here $S_\text{Wald}^{(\epsilon)}$ 
is the Wald entropy expressed in terms of the renormalized gravitational couplings and
$S^{(\epsilon)}_{\rm m}$ is a renormalized 
entanglement entropy of matter fields that is related to $S_\text{finite}$ in 
(\ref{eqn:SEE}), although the precise relation depends on the 
renormalization scheme.\footnote{A covariant 
regulator must be used to ensure that the subleading
divergences appear as  a Wald entropy.  Also, since power law divergences are not universal, 
when they are present the same renormalization scheme must be used for the 
   entanglement entropy
and the gravitational couplings.  
 }  
The identification of gravitational couplings with entanglement entropy divergences is
 therefore consistent with the renormalization group (RG)
flow in the low energy effective theory, and amounts to assuming that the bare
gravitational couplings vanish \cite{Jacobson:1994iw}.  In this picture, 
$S_\text{gen} = S_{\rm ent}$, with $S_\text{Wald}^{(\epsilon)}$ acting
as a placeholder for the UV degrees of freedom that have been integrated out.

When viewed as entanglement entropy, 
it is clear that generalized entropy can be assigned to surfaces other than black hole
horizon cross sections 
\cite{Jacobson:1999mi, Jacobson:2003wv, 
Bianchi:2012ev, Bousso:2015mna}.  For example, in holography
the generalized entropy of a minimal surface in the bulk is dual via the quantum-corrected
Ryu-Takayanagi formula \cite{Ryu:2006bv, Faulkner:2013ana} to the entanglement
entropy of a region of the boundary CFT.\footnote{The 
UV divergences in the CFT entanglement entropy have no relation to the Planck length
in the bulk, but instead are related to the infinite area of the minimal surface in AdS, courtesy
of the UV/IR correspondence.}
Even without assuming  holographic duality, 
the generalized entropy  provides a link between the geometry of  surfaces and  
entanglement entropy.   When supplemented with thermodynamic 
information, this link can give rise to dynamical equations for gravity.  
The first demonstration of this was Jacobson's derivation of the Einstein equation as an 
equation of state for local causal horizons possessing an entropy proportional to their area 
\cite{Jacobson:1995ab}.  Subsequent work using entropic arguments 
\cite{Verlinde:2010hp, Verlinde:2016toy} and holographic entanglement entropy \cite{Lashkari:2013koa, Faulkner:2013ica, Swingle:2014uza} confirmed that entanglement thermodynamics is connected to 
gravitational dynamics.

\begin{figure}[tp!]
\centering
\includegraphics[width=.5\textwidth]{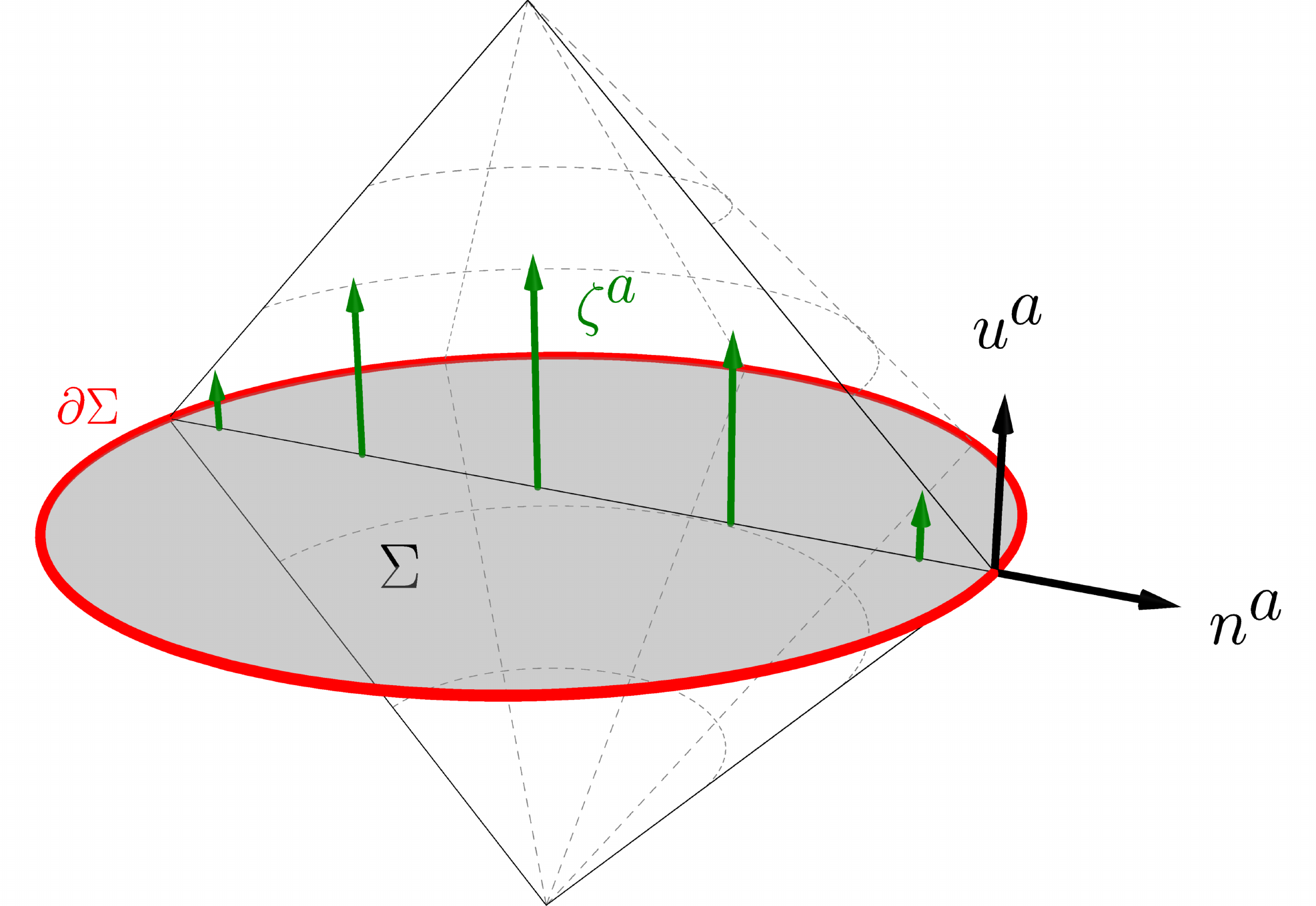}
\caption{\small The causal diamond consists of the future and past domains
of dependence of a spatial sphere $\Sigma$ in a MSS. 
$\Sigma$ has a unit normal $u^a$, induced metric $h_{ab}$, and volume form $\eta$.  
The boundary  $\partial\Sigma$ has a spacelike unit normal $n^a$, defined
to be orthogonal to $u^a$,  and volume form $\mu$.  
The conformal Killing vector $\zeta^a$ generates a flow within the causal diamond, and 
 vanishes on  the bifurcation surface $\partial\Sigma$.} \label{fig:diamond}
\end{figure}

Recently, Jacobson has advanced a new viewpoint on the relation between geometry 
and entanglement that has been dubbed ``entanglement equilibrium'' \cite{Jacobson:2015hqa}.
This proposal considers spherical, spatial subregions in geometries that are a 
perturbation of a maximally symmetric spacetime (MSS).  Each such subregion $\Sigma$
 in the maximally
symmetric background defines a causal diamond, which admits a conformal Killing vector 
$\zeta^a$
whose flow preserves the diamond (see Figure \ref{fig:diamond}).  
The entanglement equilibrium hypothesis states that any perturbation of the matter
fields and geometry inside the ball leads to a decrease in entanglement, i.e.  the vacuum
is a maximal  entropy state.  
This hypothesis applies holding the volume of $\Sigma$ fixed; even so, 
the
introduction of curvature from the geometry variation 
can lead to a decrease in the area of the boundary $\partial\Sigma$. 
This affects the divergent terms in the entanglement entropy by changing Wald entropy,
which at leading order is simply $A/4G$. The variation of the quantum
state  contributes a piece $\delta S_{\rm m}$, and maximality implies the   total
variation of the entanglement entropy vanishes at first order,\footnote{The separation
of the entanglement entropy into a divergent Wald piece and a finite matter piece is 
scheme dependent, and can change under the RG flow \cite{Jacobson:2012ek}.  
Also the matter variation can sometimes produce state-dependent divergences
\cite{Marolf:2016dob}, which appear as a variation of the Wald entropy.  Since
we only ever deal with total variations of the generalized entropy, these subtleties do not affect 
any results.  For simplicity, we will refer to $\delta S_\text{Wald}$ as coming from
the geometry variation, and $\delta S_{\rm m}$ from the matter state variation. } 
\beq \label{eqn:dSEEV}
\delta S_{\rm ent}\big|_V = \frac{\delta A\big|_V}{4G} + \delta S_{\rm m} = 0 \, .
\eeq
When applied to small spheres, this maximal entropy 
condition was 
shown to be equivalent to imposing the Einstein equation at the center of the ball.

Taken as an effective field theory, gravity is expected to contain higher curvature corrections 
that arise from matching to its UV completion.  An important test 
of the entanglement equilibrium hypothesis is whether it can consistently accommodate these 
corrections.  It is the purpose of this Chapter to demonstrate that a generalization
to higher curvature theories is possible, 
 and relates to  
the subleading divergences appearing in (\ref{eqn:SEE}).

\subsubsection{Summary of results and outline}
 It is not {\it a priori} clear what the precise statement of the entanglement equilibrium condition
should be for a higher curvature theory, and in particular what replaces the fixed-volume
constraint. 
The  formulation we propose here  is advised by
the {\it first law of causal diamond mechanics}, a purely geometrical identity that holds independently of 
any entanglement considerations.  
It was derived  for Einstein gravity
in the supplemental materials of \cite{Jacobson:2015hqa}, and one of the 
main results of this Chapter is to extend it to arbitrary, higher derivative theories.  
As we show in Section \ref{sec:firstlaw}, the first law is related to the off-shell identity 
\beq\label{titis}
\frac{\kappa}{2\pi} \delta S_{\text{Wald}}\big|_W + \delta H^{\rm m}_\zeta = \int_\Sigma \delta C_\zeta \, ,
\eeq   
where $\kappa$ is the surface gravity of $\zeta^a$ \cite{Jacobson:1993pf}, $S_\text{Wald}$ is the 
Wald entropy of $\partial \Sigma$ given in equation (\ref{eqn:SWald}) \cite{Wald:1993nt, Iyer:1994ys},
$H_\zeta^{\rm m}$ is the matter Hamiltonian for flows along $\zeta^a$, defined in equation 
(\ref{potati}), and $\delta C_\zeta=0$ are the linearized constraint equations of the higher
 derivative  theory.  The Wald entropy is varied holding fixed a local geometric 
functional
\beq \label{eqn:Wolume}
W = \frac{1}{(d-2)E_0} \int_\Sigma {\eta}\left(E^{abcd} u_a h_{bc} u_d - E_0\right) \, ,
\eeq
with $\eta$, $u^a$ and $h_{ab}$ defined in Figure \ref{fig:diamond}.  
$E^{abcd}$ is the variation of the gravitational Lagrangian scalar with respect to 
$R_{abcd}$, and $E_0$ is a constant determined by the value of $E^{abcd}$ in a MSS via
$E^{abcd}\overset{\text{MSS}}{=}E_0(g^{ac}g^{bd}-g^{ad}g^{bc})$.
We refer to $W$ as the ``generalized volume'' since it reduces to the volume for
Einstein gravity.

The Wald formalism contains ambiguities identified by Jacobson, Kang and Myers (JKM)
\cite{Jacobson:1993vj}  that modify the  Wald entropy and the 
generalized volume by the terms $S_\text{JKM}$ and $W_\text{JKM}$ 
given in (\ref{eqn:SJKM}) and (\ref{eqn:WJKM}).  Use a modified generalized
volume defined by
\beq \label{eqn:Wp}
W' = W + W_\text{JKM} \, ,
\eeq
the identity (\ref{titis}) continues to hold with $\delta (S_\text{Wald}+S_\text{JKM})\big|_{W'}$
replacing $\delta S_\text{Wald}\big|_W$\,.  As discussed in Section \ref{sec:subleading}, the subleading divergences for the 
entanglement entropy involve a particular resolution of the JKM ambiguity, while 
Section \ref{sec:fixedflux} argues that
the first law of causal diamond mechanics applies for \emph{any} resolution, as long as the 
appropriate generalized volume is held fixed.

Using the resolution of the JKM ambiguity required for the entanglement entropy calculation,
the first law leads to the following statement of entanglement
equilibrium, applicable to higher curvature theories:\\

\begin{hypothesis*}[Entanglement Equilibrium]
In a quantum gravitational theory, the entanglement entropy of a spherical 
region with fixed generalized volume $W'$ is maximal in vacuum.
\end{hypothesis*}

This modifies the original equilibrium condition (\ref{eqn:dSEEV}) by replacing the area
variation with 
\beq
\delta(S_\text{Wald} + S_\text{JKM})\big|_{W'} \, .
\eeq
In Section \ref{sec:equilibrium}, this equilibrium condition is shown
to be equivalent to   the linearized higher derivative field equations 
in the case that the matter fields are conformally invariant.\footnote{There is a proposal
for including nonconformal matter that involves varying a local cosmological constant
\cite{Jacobson:2015hqa, Casini:2016rwj, Speranza:2016jwt}.  If valid, this proposal applies 
in the higher curvature case as well, since it deals only with the matter variations.}  
Facts about  
entanglement entropy divergences and 
the reduced density matrix for a sphere in a CFT are used to relate the total variation of 
the entanglement entropy to the left-hand 
side of (\ref{titis}).  Once this is done, it becomes clear that imposing the linearized 
constraint equations is 
equivalent to the entanglement equilibrium condition.

In \cite{Jacobson:2015hqa}, this condition was applied in the small
ball limit, in which {\it any} geometry looks like a perturbation of a MSS.  Using Riemann
normal coordinates (RNC), the linearized equations were shown to impose the fully nonlinear
equations for the case of Einstein gravity.  
We will discuss this argument in Section \ref{sec:equations} for higher
curvature theories, and show that the nonlinear equations can \emph{not} be obtained from the 
small ball limit, making general relativity unique in that regard. 
Finally, in Section \ref{sec:comparisonto}  we describe how this work   compares to other approaches  connecting geometry and entanglement.  

\section{First law of  causal diamond mechanics} \label{sec:firstlaw}

Jacobson's 
entanglement equilibrium argument \cite{Jacobson:2015hqa} 
compares the surface area of a 
small spatial ball $\Sigma$ in a curved spacetime to the one that would be obtained in a MSS.  
The comparison is made using balls of equal volume $V$, a choice justified 
by an Iyer-Wald variational identity \cite{Iyer:1994ys} 
for the conformal Killing vector $\zeta^a$ of the causal diamond in the maximally symmetric background. When the Einstein equation holds, this identity implies the 
{\it first law of causal diamond mechanics} \cite{Jacobson:2015hqa, Jacobson:2018ahi}
\beq \label{barbecue}
- \delta H_\zeta^{\rm m} = \frac{\kappa}{8\pi G} \delta A   -\frac{\kappa k}{8\pi G}  \delta V \, ,
\eeq
where $k$ is 
  the trace of the extrinsic curvature of $\partial \Sigma$ embedded in $\Sigma$,
and the matter conformal Killing energy $H_\zeta^{\rm m}$ is constructed from the  stress 
tensor $T_{ab}$ by
\begin{equation}\label{potati}
 H_{\zeta}^{\rm m}=\int_\Sigma\eta\, u^a \zeta^bT_{ab}\, .
 \end{equation}
 The purpose of this section is to generalize the variational identity to higher derivative
 theories, and to clarify its relation to the equations of motion.  
This is done by focusing on an off-shell version of the identity, which reduces to the first law
when the  linearized  constraint equations for the theory are satisfied.  We begin by reviewing the 
Iyer-Wald formalism in Section \ref{sec:IyerWald}, which also serves to establish notation.
After describing the geometric setup in Section \ref{subsec:setup}, we show in Section
\ref{sec:localgeo} how the quantities appearing in the identity can be written as variations of 
local geometric functionals of the surface $\Sigma$ and its boundary $\partial \Sigma$.  As
one might expect, the area is upgraded to the Wald entropy $S_\text{Wald}$, and we derive 
the generalization of the volume  given in equation (\ref{eqn:Wolume}). 
Section \ref{sec:fixedflux} describes how the variational identity can instead be viewed 
as a variation at fixed generalized volume $W$, as quoted in equation (\ref{titis}), 
and describes the effect that JKM
ambiguities have on the setup.

\subsection{Iyer-Wald formalism} \label{sec:IyerWald}
We begin by recalling the Iyer-Wald formalism \cite{Wald:1993nt, Iyer:1994ys}.  A general diffeomorphism invariant theory may 
be defined by its Lagrangian $L[\phi]$, a spacetime $d$-form locally constructed 
from the dynamical fields $\phi$, which include the metric and matter fields.  A variation of this
Lagrangian takes the form  
\beq \label{eqn:dL}
\delta L = E   \delta\phi + \dd\theta[\delta\phi] \, ,
\eeq
where  $E$ collectively denotes the equations of motion for the dynamical fields, and $\theta$
is the symplectic potential $(d-1)$-form.  Taking an antisymmetric variation of $\theta$ yields
the symplectic current $(d-1)$-form
\beq \label{eqn:symplectomega}
\omega[\delta_1\phi, \delta_2\phi] = \delta_1 \theta[\delta_2\phi] - \delta_2\theta[\delta_1\phi] \,,
\eeq
whose integral over a Cauchy surface $\Sigma$ gives the symplectic form for the phase space description
of the theory.  Given an \emph{arbitrary} vector field $\zeta^a$, 
   evaluating the symplectic
form on the Lie derivative  $\lie_\zeta \phi$  
gives the variation of the Hamiltonian $H_\zeta$ that generates the flow of  $\zeta^a$ 
\beq \label{eqn:hamilton}
\delta H_\zeta = \int_\Sigma \omega[\delta\phi, \lie_\zeta\phi] \,.
\eeq
Now consider   a ball-shaped region $\Sigma$, and take $\zeta^a$ to be any future-pointed,
timelike vector that vanishes on the boundary $\partial\Sigma$. Wald's variational identity then
 reads 
\beq \label{eqn:dHz}
\int_\Sigma \omega[\delta\phi, \lie_\zeta\phi] =  \int_\Sigma \delta j_\zeta \,,
\eeq
where the Noether current $j_\zeta$ is defined by
\beq
j_\zeta = \theta[\lie_\zeta \phi]-i_\zeta L \,.
\eeq
Here $i_\zeta$ denotes contraction of the vector $\zeta^a$ on the first index of the differential
form $L$. The identity (\ref{eqn:dHz}) holds when the background geometry satisfies
the field equations $E=0$, and it assumes that $\zeta^a$ vanishes on $\partial \Sigma$.  
Next we note that the 
Noether current can always be expressed as \cite{Iyer:1995kg}
\beq  \label{eqn:noethercurrent}
j_\zeta = \dd Q_\zeta + C_\zeta,
\eeq
where $Q_\zeta$ is the Noether charge $(d-2)$-form and $C_\zeta$ are the constraint
field equations, which arise as a consequence of the diffeomorphism gauge symmetry. 
For non-scalar matter, these constraints are a combination of the metric and matter field
equations \cite{Seifert:2006kv, Jacobson:2011cc}, but, 
assuming the matter equations  are imposed, 
we can take $C_\zeta = -2 \zeta^a E\indices{_a^b}\epsilon_b$, where $E^{ab}$ is the variation
of the Lagrangian density with respect to the  metric.  
By combining equations (\ref{eqn:hamilton}), (\ref{eqn:dHz}) and (\ref{eqn:noethercurrent}), one finds that 
\beq \label{eqn:FLDM}
-\int_{\partial\Sigma} \delta Q_\zeta+\delta H_\zeta =   \int_\Sigma \delta C_\zeta \, .
\eeq
When the linearized constraints hold, $\delta C_\zeta = 0$, the variation of the Hamiltonian
is a boundary integral of $\delta Q_\zeta$.  This on-shell identity forms the basis for 
deriving the first law of causal diamond mechanics.  Unlike the situation encountered
in black hole thermodynamics, $\delta H_\zeta$ is not zero because below we take $\zeta^a$ to be a conformal Killing vector as opposed to a true Killing vector.

\subsection{Geometric setup}
\label{subsec:setup}

Thus far, the only restriction that has been placed on the vector field $\zeta^a$ is that it 
vanishes on $\partial\Sigma$.  
As such, the quantities $\delta H_\zeta$ and $\delta Q_\zeta$ appearing 
in the identities depend rather explicitly on the fixed vector $\zeta^a$, and therefore these
quantities are not written in terms of only the geometric properties of the surfaces
$\Sigma$ and $\partial\Sigma$.  A purely geometric description is desirable if
the Hamiltonian and Noether charge are 
to be interpreted as  
thermodynamic state functions, which ultimately may be used to define the 
ensemble of geometries in any proposed quantum description of the microstates.  
This situation may be remedied by choosing the vector $\zeta^a$ and the surface
$\Sigma$ to have special properties in the background geometry.  In particular, by choosing
$\zeta^a$ to be a conformal Killing vector for a causal diamond in the MSS, and picking
$\Sigma$ to lie on the surface where the conformal factor vanishes, one finds that 
the perturbations $\delta H_\zeta$ and $\delta Q_\zeta$    have expressions in terms of local
geometric functionals on the surfaces  $\Sigma$ and $\partial\Sigma$, respectively.

Given a causal diamond in a MSS,
there exists a conformal Killing vector $\zeta^a$ which generates a flow within the diamond and 
vanishes at the bifurcation surface $\partial\Sigma$ (see Figure \ref{fig:diamond} on p. \pageref{fig:diamond}).  
The metric satisfies the conformal Killing equation
\beq  \label{eqn:lie}
\lie_\zeta g_{ab} =    2\alpha g_{ab} \quad \text{with} \quad \alpha = \frac{1}{d}  \nabla_c \zeta^c  \,.
\eeq
and the conformal factor $\alpha$ vanishes on the spatial ball $\Sigma$.  
The gradient of $\alpha$ is hence proportional to the unit normal to $\Sigma$, 
\beq
u_a = N \nabla_a \alpha  \quad \text{with} \quad N =  \lVert \nabla_a \alpha \rVert^{-1} .
\eeq
Note the vector $u^a$ is future pointing since the conformal factor $\alpha$ decreases to the 
future of $\Sigma$. 
In a MSS, the normalization 
function $N$ has the curious property that it is constant over   $\Sigma$, and is
given by \cite{Jacobson:2018ahi}\footnote{Notice that   $-1/N$ is equal to the quantity $\dot \alpha |_{\Sigma}$,     introduced in Section \ref{sec:diamond}.}
\beq \label{eqn:N}
N =   \frac{d-2}{\kappa  k},
\eeq
where $k$ is the trace of the extrinsic curvature of $\partial \Sigma$ embedded in $\Sigma$, and $\kappa$ is the surface gravity of the conformal Killing horizon, defined momentarily.  
This constancy ends
up being crucial to finding a local geometric functional for $\delta H_\zeta$. 
Throughout this work, $N$ and $k$ will respectively
denote  constants equal to the normalization
function and extrinsic curvature trace, both evaluated in the background spacetime.  

Since $\alpha$ vanishes on $\Sigma$, $\zeta^a$ is instantaneously a Killing vector.  On the other hand, the covariant derivative of $\alpha$  is nonzero, so
\beq \label{eqn:covlie}
\nabla_d (\lie_\zeta g_{ab}) \big |_\Sigma = \frac{2}{N} u_d g_{ab} \, .
\eeq
The fact that the covariant derivative is nonzero on  $\Sigma$
is responsible for making   $\delta H_\zeta$ nonvanishing. 
 A conformal Killing vector with a horizon has a well-defined surface gravity 
$\kappa$ \cite{Jacobson:1993pf},
and since $\alpha$ vanishes on $\partial\Sigma$, we can conclude that 
\beq\label{eqn:dz}
\nabla_a \zeta_b  \big  |_{\partial \Sigma}= \kappa n_{ab}\, , 
\eeq
where $n_{ab} = 2u_{[a} n_{b]}$ is the binormal for the surface $\partial\Sigma$,  
and $n^b$ is the outward
pointing spacelike unit normal to $\partial\Sigma$.  
Since $\partial\Sigma$ is a bifurcation surface of a conformal Killing horizon, 
$\kappa$  is  constant everywhere on 
it.
We provide an example of these constructions in Appendix \ref{appkill} where
we discuss the conformal Killing vector for a causal diamond in flat space.

\subsection{Local geometric expressions} \label{sec:localgeo}

In this subsection we    evaluate the Iyer-Wald identity (\ref{eqn:FLDM})   for an arbitrary higher derivative theory of gravity and for the geometric setup described above. The final on-shell 
result is given in (\ref{firstlawhigher}), which is the first law of causal diamond mechanics for higher derivative gravity.

Throughout the computation we   assume that the  matter fields are minimally coupled, so that the Lagrangian splits into a metric and matter piece $L = L^{\rm g}+L^{\rm m}$, 
 and we take $L^{\rm g}$ to be an \emph{arbitrary}, 
diffeomorphism-invariant function of the metric, Riemann tensor, and 
its covariant derivatives.  The symplectic potential and variation of the Hamiltonian  then   exhibit a similar  separation, $\theta  = \theta^{\rm g} +\theta^{\rm m}$ and $\delta H_\zeta = \delta H_\zeta^{\rm g}+ \delta H_\zeta^{\rm m}$, and so we can write equation (\ref{eqn:FLDM}) as 
\beq \label{newvarid}
-\int_{\partial \Sigma} \delta Q_\zeta+ \delta H_\zeta^{\rm g} + \delta H_\zeta^{\rm m} =  \int_\Sigma \delta C_\zeta \,.
\eeq
Below, we explicitly    compute the two terms $\delta H_\zeta^{\rm g}$ and $\int_{\partial \Sigma} \delta Q_\zeta$ 
for the present geometric context.

 \subsubsection{Wald entropy}

By virtue of equation (\ref{eqn:dz})  and the fact that $\zeta^a$ vanishes on $\partial\Sigma$, one can show that the integrated Noether charge
is simply related to the Wald entropy \cite{Wald:1993nt, Iyer:1994ys}
\begin{equation}
 -\int_{\partial\Sigma} 
Q_\zeta  =   \int_{\partial\Sigma}   \, E^{abcd}  \, \epsilon_{ab} \nabla_c \zeta_d =\frac{\kappa}{2\pi}S_\text{Wald} \,, \label{eqn:Sbar}
\end{equation}
where the Wald entropy is defined as
\beq \label{eqn:SWald}
S_\text{Wald} =  -  2\pi \int_{\partial\Sigma} \mu  \, E^{abcd} n_{ab} n_{cd} \,.
\eeq
 $E^{abcd}$ is the variation of the Lagrangian scalar   with respect to the 
Riemann tensor $R_{abcd}$  taken as an independent field, given in (\ref{defEtensor}), and $\mu$ is the 
volume form on $\partial \Sigma$, so that $\epsilon_{ab}
=-n_{ab}\wedge \mu$ there.  
The  equality (\ref{eqn:Sbar}) continues to hold at first order in perturbations, which 
can be shown following the same arguments as   given in \cite{Iyer:1994ys}, hence,
\begin{equation} \label{QWald}
\int_{\partial \Sigma} \delta Q_\zeta = -  \frac{\kappa}{2\pi } \delta S_\text{Wald} \,.
\end{equation}
 The minus sign is opposite to the black hole case \cite{Iyer:1994ys}, since the conformal Killing field   points in the opposite direction compared to the horizon Killing field in a black hole.  The conformal Killing field  is past pointing on the outside of the diamond, whereas the horizon Killing field is future pointing on the outside of the black hole horizon.

\subsubsection{Generalized volume} The gravitational part of  $\delta H_\zeta$ is related to the symplectic current $\omega[\delta g, \lie_\zeta g]$   via (\ref{eqn:hamilton}).  The symplectic
form has been computed 
 on an arbitrary background 
for any higher curvature gravitational theory whose Lagrangian is a function 
of the Riemann tensor, but not its covariant derivatives \cite{Bueno:2016ypa}.  Here,
we  take advantage of the maximal symmetry of the background to compute the symplectic
form and Hamiltonian for the causal diamond in any higher order theory, including those 
with  
derivatives of the Riemann tensor.

Recall that the symplectic current $\omega$ is defined in terms of  the symplectic potential~$\theta$ through (\ref{eqn:symplectomega}).
For a     Lagrangian that depends on the Riemann tensor and its covariant derivatives, the symplectic potential $\theta^{\rm g}$ is given in Lemma 3.1 of \cite{Iyer:1994ys} 
\begin{equation}
\theta^{\rm g}  = 2 E^{bcd}\nabla_d\delta g_{bc}  +S^{ab}\delta g_{ab}  +\sum_{i=1}^{m-1} 
T_i^{abcd a_1\ldots a_i} \delta \nabla_{(a_1}\cdots\nabla_{a_i)} R_{abcd} \, ,
\end{equation}
where 
$
E^{bcd} = \epsilon_a E^{abcd}
$
and
the tensors $S^{ab}$ and $T_i^{abcda_1\ldots a_i}$ are locally constructed from the metric, 
its curvature, and covariant derivatives of the curvature.  Due to the antisymmetry of 
$E^{bcd}$ in $c$ and $d$, the symplectic current takes the form 
\begin{align}
&\omega^{\rm g} = 2\delta_1 E^{bcd}\nabla_d\delta_2 g_{bc}-2E^{bcd}\delta_1\Gamma^{e}_{db}
\delta_2 g_{ec} +\delta_1 S^{ab}\delta_2 g_{ab}   \nonumber\\ 
 &+ \sum_{i=1}^{m-1} 
\delta_1T_i^{abcd a_1\ldots a_i} \delta_2 \nabla_{(a_1}\cdots\nabla_{a_i)} R_{abcd} - 
(1\leftrightarrow2) \label{eqn:og}  .
\end{align}
Next we specialize to the geometric setup described in Section \ref{subsec:setup}. We may thus employ the fact that we are perturbing around a maximally symmetric background.
This means the background curvature tensor takes the form
\beq\label{msb}
R_{abcd} = \frac{R}{d(d-1)}(g_{ac}g_{bd}-g_{ad}g_{bc})
\eeq
with a constant Ricci scalar $R$, so that $\nabla_e R_{abcd}=0$, and also $\lie_\zeta R_{abcd} \big|_\Sigma = 0$.
Since the tensors $E^{abcd}$, $S^{ab}$, and $T_i^{abcda_1\ldots a_i}$ are 
all constructed from the  metric and  
curvature, they will also have vanishing Lie derivative along
$\zeta^a$ when evaluated on $\Sigma$. 

Replacing $\delta_2 g_{ab}$ in equation (\ref{eqn:og}) with $\lie_\zeta g_{ab}$ and using (\ref{eqn:covlie}),
we obtain
\begin{equation} \label{eqn:olzg}
 \omega^{\rm g}[\delta g, \lie_\zeta g]  \big |_\Sigma =  
 \frac2N\left[2 g_{bc}u_d \delta E^{bcd} + E^{bcd}(u_d \delta g_{bc}
- g_{bd}u^e\delta g_{ec}) \right]   .
\end{equation}
We would like to write this as a variation of some scalar quantity.  To do so, 
we split off the background value of $E^{abcd}$ by writing
\beq\label{eqn:Fabcd}
F^{abcd} = E^{abcd} - E_0(g^{ac} g^{bd}-g^{ad}g^{bc}) \,.
\eeq
The second term in this expression is the background value, and, due to maximal symmetry, 
the scalar $E_0$ must be a constant determined by the parameters appearing in the Lagrangian.
By definition, $F^{abcd}$ is zero in the background, so any term in (\ref{eqn:olzg}) that depends
on its variation 
may be immediately written as a total variation, since variations of other tensors appearing
in the formula would multiply the background value of $F^{abcd}$, which vanishes.  Hence,
the piece involving $\delta F^{abcd}$ becomes
\beq\label{eqn:dF}
\frac{4}{N} g_{bc}u_d \delta(F^{abcd}\epsilon_a) = \frac{4}{N} 
\delta(F^{abcd} g_{bc}u_d\epsilon_a) \, . 
\eeq
The remaining terms simply involve replacing $E^{abcd}$ in (\ref{eqn:olzg}) with $E_0
(g^{ac}g^{bd}-g^{ad}g^{bc})$.  These terms then take exactly the same form as the 
terms that appear for general relativity, which we know from 
the appendix of \cite{Jacobson:2015hqa}
combine to give an overall variation of the volume.  The precise form of this variation when
restricted to $\Sigma$ is 
\beq
-\frac{4(d-2)}{N}\delta \eta \, ,
\eeq
where $\eta$ is the induced volume form on $\Sigma$.  Adding this to (\ref{eqn:dF}) 
produces
\beq \label{eqn:symplform}
\omega[\delta g, \lie_\zeta g]  \big |_\Sigma   = -\frac{4}{N} \delta\left[ \eta(E^{abcd} u_a u_d h_{bc} - E_0) \right] \,,
\eeq
where we used that $\epsilon_a = - u_a \wedge \eta$ on $\Sigma$. This leads us to define a generalized volume functional 
\beq \label{eqn:W}
W = \frac{1}{(d-2)E_0} \int_\Sigma{\eta}(E^{abcd}u_a u_d h_{bc} - E_0) \,,
\eeq
and the variation of this quantity is related to the variation of the gravitational Hamiltonian by 
\beq \label{eqn:dJz}
\delta H_\zeta^{\rm g} = -4E_0 \kappa k\, \delta W \,,
\eeq
where we have expressed $N$ in terms of $\kappa$ and $k$ using (\ref{eqn:N}).
We have thus succeeded in writing $ \delta H^{\rm g}_\zeta$ in terms of a 
local geometric functional defined on the surface $\Sigma$. 

It is worth emphasizing that $N$ being constant over the ball was crucial to this derivation, 
since otherwise it could not be pulled out of the 
integral over $\Sigma$ and would define a non-diffeomorphism invariant structure on the surface.
Note that the overall normalization of $W$ is arbitrary, since a different
normalization would simply change the coefficient in front of $\delta W$ in (\ref{eqn:dJz}).  As
one can readily check, the normalization in (\ref{eqn:W}) 
was chosen so that $W$  reduces to the volume in the case of Einstein gravity. In Appendix \ref{app:W} we provide   explicit expressions for the generalized volume in $f(R)$ gravity and quadratic gravity.

Finally, combining (\ref{QWald}), (\ref{eqn:dJz}) and (\ref{newvarid}), 
we arrive at the off-shell variational identity in terms of local geometric quantities
\beq \label{eqn:offshelllocalgeo}
\frac{\kappa}{2\pi} \delta S_\text{Wald} -4E_0 \kappa k \delta W + \delta H_\zeta^{\rm m} = 
\int_\Sigma \delta C_\zeta \, .
\eeq
By imposing the linearized constraints $\delta C_\zeta = 0$, this becomes 
the first law of causal diamond mechanics for higher derivative gravity 
\beq \label{firstlawhigher}
- \delta H_\zeta^{\rm m} = \frac{\kappa}{2 \pi} \delta S_{\text{Wald}} - 4 E_0 \kappa k  \delta W  \, .
\eeq
This reproduces (\ref{barbecue}) for Einstein gravity with Lagrangian $L = \epsilon R/16\pi G$,
for which $E_0 = 1/32\pi G$.

\subsection{Variation at fixed $W$} \label{sec:fixedflux}
We now show that the first two terms in (\ref{eqn:offshelllocalgeo}) can be written in 
terms of the variation of the Wald entropy at fixed $W$, defined as
\beq \label{eqn:dXbarY}
\delta S_\text{Wald}\big|_{W} = \delta S_\text{Wald} - \frac{\partial S_\text{Wald}}{\partial W}  \delta W \, .
\eeq
Here we must specify what is meant by $\frac{\partial S_\text{Wald}}{\partial W}$.  
We will take this partial
derivative to refer to the changes that occur in both quantities when the size of the ball is 
deformed, but the metric and dynamical fields are held fixed.  Take a vector $v^a$ that is 
everywhere tangent to $\Sigma$ that defines an infinitesimal change in the shape of $\Sigma$.
The first order change this produces in $S_\text{Wald}$ and $W$ can be computed by 
holding $\Sigma$ fixed, but varying the Noether current and Noether charge as $\delta j_\zeta
= \lie_v j_\zeta$ and $\delta Q_\zeta = \lie_v Q_\zeta$.  
Since the background field equations are satisfied  and $\zeta^a$ vanishes on
$\partial \Sigma$, we have there that $\int_{\partial \Sigma}   Q_\zeta = \int_\Sigma j_\zeta^{\rm g}$,
without reference to the matter part of the Noether current.  
Recall that $\delta W$ is related to the variation of the gravitational Hamiltonian, which can be expressed in terms of $\delta j^{\rm g}_\zeta$ through (\ref{eqn:hamilton}) and (\ref{eqn:dHz}).
Then using the relations (\ref{eqn:Sbar}) and 
(\ref{eqn:dJz}) and the fact that the Lie
derivative commutes with the exterior derivative, we may compute
\beq
\frac{\partial S_\text{Wald}}{\partial W} = \frac{-\frac{2\pi}{\kappa}\int_{\partial\Sigma} 
\lie_v Q_\zeta}{-\frac{1}{4E_0 \kappa k}\int_\Sigma \lie_v j_\zeta^{\rm g}}
=  8 \pi E_0 k \,.
\eeq
Combining this result with equations  (\ref{firstlawhigher}) and (\ref{eqn:dXbarY}) we arrive at the off-shell variational identity for higher derivative gravity quoted in the introduction
\beq \label{monkey}
\frac{\kappa}{2\pi} \delta S_\text{Wald}\big|_W +   \delta H_\zeta^{\rm m}
 = \int_\Sigma \delta C_\zeta \,.
\eeq
Finally, we comment on how JKM ambiguities \cite{Jacobson:1993vj} 
 affect this identity.  The particular ambiguity we are concerned with
comes from the fact that the symplectic potential $\theta$ in equation (\ref{eqn:dL}) 
is defined only up to 
addition of an exact form $\dd Y[\delta \phi]$ that is linear in the field variations and their 
derivatives. This has the effect of changing the Noether current and Noether charge by
\begin{align}
j_\zeta &\rightarrow j_\zeta + \dd Y[\lie_\zeta \phi] \, ,\\
Q_\zeta &\rightarrow Q_\zeta + Y[\lie_\zeta \phi] \,. \label{eqn:QJKM}
\end{align}
This modifies both the entropy and the generalized volume by  surface terms on $\partial\Sigma$
given by
\begin{align}
   S_{\text{JKM}} &= - \frac{2 \pi}{\kappa} \int_{\partial \Sigma} Y[\lie_\zeta \phi]  \, , \label{eqn:SJKM}\\
  W_\text{JKM} &= - \frac{1}{4 E_0 \kappa k} \int_{\partial\Sigma}  Y[\lie_\zeta \phi] \label{eqn:WJKM} \,.
\end{align}
However, it is clear that this combined change in $j_\zeta$ and $Q_\zeta$ leaves the 
left-hand side of (\ref{monkey}) unchanged, since the $Y$-dependent terms cancel out.  
In particular,
\beq
\delta S_\text{Wald}\big|_{W} = \delta(S_\text{Wald}+S_\text{JKM})\big|_{W+W_\text{JKM}} \,,
\eeq
showing that any resolution of the JKM ambiguity gives the same first law, provided that 
   the Wald entropy   and generalized volume   are modified by the terms (\ref{eqn:SJKM}) and 
(\ref{eqn:WJKM}).    This should be expected, because the right-hand side of
(\ref{monkey}) depends only on the field equations, which are unaffected by JKM 
ambiguities.


\section{Entanglement Equilibrium}\label{sec:equilibrium}

The original entanglement equilibrium argument for Einstein gravity stated that 
the total variation away from the vacuum of the entanglement of a region at fixed volume is zero.  
This statement is encapsulated in equation (\ref{eqn:dSEEV}), which 
shows both an   area variation due to the change in geometry, and a matter piece
from varying the quantum state.  The area variation at fixed volume can 
equivalently be written  
\beq
\delta A\big|_V = \delta A - \frac{\partial A}{\partial V} \delta V 
\eeq
and the arguments of Section \ref{sec:fixedflux} relate this combination to the 
terms appearing in the first law of causal diamond mechanics (\ref{barbecue}).  Since $\delta H_\zeta^{\rm m}$
in (\ref{barbecue})
is related to $\delta S_{\rm m}$ in (\ref{eqn:dSEEV}) for conformally invariant 
matter, the first law
may be interpreted entirely in terms of entanglement entropy variations.  

This section discusses the extension of the argument to higher derivative theories
of gravity.  Section
\ref{sec:subleading} explains how  subleading divergences in the entanglement entropy 
are related to a Wald
entropy, modified by a particular resolution of the JKM ambiguity.  Paralleling the 
Einstein gravity derivation, we seek to relate variations of the subleading 
divergences to the higher derivative
first law of causal diamond mechanics (\ref{firstlawhigher}).  Section 
\ref{cons} shows that this can be done as long as the generalized volume $W'$ 
[related to 
$W$ by a boundary JKM term as in (\ref{eqn:Wp})] is held fixed.  Then, using the relation of the 
first law to the off-shell identity (\ref{monkey}), we discuss how the entanglement 
equilibrium condition is equivalent to imposing the linearized constraint equations.

\subsection{Subleading entanglement entropy divergences} \label{sec:subleading}
The structure of  divergences in entanglement entropy is reviewed in \cite{Solodukhin:2011gn}
and the appendix of \cite{Bousso:2015mna}.  It is well known that the leading divergence 
depends on the area of the entangling surface.  More surprising, however, is the fact that 
this divergence precisely matches the matter field divergences that renormalize Newton's 
constant.  This ostensible coincidence arises because the two divergences have a 
common origin in the gravitational effective action $I_\text{eff}$, which
includes both gravitational and matter pieces.  Its relation to 
entanglement entropy comes from the replica trick, which defines the entropy as
\cite{Callan:1994py,
Calabrese:2009qy} 
\beq \label{eqn:replica} 
S_{\rm ent} = (n\partial_n - 1)I_\text{eff}(n)\big|_{n=1} \,,
\eeq
where the effective action $I_\text{eff}(n)$ is evaluated on a   manifold with a conical singularity 
at the entangling surface whose excess angle is $2\pi(n-1)$. 

As long as a covariant regulator is used to define the theory, 
the effective action will consist of terms that are local, diffeomorphism invariant
integrals over the manifold, as well as nonlocal contributions. 
All UV matter divergences must appear in the local piece of the effective action, 
and each combines with terms in the classical gravitational part of the action,
renormalizing  the gravitational coupling constants.  Furthermore, each such local
term contributes to the entanglement entropy in (\ref{eqn:replica}) only at the conical 
singularity, giving a local integral over the entangling surface \cite{Fursaev:1994ea,
Larsen:1995ax, Cooperman:2013iqr}.  

When the entangling surface is the bifurcation surface
of a stationary horizon, this local integral is simply the Wald entropy 
\cite{Nelson:1994na, Iyer:1995kg}. On nonstationary entangling surfaces, 
the computation can be done using the squashed cone techniques of \cite{Fursaev:2013fta},
which yield terms involving extrinsic curvatures that modify the Wald entropy.  
In holography, the
squashed cone method plays a key role in the proof of the Ryu-Takayanagi
formula \cite{Ryu:2006bv, Lewkowycz:2013nqa}, and its higher curvature generalization
\cite{Dong:2013qoa, Camps:2013zua}.  The entropy functionals obtained
in these works seem to also apply outside of holography, giving the 
extrinsic curvature terms in the entanglement entropy for general theories \cite{Fursaev:2013fta,
Bousso:2015mna}.\footnote{For 
terms involving four or more powers of extrinsic curvature, there are additional
subtleties associated with the so-called ``splitting problem'' \cite{Miao:2014nxa, Miao:2015iba, Camps:2016gfs}. }

The extrinsic curvature modifications to the Wald entropy in fact take the form of a 
JKM Noether charge ambiguity \cite{Jacobson:1993vj, Sarkar:2013swa, Wall:2015raa}.  
To see this,  note the vector $\zeta^a$  used to define the Noether charge 
vanishes at the entangling surface and its covariant derivative is antisymmetric and 
proportional to the binormal as in equation (\ref{eqn:dz}).  This means it  
acts like a boost on the normal bundle at the entangling surface.  General covariance
requires that any extrinsic curvature contributions can be written as a sum of boost-invariant
products,
\beq
S_\text{JKM} = \int_{\partial\Sigma} \mu \sum_{n\geq 1} B^{(-n)} \cdot C^{(n)}
\eeq
where the superscript $(n)$ denotes the boost weight of that tensor, so that at the surface:
$\lie_\zeta C^{(n)} = n C^{(n)}$.  
It is necessary that the terms consist of two pieces, each of which has nonzero boost weight,
so that they can be written as
\beq
S_\text{JKM} = \int_{\partial\Sigma} \mu \sum_{n\geq1}\frac1n B^{(-n)}\cdot \lie_\zeta C^{(n)} \, .
\eeq
This is of the form of a Noether charge ambiguity from equation (\ref{eqn:QJKM}), 
with\footnote{This formula defines $Y$ at the entangling
surface, and allows for some arbitrariness in defining it off the surface.  It is not clear that $Y$
can always be defined as a covariant functional of the form $Y[\delta\phi, 
\nabla_a\delta\phi,\ldots]$ without reference to additional structures, such as the normal
vectors to the entangling surface.  It would be interesting to understand better if and when 
$Y$ lifts to such a spacetime covariant form off the surface. }
\footnote{We thank
Aron Wall for this explanation of JKM ambiguities.} 
\beq
Y[\delta \phi] = \mu \sum_{n\geq1}\frac1n B^{(-n)} \delta C^{(n)} \,.
\eeq
The upshot of this discussion is that all terms in the entanglement entropy that are local on the
entangling surface, including all divergences, are given by a Wald entropy modified by
specific JKM terms.  
The couplings for the Wald entropy are determined by 
matching to the UV 
completion, or, in the absence of the UV description, these are simply parameters
characterizing the low energy effective theory. 
In induced gravity scenarios, the divergences are determined  by 
the matter content of the theory, and the matching
to gravitational couplings  
has 
been borne
out in explicit examples \cite{Frolov:1996aj, Myers:2013lva, Pourhasan:2014fba}.

\subsection{Equilibrium condition as gravitational constraints}\label{cons}

We can now relate the variational identity (\ref{monkey}) to entanglement entropy.  
The reduced density matrix for the ball in vacuum takes the form
\beq
\rho_\Sigma = \frac{1}{Z}e^{- K / T_{\rm H}}  \,,
\eeq
where $K$ is the modular Hamiltonian, $T_{\rm H} = \kappa/ 2\pi$ is the Hawking temperature
and $Z$ is the partition function, ensuring
that $\rho_\Sigma$ is normalized.  
Since the matter is conformally invariant, the 
modular Hamiltonian is equal to the matter Hamiltonian  
defined in (\ref{potati}) \cite{Hislop1982, Casini:2011kv}
\beq \label{eqn:Hmod}
K 
=   H_\zeta^{\rm m} \,. 
\eeq 
Next we apply the first law of 
entanglement entropy \cite{Blanco:2013joa, Bhattacharya2012}, 
which states that the first order perturbation to the
entanglement entropy is given by the change in modular Hamiltonian expectation value divided by the temperature 
\beq
\delta S_{\rm ent} = \frac{2\pi}{\kappa} \delta\vev{K } \, .
\eeq
Note that this equation holds for a fixed geometry and entangling surface, and hence
coincides with what was referred to as $\delta S_{\rm m}$ in Section \ref{sec:introhigher}.   When varying the 
geometry, the divergent part of the entanglement entropy changes due to a 
change in the Wald entropy and JKM terms of the entangling surface.  
The total variation of the entanglement entropy is therefore
\beq \label{eqn:dSEEtot}
\delta S_{\rm ent}=  \delta (S_\text{Wald}+S_\text{JKM})  + \frac{2\pi}{\kappa} \delta\vev{ K } \, .
\eeq
At this point, we must give a prescription for defining the surface $\Sigma$ in the perturbed
geometry.  Motivated by the first law of causal diamond mechanics, we  require that $\Sigma$ 
has the same generalized volume $W'$ as in vacuum, where $W'$ differs from 
the  quantity  $W$ by a JKM term, as in equation (\ref{eqn:Wp}). This provides
a diffeomorphism-invariant criterion for defining the size of the ball.  It does
not fully fix all properties of the surface, but it is enough to derive the equilibrium
condition for the entropy.  As argued in Section \ref{sec:fixedflux}, the first term in 
equation (\ref{eqn:dSEEtot}) can be written instead as $\delta S_\text{Wald}\big|_W$ 
when the variation is taken holding $W'$ fixed. 
Thus, from equations (\ref{monkey}), 
(\ref{eqn:Hmod}) and (\ref{eqn:dSEEtot}), we arrive at our main result, the equilibrium condition
\beq \label{eqn:dSEEW}
\frac{\kappa}{2\pi}\delta S_{\rm ent} \big|_{W'} = \int_\Sigma \delta C_\zeta \,,
\eeq
valid for minimally coupled, conformally invariant matter fields.

The linearized constraint equations $\delta C_\zeta=0$ may therefore be interpreted as 
an equilibrium condition on entanglement entropy for the vacuum.  
Since all first variations of the entropy vanish when the linearized gravitational constraints 
are 
satisfied, the vacuum is an extremum of entropy for 
regions with fixed generalized volume $W'$, which is necessary for it
to be an equilibrium 
state. 
Alternatively, postulating that entanglement entropy is maximal in vacuum
for all balls and in all frames would allow one to conclude that the linearized higher
derivative equations hold everywhere.

\section{Field equations from     equilibrium condition} \label{sec:equations}

The entanglement equilibrium hypothesis provides a clear connection between the 
linearized gravitational constraints and the maximality of entanglement entropy at 
fixed $W'$ in the vacuum for conformally invariant matter.  In this section, 
we will
consider whether information about the fully nonlinear 
field equations can be gleaned from the equilibrium condition. Following the approach
taken in \cite{Jacobson:2015hqa}, we employ
a limit where the ball is taken to be much smaller than all relevant scales in the problem, but
much larger than the cutoff scale of the effective field theory, which is  set by the 
gravitational coupling constants.
By expressing the linearized equations in Riemann normal
coordinates, one can infer that the full \emph{nonlinear} field equations hold in the 
case of Einstein gravity.
As we discuss here, such a conclusion can \emph{not} be reached for higher curvature
theories.  The main issue is that higher order terms in the RNC expansion are needed to 
capture the effect of higher curvature terms in the field equations, but these contribute
at the same order as nonlinear corrections to the linearized equations.

We begin by reviewing the argument for Einstein gravity.
Near any given point, the metric looks locally flat, and has an expansion in 
terms of Riemann normal coordinates that takes the form
\beq
g_{ab}(x) = \eta_{ab}-\frac13x^c x^d R_{acbd}(0) +\mathcal{O}(x^3)\, ,
\eeq
where   $(0)$ means evaluation at the center of the ball. 
At distances small compared to the radius of curvature, the second term in this expression
is a small perturbation to the flat space metric $\eta_{ab}$.  Hence, we may apply 
the off-shell identity (\ref{eqn:dSEEW}), using the first order variation
\beq \label{eqn:dgRNC}
\delta g_{ab} = -\frac13 x^c x^d R_{acbd} (0) \,,
\eeq
and conclude that the linearized constraint $\delta C_\zeta$ 
holds for this metric perturbation.  When 
restricted to the surface $\Sigma$, 
this constraint in Einstein gravity is \cite{Seifert:2006kv}
\beq
C_\zeta\big|_\Sigma=-u^a \zeta^b\left(\frac1{8\pi G} G_{ab}-T_{ab}\right) \eta \,.
\eeq 
Since the background constraint is assumed to hold, the perturbed constraint is 
\beq
\delta C_\zeta\big|_\Sigma = -u^a \zeta^b\left(\frac{1}{8\pi G} \delta G_{ab}- \delta T_{ab}\right)\eta\, ,
\eeq
but in Riemann normal coordinates, we have that the linearized perturbation to the curvature is
just $\delta G_{ab} = G_{ab}(0)$, up to terms suppressed by the ball radius.  
Assuming that the ball is small enough so that the stress tensor
may be taken constant over the ball, one concludes that the vanishing constraint implies the 
nonlinear field equation  at the center of the ball\footnote{In this equation, 
$\delta T_{ab}$ should be thought of as a quantum expectation value of the 
stress tensor.  Presumably, for sub-Planckian energy densities and in the small ball limit, this first order
variation approximates the true energy density.  However, 
there exist states for which the change in stress-energy is zero at first order in perturbations
away from the vacuum, most notable for coherent states \cite{Varadarajan:2016kei}.  Analyzing how these states can be incorporated into the entanglement equilibrium 
story deserves further attention. }
\beq
u^a \zeta^b(G_{ab}(0) - 8\pi G \delta T_{ab}) = 0\,.
\eeq
The procedure outlined above applies at all points and all frames, allowing us to obtain the full tensorial Einstein equation.

The situation in higher derivative theories of gravity is much different.
 It is no longer the case that the linearized equations evaluated in RNC imply the full nonlinear 
field equations   in a small ball.
To see this,  consider an $L[g_{ab},R_{bcde}]$  
higher curvature theory.\footnote{Note that an analogous argument should hold for general higher derivative theories, which also involve covariant derivatives of the Riemann tensor.} The equations of motion read
\begin{equation}\label{eomhigh}
- \frac{1}{2} g^{ab} \mathcal L  + E^{aecd} \tensor{R}{^b_{ecd}} - 2  \nabla_c \nabla_d E^{acdb}
= \frac{1}{2} T^{ab} \,.
\end{equation}
In Appendix \ref{app:FLDMRNC} we show that linearizing these equations
around a Minkowski background leads to
\begin{align}\label{lineomhigh}
 \frac{\delta G^{ab}}{16\pi G} - 2  \partial_c \partial_d \delta E^{a c d b}_{\text{higher}} = \frac{1}{2} \delta T^{ab}\, ,
\end{align}
where we split $E^{abcd}=E^{abcd}_{\text{Ein}}+E^{abcd}_{\text{higher}}$ into its Einstein piece, which gives rise to the Einstein tensor, and a piece coming from higher derivative terms.
As noted before, the variation of the Einstein tensor evaluated in RNC gives the nonlinear Einstein tensor, up to corrections that are suppressed by the ratio of the ball size to the radius of curvature. 
However, in a higher curvature theory of gravity, the equations of motion 
\eqref{eomhigh} contain terms that are nonlinear in the curvature.
Linearization around a MSS background of these terms would 
produce, schematically, $\delta ( R^n ) = n \bar{R}^{n-1} \delta R$, where $\bar{R}$ denotes evaluation in the MSS background.
In Minkowski space, all such terms would vanish.
This is not true in a general MSS, but evaluating the curvature tensors in the 
background still leads to a significant loss of information about the tensor structure 
of the equation.  
We conclude that the linearized equations cannot reproduce the full nonlinear 
field equations for higher curvature gravity, and it is only the linearity of the Einstein equation
in the curvature that allows the nonlinear equations to be obtained for general relativity.

When linearizing around flat space, the higher curvature corrections to the Einstein equation  are entirely captured by the  second term in \eqref{lineomhigh}, which features  
four derivatives acting on the metric, since $E_\text{higher}^{abcd}$ is constructed from 
curvatures that already contain two derivatives of the metric.
Therefore, one is insensitive to higher curvature corrections unless at least $\mathcal{O}(x^4)$ corrections \cite{Brewin:2009se} are added to the Riemann normal coordinates expansion \eqref{eqn:dgRNC}
\beq\label{eqn:dg4}
\delta g^{(2)}_{ab} = x^c x^d x^e x^f\bigg(\frac2{45}R\indices{_a _c_d^g}R\indices{_b_e_f_g} -\frac1{20}\nabla_c\nabla_d R_{aebf}\bigg) \,.
\eeq
Being quadratic in the Riemann tensor, this term contributes at the same order as 
the nonlinear corrections to the linearized field equations. Hence, linearization based on 
the RNC expansion up to $x^4$ terms is not fully self-consistent.
This affirms the claim that for higher curvature theories, the nonlinear equations at a point cannot be derived by only imposing the linearized equations.

\section{Comparison to other ``gravitational dynamics from entanglement'' approaches}
\label{sec:comparisonto}

Several proposals have been put forward to understand gravitational dynamics
in terms of thermodynamics and entanglement.  Here we will compare the 
entanglement equilibrium program considered in this chapter  
to two other approaches: the equation of state for
local causal horizons, and gravitational dynamics from holographic entanglement
entropy  (see \cite{Carroll:2016lku} for a related discussion).

\subsubsection{Causal horizon equation of state}

By assigning an entropy proportional to the area of local causal horizons, Jacobson
showed that the Einstein equation arises as an equation of state  
\cite{Jacobson:1995ab}.  This approach employs a physical process first law for the local 
causal horizon, defining a heat $\delta Q$ as the flux of local boost energy across the horizon.  
By assigning an entropy $S$ to the horizon proportional to its area, one finds that the 
Clausius relation $\delta Q = Td S$ applied to all such horizons is equivalent to the Einstein
equation.

The entanglement equilibrium approach differs in that it employs an equilibrium state first 
law [equation (\ref{firstlawhigher})], instead of  a physical process one \cite{Wald1994}.  
It therefore represents 
a different perspective that focuses on the steady-state behavior, as opposed to dynamics
involved with evolution along the causal horizon.
It is consistent therefore that we obtain constraint
equations in the entanglement equilibrium setup, since one would not expect evolution
equations to arise as an equilibrium condition. 
  That we can infer dynamical equations
from the constraints is related to the fact that the dynamics of diffeomorphism-invariant 
theories is entirely determined by the constraints evaluated in all possible Lorentz frames.  

Another difference comes from the focus on spacelike balls as opposed to local causal 
horizons.  Dealing with a compact spatial region has the advantage of providing an IR 
finite entanglement entropy, whereas the entanglement associated with local causal
horizons can depend on fields far away from the point of interest.  This allows us to give 
a clear physical interpretation for the surface entropy functional as entanglement entropy, 
whereas such an interpretation is less precise in the equation of state approaches.  

Finally, we note that both approaches attempt to obtain fully nonlinear equations by
considering ultralocal regions of spacetime.  In both cases the derivation of the field equations
for Einstein gravity is fairly robust, however higher curvature corrections present some problems.
Attempts have been made in the local causal horizon approach
that involve modifying
the entropy density functional for the horizon
\cite{Eling:2006aw, Elizalde:2008pv, Chirco:2010sw,
 Brustein:2009hy, Parikh2016, Dey2016, Padmanabhan:2009ry, Padmanabhan:2009vy,
 Guedens:2011dy}, but they meet certain challenges.  
 These include a need for a physical interpretation of the chosen entropy density functional, 
 and dependence of the entropy on arbitrary features of the local
Killing vector in the vicinity of the horizon
 \cite{Guedens:2011dy, Jacobson:2012yt}.  
While the entanglement equilibrium argument avoids these problems, it fails to get beyond
linearized higher curvature equations, even after considering the small ball limit.  The 
nonlinear equations in this case appear to involve information beyond first order perturbations,
and hence may not be accessible based purely on an equilibrium argument.

 \newpage

\subsubsection{Holographic entanglement entropy }
A different approach  comes from holography and the 
Ryu-Takayanagi formula \cite{Ryu:2006bv}.  By demanding that areas of
minimal surfaces
in the bulk match the entanglement entropies of spherical regions in the boundary CFT, one
can show that the linearized gravitational equations must hold \cite{Lashkari:2013koa, Faulkner:2013ica,
Swingle:2014uza}.  The argument employs an equilibrium state first law for the bulk geometry,
utilizing the Killing symmetry associated with Rindler wedges in the bulk.

The holographic approach is quite similar to the entanglement equilibrium argument since
both use equilibrium state first laws.  One difference is that the holographic argument must 
utilize minimal surfaces in the bulk, which extend all the way to the boundary of AdS.  This 
precludes using a small ball limit as can be done with the entanglement 
equilibrium derivation, and is the underlying reason that entanglement equilibrium can derive fully
nonlinear field equations in the case of Einstein gravity, 
whereas the holographic approach has thus far only obtained linearized equations.  Some progress has been made to go beyond 
linear order in the holographic approach by considering higher order perturbations
in the bulk \cite{Faulkner:2014jva, Lashkari:2015hha, Beach:2016ocq}.  Higher order 
perturbations will prove useful in the entanglement equilibrium program as well,
and has the potential to extend the higher curvature derivation to fully nonlinear equations.
Due to the similarity between the holographic and entanglement equilibrium approaches, 
progress in one will complement and inform  the other.


 \section{Discussion} \label{sec:conclusion5}
 Maximal entanglement of the vacuum state was proposed in \cite{Jacobson:2015hqa} as a new 
 principle in quantum gravity.  It hinges on the assumption that divergences in the 
 entanglement entropy are cut off  at short distances, so it
 is ultimately a statement about the UV complete quantum gravity theory.  However, the 
 principle can be phrased in terms of the generalized entropy, which is intrinsically UV 
 finite and well defined within the low energy effective theory.  
 Therefore, if true, maximal vacuum entanglement provides a 
 low energy constraint on any putative
 UV completion of a gravitational effective theory.  
 
 Higher curvature terms arise generically in any such effective  field theory.  Thus, it
 is important to understand how the entanglement equilibrium argument is modified by 
 them.  
As explained in Section \ref{sec:firstlaw}, the precise characterization of the entanglement equilibrium hypothesis  
relies on a classical variational identity for causal diamonds in maximally symmetric spacetimes.
This identity leads to equation (\ref{monkey}), which relates variations of the Wald entropy and 
matter energy density of the ball to the linearized constraints. The variations
are taken holding fixed a new geometric
functional $W$, defined in (\ref{eqn:W}), which we call the ``generalized volume.''

We connected this identity to entanglement equilibrium in Section \ref{sec:equilibrium},
invoking the   fact that subleading entanglement entropy divergences are given by a Wald
entropy, modified by specific JKM terms, which also modify $W$ by the boundary
term (\ref{eqn:WJKM}).  
  With the additional assumption that matter is conformally invariant, we arrived at our main result \req{eqn:dSEEW}, showing that the equilibrium condition $\delta S_{\rm ent}\big|_{W'}=0$ 
  applied to small balls is
  equivalent to imposing the linearized constraints $\delta C_\zeta = 0$.

In Section \ref{sec:equations}  we reviewed the argument that
in the special case of Einstein gravity, 
 imposing the linearized equations within small enough balls is equivalent to 
requiring that the fully nonlinear equations hold within the ball \cite{Jacobson:2015hqa}. Thus by considering spheres
centered at each point and in all Lorentz frames, one could 
conclude that the full Einstein equation  holds everywhere. Such an argument cannot be
made for a theory that involves higher curvature terms.  One finds that higher order terms
in the RNC expansion are needed to detect the higher 
curvature pieces of the field equations, but these terms enter at the same order as 
the nonlinear corrections to the linearized equations. This signals 
a breakdown of the perturbative
expansion unless the curvature is small.   

 There
is a subtlety associated with whether the solutions within each small
ball can be consistently glued together to give a solution over all of spacetime.  
One must solve for the gauge transformation relating the Riemann normal 
coordinates at different nearby points, and errors in the linearized approximation
could accumulate as one moves from point to point.  The question of whether the ball
size can be made small enough so that the total accumulated error goes to zero
deserves further attention.

The fact that we obtain only linearized equations for the higher curvature theory 
is consistent with the effective field theory standpoint.  One could take 
the viewpoint that higher curvature corrections are 
 suppressed by powers of a UV scale, and the effective field theory is valid 
only when the curvature is small compared to this scale.  This suppression would suggest
that the linearized equations largely capture the effects of the higher curvature corrections
in the regime where effective field theory is reliable.

%% file: chapter4.tex
\chapter{Towards non-AdS Holography}
\label{ch4}


\section{Introduction}

In the last 20 years considerable progress has been made on the holographic description of Anti-de Sitter space \cite{Maldacena:1997re,Witten:1998qj,Aharony:1999ti}. An important open question concerns the reconstruction of the bulk spacetime from  the   boundary theory. In addition, little is known about the microscopic theories for more general spacetimes, such as Minkowski or de Sitter space. 

One of the general lessons from the AdS/CFT correspondence is the concept of holographic renormalization \cite{Henningson:1998gx, Akhmedov:1998vf,Alvarez:1998wr, Balasubramanian:1999jd, Skenderis:1999mm,deBoer:1999tgo}.  If one moves  the holographic boundary into the bulk one introduces a UV cutoff in the conformal field theory. Going further into the bulk increases  the cutoff length scale and reduces the number of microscopic degrees of freedom of the holographic theory.  This  holographic RG description of AdS spacetimes works well near the boundary at scales large compared to the curvature radius. The microscopic theory is, however, not so well understood if the boundary reaches the AdS scale and breaks down at sub-AdS scales.  Here the curvature radius of AdS is irrelevant, and the geometry becomes approximately that of flat space.  

A crucial hint about  the microscopic theory for non-AdS geometries comes from the holographic principle, motivated by  the Bekenstein-Hawking entropy formula \cite{Bekenstein:1973ur,Hawking:1974sw}:
\\[-2mm]
\begin{equation}  
S= {A\over 4G}   \label{BHformula} \, .
\end{equation}
The holographic principle states that the maximal number of microscopic degrees of freedom associated to a spacelike region is proportional to the area $A$ of its boundary in Planckian units  \cite{tHooft:1993dmi, Susskind:1994vu,Bousso:1999xy,Bousso:2002ju}. However,
an important indication that the microscopic  
holographic descriptions for super-AdS scales and those for sub-AdS scale and dS and 
Minkowski spacetimes are qualitatively different is given by the properties of black holes in these geometries:
large AdS black holes are known to have a positive specific heat, whereas the specific heat of black holes in sub-AdS, flat or dS is negative.  A complete understanding of this feature from a microscopic holographic perspective is still lacking, which is illustrative of our poor understanding of holography for non-AdS spacetimes. 

In this chapter we will   make   modest steps towards answering some of these questions. Our strategy is to follow the same line of reasoning as in the original paper \cite{Susskind:1998dq}, which clarified the holographic nature of the AdS/CFT correspondence by showing that it obeys all the general properties which are expected to hold in a microscopic theory that satisfies the holographic principle.  In their work Susskind and Witten established the UV-IR correspondence that is underlying AdS/CFT by relating the UV cutoff of the microscopic theory to the IR cutoff in the bulk. They furthermore showed that the number of degrees of freedom of the cutoff CFT is proportional to the area of the holographic surface  in Planckian units. They also pointed out that if the temperature of the CFT approaches the cutoff scale all the microscopic degrees of freedom become excited  and produce a state whose entropy is given by the Bekenstein-Hawking entropy for a black hole horizon  which coincides with the holographic boundary.

Following this same logic we focus on general features of the holographic theory for sub-AdS geometries and Minkowski and de Sitter space,  such as the number of microscopic degrees of freedom and the typical energy that is required to excite these degrees of freedom.  We will assume that   these   non-AdS geometries are also described by an underlying microscopic quantum theory, that obeys the general principles of holography.  
 A logical assumption is that the number of microscopic degrees of freedom of the cutoff boundary theory for general spacetimes is also determined by the area of the holographic boundary in Planckian units.  Here, by a `cutoff boundary' we mean a holographic screen located inside the spacetime at a finite distance from its `center'.  In this chapter we will for definiteness and simplicity only consider  spherically symmetric spacetimes, so that we can choose the center at the origin.  Our main cases of interest are empty (A)dS and Minkowksi space, but we will also study  Schwarzschild geometries. 

  In addition to the number of degrees of freedom we are interested in the excitation energy per degree of freedom. In sub-AdS, flat and de Sitter space we find that this excitation energy decreases with the distance from the center. One of our main conclusions is that the UV-IR correspondence, familiar from AdS/CFT,  is inverted in these spacetimes: long distances (=IR) in the bulk   correspond  to   low energies  (=IR)   in the microscopic theory. And contrary to AdS/CFT the number of degrees of freedom increases towards the IR of the microscopic theory. Hence, we are dealing with a  holographic quantum theory whose typical excitation energy decreases if the number of degrees of freedom increases.  This fact is directly related to the negative specific heat of black holes. 

   Our aim is to find an explanation of this counter-intuitive feature of the microscopic theory. 
  For this purpose we employ a conformal map between  three non-AdS geometries and spacetimes of the form $AdS_3\!\times\! S^{q}$.  This conformal map relates general features of the microscopic theories on holographic screens in both spacetimes, and allows us to identify the mechanism responsible for the inversion of the energy-distance relation compared to AdS/CFT.  We find that it is a familiar mechanism, often invoked in the microscopic description of black holes \cite{Strominger:1996sh, Maldacena:1996ds, Dijkgraaf:1996xw}, known as the `long string phenomenon'. This mechanism operates on large symmetric product CFTs, and identifies a twisted sector consisting of `long strings' whose typical excitation energy is considerably smaller than that of the untwisted sector. Our conclusion is that this same long string mechanism reduces the excitation energy at large distances in the bulk and towards the IR of the microscopic theory, and therefore explains the negative specific heat of non-AdS black holes.  Furthermore, it clarifies the   value of the vacuum energy of (A)dS, which, contrary to most expectations, differs from its natural value set by the Planck scale. 

The outline of this chapter is as follows.
 In Section \ref{sec:lessons} we use lessons from AdS/CFT to give a geometric definition of the number of holographic degrees of freedom and their excitation energy. In Section \ref{sec:towardsholographynonads} we present a conjecture relating the microscopic theories for two Weyl equivalent spacetimes and describe the conformal map from sub-AdS, Minkowski and de Sitter space to $AdS_3\!\times\! S^{q}$ type geometries. Section \ref{sec:longstring}  describes the long string mechanism and its relevance for non-AdS holography. Finally, in Section \ref{sec:physicalimplications} we discuss the negative specific heat of black holes and   vacuum energy of  (A)dS spacetimes.

\section{Lessons from the AdS/CFT correspondence}
\label{sec:lessons}

 In this chapter we are interested in obtaining a better understanding of the microscopic description of  spacetime. For definiteness we consider $d$-dimensional static, spherically symmetric spacetimes with a metric of the form
\begin{equation} \label{metric}
ds^2 = -  f(R) dt^2 + \frac{d R^2}{f(R)} + R^2 d \Omega_{d-2}^2 \, ,
\end{equation}
where $d\Omega_{d-2}$ is the line element on a $(d-2)$-dimensional unit sphere.
This class of metrics allows us to study (Anti-)de Sitter space, flat space and (A)dS-Schwarzschild solutions.  In these geometries we consider a $(d-2)$-surface $\mathcal S$  located at a finite radius $R$, corresponding to the boundary of a spacelike ball-shaped region $\mathcal B$ centered around the origin.  We will call $\mathcal S$ the `holographic screen' or `holographic surface'.   We will study     general features of the microscopic description of these spacetimes, where we imagine that   the quantum theory lives on the holographic screen $\mathcal S$.   We will start with discussing the familiar case of Anti-de Sitter space, where we have a good qualitative understanding of the holographic theory, on boundaries at a finite radius $R$.

\subsection{General features of a  microscopic holographic theory}
\label{sec:generalfeatures}

Our first goal is to present a number of general features of the holographic description of asymptotically AdS spacetimes in a way that is generalizable to other  spacetimes. 
 Motivated by \cite{Susskind:1998dq}, we focus on the following aspects of the microscopic holographic  description:\footnote{ The quantities $\mathcal C$ and $1/\epsilon$ correspond to $N_{dof}$ and $\delta$ in \cite{Susskind:1998dq}.   
 } 
\begin{eqnarray}
  \mathcal C &=&  \text{number of  UV  degrees of freedom of the holographic theory}  \nonumber \\
   \epsilon &=& \text{excitation energy per UV degree of freedom}  \nonumber \\
  \mathcal N &=& \text{total energy measured in terms of the cutoff energy $\epsilon$}. \nonumber 
\end{eqnarray}  
 More precisely, by $\epsilon$ we mean the total energy of the maximally excited state  divided by the number of UV degrees of freedom. The maximally excited state corresponds to a black hole in the bulk description. Below we will give a definition of each of these quantities purely in terms of the geometry in the neighbourhood of the holographic surface.  We will motivate and verify these definitions for the case of AdS/CFT, but afterwards we will apply those same definitions to other geometric situations.

The number of degrees of freedom $\mathcal C$ is in the case of AdS/CFT directly related to the central charge of the CFT.  According to the holographic dictionary the central charge $c$ of a CFT dual to Einstein gravity is given by \cite{Myers:2010tj}
 \begin{equation} \label{centralcharge}
\frac{c}{12}=\frac{A(L)}{16\pi G_d}  \qquad \text{with} \qquad A(L) = \Omega_{d-2} L^{d-2} \, ,
\end{equation}
where $L$ is the AdS radius and $G_d$ is Newton's constant in $d$ dimensions. The central charge $c$ is defined in terms of the normalization of the 2-point function of the stress tensor \cite{Osborn:1993cr}.  It measures the number of field theoretic degrees of freedom of the CFT. The central charge $c$ is normalized so that it coincides with the standard  central charge in  2d CFT. In three dimensions it reduces  to the Brown-Henneaux formula \cite{Brown:1986nw}. 

The number of degrees of freedom of the microscopic   theory that lives on a holographic screen $\mathcal S$ at radius $R$ is   given by
\begin{equation} \label{nodof}
\mathcal{C}=\frac{c}{12} \left ( \frac{R}{L}  \right)^{\! \!  d-2} \!\!  =\frac{A(R)}{16\pi G_d} \qquad \text{with} \qquad A(R) = \Omega_{d-2} R^{d-2} \, .
\end{equation}
This result can be interpreted as follows. We imagine that the  CFT lives on   $\mathbb R \times S^{d-2}$, where the radius of the sphere is given by $L$ and $\mathbb R$ corresponds to the time $t$. The sphere is now partitioned in cells of size $\delta$, where the lattice cutoff is  related to the radius $R$ 
through the UV-IR correspondence via $\delta = L^2/R$. Hence the number of cells on the sphere is given by: $ \left ( L/\delta \right)^{d-2} = \left ( R/L \right)^{d-2}$.\footnote{The UV-IR relation between the short distance   cutoff in the CFT and the large distance regulator in the bulk  is central to the AdS/CFT correspondence. Susskind and Witten  \cite{Susskind:1998dq} inferred this relation by comparing two-point correlators of local operators in the CFT to massive propagators in AdS. 
They also pointed out that it is implied by   the requirement   that the temperature of a thermal state in a cutoff  CFT  has a maximum value of $1/\delta$, which is dual  to the  Hawking temperature $T_{\rm H} \sim R_{\rm h}/ L^2$ of the largest  black hole    in AdS. Peet and Polchinski \cite{Peet:1998wn} showed that  the characteristic transversal perturbation length scale of a massless field in AdS    is inversely proportional to the large distance cutoff in the bulk. 
See   Section 3.1.3 of \cite{Aharony:1999ti} for a further discussion of the UV/IR correspondence. 
}
  Further, the factor $c/12$  counts  the number of quantum mechanical degrees of freedom contained in each cell. This simple argument  due to \cite{Susskind:1998dq} holds in any number of dimensions for   generic holographic CFTs.

The second quantity $\epsilon$  determines the energy that is required to excite one UV degree of freedom and is inversely related to the UV regulator $\delta$ in the boundary theory: $\epsilon\sim 1/\delta$.   Before determining its precise value, let us first discuss  the third quantity $\mathcal N$.  
The total energy of a CFT state is through the operator-state correspondence determined by the scaling dimension $\Delta$ of the corresponding operator. The holographic dictionary relates the dimension $\Delta$ to the mass of the dual field in the bulk: for large scaling dimensions $\Delta\gg d$ the relationship is $\Delta \sim M L$.  Hence $\Delta$ counts the energy in terms of the IR cutoff scale $1/L$.  
The quantity $\mathcal N$ can be viewed as the UV analogue of $\Delta$: it counts the energy $E$ in terms of the UV cutoff $\epsilon$
\begin{equation} 
\label{defN}
E=\mathcal N  \epsilon \, .
\end{equation}
Here $E$ represents the energy of the microscopic theory.   
The excitation number $\mathcal N$ is linearly related to the conformal dimension, where the linear coefficient is given  by the ratio of the IR and UV energy scales. 

By increasing the energy $E$ one starts to excite more degrees of freedom in the microscopic theory and eventually all UV degrees of freedom are excited on the holographic surface at radius $R$. This corresponds to the creation of a black hole of size $R$.  We will choose to normalize $\epsilon$ so that for a black hole we precisely have ${\mathcal N}={\mathcal C}$.   The asymptotic form of the AdS-Schwarzschild metric is given by (\ref{metric}) with blackening factor\footnote{Here we consider  large black holes with horizon size $R_{\rm h}\gg L$, so that we can ignore the constant term in $f(R)$. We will include this term later in Section \ref{sec:physicalimplications} and discuss its microscopic interpretation.}
\begin{equation}
\label{blackening}
f(R) = \frac{R^2}{L^2} - \frac{16 \pi G_d E }{(d-2) \Omega_{d-2} R^{d-3}} \, .
\end{equation}
 It is now easy to deduce the normalization of $\epsilon$ for which ${\mathcal N}= {\mathcal C}$ when $f(R)=0$.  We find that for super-AdS scales the excitation energy $\epsilon$ of the UV degrees of freedom equals
\begin{equation} \label{epsilonads}
\epsilon 
= (d-2)\frac{R}{L^2 } \  \qquad  \text{for}  \qquad R\gg L \, .
\end{equation}  
As we will discuss in Section \ref{sec:geometric}, all the three quantities, $\mathcal C$, $\mathcal N$ and $\epsilon$ can be defined geometrically, where the latter two make use of an appropriately chosen reference metric.  Before discussing these geometric definitions, let us first consider the specific example of AdS$_3$/CFT$_2$, where our definitions will become more transparent. This case will also be  of crucial importance to   our study of holography in other spacetimes.

\subsection{An example: \texorpdfstring{AdS$_3$/CFT$_2$}{AdS(3)/CFT(2)}}
\label{sec:BTZ}
 
 To illustrate the meaning of the various quantities introduced in the previous section, let us consider the  AdS$_3$/CFT$_2$ correspondence. In particular we will further clarify the relation between $\mathcal C$ and $\mathcal N$, on the one hand, and the central charge $c$ and scaling dimensions $\Delta$ of the 2d CFT, on the other hand.  The metric for a static asymptotically  AdS$_3$ spacetime can be written as
  \begin{equation} \label{BTZmetric}
  ds^2 = \left ( \frac{r^2}{L^2} - \frac{\Delta - c/12}{c/12} \right) dt^2 + \left ( \frac{r^2}{L^2}  - \frac{\Delta - c/12}{c/12} \right)^{-1} dr^2 + r^2 d \phi^2 \, ,
  \end{equation}
    where $c$ and $\Delta$ are the  central charge and the scaling dimension in the dual 2d CFT.   $\Delta = 0$ corresponds to empty AdS; $\Delta \leq c/12$ is dual to a conical defect in AdS$_3$; and for $\Delta \ge c/12    $ the metric represents a BTZ black hole \cite{Banados:1992wn}. 
The holographic dictionary between AdS$_3$  and CFT$_2$ is well understood and states that  (see \cite{Kraus:2006wn} for a review)
 \begin{equation}
 \begin{aligned}
&\text{central charge:} \qquad    \quad  \,      \frac{c}{12} = \frac{2\pi L}{16 \pi G_3}  \,,\label{BrownHen}\\
&\text{scaling dimension:} \qquad      \Delta  -   \frac{c}{12 }  =  E L         \,.
 \end{aligned}
 \end{equation}
The first equation is just the Brown-Henneaux formula, and the second equation is the standard relation between the energy on the cylinder  and the scaling dimension $\Delta$.
The energy $E$ corresponds to the ADM energy of the bulk spacetime.   The holographic quantities $\mathcal C$ and $\mathcal N$ for AdS$_3$ are easily determined from  the   expressions    (\ref{nodof}) and (\ref{defN}) 
\begin{equation}
 \begin{aligned}
&\text{number of d.o.f.:}    & \mathcal C &= \frac{2\pi r }{16 \pi G_3} = \frac{c}{12} \frac{r}{L}  \, ,  \\
&\text{excitation number:}    &\mathcal N &=  \,  E\, \frac{L^2}{r} \, \,   \, \,  =  \left (   \Delta - \frac{c}{12} \right) \frac{L}{r}   \,.
\label{excnumber1}
 \end{aligned}
 \end{equation}
 Note that excitation number can be negative, because $E$ is negative for empty AdS and the conical defect spacetime.
The reason for the increase in the number of degrees of freedom by $r/L$ is that the cutoff CFT at radius $r$ allows each field theoretic degree of freedom to have $r/L$ modes. Holographic renormalization (or the UV-IR connection) now tells us that for larger distances in the bulk the modes in the cutoff CFT carry a higher energy, given by $r/L^2$.  This means that at larger distances fewer UV degrees of freedom are excited for a state with fixed energy. Therefore, the excitation number at a radius $r>L$ decreases  by a factor  $L/r$ with respect to the value at the AdS radius. 

We see that $\mathcal C$ and $\mathcal N$  are rescalings of the central charge and the scaling dimension, respectively.   Since the rescaling is exactly opposite, the Cardy formula in 2d CFT remains invariant, and   can hence also be expressed in terms of $\mathcal C$ and $\mathcal N$ 
    \begin{equation}
  \label{CHR-formula}  
  S = 4 \pi \sqrt{\mathcal C \mathcal N} = 4 \pi \sqrt{\frac{c}{12} \left ( \Delta - \frac{c}{12} \right) } \, .
  \end{equation}
 To see that this  formula correctly reproduces the Bekenstein-Hawking entropy (\ref{BHformula}) \cite{Strominger:1997eq} one can use that  the following relations hold at the horizon of the BTZ black hole
\begin{equation}
   \Delta - \frac{c}{12} = \frac{r_{\rm h}^2}{L^2} \frac{c}{12} \qquad   \text{hence} \qquad
\, .\mathcal N = \mathcal C  .
\end{equation}
Hence our definition of $\cal N$ indeed equals $\cal C$ on the horizon for this 3d situation.

\subsection{Geometric definition and generalization to sub-AdS}
\label{sec:geometric}

The number of degrees of freedom, UV cutoff energy and excitation number, as defined above, are general notions that in principle apply to any microscopic theory. A natural question is whether these concepts can be generalized to the microscopic theories on other holographic screens than those close to the AdS boundary. Our reason for introducing the quantities $\mathcal C$, $\mathcal N$ and $\mathcal \epsilon$ is that they can be defined in terms of the local geometry near the holographic surface $\mathcal S$.  In this subsection we will present this geometric definition and verify that it holds for large holographic screens.  Our next step is to postulate that the same geometric definition holds for other situations, in particular for the microscopic theory that lives on holographic screens at sub-AdS scales.

\begin{figure}
	\centering
	\includegraphics
		[width=.35\textwidth]
		{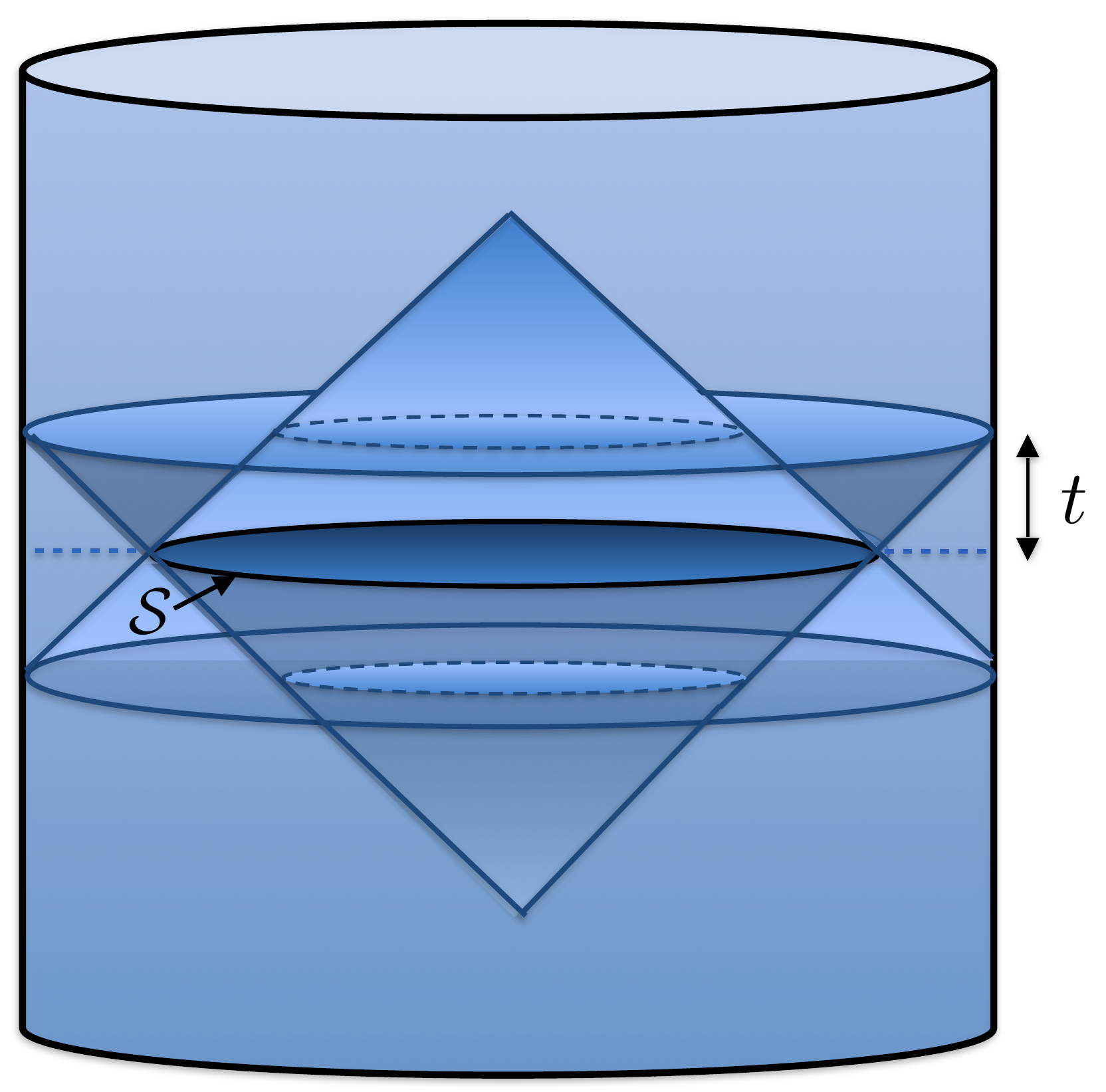}
\caption{\small  A large causal diamond in AdS consisting of the domain of dependence of the ball  bounded by the holographic screen $\mathcal S$. The ball and the screen lie in the $t=0$ time slice, and are centered around the origin. The distance of $\mathcal S$ to the AdS boundary can be characterized by the time $t$ at which the outward future lightsheet reaches the AdS boundary.}
	\label{fig:largediamond}
\end{figure}

Our geometric definition makes use of the causal diamond that can be associated to the holographic screen. Causal diamonds play an important role in the literature on holography because of their 
invariant light-cone structure \cite{Bousso:2002ju,Bousso:1999xy,Banks:2011av,Banks:2013qpa}. Given a spherical holographic screen $\mathcal S$ with radius $R$ on a constant time slice of a static spherically symmetric spacetime,   the  associated causal diamond consists of the future and past domain of dependence of the ball-shaped region  contained within $\mathcal S$. For a large holographic screen in an asymptotically AdS spacetime, the corresponding causal diamond is depicted in Figure \ref{fig:largediamond}. As shown in this figure, the distance to the AdS boundary can be parametrized by the time $t$ at which the extended lightsheets of the   diamond intersect the boundary.  The coordinate~$t$ corresponds to the global AdS time   and also gives a normalization of the local time coordinate near the screen $\mathcal S$. It is with respect to this time coordinate that we define the energy $E$. 

\begin{figure}
	\centering
	\includegraphics
		[width=.35\textwidth]
		{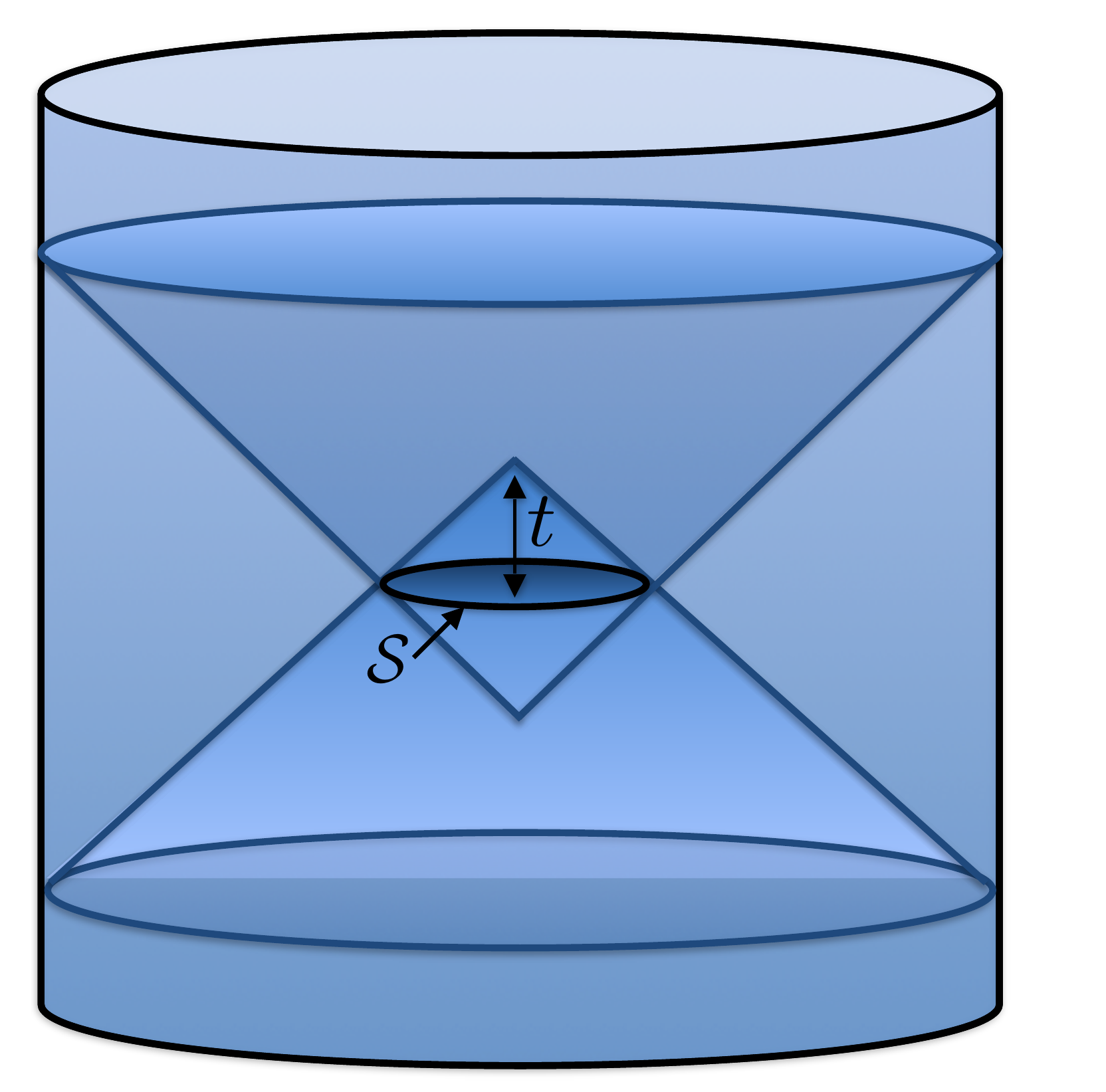}
\caption{\small   A small causal diamond in AdS consisting of the domain of dependence of the ball  bounded by the holographic screen $\mathcal S$. The ball lies in the $t=0$ time slice, and is centered around the origin. The location of $\mathcal S$ with respect to the origin can be parametrized by the time $t$ at which the inward future lightsheet arrives at the origin.}
	\label{fig:smalldiamond}
\end{figure}

For   holographic screens at sub-AdS scales we will   introduce  a similar causal diamond. Except in this situation we define the time coordinate $t$ with respect to the local reference frame in the origin. For empty AdS this time coordinate is again given by the global AdS time $t$. The  causal diamond associated to a holographic screen at a small radius $R\ll L$ is depicted in Figure \ref{fig:smalldiamond}.

Let us consider the rate of change of the number of degrees of freedom $\mathcal C$ along a null geodesic  on the future horizon of the diamond. 
The time  $t$ can be used as a (non-affine) parameter along the null geodesic. For   metrics of the form (\ref{metric}) it is related to the radius $R$ by $dt = \pm dR /f(R)$, where the sign is determined by whether the null geodesic is outgoing (=plus) or ingoing (=minus). The rate of change with respect to $t$ can   thus be positive or negative depending on whether the time $t$ is measured with respect to  the origin or infinity, respectively. 
Our definitions for the excitation number $\mathcal N$ and excitation energy $\epsilon$ are chosen such that the following identity holds
\begin{equation}  \label{derivative2}
\left|\frac{d \mathcal C}{dt}\right|   =  \left (  \mathcal C - \mathcal N   \right) \epsilon   \,   ,
\end{equation}
where the absolute value is taken to ensure that $\epsilon$ is positive. This definition is motivated by the fact that the quantity $d\mathcal C/dt$ vanishes on the horizon of a black hole, if the horizon size is equal to $R$.  In this way it follows that on the horizon $\mathcal N = \mathcal C$.

The identity (\ref{derivative2}) is not yet sufficient   to fix the values of $\mathcal N$ and $\epsilon$. We need to specify for which geometry the excitation number $\mathcal N$ is taken to be zero. In other words, we need to introduce a reference metric that defines the state of zero energy. One could take this to be the empty AdS geometry. However, to simplify the equations  and clarify the discussions in the   subsequent sections
we will take a different choice for our reference geometry. For    super-AdS regions  the reference metric can be found by only keeping the leading term for large $R$ in the function $f(R)$. Whereas for   sub-AdS regions with $R\ll L$ we   take  the Minkowski metric to be the reference metric. Thus for these two cases the reference geometry has the form (\ref{metric}) where the function $f(R)$ is given by
\begin{equation}
\label{reference}
f_0(R) =\left\lbrace 
\begin{array}{c}    \!\!
\quad 1\ \, \qquad \text{for} \quad R\ll L\, ,\\
\!\! {R^2/L^2} \ \quad \text{for} \quad R\gg L\, . 
\end{array}\right.
\end{equation}
This geometry defines the state with vanishing   energy. We also take it to be the geometry where $\mathcal N$ is equal to zero. From (\ref{derivative2}) we thus conclude that the excitation energy is defined in terms of the reference metric via
\begin{equation}\label{excenergy}
 \epsilon = \left| \frac{1}{\mathcal C}\frac{d \mathcal C}{dt}  \right|_{0}      \, .
\end{equation}
We can now use the fact that in the reference geometry $dt = \pm dR /f_0(R)$ to compute the 
excitation energy explicitly
\begin{equation} \label{eq:newdefepsilon}
\epsilon = {f_0(R)\over \mathcal C}	{d\mathcal C\over dR} =\ 
\left\lbrace \begin{array}{c}  \!
 (d-2)/R\ \, \qquad \text{for} \quad R\ll L\, , \\
  \!\!(d-2){R/L^2} \ \quad \,\text{for} \quad R\gg L\, . \end{array}\right.
\end{equation}
For small causal diamonds the dependence on $R$ for the excitation energy is quite natural, because   the radius of the holographic screen is effectively the only scale there is.  
What is remarkable, though, is that the excitation energy increases when the size of the screen decreases. This is opposite to the situation at super-AdS scales, because in that case the excitation energy increases with the size of the screen.

Another way to arrive at this identification is to use the fact that the metric outside a mass distribution at sub-AdS ($R \ll L$) as well as   super-AdS scales ($R \gg L $) takes the form (\ref{metric}) where the blackening function is given by 
 \begin{equation}  \label{blackeningfactorfull}
 f(R) = f_0(R)    - \frac{ 16 \pi G_d E  }{(d-2) \Omega_{d-2} R^{d-3}} \,. 
 \end{equation} 
This equation allows us to verify     our geometric definition (\ref{derivative2}) of the excitation energy $\epsilon$ per degree of freedom and excitation number $\mathcal N$, and show that it is consistent with the identity (\ref{defN}) that expresses the total excitation energy as $E={\mathcal N}\epsilon$.  Using the fact that along a null geodesic $dt =\pm dR/f(R)$ one can derive the following relation\footnote{A similar observation was made by Brewin \cite{Brewin}, who noted that the ADM  mass is proportional to the rate of change of the area of a closed ($d-2$)-surface with respect to the geodesic distance. } 
\begin{equation} \label{derivative1}
\left| \frac{d \mathcal C}{dt} \right|   =   
\left| \frac{d \mathcal C}{dt} \right|_{0}  - E \, .
\end{equation}    
The first term on the right-hand side is the contribution of the reference spacetime with blackening factor $f_0(R)$. By inserting the geometric definition (\ref{excenergy}) for $\epsilon$ and the definition  (\ref{defN}) of $\mathcal N$  into the equation above it is easy to check that this  reproduces the relation (\ref{derivative2}). In the following section we will provide further evidence for these relations for $\mathcal C$, $\mathcal N$ and $\epsilon$ by showing that the super-AdS and sub-AdS regions can be related through a conformal mapping that preserves the number of degrees of freedom $\cal C$ as well as the excitation number $\mathcal N$.

\section{Towards holography for non-AdS spacetimes}
\label{sec:towardsholographynonads}

We start this section by presenting two related conjectures that allow us to connect the physical properties of the microscopic theories that live on holographic screens in different spacetimes. In particular, we argue that the holographic theories for
 two spacetimes that are related by a Weyl transformation  have identical microscopic properties when the Weyl factor equals one on the corresponding holographic screens. We will apply this conjecture to obtain insights into the holographic theories for non-AdS spacetimes by relating them to the familiar case of AdS holography. 
We are especially interested in the holographic properties of sub-AdS regions, Minkowski  and  de Sitter space. We will describe a conformal mapping between AdS space (at super-AdS scale) on the one hand and AdS space (at sub-AdS scale), dS space or Minkowski space on  the other hand. We will use this mapping to derive a correspondence between the holographic descriptions of these spaces. Specifically, we will identify the quantities $\mathcal N$ and $\mathcal C$ in the two conformally related spacetimes.  

Other approaches towards non-AdS holography which also invoke holographic screens include the early work \cite{Bousso:1999cb} and more recently the Holographic Space Time framework \cite{Banks:2011av, Banks:2013qpa}.
A separate line of research has focused on the generalization of the Ryu-Takayanagi proposal in AdS/CFT \cite{Ryu:2006bv,Ryu:2006ef} to more general spacetimes. 
In particular, suitable holographic screens may be used to anchor the bulk extremal surfaces, whose areas are conjectured to provide a measure for the entanglement entropy of the holographic dual theory.  
See for example \cite{Li:2010dr,Shiba:2013jja,Miyaji:2015yva} for both bulk and boundary computations of this proposal.
In addition, the work of \cite{Sanches:2016sxy,Nomura:2016ikr,Nomura:2017fyh,Nomura:2018kji}  uses geometric aspects of carefully defined holographic screens to both establish and infer properties of the holographic entanglement entropy.
In the conclusion, we will point out how some of the conclusions of these works overlap with our own.

\subsection{A conjecture on  the microscopics  of conformally related spacetimes}
\label{sec:conjecture-on-micro}

In the previous section we showed that the number of holographic degrees of freedom $\mathcal C$ is given by the area of the holographic screen, and hence is purely defined in terms of the induced metric on $\mathcal S$. 
    Our definitions of $\mathcal N$ and $\epsilon$, on the other hand,  depend in addition on the geometry of the reference spacetime. Schematically  we have
\begin{equation}
\begin{aligned}
\mathcal C = \mathcal C \,  (  g , \mathcal S)  \,,  \qquad 
\epsilon = \epsilon \, (  g,   g_0 , \mathcal S) \, , \qquad 
\mathcal N = \mathcal N \,  (  g,   g_0, \mathcal S) \, , 
\end{aligned}
\end{equation}
where $g$ is the metric of the spacetime under consideration, and $g_0$ is the metric of the 
reference spacetime.  The presented definitions can in principle be applied to any static, spherically symmetric spacetime. The goal of this section is to gain insight into the nature and properties of these holographic degrees of freedom by comparing holographic screens in different spacetimes with the same local geometry. For this purpose it is important to note that the quantities $\mathcal C$ and $\mathcal N$ are purely expressed in terms of the (reference) metric on the holographic screen $\mathcal S$, without referring to any derivatives. One has for the case of spherically symmetric spacetimes
\begin{equation} \label{metricdef}
\mathcal C = \frac{1}{16 \pi G_d} \int_{\mathcal S} R^{d-2} \,  d\Omega_{d-2} \qquad \text{and} \qquad  \frac{ \mathcal N}{\mathcal C} = 1 - \frac{f(R)}{f_0(R)} \Big |_{\mathcal S}\, ,
\end{equation}
where $R$ is the radius of the spherical holographic screen. Here the second equation follows from   combining (\ref{derivative2}), (\ref{eq:newdefepsilon})  and   (\ref{derivative1}).  

The fact that $\mathcal C$ and $\mathcal N$ can be expressed purely in terms of the metric and not its derivative, suggests that  for holographic screens in two different spacetimes these quantities are the same if the local metrics on these holographic screens coincide. 

\begin{quote} \label{conjecture1}
\textbf{Conjecture I}: The holographic quantum systems on two holographic screens $\mathcal S$ and $\tilde{\mathcal S}$ in two different spacetime geometries $g$ and $\tilde g$, and with reference metrics $g_0$ and $\tilde{g}_0$, have the same number of (excited) degrees of freedom if the (reference) metrics are identical on the holographic screens $\mathcal S$ and $\tilde{\mathcal S}$:  
\begin{equation}
 g|_{\mathcal {S}}=\tilde{g}|_{\tilde{\mathcal {S}}},  \quad g_0|_{\mathcal{S}}=\tilde{g}_0|_{\tilde{\mathcal S}} \quad    \Rightarrow  \quad  {\mathcal C}  (g, \mathcal S) =  {\mathcal C} (\tilde g, \tilde{\mathcal S}), \quad {\mathcal N} (g,g_0, \mathcal S) = {\mathcal N} (\tilde g, \tilde g_0, \tilde{\mathcal S})   \, . \nonumber
\end{equation}
\end{quote}
The excitation energy $\epsilon$ will in general not be the same. In the specific cases discussed below, we find that $\epsilon$ is of the same order of magnitude in the two spaces, but in general differs by a (dimension dependent) constant  factor of order unity.  This motivates us to add to the conjecture that, in the specific cases we study, the excitation energies in the dual quantum systems differ only by an order unity constant: $   \epsilon (g,g_0, \mathcal S) \sim \epsilon (\tilde g, \tilde g_0, \tilde{\mathcal S})$.

Our goal is to study   the general properties of the microscopic theories that live on not just one, but a complete family of holographic screens in a given spacetime with metric $g$. By a complete family we mean that the holographic screens form a foliation of the entire spacetime. For a spherically symmetric spacetime one can choose these to be all spherical holographic screens centered around the origin. A particularly convenient way to find a mapping for all holographic screens is if the spacetime with metric $g$ is conformally related to $\tilde{g}$ via a Weyl rescaling\\[-2mm]
\begin{equation}
g=\Omega^2 \tilde{g}	 \, , 
\end{equation}
in such a way that the holographic screens $\mathcal S$ inside the spacetime geometry $g$ precisely correspond to the loci at which the conformal Weyl factor $\Omega$ takes a particular value. In other words, the holographic screens are the constant-$\Omega$ slices.  
 We denote this constant with $\Omega_{\mathcal S}$, so that $\mathcal S$ corresponds to the set of points for which $ \Omega=\Omega_{\mathcal S}$.  We thus have
 \begin{equation}
 \label{OmegaS}
  \left(\Omega - \Omega_{\cal S} \right)_{|{\mathcal S}} =0\\[-2mm]
 \end{equation}
 
 \begin{figure}[t] 
	\centering
	\includegraphics
		[width=0.7\textwidth]
		{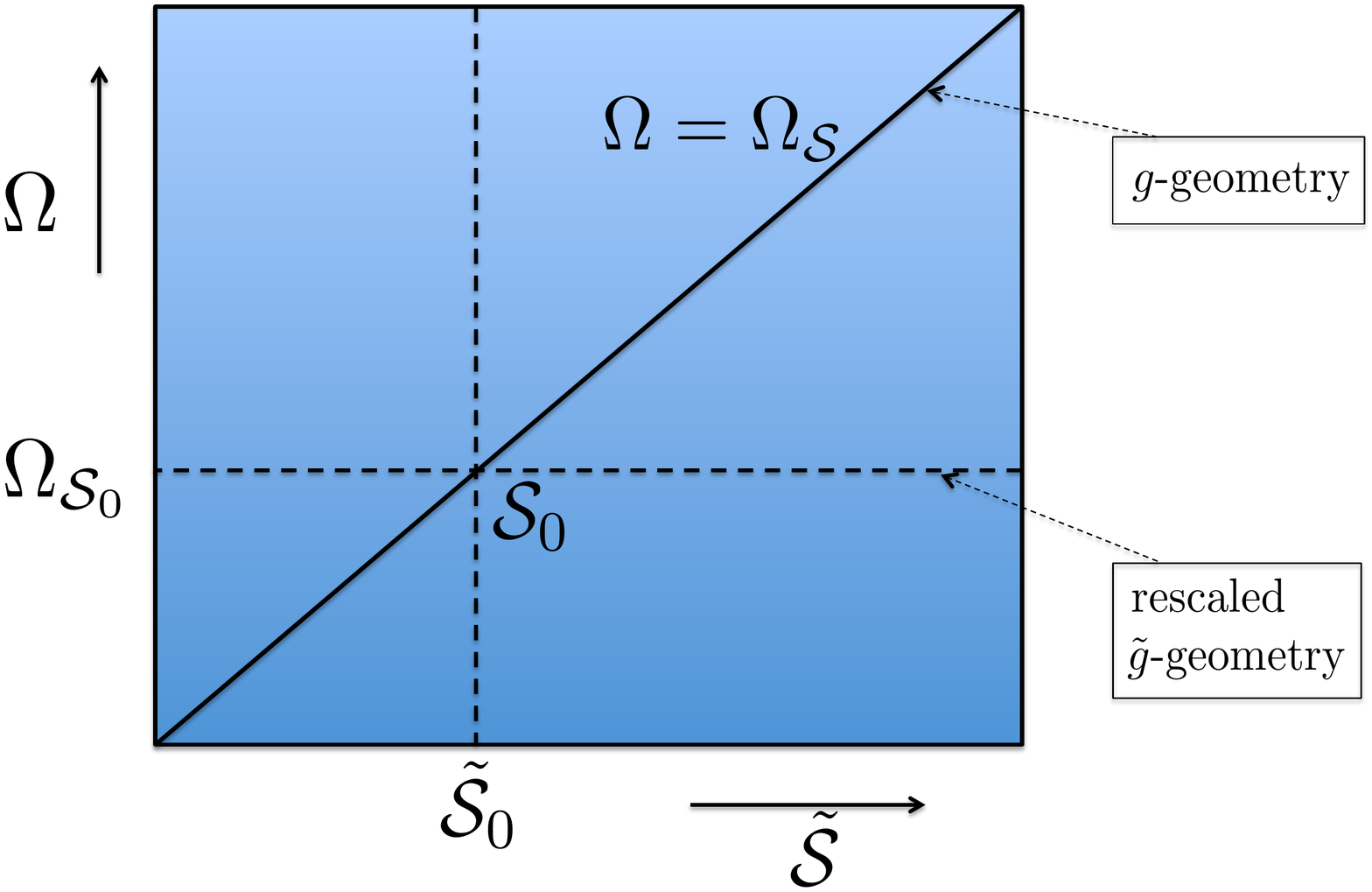}{}\\[-8mm]
	\caption{ \small  The family of rescalings of the geometry $\tilde g$ contains the geometry $g$ as a `diagonal' described by the equation $\Omega=\Omega_{\mathcal S}$. A horizontal line corresponds to a constant rescaling of $\tilde g$:  $\Omega_{\mathcal S_0} \tilde g$. On the holographic screen ${\mathcal S}_0$ the two geometries coincide: $g_{|\mathcal S_0} = \left(\Omega^2_{\mathcal S_0} \tilde g\right)_{|\mathcal S_0}$. By repeating this process for all values of $\Omega$ a foliation of the geometry $g$ is constructed and represented by the diagonal.}
	\label{fig:family-of-geometries}
\end{figure}

\noindent The screens $\tilde{\mathcal S}$ inside the geometry $\tilde g$ are mapped onto a family of rescaled screens inside the rescaled geometries $\Omega^2 \tilde{g}$, where here $\Omega$ is taken to be constant on the $\tilde{g}$ geometry.  We will collectively denote the family of these screens by $\tilde{\mathcal S}$. The conformal map relates the screen $\mathcal S$ inside the geometry $g$ to one representative of this family, namely its image in the geometry $\Omega_{\mathcal S}^2 \tilde{g}$.  We will use the same notation for the holographic screen $\mathcal S$ and its image under the conformal mapping, and reserve the notation $\tilde{\mathcal S}$ for the family of screens inside the complete family of rescaled geometries $\Omega^2 \tilde{g}$.  
In fact, one can view $\tilde{\mathcal S}$ as the family (or equivalence class) of rescalings of the representative $\mathcal S$.  Together the set of all families $\tilde{\mathcal S}$ forms a foliation of the family of rescaled $\tilde{g}$-geometries, while the spacetime with metric $g$ is foliated by the particular representative $\mathcal S$.  In this way the latter spacetime can be viewed as a `diagonal' inside the family of rescaled geometries $\Omega^2 \tilde {g}$ defined by taking $\Omega =\Omega_{\mathcal S}$.  This situation is illustrated in Figure \ref{fig:family-of-geometries}.

An additional requirement on the conformal relation between the two spacetimes is that the reference metrics $g_0$ and $\tilde{g}_0$ are related by the conformal transformation\\[-2mm] \begin{equation}
	g_0=\Omega^2 \tilde{g}_0
\end{equation}
with the same Weyl factor $\Omega$. 
In this situation  each screen $\mathcal S$ in the spacetime with metric $g$ has a corresponding holographic screen   
in the spacetime with metric $\Omega^2_{\mathcal S} \tilde{g}$ satisfying the requirements of our first conjecture.  Namely, one has  
\begin{equation}
g _ {|\mathcal S} =\left(\Omega^2 \tilde{g}\right) \!_{|\mathcal S}	
= \left(\Omega^2_{\mathcal S} \, \tilde{g}\right) \! _{|\mathcal S}
\end{equation}
and a similar relation holds for the reference metrics.   
The rescaling with $\Omega_{\mathcal S}$ changes the area of the holographic screen in the $\tilde{g}$ geometry, and hence its number of degrees of freedom $\mathcal C$, so that it precisely matches the number on the screen $\mathcal S$ in the geometry $g$.  

We are now ready to state our second conjecture

\begin{quote} \label{conjecture2}
\textbf{Conjecture II}:  For  two conformally related spacetimes $g$ and $\tilde g$ with $g=\Omega^2 \tilde{g}$ the holographic quantum systems on a holographic screen $\mathcal S$ in the spacetime with metric $g$ and its image in the spacetime with metric $\Omega^2_{{\mathcal S}}\tilde{g}$ have the same number of (excited) degrees of freedom:   
\begin{equation}
g=\Omega^2 \tilde {g}  \quad \Rightarrow  \quad  \mathcal C  (g, \mathcal S) =  \mathcal C (\Omega^2_{\mathcal S} \, \tilde g, \mathcal S), \quad \mathcal N (g,g_0, \mathcal S) = \mathcal N (\Omega^2_{\mathcal S}\,\tilde g, \Omega^2_{\mathcal S}\tilde g_0, \mathcal S)   \, . \nonumber
\end{equation}
\end{quote}
In the rest of this chapter we will postulate that conjecture II is true and apply it to gain insights into the microscopic features of the holographic theories for a number of spacetimes, including de Sitter and Minkowski space.  Since the metrics on the holographic screens $\mathcal S$ coincide, one can imagine cutting the two spacetimes with metric $g$ and $\Omega^2_{\mathcal S_0}\tilde{g}$ along the slice $\mathcal S_0$ and gluing them to each other along $\mathcal S$. In Figure \ref{fig:family-of-geometries} this amount to first going along the diagonal $\Omega =\Omega_{\mathcal S}$ and then continue horizontally along $\Omega =\Omega_{\mathcal S_0}$.  The resulting metric will be continuous but not differentiable. Hence, to turn this again into a solution of the Einstein equation, for instance, one should  add a mass density on $\mathcal S$, where we assume that the geometries $g$ and $\Omega^2_{\mathcal S} \tilde g$  are both solutions.

\subsection{An example: \texorpdfstring{$AdS_{d}\times  {S}^{p-2}  \cong AdS_{p}\times  {S}^{d-2}$}{AdS(d)xS(p-2)=AdS(p)xS(d-2)}} 
\label{sec:anexample5}

Let us illustrate the general discussion of the previous subsection with an example.  In  Appendix \ref{embedding} we discuss a class of spacetimes that are all  conformally related to (locally) AdS spacetimes. One particular case is the conformal equivalence 
\begin{equation}
  \textAdS_{d}\times S^{p-2}  \cong \textAdS_{p}\times S^{d-2}  \, . 
\end{equation}
The holographic screens $\mathcal S$ have the geometry of $S^{d-2}\times S^{p-2}$. Hence, the conformal map exchanges the spheres inside the  AdS spacetime with the sphere in the product factor. We identify the left geometry with 
$g$ and the right with $\tilde g$,  and denote the corresponding radii by $R$ and $\tilde{R}$. The AdS radius and the radius of the spheres are all assumed to be equal to $L$.  The Weyl factor $\Omega$ is a simple function of $R$ or $\tilde R$. It is easy to see that
\begin{equation}
\label{OmegaR}
\Omega = {R\over L} = {L\over \tilde{R}}	
\end{equation}
 since it maps the $(d\!-\!2)$-sphere of radius $L$  onto one of radius $R$ and the $(p\!-\!2)$-sphere with radius $\tilde R$ to one with radius $L$. This equation is the analogue of (\ref{OmegaS}).  Note that the conformal map identifies the holographic screens with radius $L$ inside both AdS-factors.
 
 We now come to an important observation: the conformal map relates sub-AdS scales on one side to super-AdS scales on the other side. Hence it reverses the UV and IR of the two spacetimes. 
 In particular holographic screens with $\tilde{R}\gg  L$ are mapped onto screens with $R\ll L$. This means that we can hope to learn more about the nature of the microscopic holographic theory for holographic screens in a sub-AdS geometry
by relating it to the microscopic theory on the corresponding screens in  the geometry with metric $\Omega^2 \tilde{g}$, which live  at super-AdS scales. 
First let us compare the values of the cutoff energies  $\epsilon$ and $\tilde \epsilon$. For the situation with $R\ll L$ and $\tilde{R}\gg L$ we found in (\ref{eq:newdefepsilon}) that
\begin{equation}\label{eq:epsilon-discont}
\epsilon = (d-2){1\over R} \qquad \mbox{and} \qquad \tilde \epsilon = (p-2){\tilde{R}\over L^2} \, , \qquad \mbox{hence}\qquad   {\tilde{\epsilon}\over \epsilon } = {p-2 \over d-2} \, .  
\end{equation}
We have thus verified that the energy cutoffs $\tilde{\epsilon}$ and $\epsilon$, before and after the conformal map are of the same order of magnitude, but differ by a dimension dependent factor. The inversion of the dependence on the radial coordinate is qualitatively explained by the fact that the conformal mapping reverses the UV and IR of the two AdS geometries. 

 Next let us compare the central charges of the two CFTs.  Since we imposed that the number of degrees of freedom of the microscopic theories are the same on corresponding holographic screens,  it follows that the central charges of the two sides must be the same at $R=\tilde{R} =L$. Namely, for this value of the radius the number of holographic degrees of freedom is equal to the central charge of the corresponding CFT.   Since we keep the value of the $d+p-2$ dimensional Newton's constant $G_{d+p-2}$ before and after the conformal map fixed, one can indeed verify that the central charges $c$ and $\tilde{c}$ agree when the AdS radius is the same on both sides:
 \begin{equation}
{c\over 12} = {\Omega_{d-2} L^{d-2}\over 16\pi G_d} =  {\Omega_{d-2}\Omega_{p-2} L^{d+p-4} \over 16\pi G_{d+p-2}} =  {\Omega_{p-2} L^{p-2} \over 16\pi G_p}  = {\tilde{c}(L)\over 12} . 
\end{equation}
Here $G_d$ denotes Newton's constant on AdS$_d$, while $G_p$ equals Newton's constant on AdS$_p$.  We indicated that the central charge $\tilde{c}$ is computed for the $R=\tilde{R}=L$ slice.   But how does the number of degrees of freedom change as we move to  say $R=R_0\ll L$?  This leads to a rescaling of the geometry on the right-hand side with $\Omega_0 = R_0/L$, and hence it changes the curvature radius of the AdS$_p$ geometry and correspondingly the value of the central charge $\tilde{c}$. Note that the radius of the $S^{d-2}$ is rescaled as well and now equals $R_0$: this affects the relationship between the Newton constants $G_{d+p-2}$ and $G_p$.  The radius of the $S^p$ on the left-hand side is unchanged, however, so the relation between $G_{d+p-2}$ and $G_d$ is still the same. In this way one finds that the central charge of the CFT corresponding to the rescaled $AdS_p\times S^{d-2}$ geometry becomes
\begin{equation}\label{eq:relation-c-tildec}
{\tilde{c} (R_0)\over 12} 	 = {\Omega_{p-2} R_0^{p-2} \over 16\pi G_p}   = {\Omega_{d-2}\Omega_{p-2} R_0^{d+p-4} \over 16\pi G_{d+p-2}} = {\Omega_{d-2} R_0^{d-2} \over 16\pi G_{d}} \left({R_0 \over L}\right)^{p-2}.  
\end{equation}
Thus the effective central charge of the CFT corresponding to the rescaled $\tilde{g}$ geometry depends on the radius $R_0$.  Note that in the left geometry $g$ we are at sub-AdS scales, while on the right $\tilde{R}_0= L^2/R_0\gg L$. This means that on the right we can use our knowledge of AdS/CFT to describe the microscopic holographic degrees of freedom.  Our conjecture II states that the general features of the microscopic theories on both sides agree. In this way we can learn about the microscopic theories at sub-AdS scales. 
 
As shown in Appendix \ref{embedding}, one can construct more general conformal equivalences that instead of AdS$_d$ contain Mink$_d$ or dS$_d$. All these geometries can, after taking the product with  $S^{p-2}$,  be conformally related to again a product manifold of a locally AdS$_p$ geometry with a $S^{d-2}$. 
The required conformal mappings may be obtained via an embedding formalism, as explained in detail in the Appendix.  

In the rest of this chapter  we will focus on the particular case $p=3$.  For this situation we have even more theoretical control, because of the AdS$_3$/CFT$_2$ connection.  Note that in this case the central charge $\tilde{c}(R_0)$ in \eqref{eq:relation-c-tildec}  grows as $R_0^{d-1}$ and hence as the volume. Another important reason for choosing $p=3$ is that the $(p\!-\!2)$-sphere becomes an $S^1$, whose size can be reduced by performing a $Z_N$ orbifold with large $N$, while keeping our knowledge about the microscopic degrees of freedom. The latter construction, as well as the significance of the volume law for the central charge, will be explained in Section \ref{sec:longstring}.

\subsection{Towards holography for sub-AdS, Minkowski and de Sitter space}
\label{sec:three-ex}

In Section \ref{sec:conjecture-on-micro} we explained how to foliate a spacetime metric $g$ in holographic screens by using a family of Weyl rescaled metrics $\tilde{g}$.   We will now apply this construction to the cases of sub-AdS$_d$, Mink$_d$ and dS$_d$ with a Kaluza-Klein circle,  by making use of the following conformal equivalences (see Appendix \ref{embedding})
\begin{align}\label{eq:global-ads/ds-weyl-equivalence-p=3}
\begin{split}
  \textAdS_{d}\times S^{1} &\, \cong \, \textAdS_{3}\times S^{d-2}\, ,\\
   \textMink_d  \times S^{1} &\,\cong \,  \textBTZ_{E=0} \times S^{d-2} \, ,  \\
 \textdS_{d}\times S^{1} &\, \cong \,   \textBTZ  \times S^{d-2} \, . 
\end{split}
\end{align} 
For all these examples, the Weyl factor $\Omega$  is given by (\ref{OmegaR}). In this and the following sections we will denote the coordinate radius of AdS$_3$ by $r$ instead of $\tilde R$, while we keep $R$  as the radius in the spaces on the left-hand side of (\ref{eq:global-ads/ds-weyl-equivalence-p=3}). 
Earlier versions of the conformal map between $dS_{d}\times S^{1}$ and  $BTZ \times S^{d-2}$ appeared in  \cite{Anninos:2011af, Hubeny:2009rc}. 

The conformal equivalences can be verified easily using the explicit metrics. One can represent the metrics of the spacetimes on the right-hand side of \eqref{eq:global-ads/ds-weyl-equivalence-p=3} as
\begin{equation}\label{confmap1}
d \tilde{s}^2 = -\left({r^{2}\over L^2} - \kappa\right) dt^2  +\left( {r^{2}\over L^2}  - \kappa  \right)^{-1} dr^{2} + r^{2} d\phi^{2} + L^2d\Omega_{d-2}^2 \, ,
\end{equation}
with  $\kappa = -1, 0$ or $+1$. Here   $\kappa = -1$ describes pure $AdS_3\times S^{d-2}$ in global coordinates;  $\kappa=0$ corresponds to the so-called massless BTZ black hole (for which $E=0$ and $r_{\rm h}=0$); and $\kappa = +1$ represents the metric of a BTZ black hole with horizon radius $r_{\rm h} = L$.

We now rescale the metric by a factor $L^2/r^2$ and subsequently perform the coordinate transformation $R=L^2/r$:
\begin{equation}  \label{conformafactor1}
d  s^2 = \Omega^2  d \tilde {s}^2 \qquad \qquad \Omega = \frac{R}{L} = \frac{L}{r} \, .
\end{equation}
This leads to the metrics
\begin{equation}\label{confmap2}
d {s}^2 = -\left( 1 -  \kappa {R^{2}\over L^2}\right) dt^2  +\left( 1 - \kappa {R^{2}\over L^2}\right)^{-1} dR^{2} + R^{2}d\Omega_{d-2}^2  + L^2d\phi^{2} \, , 
\end{equation}
with
\begin{equation} \label{kappa1}
\kappa =
\begin{dcases*}
 \ -1 &  for  $\ \textAdS_{d}\times S^{1} $\\
\ \ \ 0 & for $\ \textMink_d  \times S^{1}$ \\
\ +1  & for $\ \textdS_{d}\times S^{1}$  
\end{dcases*}
\, .
\end{equation}
An important property of these conformal equivalences is that the radius is inverted, i.e. $R= L^2/r$.  The inversion of the radius means, for example,  that asymptotic infinity  in AdS$_3$ is mapped to the origin in AdS$_d$, and vice versa.
Note also that the horizon of the BTZ black hole ($r=L$) is mapped onto the horizon of dS$_d$ space ($R=L$). 

On the AdS$_3$/BTZ side the different values of $\kappa$ correspond to different states in the dual two-dimensional CFT. One can read off the scaling dimensions  by comparing the metric (\ref{confmap1}) with the asymptotically AdS$_3$ metric (\ref{BTZmetric}): $\Delta =c/12 \left (  1 + \kappa \right)$. Using this, we rewrite the conformally rescaled metric  as
\begin{equation}  \label{eq:masterformula}
d  {s}^2=\frac{R^2}{L^2}\!\!\left[-\!\left(\frac{r^2}{L^2}\!-\!\frac{\Delta-c/12}{c/12}\right)  dt^2\!+\!\left(\frac{r^2}{L^2}\!-\!\frac{\Delta-c/12}{c/12}\right)^{-1}\!\!\! \!dr^2+r^2d\phi^2\!+\!L^2d\Omega^2_{d-2}\right]\!.
\end{equation}
This metric turns into (\ref{confmap2}) for the following values of the scaling dimension:
\begin{equation}
\begin{aligned}
&  \Delta = 0 \,\,\,\, : \quad &&  \text{Anti-de Sitter space} \, ,        \\
&  \Delta  =   \frac{c}{12} : \quad   &&\text{Minkowski space} \, ,    \\
& \Delta =  \frac{c}{6} \,\,\,: \quad && \text{de Sitter space} \, . 
\end{aligned}
\end{equation}
The physical implications of these observations will be discussed further below. 

\begin{figure}[t]
	\centering
	\includegraphics
		[width=1\textwidth]
		{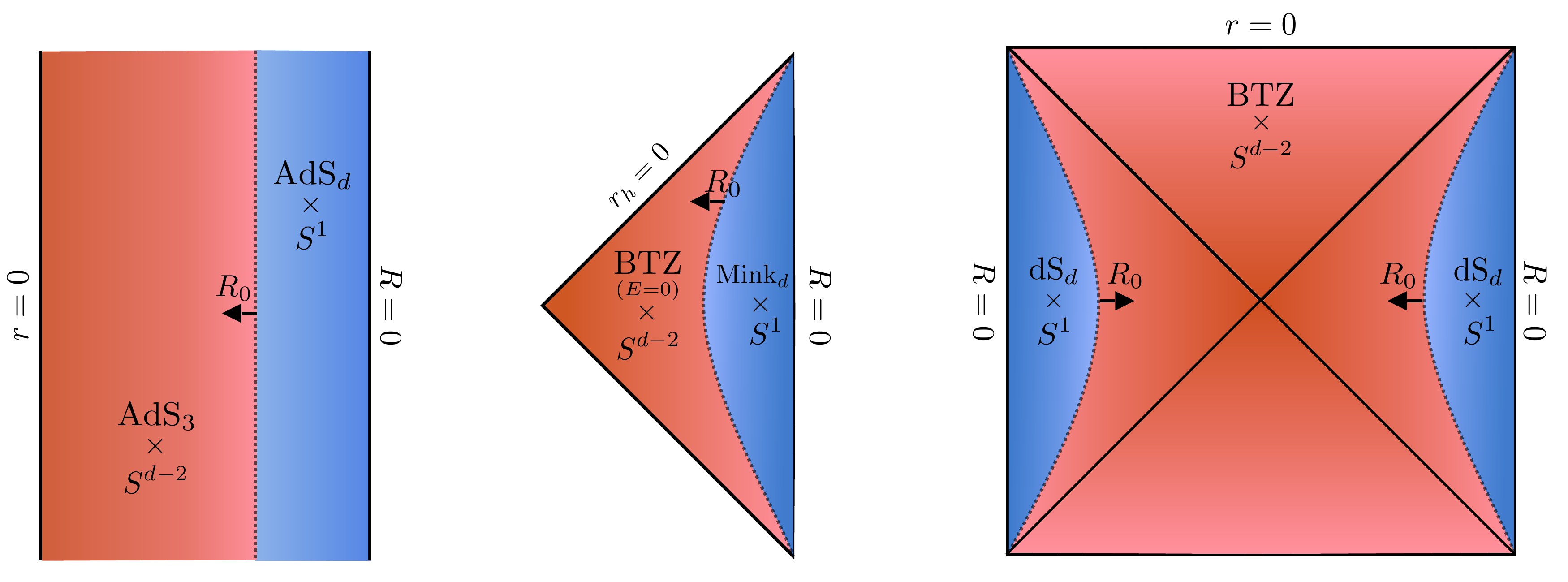}
	\caption{ \small $\!\!\!$ The first figure depicts, from left to right, the gluing of the $\left[0,r_0\right]$ region  of  Weyl rescaled $AdS_3\times S^{d-2}$ to the $[0,R_0]$ region of $AdS_d \times S^1$. The second figure illustrates, from left to right, the gluing of the $[0,r_0]$ region of massless (Weyl rescaled) $BTZ \times S^{d-2}$ to the $[0,R_0]$ region of $Mink_d \times S^1$. Finally, the third figure shows  the gluing of the $\left[0,r_0\right]$ region   of (Weyl rescaled) $BTZ\times S^{d-2}$ to the $[0,R_0]$ region of $dS_d\times S^1$. }
	\label{fig:penrose-ads-ds-inside-out}
\end{figure}

To connect to our discussion in the previous section, consider our family of Weyl rescaled AdS$_3$ spacetimes obtained by taking a constant value of $R=R_0$. 
The AdS$_d$ slice at $R=R_0$ corresponds to the $r=r_0\equiv L^2/R_0$ slice in the Weyl rescaled AdS$_3$ geometry.
We now glue the region $0\leq r\leq r_0$ of the $AdS_3/BTZ\times S^{d-2}$ geometry to the $(A)dS_d/Mink_d\times S^1$ geometry, as illustrated in Figure \ref{fig:penrose-ads-ds-inside-out}.
We take $R_0<L$, so that the position at which the spacetimes are glued is at super-AdS$_3$/BTZ and sub-(A)dS$_d$ scales.
This allows us to interpret the microscopic description of sub-(A)dS$_d$ slices from a super-AdS$_3$ perspective through our conjecture. For this purpose we just need to give an interpretation to the metric \eqref{eq:masterformula} for $\ell\leq R_0\leq L$. This will be the main approach that we employ in Section \ref{sec:longstring}.

However, already at this level we will make a couple of general remarks.
First of all, the metric within brackets is a locally AdS$_3$ spacetime. At super-AdS$_3$ scales we can thus use our knowledge of $\text{AdS}_3/\text{CFT}_2$ to interpret the geometry in terms of a microscopic theory. As discussed in Section \ref{sec:anexample5},  the rescaling of the metric with the scale factor $R_0^2/L^2$ effectively changes the curvature radius of the spacetime.  
In the microscopic quantum system this implies that the central charge of the $\text{CFT}_2$ now depends on the radius $R_0$ in the $d$-dimensional spacetime.  This $R_0$-dependence is given by the $p=3$ case of the general formula \eqref{eq:relation-c-tildec}. In fact, the central charge scales with the volume of the ball inside the holographic surface in AdS$_d$. As we will explain, the number of degrees of freedom  still grows like the area, as it should according to the holographic principle. 

Our goal in the remaining sections is to illuminate the nature of the holographic degrees of freedom that describe non-AdS spacetimes, by exploiting our conjecture and the conformal equivalence with the AdS$_3$ spacetimes. In particular we like to explain the origin of the reversal of the UV-IR correspondence at sub-(A)dS scales.  The usual concept of holographic renormalization takes us from the UV to the IR. In the sub-(A)dS case this means we have to start from a small radius and increase the radius as we go from UV to IR. The holographic principle then tells us that the number of degrees of freedom actually grows towards the IR. This is a rather unusual property compared with the familiar case of AdS/CFT, where as one moves from the UV to the IR one is integrating \emph{out} degrees of freedom.  In the case of sub-AdS or de Sitter holography, our conformal map suggests that we integrate \emph{in} degrees of freedom as we move from the UV to the IR  (a similar point of view was proposed  in \cite{Nomura:2017fyh,Anninos:2011af}). In the following section, we will describe a mechanism by which UV holographic degrees of freedom at the center of sub-AdS or de Sitter can be embedded in the IR holographic degrees of freedom at the (A)dS radius, thus still realizing the holographic principle.

\section{A long string  interpretation}
\label{sec:longstring}

In this section we will give a microscopic interpretation of the degrees of freedom of the holographic quantum system describing the metric (\ref{eq:masterformula}).
The identification of the correct degrees of freedom   enables us, through  our  conjecture II, to make some precise quantitative and qualitative statements about the holography for sub-AdS scales, Minkowski space and de Sitter space.
It will turn out that the microscopic quantum system underlying the metric in \eqref{eq:masterformula} has an interpretation in terms of so-called  `long strings'.
We will constructively arrive at these long strings as follows. We begin in the UV with a small number of degrees of freedom described by a seed CFT with a small central charge. As we go to larger distances we start taking symmetric products of this seed CFT. To arrive at the correct value of the number of degrees of freedom and excitation energy, we apply a so-called long string transformation. In this way we are able to build up the non-AdS spacetimes from small to large distances.  

In our construction we make use of three different length scales: a UV scale, an IR scale and an intermediate scale. These different scales are denoted by: \\[-4mm]
\begin{table}[H]
\centering
\begin{tabular}{lll}
 $\ell$: & \quad   \qquad UV scale & $=\quad $ short string length   \\ \\
$R_0$: & \quad    \qquad intermediate scale&  $=\quad $ fractional string length \\ \\
$L$: &   \quad   \qquad  IR scale & $=\quad $ long string length  
\end{tabular}
\label{tab:string-scales}
\end{table}
\vspace*{-5mm}
\noindent 
The terminology `short', `fractional' and `long' strings will be further explained below in our review of the long string phenomenon. We will take  $R_0$ and $L$ to be given by integer multiples of the short string length $\ell$ \begin{equation}
R_0=k\ell \qquad \text{and} \qquad L = N \ell  \, .
\end{equation}
The microscopic holographic theory can be described from different perspectives, and depends on which degrees of freedom one takes as fundamental: the short, the fractional or the long strings.  It turns out that the value of the central charge depends on which perspective one takes. The metric that was written in Section \ref{sec:three-ex} will arise in the long string perspective. However, the short and fractional string perspective will turn out to give useful insights as well. In the following we will therefore use the string length as a subscript on the central charge to indicate in which perspective we are working.  For instance, the central charge in the fractional string perspective is written as $c_{R_0} (\cdot)$, where the value between brackets is a measure for the size of the symmetric product under consideration.  

As we will review in Section \ref{sec:review-long-string},  the long string phenomenon relates the central charges and spectra of these differently sized strings according to
\begin{equation}
c_L (R_0)= \frac{k}{N} c_{R_0} (R_0)= \frac{1}{N} c_{\ell} (R_0) \, .
\end{equation}
 The relevance of the fractional string perspective consists of the fact that its central charge   always equals the total number of microscopic degrees of freedom $\mathcal C$ associated to a holographic screen at radius $R_0$. 
Nevertheless, we will argue below that the long string picture is more fundamental.
We will now start with explaining the long string phenomenon and these formulas in more detail, and we will also discuss aspects of the corresponding dual AdS$_3$ geometries.

\subsection{The long string phenomenon}\label{sec:review-long-string}
The long string phenomenon was originally discovered in \cite{Maldacena:1996ds}  and developed in detail in \cite{Dijkgraaf:1996xw}.
The starting point is a so-called `seed CFT' with central charge $c_{\ell}$. 
Consider now the CFT that is constructed by taking a (large) symmetric product of the seed CFT, i.e.
$$
\textCFT^M/S_M \, .
$$
This symmetric product CFT has central charge $c_{\ell}(M)=Mc_{\ell}$.
Operators in this theory may now also have twisted boundary conditions in addition to ordinary periodic ones.
The resulting twisted sector, labeled by a conjugacy class of $S_M$, gives rise to long string CFTs.
The word `{long}' refers to the fact these sectors behave as if they were quantized on larger circles than the original seed CFT.
For instance, the twisted sector that corresponds to the conjugacy class consisting of $M$-cycles gives rise to a single long string that is $M$ times larger than the seed  (or short) string.
 As a result, the spectrum of modes becomes fractionated because the momenta are quantized on a circle of larger radius. Moreover, the central charge is reduced since the twisted boundary condition sews together  independent short degrees of freedom into  a single long degree of freedom.  
For consistency, the spectrum of the long string is subjected to a constraint
$$
P=L_0-\bar{L}_0=0\mod M  .
$$
This implies that the total momentum of a state on the long string should   be equal to the momentum of some state on the short string. 
However, due to its fractionated spectrum there are many more states on the long string for any given total momentum.
In fact,   in a large symmetric product CFT  the dominant contribution to the entropy of a certain macrostate comes from the longest string sector \cite{Maldacena:1996ds}.

In more detail, to project onto a long string sector of size $N$, one inserts degree $N$ twist operators in the symmetric product CFT
\begin{equation}\label{eq:twist-operator}
(\sigma_N)^{M/N}|0\rangle \, ,
\end{equation}
where $N$ is assumed to be a divisor of $M$. $\!$The twist operator $\sigma_N$ has conformal dimension
\begin{equation}\label{eq:twist-op-dim-short}
\Delta_\ell=\frac{c_\ell(M)}{12}\left(1-\frac{1}{N^2}\right).
\end{equation}
The insertion of the twist operators has, as mentioned above, two important effects: it reduces the number of UV degrees of freedom and furthermore lowers their excitation energy. The reduction of the number of degrees of freedom is due to the fact that the spectrum becomes fractionated, which lowers the number of degrees of freedom by a factor $N$. The long string phenomenon thus operates as
\begin{align}\label{eq:long-string-pheno}
\begin{split}
\Delta_L-\frac{c_L(M)}{12}&=N\left(\Delta_{\ell}-\frac{c_\ell(M)}{12}\right) \, ,\\
c_L(M)&= \frac{1}{N}  \,  \, c_\ell(M)  \, .
\end{split}
\end{align}
Note that the conformal dimension and the central are rescaled in opposite direction. This implies in particular that the Cardy formula (\ref{CHR-formula}) is invariant under the long string transformation \eqref{eq:long-string-pheno}. Since the long string central charge is smaller, the vacuum (or Casimir) energy in the CFT$_2$ is lifted to a less negative value.
Moreover, the   state  \eqref{eq:twist-operator} in the short string perspective coincides with the ground state in the long string CFT, as can  easily be verified by inserting \eqref{eq:twist-op-dim-short} into \eqref{eq:long-string-pheno} 
\begin{equation}
\Delta_L=N\left(\Delta_\ell-\frac{c_\ell(M)}{12}\right)+\frac{c_L(M)}{12}=0 \, .
\end{equation}
After having introduced the long and short string perspective, let us go to the intermediate or fractional string perspective. Instead of applying the long string transformation, one could  also consider an intermediate transformation, replacing $N$ in \eqref{eq:long-string-pheno} by $k\!<\!N$. This would only partially resolve the twist operator, which means that there still remains a non-zero conical deficit. 
The resulting fractional strings have size $R_0= kL/N$.
In this case, a twist operator will remain, whose conformal dimension is smaller than (\ref{eq:twist-op-dim-short}). Its presence indicates that the fractional string of length $R_0$ does not close onto itself and should be thought of as a fraction of a long string of length $L$.
The spectrum and central charge of the fractional string are related to those of the short string by
\begin{align}\label{eq:fractional-string-pheno}
\begin{split}
\Delta_{R_0}-\frac{c_{R_0}(M)}{12} & =k\left(\Delta_\ell-\frac{c_{\ell}(M)}{12}\right),\\
c_{R_0}(M)&=\frac{1}{k} \, \, c_{\ell}(M) \, .
\end{split}
\end{align}
The dimension of the remaining twist operator is obtained by inserting (\ref{eq:twist-op-dim-short})  into   the equation above
\begin{equation}\label{eq:twist-op-dim-fractional}
\Delta_{R_0}=\frac{c_{R_0}(M)}{12}\left(1-\frac{k^2}{N^2}\right) \, .
\end{equation}
We now turn to the AdS side of this story.
The  state   \eqref{eq:twist-operator} is   dual to a conical defect of order $N$ in AdS$_3$ \cite{Martinec:1998wm}.
The conical defect metric is simply given by the metric for empty AdS$_3$  with the following identification for the azimuthal angle
\begin{equation}
\phi\equiv \phi +2\pi/N \, .
\end{equation}
where $N = L /\ell$. For now we take $M=N$ so that $L$ becomes the size of the longest string.   Also, it plays the role of the AdS radius, since the symmetric product central charge is related to the AdS radius through the Brown-Henneaux formula:
$$
c_\ell(L)=\frac{2L}{3G_3} \, .
$$
Here we introduced a slightly different notation for the symmetric product central charge by replacing $M$ with the corresponding AdS radius. This notation will be used in the rest of this chapter.

We can rewrite the AdS$_3$ metric with conical defect in the following way 
\begin{align}\label{eq:metric-long-string-from-seed-no-sphere}
\begin{split}  
ds^2& =-\left ({r^{\,2}\over L^2}+1\right) dt^2  +\left ({r^{\,2}\over L^2}+1\right)^{-1} \! \!dr^{2} +r^{2} d\phi^{2} \\
&=N^2\left[-\left ({\hat{r}^{\,2}\over \ell^2}+\frac{1}{N^2}\right) dt^2  +\left ({\hat{r}^{\,2}\over \ell^2}+\frac{1}{N^2}\right)^{-1} \! \!d\hat{r}^{2} +\hat{r}^{2} d\hat{\phi}^{2} \right],
\end{split}
\end{align}
where $\hat \phi \equiv \hat\phi +2\pi$, and
\begin{align}\label{eq:coord-transf-ell-to-L-no-sphere}
\begin{split}
\hat{r}&= r/N^2  \qquad  \text{and} \qquad \hat{\phi}=N\phi \, .
\end{split}
\end{align}
This rewriting illustrates the geometric analog  of taking an $N^{\text{th}}$ symmetric product and projecting to a long string sector of size $N$.
Indeed, the metric with curvature radius $\ell$ can be interpreted as the dual of the seed CFT. 
The multiplication by $N^2$ scales up the curvature radius to $L$, which is the geometric analog of the symmetric product. 
Moreover, the $1/N^2$ term in the $g_{tt}$ and $g_{\hat{r}\hat{r}}$ components is the analog of the insertion of the twist operator in the short string perspective.
Finally, the coordinate transformation leads us to the first metric in (\ref{eq:metric-long-string-from-seed-no-sphere}) whose covering space is dual to the long string CFT \cite{Balasubramanian:2014sra}.

The analog of \eqref{eq:metric-long-string-from-seed-no-sphere} and \eqref{eq:coord-transf-ell-to-L-no-sphere} for a fractional string transformation is given by
\begin{align}\label{eq:metric-fract-string-from-seed-no-sphere}
\begin{split}
ds^2   &=  \frac{N^2}{k^2} \left[ -\left ({\tilde r^{\,2}\over R_0^2}+\frac{k^2}{N^2}\right) dt^2  +\left ({\tilde r^{\,2}\over R_0^2}+\frac{k^2}{N^2}\right)^{-1} \! \!d\tilde r^{2} +\tilde r^{2} d\tilde\phi^{2}\right]     \\
&=N^2 \left[-\left ({\hat{r}^{\,2}\over \ell^2}+\frac{1}{N^2}\right) dt^2  +\left ({\hat{r}^{\,2}\over \ell^2}+\frac{1}{N^2}\right)^{-1} \! \!d\hat{r}^{2} +\hat{r}^{2} d\hat{\phi}^{2}  \right],
\end{split}
\end{align}
where $\tilde{\phi}=\tilde{\phi}+2\pi /k$, and 
\begin{align}\label{eq:coord-transf-ell-to-R0-no-sphere}
\begin{split}
\hat{r}&= \tilde r/k^2 \qquad \text{and}  \qquad \hat{\phi}=k\tilde{\phi} \, .
\end{split}
\end{align}
The presence of the $k^2/N^2$ is the geometric manifestation of the fact that we have not fully resolved the twist operator.
In particular, it indicates that the fractional string does not close onto itself.
Since $\tilde{\phi}$ is periodic with $2\pi/k$,   a fractional string precisely fits on the conformal boundary $\tilde{r}=R_0$ in the covering space. 

 In the next section, we will also be interested in changing the size of the symmetric product. In particular, instead of multiplying the metric by $N^2$ we will also consider multiplication by $k^2$.  In that case, the conformal  factor in front of the first metric in \eqref{eq:metric-fract-string-from-seed-no-sphere}  is one. This describes the AdS dual of a single fractional string. 
This metric and its dual CFT interpretation will play an important role in our discussion of sub-(A)dS$_d$.

\subsection{Sub-AdS scales}\label{sec:sub-ads}
In this section we will put our conjecture II in Section \ref{sec:conjecture-on-micro} to use.
By employing the long string mechanism explained in the previous section  we will give an interpretation of the holographic degrees of freedom relevant for sub-AdS scales.

We are interested in the slice $R=R_0 \le L$  in the $AdS_d\times S^1$ metric  
\begin{equation}\label{eq:metric-adsd-cirlce}
ds^2=-\left (1+{R^{\,2}\over L^2}\right) dt^2  +\left (1+{R^{\,2}\over L^2}\right)^{-1} \! \!dR^{2} +R^{2}d\Omega_{d-2}^2+ \ell^2 d\Phi^{2} \, ,
\end{equation}
with $\Phi\equiv \Phi +2\pi$.
In contrast to the metric (\ref{confmap2}),  here we have   adjusted the size of the transverse circle to $\ell\ll L$ in order to compactify to AdS$_d$ even at sub-AdS scales.
In the AdS$_3$ spacetime this can be achieved by the insertion of a conical defect, as in the first metric of (\ref{eq:metric-long-string-from-seed-no-sphere}). Combining the conformal map of Section \ref{sec:three-ex} with equation (\ref{eq:metric-long-string-from-seed-no-sphere}), we can rewrite the metric  above as  
\begin{align}\label{eq:metric-long-string-from-seed}
\begin{split}
ds^2 
&=\frac{R^2}{\ell^2}\left[-\left ({\hat{r}^{\,2}\over \ell^2}+\frac{1}{N^2}\right) dt^2  +\left ({\hat{r}^{\,2}\over \ell^2}+\frac{1}{N^2}\right)^{-1} \! \!d\hat{r}^{2} +\hat{r}^{2} d\hat{\phi}^{2} + \ell^2d\Omega^2_{d-2}\right],
\end{split}
\end{align}
with $\hat{\phi}\equiv \hat{\phi}+2\pi$. Note that we have made the following identification between the radial and angular coordinates
\begin{equation}
R   = \frac{\ell^2}{\hat r} \,  \qquad \text{and} \qquad \Phi = \hat \phi \, . 
\end{equation}
For any fixed $R=R_0$  we obtain an equivalence between slices in $AdS_d\times S^1$ and in a conformally rescaled $AdS_3\times S^{d-2}$ spacetime with conical defect.
We will first consider the case $R=L$ to provide a CFT$_2$ perspective on the holographic degrees of freedom at the AdS$_d$ scale, and thereafter consider sub-AdS$_d$ scales.

Without the conformal factor and the conical defect, the $AdS_3 \times S^{d-2}$ metric with curvature radius $\ell$ is dual to the ground state of a seed CFT with central charge
\begin{equation}\label{eq:seed-central-charge}
{c_\ell(\ell)\over 12} = {2\pi \ell\over 16\pi G_3}= { A(\ell) \over 16\pi G_d} \, ,
\end{equation}
where we used  $1/G_3 = A(\ell)/ G_d$ and  $1/G_d =2\pi \ell / G_{d+1} $.
We imagine $c_{\ell}(\ell)$ to be a relatively small central charge, just large enough to be able to speak of a `dual geometry'.
Due to the presence of the transversal sphere, the multiplication of the seed metric by $N^2$ now scales up the central charge of the seed CFT by a factor $N^{d-1}$ if we choose to keep $G_{d}$ fixed:
\begin{equation}\label{eq:sym-prod-central-charge}
{c_\ell(L)\over 12} = N^{d-1}{c_\ell(\ell)\over 12}  \, . 
\end{equation}
As  explained in Section \ref{sec:review-long-string}, we may interpret this rescaling as taking an $N^{d-1}$-fold  symmetric product of the seed CFT.
Additionally, the $1/N^2$ term in the metric (\ref{eq:metric-long-string-from-seed}) signals the presence of a twist operator in the dual CFT, that puts the system in a long string sector of the symmetric product CFT.
This reduces the central charge of the system and fractionates the spectrum of the theory, as expressed in \eqref{eq:long-string-pheno}.

Our conjecture relates the holographic quantities in the microscopic dual  of AdS$_3$ to   those in the dual of AdS$_d$.
There are two apparent problems when we think of the short string degrees of freedom in the symmetric product CFT as relevant to AdS$_d$. 
First, the symmetric product central charge \eqref{eq:sym-prod-central-charge} expresses a volume law for the number of holographic degrees of freedom at $R=L$. This number should however be related to the central charge of the CFT$_{d-1}$  which obeys an area law.
Moreover, the excitation energy required to excite the degrees of freedom at $R=L$ should be of the order $1/L$, but the seed degrees of freedom have an excitation energy of the order $\epsilon_{\ell}\sim 1/\ell$.

The long string phenomenon precisely resolves both of these problems.
First, it reduces the volume law for the central charge to an area law
\begin{equation}
\frac{c_L (L)}{12} =  {1\over N} {c_\ell(L)\over 12}= \frac{A(L)}{16 \pi G_d} \, .
\end{equation}
Simultaneously, the long string phenomenon  give rises to a reduced excitation energy 
\begin{equation}\label{eq:excitation-energy-long-string}
\epsilon_L={1\over N} \epsilon_\ell = {d-2\over L} \, .
\end{equation}
The factor $(d-2)$ arises in AdS$_d$, as discussed around \eqref{eq:epsilon-discont}.
Concluding, the quantum system dual to the metric \eqref{eq:metric-long-string-from-seed} at $R=r=L$ has $c_L(L)/12$ holographic long string degrees of freedom, which may be excited with the lowest possible energy $\epsilon_L$.
This  is consistent with our expectations for the dual quantum system of AdS$_d$ at $R=L$.
Since this discussion only concerns the number of holographic degrees of freedom and their excitation energies, our conjecture allows us to give a CFT$_2$  interpretation of the AdS$_d$ holographic degrees of freedom. \\

\noindent Next, to gain access to sub-AdS$_d$ scales, we will take $R=R_0<L$.
In this case, the seed metric is multiplied by $k^2$.
Analogously to the discussion above, this is interpreted as taking a $k^{d-1}$-fold symmetric product of the seed CFT.
The central charge of the symmetric product CFT is  
\begin{equation}  \label{eq:centralcharge-shortstring}
\frac{c_{\ell}(R_0)}{12}=k^{d-1}{c_\ell(\ell)\over 12} \, . 
\end{equation}
Performing the   coordinate transformation   \eqref{eq:coord-transf-ell-to-R0-no-sphere} on the metric  \eqref{eq:metric-long-string-from-seed}, we obtain the fractional string metric:
\begin{align}\label{eq:metric-fractional-string}
\begin{split}
ds^2  &= \frac{R^2}{R_0^2}\! \left [  -\!\left ({\tilde r^{\,2}\over R_0^2}+\frac{k^2}{N^2}\right) \!dt^2  +\left ({\tilde r^{\,2}\over R_0^2}+\frac{k^2}{N^2}\right)^{-1}\! \! \!d\tilde r^{2} +\tilde r^{2} d\tilde\phi^{2} + R_0^2d\Omega^2_{d-2} \right] .   \\
\end{split}
\end{align}
The metric with curvature radius $R_0$ is dual to   fractional strings with central charge $c_{R_0}(R_0)$.
Fractional strings provide a  useful perspective on sub-(A)dS scales, since they represent the degrees of freedom that are directly related to the holographic quantities defined in Section \ref{sec:generalfeatures}.
For instance, using \eqref{eq:fractional-string-pheno} one quickly verifies that 
\begin{equation} \label{nodofsubads}
\mathcal C=\frac{c_{R_0}(R_0)}{12}  .
\end{equation}
This shows that the fractional strings can be thought of as the sub-AdS analog of the super-AdS holographic degrees of freedom, as discussed by Susskind and Witten.
Indeed, fractional strings have a larger excitation energy than long strings:
\begin{equation}
\epsilon_{R_0} = \frac{d-2}{R_0}\,. 
\end{equation}
This is the same excitation energy as defined in (\ref{eq:newdefepsilon}) for sub-AdS scales. If we interpret $\epsilon_{R_0}$ as the UV cutoff at sub-AdS scales, then the fractional strings are the corresponding UV degrees of freedom. 
Further,    the fractional string quantities are related to the number of excitations by
\begin{equation}
 \mathcal N= \Delta_{R_0} - \frac{c_{R_0}(R_0)}{12} \, . 
\end{equation}
Similarly to \cite{Susskind:1998dq} a thermal bath of these fractional strings at temperature $\epsilon_{R_0}$ creates a black hole, and $\mathcal{N}=\mathcal{C}$     translates then to fractional string quantities as $\Delta_{R_0} = c_{R_0}/6.$ We will come back to this in more detail in Section \ref{sec:bhentropy}.

Finally, we will discuss the long string perspective on sub-AdS holography.
In the long string picture the spectrum and central charge of the dual CFT are   given by
\begin{align}\label{cLR}
\begin{split}
 \Delta_L-\frac{c_L (R_0)}{12}  & = \:\frac{N}{k} \left(\Delta_{R_0}-\frac{c_{R_0}(R_0)}{12} \right)  \, ,   \\ 
c_L(R_0)& = \frac{k}{N} \, \, c_{R_0}(R_0)   \, .  
\end{split}
\end{align}
The factor $k/N$ in $c_L(R_0)$ expresses the fact that the fractional strings only carry a fraction of the long string central charge. 
Comparing with $c_L(L)$  we see that the number of long strings at $R=R_0$ is reduced from $N^{d-2}$ to $k^{d-2}$, of which only the fraction $k/N$ is accessible.

\begin{figure}[t]
	\centering
	\includegraphics
		[width=0.4\textwidth]
		{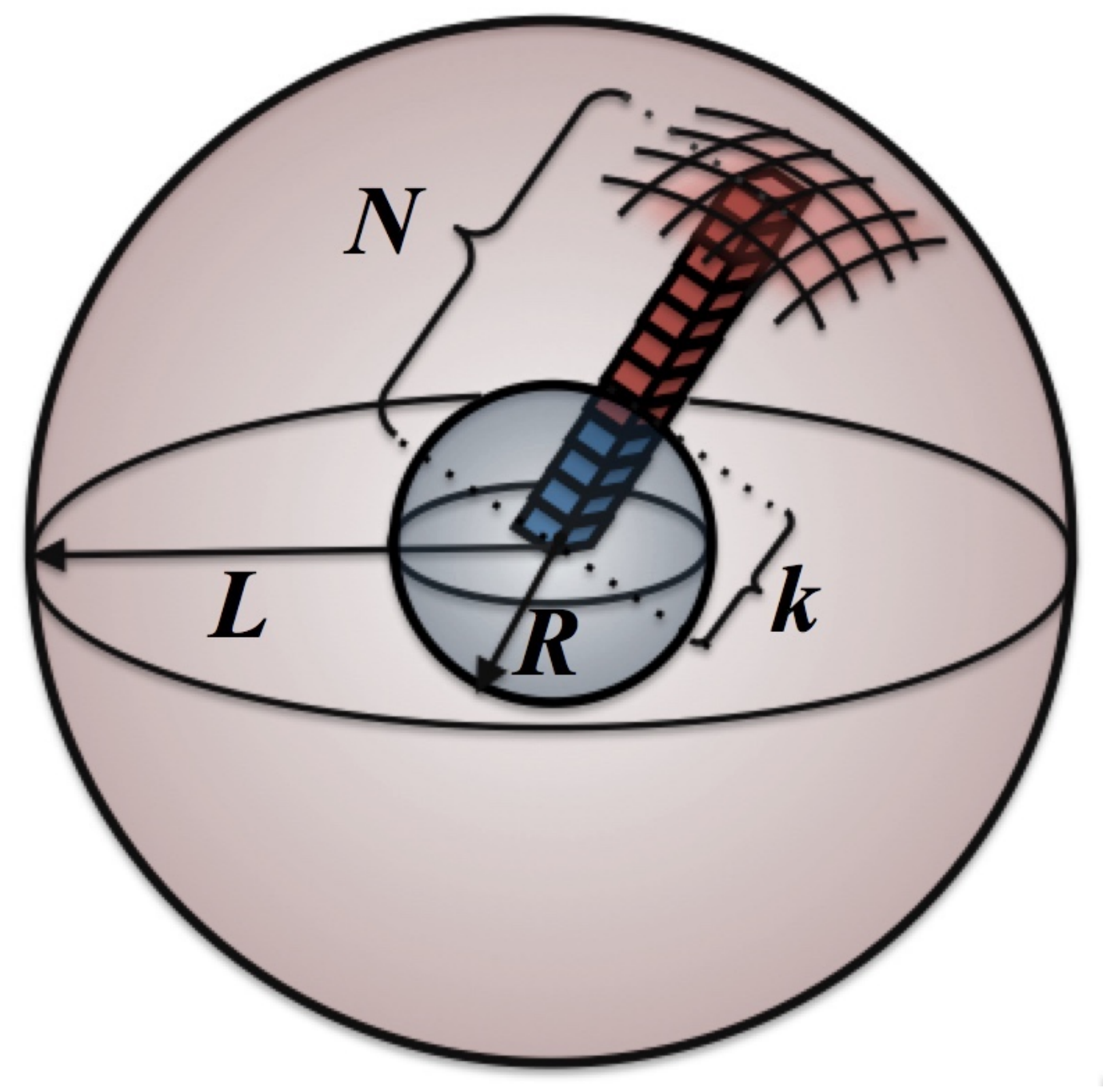}
	\caption{ \small   The degrees of freedom on the holographic screen at radius $L$ consist of single ``long strings''. The degrees of freedom on holographic screens at radius $R_0= k\ell$ with $k<N$ consist of ``fractional strings''. The total number of long (or fractional) strings is proportional to the area of the holographic screen at radius $L$ (or $R_0$). }
	\label{fig:long-strings}
\end{figure}
 
Geometrically one can arrive at the long string perspective by applying the coordinate transformation \eqref{eq:coord-transf-ell-to-L-no-sphere} on the short string metric \eqref{eq:metric-long-string-from-seed}.
In this long string metric, the super-AdS$_3$ radial slice $r=L^2/R_0$ corresponds to the sub-AdS$_d$ slice $R=R_0$. 
 From the perspective of the long string we may therefore excite more modes.  In particular, the lowest energy excitation on the fractional string corresponds to the higher excited state on the long string:
\begin{equation}
\epsilon_{R_0} = \frac{N}{k}\epsilon_L. 
\end{equation}
In this sense, one can think of the excitations on fractional strings as bound states of the lowest energy excitations on the long strings.
This also explains that in the long string perspective the total number of degrees of freedom is still given by an area law, as opposed to $c_L(R_0)$, since we may excite more modes per each of the $c_L(R_0)$ degrees of freedom.

The explicit realization of the microscopic quantum system helps in understanding  the reversal of the UV-IR correspondence  at sub-(A)dS scales. 
The process of taking symmetric products and going to a long string sector achieves to integrate in degrees of freedom while moving to the IR.
More precisely, the symmetric product is responsible for integrating in degrees of freedom, where the order of the symmetric product determines the precise amount.
The (partial) long string phenomenon then reduces the excitation energy, realizing the step towards the IR.\footnote{Note that the larger the symmetric product becomes, the further one may move to the IR.} 
Moreover, from the AdS$_d$ perspective, it ensures that the number of degrees of freedom is given by an area law, and hence is in accordance with the holographic principle. 

We conclude this section by giving a master formula for the Weyl rescaled geometries we consider in this chapter, which is a more refined version of \eqref{eq:masterformula}:
\begin{equation}\label{eq:metric-gen-conf-dim}
ds^2 =\frac{R^2}{L^2} \!\!\left[- \!\left( \!\frac{r^2}{L^2}\!-\!\frac{\Delta_{L}-\frac{c_L(R)}{12}}{\frac{c_L(R)}{12}}\right) \!dt^2+\left(\!\frac{r^2}{L^2}\!-\!\frac{\Delta_{L}-\frac{c_L(R)}{12}}{\frac{c_L(R)}{12}}\right)^{-1} \! \! \!\! \!dr^2+r^2d\phi^2+L^2d\Omega^2_{d-2}\right] \!.
\end{equation}
Here, $\phi\equiv \phi+2\pi/N$ as usual and $c_L(R)$ can be obtained by combining \eqref{nodofsubads} and \eqref{cLR}:
\begin{equation}\label{eq:cLR}
\frac{c_{L}(R)}{12} ={\Omega_{d-2} R^{d-1} \over 16\pi G_{d} L}={A(R) \over 16\pi G_{d} }{R \over L} \, .
\end{equation}
For sub-AdS, this metric is the long string version of \eqref{eq:metric-long-string-from-seed} and $\Delta_L=0$ accordingly.
However, this formula also captures all the non-AdS geometries we will consider in subsequent sections, and in addition provides them with a precise microscopic meaning.

\subsection{Minkowski space}
In this section, we briefly discuss the conformal map  between massless BTZ and Mink$_d$.
It allows us to phrase the holographic degrees of freedom relevant for Mink$_d$ in terms of  CFT$_2$ language introduced in the previous section.
Since we have discussed at length the conformal map between AdS$_3$ in the presence of a conical defect and sub-AdS$_d$, it is more convenient to think of the massless BTZ metric as the $N\to \infty$ limit of the conical defect metric.
We understand this limit as the $L\to \infty$ limit, keeping $\ell$ fixed, which on the AdS$_d$ side is of course the limit that leads to Minkowski space. 

Explicitly, the $\mathbb{R}^{1,d-1}\times S^1$ metric  
 \begin{align}  
\begin{split}
 ds^2=- dt^2  +d R^{2}+ R^2d\Omega^2_{d-2} +\ell^2 d\Phi^{2}  
\end{split}
\end{align}
is   equivalent to the $N\to \infty$ limit of the fractional string metric \eqref{eq:metric-fractional-string}, i.e.
 \begin{align}  
\begin{split}
 ds^2=\frac{R^2}{R^2_0}\left[- {\tilde r^{\,2}\over R_0^2} \,  dt^2  + { R_0^2 \over    \tilde r^{\,2}  }  \,d\tilde r^{2} +\tilde r^{2} d\tilde\phi^{2} + R_0^2d\Omega^2_{d-2}\right] \, .
\end{split}
\end{align}
In particular, at $R=\tilde{r}=R_0$ the fractional string metric is equivalent to the $\mathbb{R}^{1,d-1}\times S^1$ metric.

Our discussion on sub-AdS has taught us that the fundamental degrees of freedom are long strings of size $L$, of which we can only access a fraction at scales $R_0<L$.
In the case of Minkowski space, the long strings have an infinite length.
The infinite twist operator state in the short (or fractional) string perspective corresponds to the ground state of these infinitely long strings.
Therefore, in CFT$_2$ language Minkowski space can be thought of as the groundstate on infinitely long strings, which also has a vanishing vacuum (or Casimir) energy.
These two aspects are reflected in the equations:
$$
\Delta_{\infty}=c_\infty(R_0)=0.
$$
The vanishing of $c_L(R_0)$   in the limit $L\to \infty$ can be understood  from the fact that fractional strings of finite length carry an infinitely small fraction of the central charge of an infinitely long string.

At any finite value of $R_0$, the number of modes that can be excited on the long string is infinite as well, since the excitation energy $\epsilon_L$ goes to zero in the limit.
This balances $c_L(R_0)$ in such a way that the total number of degrees of freedom at any radius $R_0$ is still finite. This is manifested in the fractional string frame, where from  (\ref{cLR}) it follows that  $c_{R_0}(R_0)=A(R_0)/16\pi G_d \neq 0$ and  
\begin{equation}
\Delta_{R_0}-\frac{c_{R_0}(R_0)}{12}=0 \, .
\end{equation}
This implies that $\Delta_{R_0}$ corresponds to the scaling dimension of an infinite degree twist operator.

Minkowski space can also be arrived at by taking the $L\to \infty$ limit of  de Sitter space.
As will become clear in the next section, in this case one can understand  Minkowski space from the massless BTZ perspective.

 \subsection{De Sitter space}
 \label{subsec:dS}
 
In this section, we will argue that the microscopic quantum system relevant for sub-AdS$_d$ holography, as studied in detail in the Section \ref{sec:sub-ads}, plays an equally important role in the microscopic description of the static patch of dS$_d$. 
In particular, our methods identify the static patch dS$_d$ as an excited state in that quantum system, in contrast to AdS below its curvature scale which was identified as the groundstate.

As explained in Section \ref{sec:three-ex}, the de Sitter static patch metric  times  a transversal circle,
\begin{equation}\label{eq:metric-ds-circle}
ds^2=-\left(1- \frac{R^2}{L^2} \right)dt^2+\left(1- \frac{R^2}{L^2} \right)^{-1} dR^2+R^2d\Omega^2_{d-2}+\ell^2d\Phi^2,
\end{equation}
is Weyl equivalent to a Hawking-Page BTZ black hole times a transversal sphere
 \begin{align}\label{eq:metric-btz-long-string}
\begin{split}
 ds^2& =\frac{R^2}{L^2}\left[-\left ({r^{\,2}\over L^2}-1\right) dt^2  +\left ({r^{\,2}\over L^2}-1\right)^{-1} \! \!dr^{2} +r^{2} d\phi^{2} + L^2d\Omega^2_{d-2}\right], \\
\end{split}
\end{align}
where again $\Phi \equiv \Phi + 2\pi$ and  $\phi\equiv \phi +2\pi \ell/L$. 
The master formula \eqref{eq:metric-gen-conf-dim} reproduces the latter metric when $\Delta_{L}=c_L(R)/6$.

Let us start again for $R=L$.
The resulting BTZ metric arises holographically from an excited state in the CFT with conformal dimension $\Delta_L=c_L(L)/6$, as can for instance be verified by the Cardy formula \eqref{CHR-formula}:
\begin{equation}\label{eq:entropy-btz}
S= \frac{2\pi L}{4G_3}\frac{\ell}{L},
\end{equation}
This is the correct entropy for the Hawking-Page BTZ black hole with a conical defect of order $N$ \cite{deBoer:2010ac}.
The CFT state can be interpreted as a thermal gas of long strings at temperature $T \sim {1}/{L}$. Hence, the temperature is of the same order as the excitation energy $\epsilon_L$ of a long string. Our conjecture now suggests that the microscopic quantum system at the radial slice $R=L$ in dS$_d$, which coincides with the de Sitter horizon, should sit in an excited state as well. In fact, we like to interpret this state, similarly as in the CFT, as consisting of long strings, where each string typically carries only its lowest energy excitation mode. 
 
In  the gluing of $dS_d\times S^1$ and $BTZ\times S^{d-2}$, as explained in Section  \ref{sec:three-ex}, we identify the horizon in the former spacetime with the horizon in the latter.
The entropy of the de Sitter space can then be understood as the entropy of the CFT$_2$ state.
Using the relations between Newton's constants in \eqref{eq:metric-btz-long-string} at the radial slice $R=L$,
$$
\frac{1}{G_3}=\frac{A(L)}{G_{d+1}},\qquad \frac{1}{G_d}=\frac{2\pi \ell}{G_{d+1}},
$$
one quickly finds that the BTZ entropy can be rewritten as: 
\begin{equation}\label{eq:desitter-entropy}
S =\frac{A(L)}{4 G_d}.
\end{equation}
Thus, we see that the Bekenstein-Hawking entropy for a $d$-dimensional de Sitter horizon can be reproduced from the Cardy formula in two-dimensional CFT.

An important point we should stress here is the reason for why the excitation of the sub-AdS$_d$ degrees of freedom in this case does not produce a Hawking-Page AdS$_d$ black hole.
This black hole would indeed have the same entropy as in \eqref{eq:desitter-entropy}, so what is it that enables us to distinguish them?
At the AdS scale, there is in fact nothing that distinguishes them, so to answer this question we must turn to the microscopic quantum system that describes sub-AdS$_d$ scales.
The crucial difference is that the CFT$_2$ state corresponding to the AdS$_d$ black hole has a constant conformal dimension, as we will come back to in detail in Section \ref{sec:bhentropy}. 
On the other hand, the state corresponding to a sub-dS slice has $R_0$ dependent conformal dimension $\Delta_L=c_L(R_0)/6$, and therefore has a description in terms of the long strings in the $k^{d-1}$-fold symmetric product.
This is also clear from the master formula \eqref{eq:metric-gen-conf-dim}.
The dependence of the conformal dimension on $R_0$ expresses the fact that (part of) the excitations corresponding to de Sitter horizon are also present at sub-dS scales.
It is useful to move to a fractional string perspective, where we have:
\begin{equation}\label{eq:conf-dim-de-sitter-fract}
\Delta_{R_0} -  { c_{R_0}(R_0)\over 12} = \frac{R_0^2}{L^2} { c_{R_0} (R_0) \over 12} \,.
\end{equation}
This formula shows that only a part of the fractional strings at scale $R_0$ are excited, and in particular do not create a horizon, as should of course be the case for sub-dS.
It makes sense that only a fraction of the fractional strings are excited, since the de Sitter temperature $T\sim 1/L$ could only excite the longest strings with their lowest excitations.

As illustrated in Figure \ref{fig:penrose-ads-ds-inside-out}  on p. \pageref{fig:penrose-ads-ds-inside-out}, the geometric perspective on sub-dS scales glues $dS_d\times S^1$ and $BTZ\times S^{d-2}$ by replacing the outer region $R_0\!<\!R\!<\!L$ in de Sitter with the region $r_{\rm h}\!<\!\tilde{r}\!<\!R_0$ of the BTZ. 
The horizon size of this BTZ geometry is smaller than that of the Hawking-Page black hole.
This fact is expressed most clearly by the fractional string version of \eqref{eq:metric-btz-long-string}
\begin{equation}\label{eq:metric-btz-fractional-string}
  ds^2= \frac{R^2}{R_0^2}  \!\left [ - \!\left ({\tilde r^{\,2}\over R_0^2}-\frac{R_0^2}{L^2}\right) dt^2  +\left ({\tilde r^{\,2}\over R_0^2}-\frac{R_0^2}{L^2}\right)^{-1}\! \! \!d\tilde r^{2} +\tilde r^{2} d\tilde\phi^{2} + R_0^2d\Omega^2_{d-2} \right] , 
\end{equation}
where for $R=R_0<L$ the horizon radius of the BTZ is given by $r_{\rm h}=R^2_0/L<L$.
Note however that, as expected, the temperature of the black hole is not changed:
$$
T\sim \frac{r_{\rm h}}{R_0^2}= \frac{1}{L}.
$$
Since in this case it is not the horizon of the smaller BTZ but the $\tilde{r}=R_0>r_{\rm h}$ slice that is identified with the de Sitter slice $R=R_0$, one could wonder if the entropy of the BTZ can still be associated to the slice in de Sitter space.
However, in terms of the 2d CFT it is known that the entropy of a BTZ black hole is also contained in the states that live in the Hilbert space at higher energies than the black hole temperature \cite{Witten:1998zw}. Our conjecture then indeed suggests that the entropy of the smaller BTZ should also be associated to the sub-dS slice. 

The entropy of the smaller BTZ with the remaining angular deficit $\tilde{\phi}\equiv \tilde{\phi}+2\pi\ell/R_0$ in \eqref{eq:metric-btz-fractional-string} is given by:
\begin{equation}\label{eq:entropy-fract-btz}
S= \frac{2\pi R^2_0}{4G_3 L}\frac{\ell}{R_0}=\frac{A(R_0)}{4G_d}\frac{R_0}{L}.
\end{equation}
where we used the relation $1/G_3=A(R_0)/G_{d+1}$.
Hence, we see that  the entropy formula for sub-dS$_d$ scales with the volume instead of the area.
From the long string perspective, it is natural why only a fraction $(R_0/L)^{d-1}$ of the total de Sitter entropy arises at sub-dS scales. 
As we have discussed, de Sitter space corresponds to an excited state consisting of long strings at temperature $T\sim 1/L$. 
At sub-dS scales, $c_L(R_0)$ can be interpreted as $k^{d-2}$ fractions of long strings.
 It makes sense then that at the scale $R=R_0$ only a fraction of the energy and entropy associated to the long strings is accessible. 
If we apply the Cardy formula to the state $\Delta_{L}(R_0) =c_L(R_0)/6$, valid at least as long as $k^{d-1}\gg N$, we recover \eqref{eq:entropy-fract-btz}. We rewrite the entropy to make its volume dependence explicit as:
\begin{equation}\label{desitter-entropy-r0}
S=\frac{A(R_0)}{4G_d}\frac{R_0}{L} = \frac{V(R_0)}{V_0} \qquad \text{where} \qquad V_0 = \frac{4 G_d L }{d-1} \, . 
\end{equation}
Note that this entropy describes a volume law and only at the Hubble scale becomes the usual Bekenstein-Hawking entropy.
It hence seems natural to associate an entropy density to de Sitter space, which was advocated in \cite{Verlinde:2016toy}. However, the precise microscopic interpretation of this volume law remains an open question.

\subsection{Super-AdS scales revisited}
\label{sec:superadsrevisited}

In the previous sections we have given an interpretation of the microscopic holographic quantum system for sub-(A)dS$_d$ regions and flat space. To gain insight into sub-AdS$_d$ holography we used a conformal map to relate its foliation in holographic screens to the holographic screens of a family of super-AdS$_3$ screens. At this point, we have gained enough understanding of sub-AdS$_d$ to try to use the conformal map the other way around: we start with a sub-AdS$_3$ region and map it to a super-AdS$_d$ region. Even though the AdS/CFT correspondence already gives  a microscopic description of super-AdS$_d$ regions, we will argue that  this sub-AdS$_3$ perspective could still provide useful insights into the microscopic description of super-AdS$_d$ regions.
 
Let us start with the rescaled $AdS_3 \times S^{d-2}$ metric 
\begin{align}\label{eq:metric-sub-to-super}
\begin{split}
ds^2& =\frac{R_0^2}{L^2}\left[-\left ({r^{\,2}\over L^2}+1\right) dt^2  +\left ({r^{\,2}\over L^2}+1\right)^{-1} \! \!dr^{2} +r^{2} d\phi^{2} + L^2d\Omega^2_{d-2}\right] \, ,
\end{split}
\end{align}
where the angular coordinate has no deficit: $\phi\equiv \phi+2\pi$.
As should be familiar by now, this metric at slice $r_0=L^2/R_0$ is equivalent to AdS$_d$ at $R=R_0$ with a transversal circle of radius $L$.\footnote{Since we are now working at super-AdS$_d$ scales, we do not have to worry about a small size for the transversal $S^1$. However, it is perhaps not justified to ignore the transversal sphere at sub-AdS$_3$ scales.} We again glue the conformally equivalent spacetimes at $R=R_0$ and $r=r_0$, but now we take $R_0> L$ so that the sub-AdS$_3$ region is mapped to  a super-AdS$_d$ region.

The central charge for sub-AdS$_d$, as in \eqref{cLR}, together with the factor $R_0^2/L^2$ in front of the metric \eqref{eq:metric-sub-to-super} imply that the ``central charge'' associated to super-AdS$_d$ is given by
\begin{equation}\label{eq:super-adsd-central-charge-from-fract-string}
 \frac{c_L(R_0)}{12}=\frac{2\pi r_0}{16\pi G_3} \frac{r_0}{L}\left(\frac{R_0}{L}\right)^{d-1}  =\frac{A(R_0)}{16\pi G_d} \frac{L}{R_0}  \qquad \text{for} \qquad R_0>L  \, .
\end{equation}
This formula can be interpreted as the central charge of an $(R_0/L)^{d-1}$-fold symmetric product of $r_0$ sized fractional strings.
Note that this is a different quantity than the central charge of the CFT$_{d-1}$, given by formula  (\ref{centralcharge}).
Only in three dimensions the two expressions coincide and they  reproduce the   Brown-Henneaux formula for the central charge. The factor $L/R_0$ has an analogous interpretation as $R_0/L$ in the central charge at sub-AdS scales.
It expresses the fact that from the CFT$_2$ perspective, the degrees of freedom relevant at super-AdS$_d$ scales are fractions of long strings. 

Although $c_L(R_0)$ scales with $R_0^{d-3}$, the total number of quantum mechanical degrees of freedom is larger by a factor $R_0/L$. This is because the number of available modes on the long strings at energy scale $1/r_0=R_0/L^2$ is precisely $R_0/L$. As usual, the total amount of holographic degrees of freedom is reflected most clearly in the fractional string perspective:
\begin{align}
\begin{split} \label{NCsuperAdS}
\mathcal N    &=  \Delta_{R_0}-\frac{c_{R_0} (R_0)}{12}= \left (  \Delta_L-\frac{c_L (R_0)}{12}   \right) \frac{L}{R_0} \, ,  \\ 
\mathcal C   &=\frac{c_{R_0}(R_0)}{12} = \frac{c_L(R_0)}{12}  \frac{R_0}{L}   \, .  
\end{split}
\end{align}
Note that we found the same relations in (\ref{excnumber1}) for the  AdS$_3$/CFT$_2$ correspondence, but these equations hold for general $d$ and have a rather different interpretation, supplied by the sub-AdS$_3$ perspective. The  total number of degrees of freedom $\mathcal C$  is again given by an area law.

In conclusion, we see that a sub-AdS$_3$ perspective on super-AdS$_d$ identifies the degrees of freedom of the latter  with symmetric products of fractional strings. As we move outwards in AdS$_d$ the fractional string degrees of freedom become shorter and hence their excitation energy increases. 
In this way, the sub-AdS$_3$ perspective reproduces the usual UV-IR correspondence of super-AdS$_d$ holography.
We will use these results in Section \ref{sec:bhentropy}, when we discuss super-AdS$_d$ black holes.

\section{Physical implications}
\label{sec:physicalimplications}

As explained above, the long string sector of a    symmetric product CFT$_2$ gives  a detailed description of the   holographic degrees of freedom relevant for   non-AdS spacetimes. We now turn to a number of physical implications of this microscopic description. First, we will derive the Bekenstein-Hawking entropy for small and large AdS$_d$ black holes from a Cardy-like formula.
Moreover, from the CFT$_2$ point of view we will explain  why small black holes have a negative specific heat capacity and how  the Hawking-Page transition between small and large black holes  can be understood.
Finally, we will show that our long string perspective   reproduces the value of the vacuum energy for (A)dS spacetimes.

 \subsection{Black hole entropy and negative specific heat}
 \label{sec:bhentropy}
 
In this section we will apply our microscopic description   to small and large black holes in AdS$_d$.
We start with the case of small black holes, whose horizon size  $R_{\rm h}$ is smaller than the AdS scale $L$.
For this situation, we consider the metric in the master formula \eqref{eq:metric-gen-conf-dim} for
\begin{equation} \label{eq:deltablackholemass}
\Delta_L=\frac{ML}{d-2} \, .
\end{equation}
Then, one may check that the AdS-Schwarzschild metric (with a transversal circle) follows after   performing the   coordinate transformation $R=L^2/r$ and $\Phi= N\phi$:
\begin{align}\label{eq:metric-ads-schw}
ds^2 = & - \left (1+ {R^{\,2}\over L^2} - \frac{16\pi G_d M}{(d - 2)\Omega_{d-2}R^{d-3}}\right)\!  dt^2 + \left (1+{R^{\,2}\over L^2} - \frac{16\pi G_d M}{(d -  2)\Omega_{d-2}R^{d-3}}\right)^{-1} \!\! \!\!dR^{2}  \nonumber \\
& + R^{2}d\Omega^2_{d- 2} + \ell^2 d\Phi^{2}  ,
\end{align}
where $\Phi \equiv \Phi + 2\pi$. In particular, this implies that  the holographic screen at $R=R_0$ in AdS-Schwarzschild is equivalent to the $r_0=L^2/R_0$ screen in the metric \eqref{eq:metric-gen-conf-dim} for $\Delta_L$ as above.

In contrast to the previous cases, this time we do not have a clear interpretation of the AdS$_3$ metric for any value of $R$.
This is because for the particular value of $\Delta_L$ the AdS$_3$ metric between brackets in \eqref{eq:metric-gen-conf-dim} contains an $r^{(d-1)}$-dependent term.
This fact prohibits an analysis of AdS-Schwarzschild analogous to the previous cases.
However, we propose to overcome this difficulty by considering this metric only at a single constant slice $r=r_{\rm h}$, for which the $g_{tt}$ component is zero.
This happens when  
\begin{equation}\label{eq:horizon-eqn-non-btz}
\Delta_L-\frac{c_{L}(R_{\rm h})}{12}=\frac{L^2}{R_{\rm h}^2}\frac{c_{L}(R_{\rm h})}{12} \, ,
\end{equation}
where  we have used $R_{\rm h}=L^2/r_{\rm h}$.
We now interpret this slice as the horizon of an ordinary BTZ with horizon radius $r_{\rm h}$, where we make use of the fact  that horizons are locally indistinguishable \cite{Jacobson:2003wv}.
In the AdS$_d$-Schwarzschild metric, this slice becomes of course precisely the horizon $R=R_{\rm h}$ of the AdS$_d$ black hole.

Using a fractional string phenomenon we can  also express the relation above as:
\begin{equation}
\Delta_{R_{\rm h}} - \frac{c_{R_{\rm h}}(R_{\rm h})}{12} =\frac{c_{R_{\rm h}}(R_{\rm h})}{12} \qquad \text{or} \qquad \mathcal N = \mathcal C \, .
\end{equation}
Therefore, from the AdS$_3$ point of view we can think of a black hole with horizon radius $R_{\rm h}$ in AdS$_d$ as a thermal bath of $(R_{\rm h}/\ell)^{d-2}$ fractional strings at temperature $T\sim 1/R_{\rm h}$.
At the AdS$_3$ radial slice $r=r_{\rm h}$ these are all available degrees of freedom, so it is very natural that a black hole arises in the AdS$_d$ frame.

The entropy of the AdS$_d$ black hole may now be computed from a CFT$_2$ perspective.
Indeed, applying the Cardy formula to the state in \eqref{eq:horizon-eqn-non-btz} yields
\begin{equation}\label{ads-schw-entropy-r0}
S=4\pi \sqrt{\frac{c_{L}(R_{\rm h})}{6}\left(\Delta_{L}-\frac{c_{L}(R_{\rm h})}{12}\right)}=\frac{A(R_{\rm h}) }{4G_d} \, .
\end{equation}
The CFT$_2$ perspective tells us that, at the level of counting holographic degrees of freedom and their excitations, the Bekenstein-Hawking formula is a Cardy formula.
This perspective may explain the appearance of a Virasoro algebra and corresponding Cardy formula found in \cite{Majhi:2011ws,Majhi:2012tf}.  

Next we discuss large AdS$_d$ black holes, with horizon size $R_{\rm h} >L$.
For this situation we use the results from Section \ref{sec:superadsrevisited}, where we related    super-AdS$_d$ holography to sub-AdS$_3$ physics. By inserting the relations  (\ref{NCsuperAdS}) for $\mathcal N$ and $\mathcal C$ into (\ref{metricdef}) we find the following metric for sub-AdS$_3$ scales
\begin{equation}\label{eq:metric-gen-conf-dim-3}
ds^2 \!=\!\frac{R^2}{L^2} \!\!\left[-\left(\!1\!-\!\frac{r^2}{L^2}\frac{\Delta_{L}-\frac{c_L(R)}{12}}{\frac{c_L(R)}{12}}\right)\!dt^2+\left(\!1\!-\!\frac{r^2}{L^2}\frac{\Delta_{L}-\frac{c_L(R)}{12}}{\frac{c_L(R)}{12}}\right)^{-1}\!\!\!\!dr^2\!+\!r^2d\phi^2\!+\!L^2d\Omega^2_{d-2}\right].
\end{equation}
This turns into the AdS-Schwarzschild metric if one inserts formula (\ref{eq:deltablackholemass}) for $\Delta_L$ and equation \eqref{eq:super-adsd-central-charge-from-fract-string} for $c_L(R)$.
The horizon  equation now becomes
\begin{equation}
\Delta_L-\frac{ c_{L}(R_{\rm h})}{12}=\frac{R^2_{\rm h}}{L^2}\frac{c_{L}(R_{\rm h})}{12} \, .
\end{equation}
Applying the Cardy formula to this state can easily be seen to reproduce the Bekenstein-Hawking entropy for a super-AdS black hole.
This provides an explanation for the fact that the Bekensteiån-Hawking entropy for AdS$_d$ blacåk holes can be written as a Cardy-like formula for CFT$_{d-1}$ \cite{Verlinde:2000wg}.
Indeed, as for the small black holes discussed above, it shows that the AdS black hole entropy at this level of discussion \emph{is} a Cardy formula.
This result also indicates the potential usefulness of a sub-AdS$_3$ perspective on super-AdS$_d$ scales, even though the latter should be completely accessible by the CFT$_{d-1}$.

Finally, we comment on the Hawking-Page transition between small and large AdS black holes \cite{Hawking:1982dh}. 
The fact that a super-AdS$_d$ black hole has positive specific heat can be understood from the AdS$_3$ perspective in the following way. 
As we increase the size $R_{\rm h}$ of the AdS$_d$ black hole, we are decreasing the radius $r_{\rm h}$ in sub-AdS$_3$.
At the same time, we are adding degrees of freedom since the metric is multiplied by an ever growing factor $(R_{\rm h}/L)^2>1$.
Decreasing $r_{\rm h}$ at sub-AdS scales is in the direction of the UV.
In other words, the excitations have to become of larger energy since they should fit on smaller strings.
Therefore, we see that the black hole heats up as we increase its size, and hence it has a positive specific heat.
Of course, this was already well understood without referring to AdS$_3$.
The positive specific heat namely originates from the fact that the CFT$_{d-1}$ energy scale is proportional to the AdS$_d$ radial coordinate here and the number of degrees of freedom is proportional to the area of the radial slice.

On the other hand, for sub-AdS$_d$ black holes, as we make the horizon size $R_{\rm h}$ smaller, we are increasing   $r_{\rm h}$ in the super-AdS$_3$ perspective and therefore the black hole heats up.
The decrease in the number of degrees of freedom in AdS$_3$, even though we are moving outwards to larger $r_{\rm h}$, is due to the factor $(R_{\rm h}/L)^2<1$ that multiplies the metric.
Thus, the black hole heats up as we make it smaller, which establishes the negative specific heat.
There is no clear CFT$_{d-1}$ interpretation of this fact, so this is one of the main new insights from our AdS$_3$ perspective on sub-AdS$_d$ scales.

To conclude, the crucial aspect that leads to the negative specific heat is the fact that the UV-IR correspondence is reversed at sub-AdS scales.
Turning this around, one could have viewed the negative specific heat of small black holes as an important clue for the reversal of the UV-IR correspondence.

\subsection{Vacuum energy of (A)dS}
\label{sec:vacenergy}

Finally, we give an interpretation of the vacuum energy of (A)dS$_d$ spacetime from our CFT$_2$ perspective.
The vacuum energy density of (A)dS is related to the cosmological constant through $\rho_{\text{vac}} = \Lambda/8\pi G_d$ and is hence set by the IR scale. 
However, from quantum field theory one expects that the vacuum energy is instead sensitive to the UV cutoff, for example the Planck scale.
In our interpretation, the UV and IR scale correspond, respectively, to the short and long string length. 
 We will argue below that the long string phenomenon precisely explains why the vacuum energy is set by the long string scale instead of the short string scale.

The vacuum energy of $d$-dimensional (A)dS   contained in a spacelike region with volume $V(R)$ is given by 
\begin{equation}  \label{vacuumenergy}
E^{\text{(A)dS}}_{\text{vac}}= \pm \frac{(d-1)(d-2) }{16 \pi G_d L^2} V(R) \qquad \text{with} \qquad V(R) = \frac{\Omega_{d-2}R^{d-1}}{d-1} \, .
\end{equation}
The vacuum energy of AdS is negative, whereas that of dS space is positive. This expression is   obtained by multiplying the energy density $\rho_{\text{vac}} = \Lambda/8\pi G_d$ associated with the cosmological constant $\Lambda = - (d-1)(d-2)/2L^2$ with what appears to be the flat  volume of the spherical region.\footnote{A more accurate explanation of the simple expression for $V(R)$ is that in addition to describing the proper volume it incorporates the redshift factor.}  We now want to a give a microscopic interpretation of this formula  in terms of the long strings discussed in the previous section. 

In the CFT$_2$ the energy of a quantum state can be computed by multiplying the excitation number with the excitation energy of a degree of freedom. 
For long strings, this reads:
\begin{equation} \label{vacuumenergyL}
E=\left(\Delta_L-\frac{c_L(R)}{12}\right) \epsilon_L \, . 
\end{equation}
The subtraction $- c_L(R)/12$ is due to the negative Casimir energy of the CFT on a circle. 
The scaling dimension $\Delta_L$, the central charge $c_L(R)$ and the excitation energy $\epsilon_L=(d-2)/L$ are all labeled by the long string scale $L$.   
The vacuum energy of (A)dS$_d$ then follows from the expression above by imposing   specific values for the scaling dimension, $\Delta_L = 0$ (AdS) and $\Delta_L = c_L(R)/6$ (dS), i.e. 
\begin{equation}
E_{\text{vac}}^{\text{(A)dS}} = \pm \frac{c_L (R)}{12} \, \epsilon_L \, .
\end{equation}
 By inserting the values of $c_{L}(R)$ and $\epsilon_L$ into the formula above one recovers the correct vacuum energy (\ref{vacuumenergy}). This formula tells us that we can interpret the negative vacuum energy of AdS as a Casimir energy, which is what it corresponds to in the CFT$_2$. The vacuum energy of de Sitter space can be attributed to the excitations of   the lowest energy states available in the system: the long strings of size $L$.

Alternatively, if one assumes the vacuum energy is determined by the UV  or short string  degrees of freedom, one would have instead computed:
\begin{equation}
E_{\text{vac}}^{\text{UV}}   =  \pm \frac{c_\ell (R)}{12} \,   \epsilon_\ell   \, .
 \end{equation}
 By inserting the values for the short string central charge \eqref{eq:centralcharge-shortstring} and  the  excitation energy $\epsilon_\ell = (d-2)/\ell$, we find for their vacuum energy:
\begin{equation}  \label{eq:uvvacenergy}
E_{\text{vac}}^{\text{UV}}   = \pm \frac{(d-1)(d-2) }{16 \pi G_d \ell^2} V(R)   \, .
\end{equation}
This is off by a factor $1/N^2 = \ell^2/L^2$ from the true vacuum energy of (A)dS space.
The long string phenomenon precisely explains why the vacuum energy associated to long strings is $N^2$ times lower than the vacuum energy associated to short strings. It namely decreases both the central charge and the energy gap by a factor of $N$.
Thus, the identification of the long strings as the correct holographic degrees of freedom   provides a natural explanation of the value of the vacuum energy.

\section{Discussion}
\label{sec:conclusion}

In this chapter we have   proposed a new approach to  holography for non-AdS spacetimes, in particular: AdS below its curvature radius, Minkowski, de Sitter and AdS-Schwarzschild. Before summarizing our main findings, we comment on the more general lessons for non-AdS holography. First of all, we would like to make a distinction between 
  holography as manifested by the AdS/CFT correspondence and holography in general, which is only constrained by the holographic principle. The original principle states that the number of degrees of freedom in quantum gravity is bounded by the Bekenstein-Hawking formula. On the other hand, AdS/CFT is a much stronger statement,
in which the holographic degrees of freedom of quantum gravity are identified as part of a local quantum field theory in one dimension less.
We do expect (and indeed assume) the holographic principle to hold for more general spacetimes, motivated by the standard black hole  arguments \cite{tHooft:1993dmi,Susskind:1994vu,Bousso:1999xy}. However, there are indications that it is unlikely to expect a local quantum field theory dual to gravity in non-AdS spacetimes. 

One can already arrive at this conclusion for sub-AdS scales via the following reasoning. Let us assume a lattice regularization of the boundary CFT where each lattice site contains a number of degrees of freedom proportional to the central charge \cite{Susskind:1998dq}. The fact that the central charge is related to the area at the AdS radius  (cf. equation   (\ref{centralcharge}))  implies that one is left with a single lattice site   as one holographically renormalizes up to the AdS scale.
This means that the effective theory on a holographic screen at the AdS radius is completely delocalized, and can be described by  a matrix quantum mechanics.\footnote{The same conclusion is reached when considering the flat space   limit of AdS/CFT \cite{Susskind:1998vk}. In addition, generalizations of the Ryu-Takayanagi proposal to more general spacetimes suggest a non-local dual description of flat space and de Sitter space \cite{Li:2010dr,Shiba:2013jja,Miyaji:2015yva,Sanches:2016sxy,Nomura:2016ikr,Nomura:2017fyh,Nomura:2018kji}. }
It is not obvious how to further renormalize the quantum theory to probe the interior of a single AdS region. One expects, however,  that the degrees of freedom ``inside the matrix'' should play a role in the holographic description of sub-AdS regions  \cite{Susskind:1998dq,Berenstein:2005aa}. 
The reasoning also shows that one has to be careful in applying the Ryu-Takayanagi formula to sub-AdS scales, since   it typically assumes a spatial factorization of the holographic Hilbert space. For recent discussions of this issue, see for instance \cite{Balasubramanian:2014sra,Yang:2015uoa,Nomura:2018kji}. 
It is expected that similar conclusions apply to flat space holography, which should be described by the $L \to \infty$ limit of sub-AdS holography, and de Sitter static patch holography, which is connected to flat space holography via the same limit.

Even though a local quantum field theory dual may not exist for non-AdS spacetimes, we do assume that there exists a dual quantum mechanical theory which can be associated to a holographic screen. 
In this chapter we have proposed to study some general features of such quantum mechanical  theories for non-AdS geometries. These features are captured by three quantities associated to a holographic screen at radius $R$: the number of degrees of freedom $\mathcal C$, the excitation number $\mathcal N$ and the excitation energy $ \epsilon$. 
We have given a more refined interpretation of these quantities in terms of a twisted sector of a symmetric product CFT$_2$. This is achieved through a conformal map between the non-AdS geometries and locally AdS$_3$ spacetimes.  In the CFT$_2$ language, the holographic degrees of freedom are interpreted as (fractions of) long strings. 

The qualitative picture that arises is as follows. The symmetric product theory introduces a number of short degrees of freedom that scales with the volume of the non-AdS spacetime. However, the long string phenomenon, by gluing together short degrees of freedom, reduces the number of degrees of freedom to an area law, consistent with the Bekenstein-Hawking formula. We have also seen that the degrees of freedom on holographic screens at distance scales smaller than the (A)dS radius should be thought of as fractions of long strings. 
This perspective suggests that the long string degrees of freedom extend into the bulk, instead of being localized on a holographic screen. This stands in contrast with the holographic degrees of freedom that describe large AdS regions which, as suggested by holographic renormalization, are localized on their associated screens.
   
For all of the non-AdS spacetimes we studied, one of the main conclusions that arises from our proposal is that the number of degrees of freedom in the microscopic holographic theory increases towards the IR in the bulk.  
This follows from the reversal of the UV-IR relation and the holographic principle.
The familiar UV-IR correspondence thus appears to be a special feature of the holographic description of AdS space at scales larger than its curvature radius.  Our results suggest that in general the UV and IR in the spacetime geometry and in the microscopic theory are in sync.
A similar point of view has appeared in \cite{Anninos:2011af}, where the term `worldline holography' was coined (see also the review \cite{Anninos:2012qw} and their recent work \cite{Anninos:2017hhn}). This term refers to the fact that the UV observer is placed at the center of spacetime, instead of at a boundary as in AdS holography. These two observers are related by the conformal map that inverts the radius and exchanges the UV with IR.

The UV-IR correspondence in AdS/CFT is in line with the Wilsonian intuition for a quantum field theory and thus explains why the microscopic holographic theory can be described by a QFT.  On the other hand, we have argued that the general features of non-AdS holography are naturally accommodated for by symmetric products and the long string phenomenon. 
Our results then suggest that the holographic dual of non-AdS spacetimes cannot be described by (Wilsonian) quantum field theories, but  should rather be thought of as quantum mechanical systems that exhibit the long string phenomenon. 

An example of such a model is given  by matrix quantum mechanics. For instance, in the BFSS matrix model \cite{Banks:1996vh} a long string phenomenon was observed to play a role \cite{Dijkgraaf:1997vv}, inspired by the results of \cite{Taylor:1996ik}. In particular, the $N\to \infty$ limit of the matrix model corresponds to a large symmetric product, and in the far IR the long strings are the only surviving degrees of freedom. We propose that one possible way in which a smaller UV Hilbert space can be embedded in a larger IR Hilbert space, is to identify the UV degrees of freedom as excitations on fractional strings and view these excited fractional strings as bound states of the lowest energy excitations on long strings that live in the  far IR.

As we explained, the symmetric products and long string phenomenon also provide a natural framework to understand the negative specific heat of small AdS black holes.
Moreover, using a sub-AdS$_3$ perspective on super-AdS$_d$ scales, we have shown how the Hawking-Page transition between positive and negative specific heat black holes in AdS$_d$ can be understood in the CFT$_2$ language.
Even though the positive specific heat also follows from the CFT$_{d-1}$ description \cite{Witten:1998zw}, the thermal state in a strongly coupled CFT$_{\!d\!-\!1}$ is by no means a simple object to study.
It would be interesting to see whether our sub-AdS$_3$ perspective and the associated CFT$_2$ language could provide a new avenue to study strongly coupled higher dimensional CFTs.
One result we obtained in this direction is a way to understand the appearance of a Cardy-like formula in $(d\!-\!1)$-dimensional (holographic) CFTs, that describes the entropy of a $d$-dimensional AdS black hole \cite{Verlinde:2000wg,Majhi:2011ws,Majhi:2012tf}.
Namely, our conformal map relates the AdS$_d$ black hole entropy to the   entropy of a BTZ black hole, which can be derived from the Cardy formula.

We should note that the negative specific heat of small $AdS_5\times S^5$ black holes has already been studied from the $\mathcal N=4$ SYM theory in \cite{Asplund:2008xd}. 
In this paper, it is argued that a sub-matrix of the large $N$ matrix could provide a description of such ten-dimensional black holes, including the negative specific heat.
It would be   interesting to understand whether our set-up could be generalized to cover or be embedded in an $AdS_5\times S^5$ geometry.
As argued in their paper, the sub-matrix forms an essentially isolated system that consists of a dense gas of strings. 
It does not thermalize with its environment, i.e. does not spread on the $S^5$. 
Concerning the embedding, our small $S^1$, on which we also have a dense gas of strings for small AdS black holes, could be the effective geometry seen by such a localized, confined system on $S^5$.

More generally, an open question is whether the  geometries in this chapter can be embedded in string  or M-theory. 
A particularly interesting case that we leave for future study is the MSW string \cite{Maldacena:1997de}.
This string has a near-horizon geometry of the form $AdS_3\times S^2$. Applying our conformal map to a BTZ geometry in this set-up would lead to a  $dS_4\times S^1$ spacetime. 
The microscopic quantum description of the latter spacetime could be related to the $D0$-$D4$ quiver quantum mechanics theories studied by \cite{Gaiotto:2004ij}.  It is plausible that an extension of Matrix theory to $d=4$ requires the inclusion of transversal fivebranes \cite{Berkooz:1996is} and precisely leads to such a quiver QM description. 

Finally, we have argued that de Sitter space must be regarded as an excited state of the microscopic  holographic quantum system.\footnote{This is also suggested by the generalization of the Ryu-Takayanagi proposal to cosmological spacetimes, where the  holographic entanglement entropy obeys a volume law on the   holographic screen \cite{Nomura:2016ikr}.} In this sense it is similar to the BTZ spacetime to which it is conformally equivalent. 
This explains in particular the fact that the de Sitter entropy can be recovered from a Cardy formula in two-dimensional CFT.
From its description as a thermal bath of long strings at the Gibbons-Hawking temperature $T\sim 1/L$, we moreover showed that   the  vacuum energy   of de Sitter space can be reproduced from the energy carried by the long strings.

%% file: samenvatting.tex
\begin{quote}
M pulled out his pipe and started to fill it. ``And now you know as much about diamonds as I do.''

--- \textsc{Ian Fleming}, \emph{Diamonds are Forever}, book (1956)
\end{quote}
\begin{center}
   \vskip .2cm
 ***
   \vskip .2cm
\end{center}

In this thesis we have  explored thermodynamic, emergent and holographic aspects of causal diamonds. 
We conclude with a summary of the main results   and   an outlook to possible future research directions. 


\subsubsection{Gravitational thermodynamics of causal diamonds}

In   Chapter \ref{ch2} we studied the gravitational thermodynamics of causal diamonds in maximally symmetric spaces. We established a Smarr formula for such diamonds and a ``first law'' for variations to nearby solutions. The latter relates the variations of the bounding area, spatial volume of the maximal slice, cosmological constant, and matter Hamiltonian. The total Hamiltonian is the generator of evolution along the conformal Killing vector that preserves the diamond. To interpret the first law as a thermodynamic relation, it appeared necessary to attribute a negative temperature to the diamond, as has been previously suggested for the special case of the static patch of de Sitter spacetime. With quantum corrections included, for small diamonds we recovered the ``entanglement equilibrium'' result that the generalized entropy is stationary at the maximally symmetric vacuum at fixed volume, and we reformulated this as the extremization of   free conformal energy with the volume not fixed.

\subsubsection{Higher curvature gravity from entanglement equilibrium}

In   Chapter \ref{ch3} we showed that the linearized higher derivative gravitational field equations are equivalent to an equilibrium condition on the entanglement entropy of small spherical regions in vacuum. This extends Jacobson's   derivation of the Einstein equation using entanglement to include general higher derivative corrections. The corrections are naturally associated with the subleading divergences in the entanglement entropy, which take the form of a Wald entropy evaluated on the entangling surface. Variations of the Wald entropy are related to the field equations through an identity for causal diamonds in maximally symmetric spacetimes, which we derived for arbitrary higher derivative theories. If the variations are taken holding fixed a geometric functional that we call the ``generalized volume'', the identity becomes an equivalence between the linearized constraints and the entanglement equilibrium condition. The fully nonlinear higher curvature equations cannot be derived from the linearized equations applied to small balls, in contrast to the situation encountered in Einstein gravity.

\subsubsection{Towards   non-AdS holography}

In  Chapter \ref{ch4} we applied the holographic principle to the microscopic holographic theories for non-AdS spacetimes, specifically for Minkowski, de Sitter, and AdS below its curvature radius. By taking general lessons from AdS/CFT we derived the cutoff energy of the holographic theories for these non-AdS geometries. Contrary to the AdS/CFT correspondence, the excitation energy decreases towards the IR in the bulk, which is related to the negative specific heat of black holes. We constructed a conformal mapping between the non-AdS geometries and $AdS_3\times S^q$ spacetimes, and related the microscopic properties of the holographic theories for non-AdS spaces to those of symmetric product CFTs. We found that the mechanism responsible for the inversion of the energy-distance relation corresponds to the long string phenomenon. We argued  that this same mechanism naturally explains the negative specific heat for non-AdS black holes and the value of the vacuum energy in (A)dS spacetimes.

\subsubsection{Future research directions}

\emph{Non-spherical regions and non-maximally symmetric spaces.}
 The first law of causal diamonds was derived  for spherical  regions in 
 maximally symmetric spaces.  A natural question   is whether a similar first law holds for  arbitrarily shaped spatial   regions  $\S$ in any spacetime.  Such regions typically do not   admit a conformal Killing vector  which preserves its causal development; however, this may not be necessary to derive a first law.  The particular form  (\ref{firstlawcc1}) of the first law   only applies to  maximally symmetric spaces, but the   general form  (\ref{dA=-dH}), $\delta H_\z = - \kappa \delta A / 8\pi G$, may hold for every causal diamond, provided the vector field $\z$ is    locally a boost vector field near the edge of the diamond and $\kappa$ is constant on the edge.  An interesting choice of   vector field on the surface $\S$ would be the one associated to the ``York time'' flow of the constant mean curvature  (CMC) foliation of the domain of dependence of $\S$. An open question is whether    a definition of   surface gravity associated with the York time vector field exists such that it is constant on the   edge  or   on the entire  null boundary of the diamond    (i.e. a zeroth law for York time flow).  For spherically symmetric regions the constancy of $\kappa$ on the boundary is automatic, but for non-spherical regions it is not so obvious.   Since the York time   Hamiltonian is   proportional to the proper volume of   CMC slices, the gravitational Hamiltonian variation $\delta H_\z^{\rm g}$ in the first law could be equal to the variation of the volume of such slices for an appropriate choice of York time.   
 Thus, a similar first law of causal diamonds might hold for generic regions and spacetimes as for spherical regions in maximally symmetric spaces. 
 

\emph{Higher order perturbations.} In Chapters \ref{ch2} and \ref{ch3}  we restricted attention only to first order perturbations of the matter fields and the geometry.  Working to higher order in perturbation theory could yield several interesting results.  One such possibility would be proving that the vacuum entanglement 
entropy is maximal, as opposed to merely extremal.  The second order change in entanglement
entropy is no longer just the change in modular Hamiltonian expectation value.  The 
difference is given by the relative entropy, so a proof of maximality will likely invoke
the positivity of relative entropy.  On the geometrical side, a second order variational
identity would need to be derived, along the lines of the Hollands-Wald formalism \cite{Hollands2013}.  
One would expect that graviton contributions would appear at this order, and it 
would be interesting to examine how they play into the entanglement equilibrium story.  
Also, by considering small balls and using the higher order terms in the Riemann normal
coordinate expansion, in addition to   higher order perturbations,  it is possible that one could derive the fully nonlinear
field equations of any higher curvature theory.  Finally, coherent states pose a puzzle
for the entanglement equilibrium hypothesis, since they change the energy within
the ball without changing the entanglement \cite{Varadarajan:2016kei}.  However, 
their effect on the energy density only appears at second order in perturbations, so 
carrying the entanglement equilibrium argument to higher order could shed light on 
this puzzle.

\emph{CFT interpretation of first law for large diamonds in AdS.}
 Proposals exist for   holographic CFT duals of the maximal volume and boundary area in the bulk. The area  of the cutoff boundary surface 
in Planck units may be dual to the  number of degrees of freedom of the cutoff CFT \cite{Susskind:1998dq}. The proposed dual quantities for the maximal volume are  computational complexity \cite{Susskind:2014rva}, fidelity susceptibility  \cite{MIyaji:2015mia}, and  (for       variations) the symplectic form on the space of sources in the Euclidean path integral \cite{Belin:2018fxe,Belin:2018bpg}. 
One can also consider varying the cosmological constant, which is equivalent to varying the number of degrees of freedom $N$ in the CFT (if     Newton's constant is kept fixed). The thermodynamic volume is the conjugate to the cosmological constant, and in the AdS-Rindler case it is dual to the chemical potential for $N$
in a ball in the vacuum state of a CFT \cite{Kastor:2014dra}. It would be interesting to find a variational relation
in the CFT between these quantities dual to the bulk area, maximal volume, and the thermodynamic volume of the maximal slice.

 \emph{Euclidean path integral for causal diamonds.} Black hole thermodynamics, as well as de Sitter thermodynamics, has been well studied from the viewpoint of the partition function, formulated as a path integral over Riemannian geometries beginning with the work of Gibbons and Hawking \cite{Gibbons:1977mu}. Can the thermodynamics of causal diamonds similarly be formulated in terms
of the Euclidean path integral in the canonical and/or microcanonical ensemble? If so this should provide 
a foundation for determining the stability of different ensembles, 
and for a systematic treatment of the quantum corrections. 
 
 \emph{Covariant definition of holographic microscopic quantities.} In Chapter \ref{ch4} we  defined   three holographic quantities  for spherically symmetric causal diamonds  in static, spherically symmetric spacetimes. The number of degrees of freedom has a covariant definition, as the area in Planckian units of the edge of the diamond, but the definitions of the excitation energy and excitation number are coordinate dependent. In particular, the excitation energy \eqref{excenergy} depends on a time coordinate, and the definition \eqref{metricdef}  of the excitation number involves a component of the metric. It is highly desirable to find covariant definitions for these two quantities, which hold for any spatial codimension-two surface in any  (stationary) spacetime. The excitation energy could be defined as the expansion of the null generators in the reference spacetime, and the excitation number might be related to the Brown-York quasi-local energy of the codimension-two surface.

To conclude,    causal diamonds could play an important role in future studies of holographic and thermodynamic aspects of gravity. We would like to end though with a speculation about an alternative option: instead of the full diamond, the thermodynamics might be a property of the (maximal) spatial slice alone.   The diamond itself   did not play an essential role in the derivation of the first law, since all the quantities are evaluated at the maximal slice. Further, the boundary of the (maximal) spatial slice could serve as a holographic screen, and   the microscopic holographic theory could live on the timelike evolution of that boundary.  To quote Wittgenstein's \emph{Tractatus} \cite{Wittgenstein} (par. 6.54), ``[we]   must, so to speak,  throw away the ladder after [we have] climbed up it'',  that is, we must transcend causal diamonds   and instead investigate (the thermodynamics of)   spatial   and timelike surfaces.

 \begin{center}
 \vskip .5cm
  ***
 \end{center}

%% file: appendix.tex
\appendix

\chapter{Appendices}

\section{Acceleration and velocity  of the conformal Killing flow}
\label{sec:accckv}

In this appendix we show that observers, who move along the flow lines of the conformal Killing field \eqref{ckv6} that preserves a causal diamond in flat space, are uniformly accelerating inside the diamond. The upshot is that these observers are all Rindler observers, but each with a different Rindler event horizon.    We also derive the velocity and proper time for these observers. 

The line element of a diamond in flat space is
\begin{equation}
\begin{aligned} \label{xsflatdiamond}
& ds^2 = C(s,x)^2 \left [  -ds^2 + d x^2 + \rho(x)^2 d\Omega_{d-2}^2 \right], \\
   \text{with}\quad & C(s,x) =   \frac{R}{\cosh s + \cosh x} , \quad \rho(x) = \sinh x \, .
\end{aligned}
\end{equation}
The range of the (dimensionless) time and radial coordinate inside the diamond is $s\in ( - \infty, + \infty)$ and   $x\in [0, \infty)$, respectively. Further, $R$ is the proper distance between the center ($s=0,x=0$) and the edge of the diamond ($x=\infty$) or, equivalently, the proper time between the center and the future tip ($s=\infty, x=0$). See Appendix \ref{appyork} for   a derivation of this metric for   generic maximally symmetric diamonds, and see footnote \ref{fnt:Cflat} for the special case of   diamonds in flat space. 

In this   coordinate system    the  conformal Killing vector  that preserves the diamond is $\zeta = \partial_s$.
 The flow lines of $\z$ coincide at the past and future tip of the diamond, but never   cross inside the diamond (see Figure \ref{fig:causaldiamondsandx} on p. \pageref{fig:causaldiamondsandx} for   the $(x,s)$ coordinate chart of the diamond). The velocity  vector of the conformal Killing flow  is given by $u^a = \zeta^a / \sqrt{- \zeta \cdot \zeta} =  {\delta_s}^a C^{-1} $, and the acceleration  vector is  $a^b = u^a \nabla_a u^b = {\delta_x}^b C^{-3 } \partial_x C$. 
Using the expression \eqref{xsflatdiamond} for $C=C(s,x)$, we find that the proper acceleration   $a:=\sqrt{a^ba_b}$    is equal to\footnote{For a diamond in the de Sitter static patch the  acceleration is $a =\frac{\sinh x}{L \tanh(R_*/L)}$, which can be obtained by inserting the expression \eqref{Cinv} for $C(s,x)$ in de Sitter space into   \eqref{accelerationckvflow}. Note that this is  also constant on constant $x$-slices. For the entire de Sitter static patch ($R_* \to L$) the acceleration simplifies to $a = \sinh(x)/L$,
and for the Wheeler-de Witt patch of  pure AdS space ($L \to i L, R_* = L \pi/2$) it    vanishes, $a=0$.  \label{footnote:dSacceleration}} 
\beq \label{accelerationckvflow}
a (x) =  C^{-2}  \partial_x C  = \frac{\sinh x}{R} \, .
\eeq 
Note that the proper acceleration is  constant on constant $x$-slices of the diamond. This implies that the proper acceleration of observers who move along the flow lines of $\z$ is constant,   because  by construction the flow lines are constant $x$-slices (i.e. $\z \cdot dx =0$). 
Since these observers are uniformly accelerating in flat space, they are  \emph{Rindler observers} inside the diamond, who start at the past tip and end at the future tip.  However, each observer has a different Rindler event horizon, and their wordlines cross after a finite proper time from the   $s=0$ slice (namely at the past and future tip of the diamond). This stands in contrast to the  standard Rindler observers in the Rindler wedge of flat space, who all share the same Rindler horizon and whose worldlines never cross within a finite proper time  (they coincide at   infinite proper time $\tau = \pm \infty$).\footnote{The  right Rindler wedge is obtained from the finite diamond  centered at the origin in flat space coordinates, by first shifting the spatial coordinate $x^1 \to x^1 - R$ and then taking the limit $R\to \infty$. Applying the shift and   infinite size limit to the acceleration $\sinh (x)/R $, where $x = \frac{1}{2}   \ln \left[ \frac{(R + t + r)(R - t + r)}{ (R - t - r) (R + t - r)} \right]$ and $r= \sqrt{(x^1)^2 + \dots + (x^{d-1})^2}$, yields  the expression  $((x^1)^2-t^2)^{-1/2}$, which is indeed equal to the  (constant) acceleration of Rindler observers. Thus, the Rindler flow   is   a limiting case of the conformal Killing flow (see also Section \ref{largesmall:rindler}).}


As a side remark, we notice  that the surface gravity associated to $\zeta$   is equal to the redshifted proper acceleration evaluated at the bifurcation surface $\mathcal{B}$, i.e.  
$
  \kappa =  \lim_{x \to \infty} C a    = 1 \, .
$
 (Note that $C \to 0 $ and $a \to \infty$ in the limit $x\to \infty$.) 
  This formula for the surface gravity is known to be equivalent to other definitions of surface gravity for Killing horizons (see  e.g. \cite{Wald:1984rg} p. 332). For bifurcate conformal Killing horizons, following the same steps as \cite{Wald:1984rg},  the acceleration definition 
  can be shown to be equivalent to  the definition \eqref{kappa1}, but only in the limit as  the bifurcation surface is approached along the horizon, since the conformal Killing field is   instantaneously a true Killing field at  $\mathcal{B}$ (see Appendix \ref{sec:zerothlaw}).

The proper time along the flow lines of $\z$, defined as the function that   vanishes at $s=0$,  is   given by 
\beq \label{propertimediamond}
\tau (s,x) = \int_0^{s} \frac{ds'}{C(s',x)} = \frac{2 R}{\sinh x} \text{arctanh} \left [ \tanh(s/2) \tanh(x/2) \right] \, . 
\eeq
The total proper time from the $s=0$ slice to the future   tip of the diamond is finite, i.e. $\tau(s=\infty,x) = R \, x / \sinh x$, and    maximal  and equal to  $R$     at $x=0$ (for observers at rest).  Moreover, it vanishes at the horizon, i.e. $\tau (s = \infty, x= \infty) =0 $. 

Since the conformal Killing flow observers are Rindler observers, their worldlines are    described in standard Minkowski coordinates $(t,x^1)$ by the parametrization
\beq
t (\tau) = \frac{1}{a} \sinh (a \tau), \qquad x^1 (\tau) = \frac{1}{a} \cosh (a \tau) \, , 
\eeq
where $a$ is the constant proper acceleration and $\tau$ is the proper time (which vanishes at $t=0$).  The velocity of a Rindler observer with respect to an observer at rest at $x^1=0$ is
\beq
v (\tau) = \frac{dx^1}{dt} = \frac{dx^1}{d \tau}  \Big / \frac{d t}{ d \tau}=\tanh(a \tau) \, . 
\eeq
By inserting the acceleration \eqref{accelerationckvflow} and proper time \eqref{propertimediamond},     we find the velocity of the observers who follow the orbits of $\z$ as a function of $s$ and $x$
\beq
v(s,x) =  \frac{2\tanh(s/2) \tanh(x/2) }{\tanh^2(s/2) \tanh^2(x/2) + 1}  \, . 
\eeq
Note that the size $R$ of the diamond has dropped out of this expression. At the $s=0$   slice the velocity vanishes,   hence  the observers are instantaneously are rest. At the future tip, where all the observers meet, the velocity is given by: $v(s=\infty, x) = \tanh x$, which   ranges from $v=0$ for $x=0$ (an observer at rest) to $v=1$ (speed of light) for $x = \infty$ (at the null boundary of the diamond).

\section{Conformal isometry of causal diamonds in (A)dS}
\label{sec:ckv}

We seek the conformal isometry that preserves a causal diamond in de Sitter space. Since the property of being a conformal isometry is invariant under a Weyl rescaling of the metric, we will use the  form of the line element (\ref{linehyperbolic}), leaving  off the conformal factor $\text{sech}^2  (r_*/L)$. Note that any vector field of the form 
\begin{equation}
\zeta = A(u) \partial_u + B(v) \partial_v
\end{equation}
is then a conformal isometry of the $dudv$ factor of the metric: $\mathcal L_\zeta du dv = [A'(u) + B'(v)] du dv$, where $\mathcal L_\zeta$ is the Lie derivative along $\zeta$. It will be a conformal isometry of the full metric provided   $\mathcal L_\zeta \sinh^2 (r_*/L) = [A'(u) + B'(v)] \sinh^2 (r_*/L)$. Using $r_* = (v-u)/2$  we find that in fact $\mathcal L_\zeta \sinh^2 (r_*/L) = \frac{1}{L}(B-A) \sinh (r_*/L) \cosh (r_*/L)$, so $\zeta$ is a conformal Killing field if 
\begin{equation} \label{cond}
[A'(u) + B'(v)] L \tanh (r_* /L)= B(v) - A(u) \, .
\end{equation}
For $u = v$, $r_* = 0$ and thus $\tanh (r_*/L) = 0$, so this implies $B(v) = A(v)$. Then evaluating at $v=0$, we  have $r_* = - u/2$, so (\ref{cond}) becomes
\begin{equation}
[A'(u) + A'(0)] L \tanh (u/(2L))= A(u) - A(0) \, .
\end{equation}
The general solution to this equation is 
\begin{equation}
A(u) = B(u) =   a + b \sinh (u/L) + c \cosh (u/L)  \,.
\end{equation}
To map the diamond onto itself, the flow of $\zeta$ must leave invariant the boundaries $u = - R_*$ and $v = R_*$. This implies $A(\pm R_*) = 0$, and hence $A(u) = a_1 (\cosh (u/L) - \cosh (R_*/L)).$ The requirement that the surface gravity $\kappa$ of $\zeta$ be unity at the future conformal Killing horizon implies $\kappa = -B' (R_*) = - \frac{1}{L} a_1 \sinh (R_*/L) = 1$. Therefore, the conformal Killing vector that preserves a diamond in dS and has unit surface gravity at the  horizon   is   given by
\begin{equation} \label{ckv4}
 \zeta  = \frac{L}{\sinh (R_* / L)} \Big[ \left(\cosh (R_* / L) - \cosh (u/L) \right)\partial_u + \left(\cosh (R_* / L) - \cosh (v/L) \right) \partial_v    \Big]  \, . 
\end{equation}
   Expressed in terms of the standard $t$ and $r$ coordinates, $\zeta$ reads
  \begin{equation} \label{ckv3}
   \zeta  \!=\! \frac{L^2}{R} \! \!\left[    \Big ( 1 \!-\!  \frac{\sqrt{1- (R/L)^2}}{\sqrt{1- (r/L)^2}}   \cosh(t/L)  \Big)    \partial_t -   \frac{r}{L} \sqrt{\left ( 1 \!-\! (R/L)^2 \right)  \left (1 \!-\! (r/L)^2\right) }   \sinh(t/L)  \partial_r  \!     \right].
  \end{equation}
 A similar expression exists for the conformal Killing vector which preserves a causal diamond in AdS, which can be obtained by sending $L \rightarrow iL$. Moreover, in the flat space limit $L \to \infty$,   $\zeta$ reduces to  the well-known expressions \cite{Faulkner:2013ica, Jacobson:2015hqa}
 \begin{equation}
 \begin{aligned}  \label{ckvflat2}
 \zeta^{\text{flat}} &= \frac{1}{2 R} \left [ \left (R^2 - u^2 \right)\partial_u  + \left (R^2 - v^2 \right)\partial_v  \right]     \\
 &= \frac{1}{2R} \left [  \left ( R^2 - t^2 - r^2 \right)  \partial_t   - 2 rt     \partial_r  \right] \, .
 \end{aligned}
 \end{equation}

\section{Conformal Killing time and mean curvature}
\label{appyork}

In this appendix we show that on slices of constant conformal Killing time $s$ in a maximally symmetric causal diamond,
the trace $K$ of the extrinsic curvature is constant, and we establish the relation \eqref{Kands2} between $K$ and $s$. 
To this end, we first construct a coordinate system adapted to the flow of $\z$, and then we 
compute $K$ on the constant $s$ slices using this coordinate system.

Conformal Killing time is a function $s$ satisfying $\z\cdot ds=1$, but there are many such functions. Given one choice of a constant $s$ hypersurface, $s$ is determined by integration along the flow lines of $\z$. We choose the $t=0$ slice of the diamond to be the $s=0$ slice. This slice is everywhere orthogonal to $\z$, hence $\z_a$ and $\nabla_a s$ are parallel on it. The Lie derivative ${\cal L}_\z\nabla_a s$ vanishes by definition of $s$, and ${\cal L}_\z \z_a = 2\a \z_a$, since $\z$ is a conformal Killing vector satisfying ${\cal L}_\z g_{ab}=2\a g_{ab}$. The flow therefore preserves the proportionality of these two covectors, hence all of the constant $s$ slices are orthogonal to $\z$. 

Now choose a spherically symmetric coordinate $x$ on the $s = 0$ slice, such that $ |dx|=|ds| $ and $x = 0$ at $r = 0$, and extend it to the diamond by the flow of $\z$, i.e.\ so that $\z\cdot dx = 0$. Both $ds$ and $dx$ are invariant under the conformal Killing flow, and 
they are orthogonal and have equal norms at $s=0$, hence these conditions hold for all values of $s$. 
The line element \eqref{linehyperbolic} therefore takes the form 
\begin{equation} \label{confKillingmetric1}
ds^2 
=  C^2(s,x)(- ds^2 + dx^2)+ r^2 d\O_{d-2}^2 \,.
\end{equation}
The conformal Killing equation then implies that $r = C\rho$, where $\r=\r(x)$ is a function of $x$ alone, so we have
\begin{equation} \label{confKillingmetric2}
ds^2 
=  C^2(s,x)[- ds^2 + dx^2+ \r^2(x) d\O_{d-2}^2] \,.
\end{equation}
In this coordinate system the future horizon of the diamond is located at $s=~\!\!\infty$, and the past horizon is  at $s=-\infty$ (see Figure \ref{fig:causaldiamondsandx} on p. \pageref{fig:causaldiamondsandx} for an illustration of   constant $s$ and $x$ surfaces).
Further, we have $\z = \partial_s$, the surface gravity is $\kappa = - C^{-1} \partial_s C  \big |_{s \to \infty}$, and the unit timelike vector normal to the constant $s$ slices is 
$u = C^{-1}\partial_s$. The divergence $\nabla_a u^a$ is the trace $K$ of the extrinsic curvature of these slices,
\beq\label{K}
K = (d-1)C^{-2}\partial_s C = (1-d)\partial_s C^{-1} \,.
\eeq
It remains to show that $\partial_s C^{-1}$ is independent of $x$, and to evaluate it explicitly.\footnote{We note that the mean curvature can also be expressed as 
$K= (d-1) \alpha / |\zeta|$, since $\alpha = \nabla \cdot \zeta /d =  C^{-1} \partial_s C $ and $|\zeta| = C$.}

We proceed by finding the 
relation between the null coordinates $(u,v)$ defined in \eqref{uandv} and the null coordinates 
$\bar u = s-x$ and 
$\bar v= s+x$. 
The form of the metric indicates that $\bar u = \bar u(u)$ and $\bar v =\bar v(v)$, and 
the conditions $\z\cdot ds=1$ and $\z\cdot dx=0$ imply $\z\cdot d\bar u = 1 = \z\cdot d\bar v$. 
Using the expression \eqref{ckv4} for $\z$ (which is the conformal Killing vector which has unit surface gravity), we find   the coordinate relations
\beq \label{coordrelations}
e^{\bar u} = \frac{\sinh[(R_*+u)/2L]}{\sinh[(R_*-u)/2L]} \,, \quad 
e^{u/ L} = \frac{\cosh[( R_*/L +\bar u)/2]}{\cosh[( R_*/L -\bar u)/2]} \,, \quad 
\text{and} \quad (u\rightarrow v) \,.
\eeq
With the help of Mathematica we then find
\beq\label{Cinv}
C^{-1} = \frac{\cosh s + \cosh (x) \cosh(R_*/L)}{L\sinh(R_*/L)} \,,\qquad \rho =\sinh x\,.
\eeq
With this equation for $\rho$ the metric between brackets in \eqref{confKillingmetric2} becomes that of   conformal Killing time cross hyperbolic space, $ \mathbb R \times \mathbb H^{d-1}$.\footnote{The limit $L\rightarrow\infty$, $R_*\rightarrow R$ of (\ref{coordrelations}) and (\ref{Cinv}) yields the result for
a diamond in flat spacetime: $\bar u = \ln \!\left [(R+u)/(R-u)\right]$, $u = R \tanh (\bar u /2)$, and 
$C=R/({\cosh s + \cosh x})$. 
The limit $R\rightarrow L$, $R_*\rightarrow\infty$ leads to the result for the static patch of de Sitter spacetime:
$\bar u = u/L$, and $C= L/\!\cosh x$. And by   replacing $L \to i L$ and  setting $R_* =  L \pi/2$ we obtain 
  for the Wheeler-DeWitt patch of AdS:  $C = L / \cosh s$, where $L$ is the AdS radius. \label{fnt:Cflat}}
From this expression for $C^{-1}$   it follows that 
\beq
K = \frac{1-d}{L\sinh(R_*/L)}\,\sinh s = (d-1) \dot{\alpha} |_{s=0} \, \sinh s \,,
\eeq
where we used \eqref{alphadot} in the last equality. This establishes in particular that $K$ is constant on the constant $s$ slices.\footnote{A degenerate case is the dS static patch ($R_*\to\infty$), since in that case  $K=0$ for all $s$, i.e. all the CMC slices have zero mean curvature, so there is no York time flow for the dS static patch.} 
 Instead of using  \eqref{alphadot} we could  use   
$\alpha = \nabla \cdot \zeta /d =  -C \partial_s C^{-1}$ 
to write
$\dot\alpha = u^a \nabla_a \alpha = -C^{-1} \partial_s (C \partial_s C^{-1})$. 
Inserting \eqref{Cinv}  then yields  $\dot{\alpha} |_{s=0}= -1/{[L\sinh(R_*/L)]}$, which again 
establishes the fact that $\dot{\alpha} |_{s=0}$ is independent of $x$.

Finally, as a side comment, let us explain how the conformal Killing flow is related to the ``new York'' transformation of \cite{Belin:2018fxe, Belin:2018bpg}. In terms of the induced metric and the extrinsic curvature of a Cauchy surface the latter transformations reads: $\delta_Y h_{ab} =0$ and $\delta_Y K_{ab} = \tilde \alpha \, h_{ab}$. This transformation is not a diffeomorphism in general, so it is not expected to be the same as the variation induced by the conformal Killing vector: $\delta_\zeta \phi = \mathcal L_\zeta \phi$. In our setup the extrinsic curvature  is given by  $K_{ab} :=\nabla_a u_b = ( C^{-2} \partial_s  C ) h_{ab}= \alpha h_{ab}/ |\zeta|$. The Lie derivative of the induced metric $h_{ab}:= u_a u_b + g_{ab}$ on $\S$ and of the extrinsic curvature  are given by
\beq
\mathcal L_\zeta h_{ab}  = 2 \alpha h_{ab} \,  , \qquad \mathcal L_\zeta K_{ab} = ( C^{-2} \partial_s^2 C ) h_{ab}=   (\dot{\alpha} +  \alpha^2 /   |\zeta|    )  h_{ab} \, . 
\eeq
The first equation follows from the definition \eqref{lie} of a conformal Killing vector,  and from $\mathcal L_\zeta u_a = \alpha u_a$, where $u_a = \zeta_a / |\zeta|$. The second equation follows from the expression above for $K_{ab}$ and from the fact that $h_{ab} (s,x) = C^2(s,x) \sigma_{ab} (x)$, where $\sigma_{ab} (x)$ is the metric on hyperbolic space $\mathbb H^{d-1}$.
Since $\alpha=0$ on $\Sigma$, we find that on the extremal slice of the diamond
\beq
\mathcal L_\zeta h_{ab}  \big |_\Sigma =0 \,  , \qquad \mathcal L_\zeta K_{ab} \big |_\Sigma =  \dot{\alpha}   |_{s=0}  h_{ab} \, . 
\eeq
This is of the same form as the new York transformation, if we identify   $\tilde \alpha = \dot{\alpha} \big |_{s=0}$. In other words, the new York transformation and the conformal Killing transformation only coincide \emph{at} the extremal surface $\Sigma$.

\section{Zeroth law for   bifurcate  conformal Killing horizons}
\label{sec:zerothlaw}

A vector field $\zeta^a$ is a conformal Killing vector of the metric $g_{ab}$ if $ \mathcal L_\zeta g_{ab} = 2 \alpha g_{ab}$ for some function $\a$. A conformal Killing horizon $\mathcal H$ is a null hypersurface whose  null  generators are orbits of a conformal Killing field. The notion of surface gravity for Killing horizons can be extended to conformal Killing horizons \cite{Jacobson:1993pf, Dyer:1979, Dyer:2004}, however definitions that are equivalent in the former case (see e.g.  Sec.~12.5 in \cite{Wald:1984rg}) are not generally equivalent in the latter case.  
One of these definitions has two properties not shared by the other definitions: 
it is (i) Weyl invariant, and (ii) constant on the horizon. In this appendix we establish these properties,
the second one assuming  $\mathcal H$ has a bifurcation surface ${\cal B}$ where $\z^a$ vanishes.
We also show that $\nabla_a\z_b \overset{\mathcal{B}}{=}\k n_{ab}$, where $\k$ is the surface gravity and $n_{ab}$ is the binormal to ${\cal B}$. In particular, the Killing equation holds at ${\cal B}$.

The conformal Killing horizon is a hypersurface defined by the equation $\z^2=0$. The gradient $\nabla_a\z^2$ is normal 
to this surface; and, since it is a null surface, the normal is proportional to $\z_a$, so 
 \begin{equation} \label{kappa1}
 \nabla_a \! \left (  \zeta^b \zeta_b \right) \overset{\mathcal{H}}{=} - 2 \kappa \zeta_a \,
 \end{equation}
for some function $\kappa$. 
Under a Weyl transformation, $g_{ab} \rightarrow  \Omega^2 g_{ab}$,  $\zeta^a$ remains a conformal Killing field, and 
$\mathcal H$ a conformal Killing horizon. Moreover both sides of \eqref{kappa1}
transform homogeneously, acquiring a factor $\O^2$. 
This definition of the surface gravity $\kappa$ is therefore Weyl invariant \cite{Jacobson:1993pf}.

The zeroth law for  stationary black holes asserts that the surface gravity  is constant  over the entire event horizon.   This was originally proven in    \cite{Bardeen:1973gs} by assuming  the Einstein field equations and the dominant energy condition for matter.\footnote{There   exists another proof  of the zeroth law by Carter \cite{Carter:1973} and R\'{a}cz and Wald \cite{Racz:1995nh} that does not depend on the Einstein equation or an energy condition. The proof assumes the black hole horizon is a Killing horizon and  the black hole is either (i) static or (ii) stationary-axisymmetric with ``$t - \phi$'' reflection isometry. It would be interesting to see whether this proof can be generalized to conformal Killing horizons.}   
A simpler, and in a sense more general, proof of the zeroth law for black hole horizons was given by Kay and Wald in \cite{Kay:1988mu}.\footnote{Gibbons and Geroch  are acknowledged, respectively, in \cite{Kay:1988mu} for this version of the zeroth law and   for the proof.}  This proof    applies to   Killing horizons that contain a bifurcation surface where the Killing vector vanishes --- also called \emph{bifurcate Killing horizons} ---  and  it holds independently of  any gravitational field equation or energy condition.  Here we extend this proof to  bifurcate conformal Killing horizons:   \\ 
 
\noindent  {\bf Zeroth Law.} \textit{ Let $\mathcal H$ be a (connected) conformal Killing horizon with bifurcation surface $\mathcal B$. Then  the surface gravity $\kappa$,   defined in (\ref{kappa1}), is constant on $\mathcal H$.}   \\
  
\noindent \textit{Proof:} 
We  show first that $\kappa$ is constant along the generators of $\mathcal H$, and then 
that its value also does not vary from generator to generator. 

That $\k$ is constant on each generator can be expected, since 
the flow of the conformal Killing vector leaves the metric unchanged up to a Weyl transformation, 
and $\k$ is Weyl invariant. To make this into a computational proof, we take the Lie derivative 
of both sides of \eqref{kappa1} along $\z$. On the lhs we have 
\beq\label{proof1}
\mathcal L_\zeta \nabla_a \z^2=\nabla_a {\cal L}_\zeta \z^2=   \nabla_a {\cal L}_\zeta(g_{bc}\zeta^b \zeta^c) 
= \nabla_a (2\a\z^2) 
\overset{\mathcal{H}}{=} 2\a\nabla_a\z^2
\overset{\mathcal{H}}{=} -4\a\k\z_a.
\eeq
On the rhs we have 
\begin{equation} \label{proof2}
\mathcal L_\zeta (- 2\kappa \zeta_a) =  (- 2 \mathcal L_\zeta \kappa - 4 \alpha  \kappa)   \zeta_a \, . 
\end{equation}
Since (\ref{proof1}) and (\ref{proof2})  must be equal, we conclude that
\begin{equation} \label{lieofkappa}
\mathcal L_\zeta \kappa \overset{\mathcal{H}}{=} 0,
\end{equation}
i.e.\,  $\kappa$ is constant along the flow of $\z^a$ and hence along each null generator of $\mathcal H$. 

Next, to prove that $\kappa$ does not vary from generator to generator, we will show that it is constant on the bifurcation surface ${\cal B}$. We begin by 
noting that, since $\z^a$ is a conformal Killing vector, we have 
 \beq\label{sweet}
 \nabla_a\z_b = \a g_{ab} +   \o_{ab},    
 \eeq
where $\o_{ab} = -\o_{ba}$ is an antisymmetric tensor.
If  $m^a$ is tangent to $\mathcal B$, then 
the contraction of this equation with $m^am^b$ yields zero on the lhs
(since $\z_b=0$ on ${\cal B}$), and yields $\a m^2$ on the rhs. It follows that the conformal factor $\a$ vanishes on~$\mathcal B$, and thus the Killing equation $\nabla_{(a}\z_{b)}\overset{\mathcal{B}}{=} 0$ holds there.  In addition, 
the contraction of the lhs of \eqref{sweet} with a single vector $m^a$ vanishes on ${\cal B}$, so 
$m^a\o_{ab}\overset{\mathcal{B}}{=}0$ for all $m^a$ tangent to ${\cal B}$, 
which implies that $\o_{ab}$ is proportional to the binormal $n_{ab}$ to ${\cal B}$, i.e.\, 
$ \nabla_a\z_b\overset{\mathcal{B}}{=} \beta n_{ab}$ for some function $\beta$. 
To evaluate $\beta$, contract both sides of this equation with a null vector $k^b$ 
that is tangent to ${\cal H}$. On the lhs we have $k^b\nabla_a \z_b$, which
according to \eqref{kappa1} off of ${\cal B}$ 
is equal to $-\k k_a$ (since $k_a$ is proportional to $\z_a$). Taking the limit  
as ${\cal B}$ is approached along the horizon thus yields $k^b\nabla_a \z_b\overset{\mathcal{B}}{=}-\k k_a$.
On the rhs we have $\beta n_{ab}k^b=-\beta k_a$, so it follows that $\beta=\k$, 
if the sign of the binormal is chosen so that $n_{ab}k^b=-k_a$.\footnote{Note that this means that the sign of $\k$ is opposite for the two sheets of a bifurcate Killing horizon.}
This establishes the useful relation\footnote{This relation is also needed to derive the expressions \eqref{noetherarea} and \eqref{noether} for the Noether charge  (see section 7 in \cite{Jacobson:2017hks}).}
\beq\label{nice}
\nabla_a \zeta_b \overset{\mathcal{B}}{=} \kappa \, n_{ab}.
\eeq
To demonstrate that $\k$ is constant on ${\cal B}$, we act on both sides of \eqref{nice} 
with $n^{ab}m^c\nabla_c$. Since $n^{ab}n_{ab}=-2$ is constant on ${\cal B}$, we obtain
\beq\label{dnice}
n^{ab}m^c\nabla_c\nabla_a \zeta_b\overset{\mathcal{B}}{=} -2m^a\nabla_a \k    .
\eeq
Finally, a conformal Killing vector satisfies a generalization of the usual Killing identity, 
\begin{equation} \label{riemann}
\nabla_c \nabla_a \zeta_b = \zeta^d R_{dcab} + g_{ab} \nabla_c \alpha + g_{bc} \nabla_a \alpha - g_{ac} \nabla_b \alpha \,.
\end{equation}
The contraction of the rhs of this identity with $n^{ab}m^c$ vanishes on ${\cal B}$, since $\z^d$ and $\a$ vanish on ${\cal B}$, and $n^{ab}m_b=0$. It follows that the lhs of \eqref{dnice} vanishes, 
which establishes that $\k$ is constant on ${\cal B}$. $\Box$

\section{Conformal group     from    two-time embedding   formalism}
\label{sec:embedding}
It is well-known that $d$-dimensional Minkowski spacetime   can be embedded in $\mathbb R^{2,d}$ as a section of the light cone through the origin, described by the equation (see e.g. \cite{Penedones:2016voo})
  \begin{equation} \label{embconstraint}
 X \cdot X = - \left (X^{-1} \right)^2 - \left(X^0 \right)^2 +\left(X^1 \right)^2 + \dots + \left (X^{d}\right)^2  =  0 \, .
  \end{equation}
Here $X^A$ are the standard flat coordinates on   $\mathbb R^{2,d}$.  The embedding space naturally induces a metric on a  hyper-lightcone section,  for example   the     section $X^{-1} + X^{d}=1$ realizes the standard   Minkowski metric on $\mathbb R^{1,d-1}$.  In fact, any  conformally flat manifold can be embedded in $\mathbb R^{2,d}$ as a  section of the  light cone, because under the coordinate transformation $\widetilde X^A = \Omega (x) X^A $ the induced metric becomes  
\begin{equation}  \label{conftransf}
d   \widetilde  s^2 = (X d \Omega + \Omega dX)^2 =  \Omega^2  dX \cdot dX= \Omega^2 (x) ds^2 \, ,
\end{equation}
where we   used the light cone properties $X \cdot d X = 0$ and $X \cdot X  =0$. This means that the induced metrics on two different hyper-lightcone sections  are related by a Weyl transformation.

This embedding construction is particularly useful for characterizing    the  conformal group $O(2,d)$ of Minkowski space, since it corresponds to    the symmetry group that preserves a light cone in  $\mathbb R^{2,d}$. Moreover, it follows from   the observation above  that    the conformal group generators  of \emph{any} conformally flat spacetime can be obtained from the embedding space.   In this appendix we will use the two-time embedding formalism to derive  all conformal Killing vectors of dS and AdS space -- which are both conformally flat -- and,  in particular,  we will write  the conformal Killing vector   (\ref{ckv3})    in terms of   embedding coordinates.

\subsection{Conformal Killing vectors of de Sitter space}
 
The embedding coordinates for the static line element  (\ref{staticpatch})  of de Sitter space are
\begin{equation}
 \begin{aligned}
 X^{0} &=   \sqrt{L^2 - r^2}  \sinh (t/L) \,, \qquad   X^{-1} = L \, ,      \\
 X^{d} &=   \sqrt{L^2 -   r^2}  \cosh (t/L) \,, \qquad   X^i = r \, \Omega^i \, ,  \quad i = 1, \dots, d-1  \, ,  
 \end{aligned}
 \end{equation}
with  $\Omega^1 = \cos \theta_1, \Omega^2 = \sin \theta_1 \cos \theta_2, \,  ..., \,  \Omega^{d-1} =   \sin \theta_1 \sin \theta_2 \cdots \sin \theta_{d-2}$ and the condition $\sum_{i=1}^{d-1} (\Omega^i)^2 =1$.
Note that we have promoted the de Sitter length scale to a coordinate, which is a convenient trick to obtain the conformal group. If the relation $X^{-1} = L$ is inserted back into      (\ref{embconstraint}), then the embedding constraint turns into the equation for a hyperboloid in $\mathbb R^{1,d}$, which is the standard embedding of de Sitter space. 
 
 The conformal group of $d$-dimensional Lorentzian de Sitter space is $O(2,d)$, whereas the isometry group   is $O(1,d)$. Hence, de Sitter space admits   $\frac{1}{2} d (d+1) $ true Killing vectors and $d+1$ conformal Killing vectors which do not generate isometries (in total there are $\frac{1}{2} (d+1) (d+2)$ conformal generators).
The generators of $O(2,d)$ are boosts and rotations in embedding space, and   hence take the form 
 \begin{equation}
J_{AB} = i \left (     X_A \partial_{X^B} - X_B \partial_{X^A} \right) \, .
 \end{equation}
Coordinates with lower indices are defined as   $X_A = \eta_{AB} X^B$, such that for instance the generator $J_{01} = - i \left ( X^0 \partial_{X^1} + X^1 \partial_{X^0} \right)$ is a proper boost. The  generators satisfy the usual Lorentz algebra commutation relations 
 \begin{equation}
 \left [ J_{AB} , J_{CD}\right] = i \left ( \eta_{AD}    J_{BC} +   \eta_{BC} J_{AD}   - \eta_{AC} J_{BD} - \eta_{BD} J_{AC} \right)  \, .
 \end{equation}
 In embedding space the true Killing vectors correspond to boosts and rotations in the $X^0, \dots, X^d$ directions, since these generators preserve the light cone section, whereas the     extra $d+1$ conformal Killing vectors are boosts and rotations in the $X^{-1}$ direction, which do not preserve the section.  
 
  We are now ready to compute the  conformal   generators explicitly. In terms of   static coordinates    the true Killing vectors of de Sitter space are given by
  \begin{equation}
   \begin{aligned}
i J_{0d}  &=  L \partial_t  \\
i J_{0i}   &= \frac{L r \Omega_i \cosh  (t /L )}{\sqrt{L^2-r^2}} \partial_t  + \sqrt{L^2-r^2} \Omega_i  \sinh  (  t /L   ) \partial_r  + \frac{  \sqrt{L^2-r^2}}{r} \sinh
    (  t /L   ) \nabla_{ i}\\
iJ_{id}  &= \frac{L r \Omega_i \sinh  (t /L )}{\sqrt{L^2-r^2}} \partial_t  +\sqrt{L^2-r^2} \Omega_i  \cosh  (  t /L   ) \partial_r  + \frac{  \sqrt{L^2-r^2}}{r} \cosh (  t /L   ) \nabla_{ i}   \\
   i J_{ij}  &=   \Omega_j \partial_{\Omega^i} - \Omega_i \partial_{\Omega^j}  \, ,
\end{aligned}
\end{equation}
where $\nabla_{ i}= \partial_{\Omega^i}  -  \Omega_i \Omega^j \partial_{\Omega^j}$ is the covariant operator\footnote{
In terms of the angular coordinates $\theta_1, \dots, \theta_{d-2}$ the covariant operator on the unit sphere is given by 
\begin{align*}
\nabla_1 &= - \sin \theta_1 \partial_{\theta_1}, \quad   
\nabla_2 = \cos \theta_1 \cos \theta_2 \partial_{\theta_1} - \frac{\sin \theta_2}{\sin \theta_1} \partial_{\theta_2}, 
\, \,  \dots \, ,   \\
\nabla_{d-2} &= \sum_{j=1}^{d-3}   \frac{\cos \theta_j  \sin \theta_j \cdots \sin \theta_{d-3} \cos \theta_{d-2}}{\sin \theta_1 \cdots \sin \theta_j} \partial_{\theta_{j}}
- \frac{ \sin \theta_{d-2} }{\sin \theta_1 \cdots \sin \theta_{d-3}} \partial_{\theta_{d-2}} \,  ,   \\    
 \nabla_{d-1} &= \sum_{j=1}^{d-3}   \frac{\cos \theta_j  \sin \theta_j \cdots \sin \theta_{d-3} \sin \theta_{d-2}}{\sin \theta_1 \cdots \sin \theta_j} \partial_{\theta_{j}}
+ \frac{ \cos \theta_{d-2} }{\sin \theta_1 \cdots \sin \theta_{d-3}} \partial_{\theta_{d-2}}       \, .
 \end{align*}
 }
 on the unit sphere $S^{d-2}$, and the other $d+1$ conformal generators take the form
\begin{equation}
   \begin{aligned}
i J_{-10}  &= \frac{L^2 \cosh  (  t/ L  )}{\sqrt{L^2-r^2}} \partial_t   +  \frac{r}{L} \sqrt{L^2-r^2} \sinh  (t /L  ) \partial_r   \\
i J_{-1i}   &=    \frac{  L^2 -r^2}{L}    \Omega_i \partial_r + \frac{L}{r} \nabla_i       \\
iJ_{-1d}  &= - \frac{L^2 \sinh  ( t /L )}{\sqrt{L^2-r^2}} \partial_t - \frac{r}{L} \sqrt{L^2-r^2} \cosh  (t / L  ) \partial_r  \, .  
\end{aligned}
\end{equation} 
Finally, by comparing   expression  (\ref{ckv3})  for the conformal Killing vector   that preserves a causal diamond in dS with the previous conformal generators, we see that $\zeta$ can be written as  a linear combination of $J_{0d}$ and $J_{-10}$
  \begin{equation} \label{ckvdsemb}
  \zeta^{\text{dS}} = \frac{i L}{R} \left ( J_{0d} - \sqrt{1 - (R/L)^2 } J_{-1 0} \right) \, .
  \end{equation} 

\subsection{Conformal Killing vectors of Anti-de Sitter space}

Global coordinates for   $d$-dimensional  Anti-de Sitter space are  defined by
\begin{equation}
 \begin{aligned} \label{globaladscoord}
 X^{-1} &= \sqrt{L^2 + r^2} \cos (t/L)  \,, \qquad   X^{d} = L \, ,      \\
 X^{0} &=   \sqrt{L^2 + r^2} \sin (t/L)  \,, \qquad   X^i = r \, \Omega^i \, ,  \quad i = 1, \dots, d-1   \, .
 \end{aligned}
 \end{equation}
 Note that here we have set a space   coordinate $X^d$, instead of a time coordinate $X^{-1}$, equal to the curvature scale $L$. In this way the embedding constraint (\ref{embconstraint}) turns into the familiar embedding equation for AdS, which is that of a hyperboloid in $\mathbb R^{2,d-1}$. It is manifest from this embedding that the isometry group of Lorentzian AdS$_d$ is   $O(2,d-1)$, whereas the full conformal group is the same as that of dS$_d$, i.e. $O(2,d)$.
 
In terms of these coordinates, the  induced metric on the light cone reads
\begin{equation} \label{globalads}
ds^2 = - [ 1 +(r/L)^2] dt^2 + [ 1 + (r/L)^2]^{-1}  dr^2 + r^2 d \Omega_{d-2}^2 \, .
\end{equation}
For $r\ge0$ and $0\le t < 2\pi L$  this solution covers the entire hyperboloid. However, to avoid closed timelike curves   AdS is  usually defined as the universal cover of the hyperboloid. This means that the timelike cycle is unwrapped, i.e. the range of the   coordinate $t$ is extended: $-\infty < t < \infty$.

The   generators of the conformal group   take the following form in global AdS coordinates
  \begin{equation}
   \begin{aligned}
i J_{-10}  &=  L \partial_t  \\
i J_{-1i}   &= -  \frac{L r \Omega _i \sin  (t/L )}{\sqrt{L^2+r^2}} \partial_t   +  \sqrt{L^2+r^2} \Omega _i  \cos  ( t/L ) \partial_r + \frac{ \sqrt{L^2+r^2} \cos  ( t/ L )}{r} \nabla_i  \\
iJ_{0i}  &=   \frac{L r \Omega _i \cos  (t/L )}{\sqrt{L^2+r^2}} \partial_t   +  \sqrt{L^2+r^2} \Omega _i  \sin  ( t/L ) \partial_r + \frac{ \sqrt{L^2+r^2} \sin  ( t/ L )}{r} \nabla_i  \\
   i J_{ij}  &=   \Omega_j \partial_{\Omega^i} - \Omega_i \partial_{\Omega^j}  \\
   i J_{-1d}  &= - \frac{L^2 \sin  ( t/L )}{\sqrt{L^2+r^2}} \partial_t -  \frac{r}{L} \sqrt{L^2+r^2} \cos  (t/L ) \partial_r \\
   i J_{0d}  &=   \frac{L^2 \cos  ( t/L )}{\sqrt{L^2+r^2}} \partial_t -  \frac{r}{L} \sqrt{L^2+r^2} \sin  (t/L ) \partial_r \\
   i J_{id}  &= \frac{L^2 + r^2}{L} \Omega_i \partial_r + \frac{L}{r} \nabla_i      \, .
\end{aligned}
\end{equation}
The first four  generators  are true Killing vector fields, whereas the latter three are   conformal Killing fields. 
The conformal Killing vector that preserves a causal diamond centered at the origin $r=0$ of AdS can be obtained from (\ref{ckv3})  by analytically  continuing  the curvature scale  $L \rightarrow i L$. One  then   finds  the following expression in terms of the conformal generators
 \begin{equation} \label{ckvads1}
   \zeta^{\text{AdS}} 
   = \frac{i L}{R} \left (  \sqrt{1 + (R/L)^2 } J_{ 0d} - J_{-10}  \right) \, .
 \end{equation}
Notice that for a causal  diamond of infinite size, i.e. $R/L \rightarrow \infty $, the conformal Killing field $\zeta$ reduces to the conformal generator $iJ_{0d}$. The $t=0$ time slice of this diamond covers the entire AdS time slice, and at the center   of this diamond  $\zeta$ is equal to the timelike  Killing vector $L \partial_t$.   

Another useful set of coordinates   for AdS are the  Poincar\'{e} coordinates, which are defined by
\begin{equation}
  \begin{aligned}
 X^{-1} &=  \frac{1}{2z} \left ( L^2- t^2 + \vec x^2 + z^2\right), \qquad  X^{1} =\frac{1}{2z} \left ( L^2+t^2 - \vec x^2 - z^2 \right) \, ,\\
        X^0 &= L t /z \, ,   \qquad X^i = L x^i / z  \, , \qquad  X^d = L \, , \qquad i = 2, \dots, d-1 \, .
  \end{aligned}
  \end{equation}
   The induced metric in Poincar\'{e} coordinates is 
  \begin{equation}
  ds^2 = \frac{L^2}{z^2} \left ( - dt^2 + d\vec x^2 + dz^2   \right) \, ,
  \end{equation}
  where the coordinates $t$ and $x^i$ range from $-\infty$ to $\infty$. The coordinate $z$ behaves as a radial coordinate and divides the hyperboloid into two charts ($z \lessgtr 0$). The Poincar\'{e} patch of AdS is typically taken to be the chart   $z > 0$, which covers one half of the hyperboloid. The location $z=0$ corresponds to the conformal boundary of AdS.

It is convenient to introduce the following generators of the conformal group
\begin{equation}
 \begin{aligned}
D &= J_{-1   1}  \, , \qquad P_\mu =  J_{  \mu    -1 } - J_{   \mu 1}  \, , \\
 M_{\mu \nu} &= J_{\mu \nu} \,  ,  \qquad   \,  K_\mu = J_{ \mu   - 1 } + J_{   \mu 1} \, ,
 \end{aligned}
 \end{equation}
where  $\mu = 0, i , $ or $d$, corresponding to the coordinates $(t, x^i, z)$ respectively. On the conformal boundary of AdS these generators turn into   the standard conformal generators of flat space (where $D$ denotes the generator of   dilatations, $P_\mu$ of translations, $M_{\mu\nu}$ of Lorentz transformations, and $K_\mu$ of special conformal transformations). In Poincar\'{e} coordinates the conformal generators  are equal to
\begin{equation}
\begin{aligned} \label{poincconfgen}
i D &= -  \left (  t \partial_t + x^i \partial_i + z \partial_z  \right) \, , \\
  i  P_t &=- L  \partial_t \, , \quad \,\,  i P_i = - L \partial_i \, , \quad \,\,  i P_z = - L \partial_z \, ,  \\
i K_t &= \frac{-1}{L} \left[  (t^2 + \vec x^2 + z^2  ) \partial_t + 2 t (x^i \partial_i +  z \partial_z)  \right] \,  , \\
i K_i &=  \frac{1}{L} \left[   ( t^2  - \vec x^2  - z^2 ) \partial_i   + 2  x_i    (t \partial_t  + x^j \partial_{j} + z \partial_z  )  \right ] \,  , \\
i K_z &=  \frac{1}{L} \left[    ( t^2 - \vec x^2 + z^2  )\partial_z  + 2 z  (t \partial_t + x^i   \partial_i )   \right ]    \,  ,  \\
i M_{ti}&= x_i \partial_t + t \partial_i \,, \qquad  i M_{i z} = z \partial_x - x \partial_z \, , \\
i M_{tz} &= z \partial_t + t \partial_z \,, \qquad  i M_{ij} = x_j \partial_i - x_i \partial_j \, . 
\end{aligned}
\end{equation}
The  generators $P_z, K_z, M_{tz}$ and $M_{iz}$   are  conformal Killing vectors, and the other generators are true Killing vectors. With this list of conformal generators at hand, we  are able to write the conformal Killing field (\ref{ckvads1}) that preserves a   diamond in AdS in terms of Poincar\'{e} coordinates
\begin{equation}
\begin{aligned} \label{ckvpoin}
\zeta^{\text{AdS}} &= \frac{i L}{R} \left [  \sqrt{1 + (R/L)^2 } J_{ 0d} + J_{0-1}  \right ]
= \frac{i L}{R} \left [  \sqrt{1 + (R/L)^2 } M_{tz}  +\frac{1}{2} \left (  P_t + K_t \right) \right ]     \\
&= \frac{1}{2R} \left [  \sqrt{L^2 + R^2 }  (2 z \partial_t + 2 t \partial_z ) -   (L^2 + t^2 + \vec x^2 + z^2)  \partial_t  -  2 t x^i \partial_i  -  2 t z \partial_z   \right] \, .
\end{aligned}
\end{equation}
This conformal Killing vector field generates a modular flow inside a causal diamond $\mathcal D $ in the Poincar\'{e} patch of AdS. The center of the diamond $\mathcal D $  is located at $\{t=0,z=L,x^i=0 \}$, and its boundaries are given  by     
\begin{equation*}
\begin{aligned}
&\text{past and future vertices:} \quad    p, p' = \left \{  t = \pm R, \, z = \sqrt{L^2 + R^2}, \, x^i = 0 \right \} \, , \\
&\text{edge/bifurcation surface:}   \,\,\,\,\,  \mathcal B = \left \{  t= 0, \, z = \sqrt{L^2 + R^2} \pm \sqrt{R^2 - \vec x^2} \right \} \,, \\
&\text{past  null boundary:}  \qquad \,\,\,\,\,  \mathcal H_{\text{past}} = \left \{ z = \sqrt{L^2 + R^2 } \pm \sqrt{(R+t)^2 - \vec x^2} \right \}  \, ,\\
&\text{future  null boundary:}  \qquad   \mathcal H_{\text{future}} = \left \{ z = \sqrt{L^2 + R^2} \pm \sqrt{(R-t)^2 - \vec x^2} \right \} \, . \\
\end{aligned}
\end{equation*}
One can  readily  check that the conformal Killing vector (\ref{ckvpoin}) vanishes at $p, p'$ and $\mathcal B$, and acts as a null flow on $\mathcal H_{\text{past/future}} $. Hence this causal diamond is entirely contained within the $z>0$ Poincar\'{e} patch.      

One can, of course, shift the causal diamond such that its center is located at a different position. An interesting special case is the    diamond which is centered  at the boundary of AdS, i.e. $z=0$, and whose past and future null boundaries coincide with the Killing horizon of  AdS-Rindler space. This can be established by  sending $z\rightarrow z + \sqrt{L^2 + R^2}$ in (\ref{ckvpoin}). The conformal Killing vector then turns into the boost Killing vector  of AdS-Rindler space, given in  \cite{Faulkner:2013ica}
\begin{equation}  \label{adsrindlerboost}
\begin{aligned}
\xi^{\text{AdS-Rindler}} &= \frac{i L}{2R} \left [   \left ( 1 - (R/L)^2\right) J_{0 -1}  + \left ( 1 + (R/L)^2 \right) J_{01}  \right ]   = \frac{i L}{2R} \left [  K_t - (R/L)^2 P_t \right ]     \\
&= \frac{1}{2R}\left [  \left ( R^2 - t^2 - \vec x^2 - z^2 \right)  \partial_t - 2 t x^i \partial_i - 2 t z \partial_z  \right]\,  . 
\end{aligned}
\end{equation}
Note that the normalization of the boost Killing vector is different than   in  \cite{Faulkner:2013ica}, since we have set the surface gravity of the  Killing horizon   equal to one. 
The center of this diamond is located at $\{t=0,z=0,x^i=0 \}$, and  the past and future vertices are at $ \left \{ t = \pm R, z = 0, x^i =0\right\}$. Moreover,  the bifurcation surface corresponds to the hemisphere $\mathcal B = \left \{  t= 0, \, z^2 + \vec x^2 =    R^2    \right \} $. There exist two solutions to this quadratic equation, i.e. $z = \pm \sqrt{R^2 - \vec x^2}$, so there are in fact two equal-sized Rindler wedges in AdS that are preserved by this boost Killing vector (see Figure \ref{fig:adsrindler} on p. \pageref{fig:adsrindler}).

\section{Conformal transformation from Rindler space to   causal diamond}
\label{sec:conftrans}

It is well known in the literature \cite{Hislop1982, Haag:1992hx, Casini:2011kv} that there exists a special conformal transformation that maps Rindler space to  a causal diamond in flat space. In this appendix we describe this transformation and derive the conformal Killing vector that preserves the diamond from the boost Killing vector of Rindler space. In particular, we focus on the conformal transformation from the right Rindler wedge (described by coordinates $X^a$) to the diamond (described by coordinates $x^a$) whose left edge is fixed at the bifurcation surface of the Rindler wedge. (See Figure \ref{fig:mink2} on p. \pageref{fig:mink2} for a depiction of this diamond inside the right Rindler wedge.) 
The  special conformal transformation and its inverse are given by
\beq \label{sct1}
x^a =\frac{X^a -  X\cdot X  C^a }{1 - 2 X\cdot C + X\cdot X C \cdot C}   \,, \qquad X^a = \frac{x^a + x \cdot x \, C^a}{1 + 2 \,  x \cdot C + x \cdot x \,  C \cdot C} \, ,
\eeq
where $C^a = (0,-1/(2R),0,0)$, and $x^a$ and $X^a$ are the standard coordinates in Minkowski space. This is just a shift of the map mentioned in \cite{Casini:2011kv} (and we included a minus sign in $C^a$, correcting a typo in their paper).
Under the transformation $X^a \to x^a$, the metric becomes
 $\eta_{ab}dX^a   dX^b = \Omega^2 \eta_{ab} dx^a  dx^b,$ 
 where    the conformal factor is given by
\beq
\Omega = 1- 2 X \cdot C + X \cdot X C \cdot C = \left [ 1 + 2 x \cdot C + x \cdot x  \, C \cdot C \right]^{-1}  .
\eeq
 Note that the conformal transformation maps  spatial infinity $i^0=(T=0,X^1=\infty)$  to the right edge of the diamond  $ (t= 0,x^1=2R)$, and the origins are also mapped to each other. 
 
 Further, with the help of Mathematica, we find that under the coordinate transformation  \eqref{sct1} the boost Killing field 
\beq \label{Rindlerxi}
\xi = X^1 \partial_T + T \partial_{X^1}
\eeq
becomes     the conformal Killing field that preserves the diamond
\begin{equation} \label{shiftedzeta}
\zeta =\frac{1}{2R} \left [  \left (  R^2 - ( x^1  -R)^2 - (x^i)^2- t^2 \right) \partial_t - 2 t (x^1  - R) \partial_{x^1}  - 2 t x^i \partial_{i} \right] ,
\end{equation} 
where $i=2, \dots, d-1$.
By replacing     $x^1$  by $ x^1 + R$  one obtains the standard conformal Killing vector  \eqref{ckvflat2} that preserves a diamond centered at the origin of flat space.
At $T=t=0$ the norm of the two vector fields \eqref{Rindlerxi} and \eqref{shiftedzeta} is 
\beq
|\xi| \overset{(T=0)}{=}  X^1 \, , \qquad |\zeta| \overset{(t=0)}{=} \frac{X^1}{\Omega} = x^1 - \frac{\vec x^2}{2 R} \, . 
\eeq
 The former blows up at spatial infinity $i^0$, whereas the latter vanishes at the edge of the diamond $(x^1- R)^2 + (x^i)^2 =R^2$. This is because, although both $X^1$ and $\Omega$ diverge at spatial infinity, their ratio vanishes in this limit.

\section{Conformal Killing vector in flat space}  \label{appkill}
Here we make explicit the geometric quantities introduced in Section \ref{subsec:setup} in the case of a Minkowski background, whose metric we write in spherical coordinates, i.e.  $ds^2 = - dt^2 + dr^2 + r^2 d \Omega_{d-2}^2$. Let $\Sigma$ be a spatial ball of radius $R$ in the time slice $t=0$ and with center at $r=0$. The conformal Killing vector which preserves the causal diamond  of $\Sigma$ is given by \cite{Jacobson:2015hqa} 
\begin{equation}\label{zz}
\zeta=\left( \frac{R^2-r^2-t^2}{R^2}\right) \partial_t-\frac{2 r t}{R^2} \partial_r \, ,
\end{equation}
where we have chosen the normalization in a way such that $\zeta^2=-1$ at the center of the ball, which then gives the usual notion of energy for $H_\zeta^{\rm m}$ (i.e. the correct units).
It is straightforward to check that 
$
\zeta(t=\pm R,r=0)=\zeta(t=0,r=R)=0\, ,
$
 i.e.  the tips of the causal diamond and the maximal sphere $\partial \Sigma$ at its waist are fixed points of $\zeta$, as expected. Similarly, $\zeta$ is null on the boundary of the diamond. In particular, 
 $
 \zeta(t=R \pm r)=\mp 2 r(R \pm r)/R^2 \cdot (\partial_t \pm \partial_r)\, .
$
The vectors $u$ and $n$ (respectively normal to $\Sigma$ and to both $\Sigma$ and $\partial \Sigma$) read
$u=\partial_t$, $n=\partial_r$,
so that the binormal to $\partial \Sigma$ is given by 
$
n_{ab}=2\nabla_{[a} r \nabla_{b]} t \, .
$
It is also easy to check that $\lie_{\zeta}g_{ab}=2\alpha g_{ab}$ holds, where
$
\alpha\equiv \nabla_{a}\zeta^a/d=-2t/R^2 \,.
$
Hence, we immediately see that $\alpha=0$ on $\Sigma$, which implies that the gradient of $\alpha$ is proportional to the unit normal $u_a=-\nabla_a t$. Indeed, one finds $ \nabla_a \alpha=-2\nabla_a t/R^2 $, 
so in this case $N\equiv \lVert \nabla_a \alpha \rVert^{-1}=R^2/2$. It is also easy to show that $(\nabla_a \zeta_b)|_{\partial \Sigma}=\kappa n_{ab}$ holds, where the surface gravity reads $\kappa=2/R$.

As shown in \cite{Jacobson:1993pf}, given some metric $g_{ab}$ with a conformal Killing field $\zeta^a$, it is possible to construct other metrics $\bar{g}_{ab}$ conformally related to it, for which $\zeta^a$ is a true Killing field. More explicitly, if 
$\lie_{\zeta}g_{ab}=2\alpha g_{ab}$, then $\lie_{\zeta}\bar{g}_{ab}=0$ 
as long as $g_{ab}$ and $\bar{g}_{ab}$ are related through $\bar{g}_{ab}=\Phi\, g_{ab}$, where $\Phi$ satisfies
\begin{equation}
 \lie_{\zeta} \Phi+2\alpha \Phi=0\, .
\end{equation}
For the vector \req{zz}, this equation has the general solution 
\begin{equation}\label{pii}
\Phi(r,t)=\frac{\psi(s)}{r^2}\, \quad  \text{where} \quad s\equiv \frac{R^2+r^2-t^2}{r} \, .
\end{equation}
Here, $\psi(s)$ can be any function. Hence, $\zeta$ in \req{zz} is a true Killing vector for all metrics conformally related to   Minkowski's  with a conformal factor given by \req{pii}. For example, setting $\psi(s)=L^2$, for some constant $L^2$, one obtains the metric of AdS$_2\times S_{d-2}$ with equal radii, namely: 
$ds^2=L^2/r^2 (-dt^2+dr^2)+L^2 d\Omega_{d-2}^2
$. Another simple case corresponds to $\psi(s)=L^2((s^2/(4L^2)-1)^{-1}$.  Through the change of variables \cite{Casini:2011kv}: $t=L \sinh(\tau/L)/(\cosh u +\cosh (\tau/L))$, $r=L \sinh u \, /(\cosh u +\cosh (\tau/L))$, this choice leads to the $\mathbb{R}\times \mathbb{H}^{d-1}$ metric (where $\mathbb{H}^{d-1}$ is the hyperbolic plane): $ds^2= -d\tau^2+L^2(du^2+\sinh^2 u \, d\Omega^2_{d-2})$. 


\section{Generalized volume in higher order gravity} \label{app:W}

\noindent The generalized volume $W$ is defined in \eqref{eqn:W}.
We restate the expression here
\beq
W= \frac{1}{(d-2)E_0} \int_\Sigma \eta    \left (  E^{abcd} u_a u_d h_{bc}  - E_0  \right) \, ,
\eeq
where $E_0$ is   a theory-dependent constant defined by the tensor $E^{abcd}$
in a maximally symmetric solution to the field equations through
$E^{abcd}\overset{\text{MSS}}{=}E_0(g^{ac}g^{bd}-g^{ad}g^{bc})$. Moreover, $E^{abcd}$     is  the variation of the Lagrangian scalar $\mathcal L$ with respect to the Riemann tensor $R_{abcd}$ if we were to treat it as an independent field \cite{Iyer:1994ys},
\begin{equation} \label{defEtensor}
E^{abcd}  = \frac{\partial   \mathcal L}{\partial R_{abcd}}  - \nabla_{a_1} \frac{\partial \mathcal L}{\partial \nabla_{a_1} R_{abcd}} + \dots    
 + (-1)^m \nabla_{(a_1} \cdots \nabla_{a_m )} \frac{\partial \mathcal L}{\partial  \nabla_{(a_1} \cdots \nabla_{a_m )} R_{abcd}} \nonumber  \, ,
\end{equation}
where $\mathcal L$ is then  defined  through $L = \epsilon \mathcal L$. In this section we provide explicit expressions for $W$ in $f(R)$ gravity, quadratic gravity and Gauss-Bonnet gravity. Observe that throughout this section we use the bar on $\bar{R}$ to denote evaluation on a MSS. Imposing a MSS to solve the field equations of a given higher derivative theory gives rise to a constraint between the theory couplings and the background curvature $\bar{R}$. This reads \cite{Bueno:2016ypa}
\begin{equation}\label{emb}
E_0=\frac{d}{4\bar{R}} \mathcal L(\bar{R}) \, ,
\end{equation} 
where $ \mathcal L(\bar{R})$ denotes the Lagrangian scalar evaluated on the background.

\subsubsection{$f(R)$ gravity} 
A simple higher curvature gravity is obtained by replacing $R$ in the Einstein-Hilbert action by a function of $R$
\beq
L_{f(R)} = \frac{1}{16\pi G}  \epsilon f(R) \, .
\eeq
To obtain the generalized volume we need
\beq
E^{abcd}_{f(R)} = \frac{f'(R)}{32 \pi G} \left( g^{ac} g^{bd}- g^{ad} g^{bc} \right) \, , \quad E_0 = \frac{f'(\bar{R})}{32 \pi G} \, .
\eeq
The generalized volume then reads
\beq
W_{f(R)} = 
\frac{1}{d-2} \int_{\Sigma} \eta \left [ (d-1) \frac{f'(R)}{f'(\bar{R})} - 1\right] \, .
\eeq
\subsubsection{Quadratic gravity}
A general quadratic   theory of gravity is given by the Lagrangian
\begin{equation} 
  L_\text{quad}  =   \,   \epsilon \bigg [ \frac{1}{16 \pi G} \big( R -2 \Lambda \big)   + \alpha_1  R^2 + \alpha_2  R_{ab} R^{ab}   + \alpha_3   R_{abcd}R^{abcd}  \bigg ] \, .
\end{equation}
Taking the derivative of the Lagrangian with respect to the Riemann tensor leaves us with
\begin{equation} 
E^{abcd}_\text{quad} =   \, \left ( \frac{1}{32 \pi G} +  \alpha_1 R \right) 2 g^{a[c} g^{d]b} \\
 + \alpha_2 \left (   R^{a[c} g^{d]b} + R^{b[d} g^{c]a} \right) + 2 \alpha_3R^{abcd} \, ,
\end{equation}
and using \req{msb} one finds
\beq
E_0 = \frac{1}{32 \pi G} + \left( \alpha_1 + \frac{\alpha_2}{d} +  \frac{2\alpha_3}{d(d-1)} \right) \bar{R}\, .
\eeq
The generalized volume for quadratic gravity thus reads
\begin{equation}
\begin{aligned} 
&W_\text{quad}
=  \frac{1}{(d-2)E_0} \int_\Sigma \eta \bigg [ (d-1)\left(\frac{1}{32 \pi G}+\alpha_1 R \right) - E_0 \\
&+ \frac{1}{2} \alpha_2 \left (R - R^{ab} u_a u_b (d-2)\right) - 2 \alpha_3 R^{ab} u_a u_b \bigg ].
\end{aligned}
\end{equation}
An interesting instance of quadratic gravity is Gauss-Bonnet theory, which is obtained by restricting to $\alpha_1=-\frac{1}{4}\alpha_2=\alpha_3=\alpha$.
The generalized volume then reduces to
\begin{equation}  
W_{\text{GB}}  =  \, \frac{1}{(d-2)E_0}\int_\Sigma \eta \left[ \frac{1}{32 \pi G}(d-1) - E_0  \label{eq:WGB}
 + (d-3) \alpha \Big( R + 2 R^{ab} u_a u_b \Big) \right] \, ,
\end{equation}
with $E_0=1/(32 \pi G)+\alpha \bar{R} (d-2)(d-3)/(d(d-1))$.
Since the extrinsic curvature of $\Sigma$ vanishes in the background, 
the structure $R+2R^{ab}u_a u_b$ is equal to the intrinsic Ricci scalar of $\Sigma$,
in the background and at first order in perturbations.


\section{Linearized equations of motion for higher curvature gravity}
\label{app:FLDMRNC}
 
The variational identity (\ref{eqn:offshelllocalgeo}) states that the vanishing of the linearized constraint equations $\delta C_\zeta$ is equivalent to a relation between the variation of the Wald entropy, generalized volume, and matter energy density.
In \cite{Jacobson:2015hqa}, Jacobson used this relation to extract  the Einstein equation,
 making use of Riemann normal coordinates. 
Here we perform a similar calculation for the higher curvature generalization of the first law of causal diamond mechanics which will produce the linearized equations of motion.
In this appendix we will restrict to theories whose Lagrangian depends on the metric and the Riemann tensor, $L[g_{ab},R_{abcd}]$, and to linearization around flat space.

The equations of motion for such a general higher curvature theory  read
\begin{equation}\label{eomhighapp}
\begin{split}
  - \frac{1}{2} g^{ab} \mathcal L  + E^{aecd} \tensor{R}{^b_{ecd}} - 2  \nabla_c \nabla_d E^{acdb}    = \frac{1}{2} T^{ab}.
\end{split}
\end{equation}
Linearizing the equations of motion around flat space leads to
\begin{equation}
\begin{aligned}
 \label{lineomhighapp}
 & - \frac{1}{32\pi G} \eta^{ab} \delta R   + E^{aecd}_\text{Ein} \delta \tensor{R}{^b_{ecd}} - 2  \partial_c \partial_d \delta E^{a c d b}_{\text{higher}}  \\
&= \frac{\delta G^{ab}}{16\pi G} - 2  \partial_c \partial_d \delta E^{a c d b}_{\text{higher}} = \frac{1}{2} \delta T^{ab}\, ,
\end{aligned}
\end{equation}
where we split $E^{abcd}=E^{abcd}_{\text{Ein}}+E^{abcd}_{\text{higher}}$ into an  Einstein piece, which goes into the Einstein tensor, and a piece coming from higher derivative terms.
We used the fact that many of the expressions in \eqref{lineomhighapp} significantly simplify when evaluated in the Minkowski background because the curvatures vanish.
For example, one might have expected additional terms proportional to the variation of the Christoffel symbols coming from $\delta(\nabla_c\nabla_d E^{acdb})$.
To see why these terms are absent, it is convenient to split this expression into its Einstein part and a part coming from higher derivative terms.
The Einstein piece does not contribute since $E^{acdb}_{\text{Ein}}$ is only a function of the metric and therefore its covariant derivative vanishes.
The higher derivative piece will give $\partial_c \partial_d \delta E^{a c d b}_\text{higher}$ as well as terms such as $\delta \Gamma^{c}_{ce} \nabla_d E^{eadb}_{\text{higher}}$ and $\Gamma^{c}_{ce} \nabla_d \delta E^{eadb}_{\text{higher}}$.
However, the latter two terms are zero because both the Christoffel symbols
 and $E^{eadb}_{\text{higher}}$ vanish when evaluated in the Minkowski background with
 the standard coordinates.

We now want to evaluate each term in \eqref{newvarid} using Riemann normal coordinates.
Taking the stress tensor $T^{ab}$ to be  constant for small enough balls, the variation of \eqref{potati} reduces to
\begin{equation}
\delta H^{\rm m}_{\zeta} = \frac{\Omega_{d-2}\ell^{d}}{d^2-1} \kappa u_a u_b \delta T^{ab} + \mathcal{O}\left(\ell^{d+2}\right)\, ,
\end{equation}
where $\Omega_{d-2}$ denotes the area of the $(d-2)$-sphere, $\ell$ is the radius of our geodesic ball and $u_a$ is the future pointing unit normal.
As was found in \cite{Jacobson:2015hqa}, the Einstein piece of the symplectic form will combine with the area term of the entropy to produce the Einstein tensor.
Therefore, we focus on the higher curvature part of $\delta H_\zeta^{\rm g}$.
Combining \eqref{eqn:hamilton} and \eqref{eqn:symplform}, we find 
\begin{align}\label{eq:symplformhigh}
\delta H^{\rm g}_{\zeta,\text{higher}} 
&= - \frac{4\kappa}{\ell} \int d\Omega \int dr r^{d-2} u_a u_d \eta_{bc } \Big(\delta  E^{abcd}_{\text{higher}}(0)
+ \partial_i \delta  E^{abcd}_{\text{higher}}(0) r n^i  \nonumber \\
&\qquad + \frac{1}{2}\partial_i \partial_j \delta  E^{abcd}_{\text{higher}}(0) r^2 n^i n^j + \mathcal{O}\left(r^3\right)\Big)  \\
&= - 4\kappa \Omega_{d-2} \ell^{d-2} u_a u_d \eta_{bc } \bigg(\frac{\delta  E^{abcd}_{\text{higher}}(0)}{(d-1)}
 + \frac{\ell^{2}  \delta^{ij} \partial_i \partial_j \delta  E^{abcd}_{\text{higher}}(0)}{2(d^2-1)} \bigg) + \mathcal{O}\left(\ell^{d+2}\right)\, . \nonumber
\end{align}
Here, $n^i$ is the normal vector to $\partial \Sigma$ and the indices $a,b$ run over space-time directions, while the indices $i,j$ run only over spatial directions, and $\partial_i$ is the derivative operator compatible with the flat background metric on $\Sigma$.
In the first line, we simply use the formula for the Taylor expansion of a quantity $f$
in the coordinate
system compatible with $\partial_i$, 
\begin{equation}\label{Taylor}
f(x)=f(0)+\partial_a f(0) x^a + \frac{1}{2} \partial_a \partial_b f(0) x^a x^b + \mathcal{O}\left(x^3\right) \, ,
\end{equation}
where $(0)$ denotes that a term is evaluated at $r=0$.
Since we evaluate our expressions on a constant timeslice at $t=0$, we have $x^t=0$ and $x^i=r \, n^i$, where $r$ is a radial coordinate inside the geodesic ball and the index $i$ runs only over the spatial coordinates.
To evaluate the spherical integral, it is useful to note that spherical integrals over odd powers of $n^i$ vanish and furthermore
\begin{align}
\int d\Omega \, n^i n^j &= \frac{\Omega_{d-2}}{d-1}\delta^{ij} \, , \\
\int d\Omega \, n^i n^j n^k n^l &= \frac{\Omega_{d-2}}{d^2-1}\left(\delta^{ij}\delta^{kl}+\delta^{ik}\delta^{jl}+\delta^{il}\delta^{jk}\right)  \;.
\end{align}
Next, we evaluate $\delta S_\text{higher}$, the variation of the higher curvature part of the Wald entropy given in \eqref{eqn:SWald}, in a similar manner 
\begin{align}
\!\!\!\!\!\!\!\!\!\delta S_{\text{higher}} &\! = 8 \pi \Omega_{d-2}\ell^{d-2} u_a u_d \! \left(\frac{\eta_{bc}\delta E^{abcd}_{\text{higher}}(0)}{(d-1)}   
  + \frac{\ell^2\left[\eta_{bc}\delta^{ij} \partial_i \partial_j \delta E^{abcd}_{\text{higher}}(0) 
 	+ 2 \partial_b \partial_c \delta E^{abcd}_{\text{higher}}(0) \right] }{2(d^2-1)}\right) \nonumber \\
	&\,\,\,\,\, + \mathcal{O}\left(\ell^{d+2}\right) .
\end{align}
We are now ready to evaluate the first law of causal diamond mechanics \eqref{newvarid}.
Interestingly, the leading order pieces of the Hamiltonian and Wald entropy exactly cancel against each other.
Note that these two terms would have otherwise dominated over the Einstein piece.
Furthermore, the second term in the symplectic form and Wald entropy also cancel, leaving only a single term from the higher curvature part of the identity.
Including the Einstein piece, we find the first law  for higher curvature gravity reads in Riemann normal coordinates 
\begin{equation}\label{eomRNC}
\begin{split}
- \frac{\kappa \,  \Omega_{d-2}\ell^{d} }{d^2-1} u_a u_d \Big ( \frac{\delta G^{ad}(0)}{8 \pi G} - 4 \partial_b \partial_c \delta E^{abcd}_{\text{higher}}(0) - \delta T^{ad} \Big) 
 +\mathcal{O}\left(\ell^{d+2}\right) =0 \, ,
\end{split}
\end{equation}
proving equivalence to the linearized equations (\ref{lineomhighapp}).

\section{Weyl equivalent spacetimes from an embedding   formalism}

\label{embedding}

In Chapter \ref{ch4} we have studied three cases of conformally equivalent spacetimes:
\begin{align}\label{eq:conformal-equivalences-quotiented-p=3}
 AdS_{d}\times S^{1}  &\cong \text{conical} \, \,  AdS_{3}    \times S^{d-2}\, ,    \nonumber  \\
Mink_d\times S^1  & \cong \text{massless} \,\,     {BTZ}    \times S^{d-2}\, ,     \\
 dS_{d}\times S^1 & \cong \text{Hawking-Page} \, \,     {BTZ}     \times S^{d-2} \, . \nonumber
\end{align} 
The $(d+1)$-dimensional spacetimes on the left and right-hand side are  Weyl equivalent with the specific conformal factor  $\Omega=L/r = R/L.$ Notice that the spacetimes on the right-hand side are all    discrete quotients of (patches of) pure  AdS$_3$. They can  be obtained by orbifolding AdS$_3$ by an elliptic, parabolic and hyperbolic element of the isometry group, respectively \cite{Banados:1992gq}. On the left-hand side, the quotient acts on the one-dimensional space. Since taking the quotients commutes with the Weyl transformation we could also consider the conformal equivalence of the unquotiented spaces. These are in fact easier to understand and can be generalized to any dimension $p$ as follows: 
\begin{align}
 AdS_{d}\times S^{p-2} & \cong AdS_{p}\times S^{d-2},    \nonumber   \\
Mink_d\times \mathbb{R}^{p-2}  & \cong \text{Poincar\'{e}-}AdS_{p}\times S^{d-2}, \\
 dS_{d}\times \mathbb{H}^{p-2}   & \cong \text{Rindler-}AdS_{p}\times S^{d-2}.  \nonumber
\end{align} 
Note that when $\mathbb{H}^1\cong \mathbb{R}$ is quotiented by a boost, one obtains the $S^1$ in \eqref{eq:conformal-equivalences-quotiented-p=3}.
In this appendix we will   explain in detail how the conformal equivalence of these spacetimes can be understood from the embedding space perspective. 

\subsubsection{Embedding space formalism} 

The $(p+d-2)$-dimensional spacetimes above are   special in the sense that they are conformally flat. Now any $D$-dimensional conformally flat spacetime  can be embedded in $\mathbb R^{2,D-2}$, where the metric signature of the original spacetime is $(-,  +,   \dots ,+)$. This can be seen as follows. To begin with,   Minkowski spacetime can be obtained  as a section of the light cone through the origin of $\mathbb R^{2,D-2}$. The light cone equation is
\begin{equation} 
X \cdot X = -X^2_{-1}-X^2_0+X^2_1+\ldots +X^2_{D}=0.
\end{equation}
Here $X_A$ are the standard flat coordinates on $\mathbb R^{2,D-2}$. The embedding space naturally induces a metric on the light cone section. The Poincar\'e section $X_{-1} + X_{D}=1$, for example, leads to the standard   metric on Minkowski spacetime. Under the coordinate transformation, $  X_A = \Omega (x) \tilde X_A$, the induced metric becomes
\begin{equation}   
d   s^{2} =d {X}\cdot d {X}=(\Omega d\tilde X+\tilde Xd\Omega)^2=\Omega^2d\tilde X\cdot d\tilde X=\Omega^2 d\tilde s^2 \, .
\end{equation}
Here the light cone properties $\tilde X \cdot d \tilde X =0$ and $\tilde X \cdot \tilde X =0$ were used in the third equality. This means that the induced metrics on two different light cone sections are related by a Weyl transformation. 
Thus, any spacetime which is conformally flat can be embedded in $\mathbb R^{2,D-2}.$

 \subsubsection{Global AdS}
The first class of conformally equivalent spacetimes is given by
\begin{equation}\label{eq:global-ads-weyl-equivalence}
  AdS_{d}\times S^{p-2} \cong AdS_{p}\times S^{d-2} \, .
\end{equation}
Both spacetimes can be obtained as a section  of the light cone in the embedding space $\mathbb{R}^{2,p+d-2}$
\begin{equation}\label{eq:null-cone}
-X^2_{-1}-X^2_0+X^2_1+\ldots +X^2_{p+d-2}=0 \, .
\end{equation}
The scaling symmetry $X_A \to \lambda X_A$ of this equation can be fixed in multiple ways. Each choice corresponds to a different section of the light cone and realizes a different conformally flat spacetime. For instance,
\begin{equation}\label{eq:global-ads-scale-inv-fix1}
\underbrace{-\tilde X^2_{-1}-\tilde X^2_0}_{=\:- r^2- L^2}+\underbrace{\tilde X^2_1+\ldots+ \tilde X^2_{p-1}}_{=\:r^2}+\underbrace{\tilde X^2_{p}+\ldots+\tilde X^2_{p+d-2}}_{=\:L^2}=0  
\end{equation}
corresponds to  $AdS_{p}\times S^{d-2}$, where $r$ parametrizes the radial direction in AdS$_p$  and $L$ is the size of the sphere $S^{d-2}$.
On the other hand,
\begin{equation}\label{eq:global-ads-scale-inv-fix2}
\underbrace{- X^2_{-1}- X^2_0}_{=\:-L^2 -R^2}+\underbrace{ X^2_1+\ldots+  X^2_{p-1}}_{=\:L^2}+\underbrace{  X^2_{p}+\ldots+ X^2_{p+d-2}}_{=\:R^2}=0
\end{equation}
leads to $AdS_{d}\times S^{p-2}$, where $R$ is the radial coordinate in $\text{AdS}_d$. It is straightforward to see that the induced metrics on the two sections   are related by the Weyl transformation
\begin{equation}   \label{weylrelation}
d  s^2_{(d, p-2)} = \Omega^2  d \tilde s^2_{(p,d-2)}   \qquad \text{with} \qquad \Omega = \frac{L}{r} = \frac{R}{L} \, .
\end{equation}
The explicit expressions for the conformally equivalent metrics are  
\begin{equation}\label{eq:ads-sphere-pq-metric}
 d \tilde s^2_{(p,d-2)}=-\left(\frac{r^2}{L^2}+1\right)dt^2+\left( \frac{r^2}{L^2} +1\right)^{-1} dr^2+r^2d\Omega^2_{p-2}+L^2d\Omega^2_{d-2} \, , 
\end{equation}
and
\begin{equation}\label{eq:ads-sphere-qp-metric}
d  s^2_{(d, p-2)}=-\left(1 + \frac{R^2}{L^2}\right)dt^2+ \left(1+\frac{R^2}{L^2}\right)^{-1} dR^2+R^2d\Omega^2_{d-2}+L^2d\Omega^2_{p-2} \, . 
\end{equation}

\subsubsection{Poincar\'e patch}
The second class of Weyl equivalent geometries is
\begin{equation}\label{eq:ads-poinc-weyl-equivalence}
 \mathbb{R}^{1,d+p-2}  \cong \text{Poincar\'e-}AdS_{p}\times S^{d-2} \, .
\end{equation}
The Poincar\'e patch   can be understood as a flat foliation of AdS$_p$ with leaves  $\mathbb{R}^{1,p-2}$.
The choice of coordinates on the light  cone  (\ref{eq:null-cone}) that leads to the geometry on  the right-hand side is
\begin{equation}\label{eq:ads-poinc-scale-inv-fix1}
\underbrace{-\tilde X^2_{-1}+\tilde X^2_{p-1}}_{=\: - \frac{r^2}{L^2}x^{\mu}x_{\mu}- L^2 }\underbrace{-\tilde X^2_{0}+\tilde X^2_1+\ldots+ \tilde X^2_{p-2}}_{=\:\frac{r^2}{L^2}x^{\mu}x_{\mu}}+\underbrace{\tilde X^2_{p}+\ldots+\tilde X^2_{p+d-2}}_{=\:L^2}=0 \, ,
\end{equation}
where $r$ parametrizes the radial direction in the Poincar\'e patch and $x^\mu$ represent flat coordinates on $\mathbb{R}^{1,p-2}$.   The section can also simply be described by  $\tilde X_{-1} + \tilde X_{p-1} = L$.

Alternatively, we can fix the scale invariance to obtain $\mathbb{R}^{1,d+p-2}$ through
\begin{equation}\label{eq:ads-poinc-scale-inv-fix2}
\underbrace{-  X^2_{-1}+  X^2_{p-1}}_{=\: -x^{\mu}x_{\mu}-R^2}\underbrace{-  X^2_{0}+  X^2_1+\ldots+   X^2_{p-2}}_{=\:x^{\mu}x_{\mu}}+\underbrace{  X^2_{p}+\ldots+  X^2_{p+d-2}}_{=\:R^2}=0 \, .
\end{equation}
Both of these sections are so-called Poincar\'{e} sections and the latter one is described by  the equation $  X_{-1} +   X_{p-1} = R$. One can  easily verify that the induced metrics on these two sections are related by (\ref{weylrelation}), and they  explicitly  take the form
\begin{equation}\label{eq:poinc-path-metric}
d\tilde s^2_{(p, d-2)}=- \frac{r^2}{L^2} dt^2  +  \frac{L^2}{r^2}dr^2+\frac{r^2}{L^2} d \vec x^2_{p-2}  +  L^2d\Omega^2_{d-2} \, ,
\end{equation}
and
\begin{equation}\label{eq:mink-metric}
d {s}^2_{(d,p-2)}=- dt^2+dR^2 +R^2d\Omega^2_{d-2}+ d \vec x^2_{p-2}\, .
\end{equation}

\subsubsection{AdS-Rindler patch}
The third class  is given by 
\begin{equation}\label{eq:ads-rindler-weyl-equivalence}
  dS_{d}\times \mathbb{H}^{p-2} \cong  \text{Rindler-}AdS_{p}\times S^{d-2} \, ,
\end{equation}
where  $\mathbb{H}^{p-2}$ denotes $(p-2)$-dimensional hyperbolic space.
In this case, we use the fact that time slices of the AdS-Rindler patch are foliated by   hyperbolic space $\mathbb{H}^{p-2}$.
The scale fixing of the null cone that leads to  the geometry on the right-hand side is
\begin{equation}\label{eq:ads-rindler-scale-inv-fix1}
\underbrace{-\tilde X^2_0+\tilde X^2_1}_{=\:-L^2+r^2}\underbrace{-\tilde X^2_{-1}+\tilde X^2_2+\ldots+\tilde X^2_{p-1}}_{=\:-r^2}+\underbrace{\tilde X^2_{p}+\ldots+\tilde X^2_{p+d-2}}_{=\:L^2}=0 \, ,
\end{equation}
where $r$ parametrizes the radial direction in the Rindler wedge.
Furthermore,  one can  obtain $dS_{d}\times \mathbb{H}^{p-2}$ by fixing the   coordinates on the light cone as follows
\begin{equation}\label{eq:ads-rindler-scale-inv-fix2}
\underbrace{-  X^2_0+  X^2_1}_{=\:-R^2+L^2}\underbrace{-  X^2_{-1}+  X^2_2+\ldots+   X^2_{p-1}}_{=\:-L^2}+\underbrace{  X^2_{p}+\ldots+  X^2_{p+d-2}}_{=\:R^2}=0 \, .
\end{equation}
Again, one easily verifies that this implies that the induced metrics on the sections are Weyl equivalent with conformal factor $\Omega =L /r=R/L$.
Explicitly, the induced metrics are given by
\begin{equation}\label{eq:ads-rin-sphere-metric}
d\tilde s^2_{(p, d-2)}=-\left(\frac{r^2}{L^2}-1\right)dt^2+ \left( \frac{r^2}{L^2}-1\right)^{-1} dr^2+r^2\left(du^2+\sinh^2(u)d\Omega^2_{p-3}\right)+L^2d\Omega^2_{d-2} \, ,
\end{equation}
and
\begin{equation}\label{eq:ds-hyperboloid-metric}
d {s}^2_{(d,p-2)}=-\left(1-\frac{R^2}{L^2}\right)dt^2+ \left(1-\frac{R^2}{L^2}\right)^{-1} dR^2+R^2d\Omega^2_{d-2}+L^2\left(du^2+\sinh^2(u)d\Omega^2_{p-3}\right) \, .
\end{equation}
As mentioned above and used in the main text, in the case that $p=3$ we may compactify $\mathbb{H}^1$ by taking a discrete quotient by a boost. On the right-hand side, the AdS-Rindler side, this same identification produces a BTZ black hole.

%% file: samenvattingNL.tex


   \begin{center}
  {\it \Large Emergente Zwaartekracht in een Holografisch Universum}
  \end{center}
  
\vskip 0.5cm

Zwaartekracht is de belangrijkste kracht in ons universum. Zij veroorzaakt de getijden   op aarde, 
 laat alle planeten in een baan om de zon draaien,  houdt de sterren in ons Melkwegstelsel bijeen, 
 en  zorgt ervoor dat ons heelal versneld   uitdijt.    Kortom, zonder de zwaartekracht zou het universum zoals wij dat kennen niet bestaan.  
 
 
Toch is de zwaartekracht de minst begrepen fundamentele kracht van de Natuur. De zwaartekrachtstheorieën van Isaac Newton en Albert Einstein   doen goede voorspellingen op aarde en in ons zonnestelsel, maar op \emph{grote} schaal 
 stroken zij niet   met de waarnemingen van astronomen en kosmologen. Het is onduidelijk waarom sterren aan de rand  van sterrenstelsels zo snel bewegen en bovendien weten we   niet   waarom het heelal versneld uitdijt. Deze experimentele  problemen staan, respectievelijk,   bekend als  `donkere materie' en `donkere energie'. Op  \emph{kleine} schaal (lees: subatomair niveau) is het zelfs lastig om überhaupt  consistente voorspellingen\linebreak te doen  met de huidige zwaartekrachtstheorieën.   
 Het combineren van quantummechanica (de theorie van het allerkleinste) met de algemene relativiteitstheorie (Einsteins zwaartekrachtstheorie) leidt tot  allerlei wiskundige problemen. Dit  suggereert  dat zwaartekracht op quantumniveau een andere beschrijving nodig heeft dan de andere   fundamentele krachten, zoals electromagnetisme, die w\'{e}l goed door de quantummechanica    worden beschreven.  
 Dit proefschrift gaat over      \emph{quantum-\linebreak zwaartekracht} en onderzoekt ook de oorsprong van    donkere energie. 

  \subsubsection{Zwaartekracht bestaat niet (op de allerkleinste schaal)}

De afgelopen decennia     is er een nieuwe kijk op zwaartekracht ontstaan, die ik door middel van mijn onderzoek mede verder heb ontwikkeld: de theorie van \emph{emergente zwaartekracht}. Met emergentie bedoelen we dat er op grote schaal nieuwe eigenschappen voorkomen die op kleine schaal niet aanwezig zijn. 
Volgens deze kijk bestaat zwaartekracht niet op de allerkleinste schaal    (lees: de   Planckschaal $10^{-35}\,$m), maar komt zij pas tevoorschijn op grotere schaal. Hetzelfde geldt voor   ruimtetijd, omdat volgens Einstein zwaartekracht niets anders is dan de kromming van ruimtetijd. Deze emergente kijk roept de volgende vragen op die ik in mijn proefschrift behandel: wat is de microscopische beschrijving van  ruimtetijd?    En hoe  emergeert zwaartekracht precies op grote schaal uit deze beschrijving?

De   belangrijkste aanwijzingen voor de emergentie van ruimtetijd en zwaartekracht zijn   theoretisch van aard. Ten eerste ontdekten de fysici Jacob Bekenstein en Stephen Hawking (en anderen)   in de jaren zeventig van de vorige eeuw   dat zwarte gaten \emph{thermodynamische} objecten zijn, met een temperatuur, energie en entropie. Thermodynamische objecten bestaan typisch uit kleinere bestanddelen,    microscopische vrijheidsgraden genoemd. Een gas bestaat bijvoorbeeld uit grofweg $10^{23}$ individuele atomen. De thermodynamische eigenschappen van zwarte gaten suggereren dus dat zij ook uit microscopische vrijheidsgraden bestaan.  Dit vermoeden werd bevestigd  in 1995 door een belangrijk resultaat uit de snaartheorie van Andrew  Strominger en Cumrun Vafa: zij telden     het aantal microtoestanden voor specifieke (supersymmetrische) zwarte gaten     en  toonden aan dat dit   overeenkomt  met de Bekenstein-Hawking formule voor de entropie van zwarte gaten in   Einsteins zwaartekrachtstheorie. Dit betekent dat de (gekromde ruimte van) zwarte gaten    die Strominger and Vafa onderzochten    emergeren uit snaartheoretische toestanden.  

 Ten tweede wijst theoretisch onderzoek uit dat zwaartekracht een \emph{holografisch} fenomeen  is. Dat betekent   dat  zij net zo goed kan worden beschreven  door een   quantumtheorie zonder zwaartekracht die leeft in een lagerdimensionale ruimte,   \'e\'en dimensie lager dan de ruimte waarin de zwaartekracht   werkt. Volgens het holografisch principe is een zwaartekrachtstheorie in een driedimensionale ruimte, zoals ons universum,  equivalent aan een quantumtheorie op een tweedimensionaal `hologram'.  Aangezien de  theorie   op het hologram   quantummechanisch is en geen zwaartekracht  bevat, zou zij fundamenteler kunnen zijn dan de zwaartekrachtstheorie  (en dat is ook hoe veel fysici holografie interpreteren). In dat geval emergeert zwaartekracht    vanuit een microscopische  beschrijving in \'e\'en dimensie lager.

 \subsubsection{Thermodynamica van causale diamanten}

In dit proefschrift hebben we   een nieuw principe voor emergente zwaartekracht onderzocht, dat Ted Jacobson heeft voorgesteld in 2015. Hij leidde de zwaartekrachts-vergelijking van Einstein af vanuit een \emph{equilibriumconditie} voor de  entropie van kleine causale diamanten. Een \emph{causale diamant} is het grootste gebied in ruimtetijd dat causaal  is verbonden met een waarnemer gedurende zijn hele leven. Jacobsons argument is gebaseerd op een aantal   aannames, die in dit proefschrift onder de loep  worden genomen, zoals het vasthouden van het    volume van het maximale constante-tijdoppervlak  van de diamant   en een hypothese over de modulaire Hamiltoniaan van niet-conforme veldentheorieën in kleine vlakke causale diamanten. \textsc{Hoofdstuk \'{e}\'{e}n} bevat een introductie tot deze afleiding van   Einsteins vergelijking en tot andere concepten die belangrijk zijn voor dit proefschrift, zoals emergente zwaartekracht en holografie. 

   In \textsc{hoofdstuk twee} bestuderen we   de thermodynamische eigenschappen van causale diamanten in maximaal symmetrische ruimtes. 
   Door een   conforme Killing-symmetrie te gebruiken   kunnen we  een thermodynamische relatie afleiden tussen de  temperatuur, energie en entropie van deze diamanten. We beargumenteren dat deze   relatie alleen consistent is als de (absolute) temperatuur negatief is, dat wil zeggen een lagere temperatuur dan nul graden Kelvin. 
 We laten verder  zien hoe   de equilibriumconditie voor entropie volgt uit de thermodynamische relatie (waarin het   volume de rol speelt van    gravitationele energie).   Een ander   resultaat is dat deze conditie \emph{equivalent} is aan een equilibriumconditie voor de vrije energie van causale diamanten, die dus evengoed als input   kan worden gebruikt in de afleiding van      Einsteins vergelijking. Het voordeel van de equilibriumconditie voor vrije energie is dat zij niet is gebaseerd op de aannames over het volume en de modulaire Hamiltoniaan, en dus verdient zij de voorkeur   volgens  Ockhams scheermes. 


In  \textsc{hoofdstuk drie} generaliseren we Jacobsons originele argument naar andere zwaartekrachtstheorieën, die een uitbreiding zijn op Einsteins theorie (in de zin dat de Lagrangiaan extra krommingstermen bevat) en die belangrijk zijn bij hoge energieën.
We leiden de zwaartekrachtsvergelijking  voor deze theorieën  af vanuit een equilibriumprincipe voor entropie. Hiervoor leiden we  eerst  een  thermodynamische relatie af voor   causale diamanten in de context van deze theorieën. Het verschil met Jacobsons resultaat is dat voor deze  theorieën  alleen de gelineariseerde zwaartekrachtsvergelijking kan worden afgeleid, terwijl voor Einsteins theorie de volledige niet-lineaire   vergelijking volgt.  
 

 \subsubsection{Een holografische kijk op het universum}

In \textsc{hoofdstuk vier} van het proefschrift hebben we de holografische eigenschappen van causale diamanten    onderzocht. We nemen aan dat causale diamanten  aan het holografisch  principe voldoen, dat wil zeggen  dat er een microscopische    theorie leeft op de rand van   diamanten.  Voor diamanten in sferisch symmetrische ruimtes definiëren we drie microscopische grootheden: het   aantal vrijheidsgraden, het aantal excitaties en de energie per vrijheidsgraad.  We motiveren deze grootheden vanuit de AdS/CFT correspondentie en relateren ze voor non-AdS ruimtes aan lange snaren in symmetrisch-producttheorieën. Aangezien holografie niet goed  is begrepen voor non-AdS ruimtes  --- zoals de  ruimte  die lijkt op ons universum, genaamd de Sitter ruimte --- helpen deze grootheden bij het vinden van de juiste microscopische theorie voor deze ruimtes  (en dus ook voor ons universum). 







%% file: dankwoord.tex
 
 
 \vspace*{-1cm}
In de eerste plaats dank ik mijn promotor en leermeester Erik Verlinde   dat hij mij heeft betrokken bij zijn inspirerende  onderzoek. 
 Het meeste wat ik   over zwaartekracht    weet,   heb ik van hem geleerd.  Door zijn diepe en brede kennis, verbeeldingskracht, bevlogenheid en  voorliefde voor wiskundige eenvoud is hij een voorbeeld voor mij  voor hoe een   natuurkundige hoort te zijn. 
  Ik prijs mijzelf   gelukkig dat ik met hem heb mogen samenwerken, en heb genoten van onze urenlange, stimulerende discussies. Ik wil hem ook bedanken voor het mogelijk maken van mijn vele reizen, vooral  onze reis   samen  naar Princeton.  

Dit proefschrift is het resultaat van intensieve samenwerking met collegae. Ik ben   veel dank verschuldigd aan al mijn co-auteurs voor het delen van hun inzichten en   hulp in de zoektocht naar de bouwstenen van ons universum: Margot Brouwer, Pablo Bueno, Pablo Cano,   Sam van Leuven, Niels Linnemann, Vincent Min, Juan Pedraza, Antony Speranza and  Watse Sybesma. Ik wil ook Jorrit Kruthoff en Lars Aalsma bedanken voor de vele discussies over de Sitter ruimte. Moreover,  I had the privilege to work with the   American physicist Ted Jacobson, who was like a  second supervisor to me.  I admire him as a theoretical physicist for   his penetrating thoughts, slick proofs, drive to understand and wordsmithing. I would like to thank him for our  two-year journey through the landscape of causal diamonds. 


  Verder dank ik het gehele Instituut voor Theoretische Fysica  voor de plezierige onderzoeksomgeving. 
 Ik heb mij zowel bij de snaartheoriegroep als bij   de HPS groep op mijn gemak gevoeld. In het laatste verband dank ik  Jeroen van Dongen, Sebastian de Haro en Jeremy Butterfield    voor de fijne samenwerking. Ook    noem ik graag       Jaco de Swart, mijn handlanger in het  combineren van  filosofie en natuur-\linebreak kunde. 
   Verder denk ik met    plezier terug aan de   filosofische discussies met Niels Linnemann  over emergente zwaartekracht. Ik wil hem en mijn oudste natuurkunde-\linebreak vriend  Gerben Oling  tevens  bedanken voor hun commentaar op het manuscript.    

  
Tot slot bedank ik mijn (schoon)familie, vooral mijn ouders, dat ze mij de vrijheid en de nodige bagage hebben gegeven om  mijn eigen weg te kunnen bewandelen.\linebreak Mijn grote liefde,  en inmiddels vrouw, Charlotte heeft veel moeten doorstaan gedurende mijn promotietraject en ik dank haar voor haar  steun en   vertrouwen. Dit proefschrift draag ik op aan haar en onze allerliefste Elias. 
  

